\documentclass[11pt,a4paper]{report}      
\usepackage{warwickthesis,setspace,graphicx,amsmath,bm,amssymb,mathabx,cite}    
\usepackage[vcentermath]{youngtab}
\usepackage{subfig}
\usepackage[usenames,dvipsnames]{color}

\def\listofsymbols{
\begin{tabbing}
$\Uparrow,\Downarrow$~~~~~~~~~~~\=\parbox{5in}{valley pseudospin projections\dotfill \pageref{sym:thickarrow}}\\
\addsymbol \uparrow,\downarrow: {spin projections}{sym:thinarrow}
\addsymbol a: {spacing between nearest neighbour carbon atoms in graphene
lattice}{sym:a}
\addsymbol a^\dagger(a): {raising (lowering) bosonic ladder operators
for $n$ in the symmetric gauge}{sym:adag}
\addsymbol a_0: {Bohr radius}{sym:a0}
\addsymbol a_n: {$2^{\frac{1}{2}(\delta_{n,0} -1)}$}{sym:an}
\addsymbol a_{nk}: {coefficients in FSS expansion of $\tilde{f}_n$}{sym:ank}
\addsymbol \mathbf{A}: {magnetic vector potential}{sym:boldA}
\addsymbol A^-: {acceptor Coulomb impurity}{sym:Aminus}
\addsymbol A,B: {two distinct sublattices in graphene}{sym:AB}
\addsymbol \mathbf{A}_\mathrm{L}: {magnetic vector potential in the
Landau gauge}{sym:boldAL}
\addsymbol \mathbf{A}_\mathrm{s}: {magnetic vector potential in the
symmetric gauge}{sym:boldAs}
\addsymbol b^\dagger(b): {raising (lowering) bosonic ladder operators
for the $m$ quantum number}{sym:bdag}
\addsymbol b_m: {coefficients in expansion of $\chi_\mathrm{r}$}{sym:bm}
\addsymbol \mathbf{B}: {magnetic field}{sym:boldB}
\addsymbol c: {spin/pseudospin flavour $\Uparrow \uparrow$}{sym:cflavor}
\addsymbol \hat{c}: {charm operator}{sym:chat}
\addsymbol c^\dagger_i(c_i): {creation (annihilation) operator for an electron
at the $i^\mathrm{th}$ lattice site}{sym:cdag}
\addsymbol c_m: {coefficients in expansion of $\chi_\mathrm{i}$}{sym:cm}
\addsymbol c^\dagger_{n \tau s m}: {creation operator for an electron in graphene (symmetric gauge)}{sym:cdaggergrap}
\addsymbol c^\dagger_{n \tau s X}: {creation operator for an electron in graphene (Landau gauge)}{sym:cdagntausX}
\addsymbol C: {real space path around solenoid}{sym:C}
\addsymbol \bm{\mathrm{C}}_{i}: {connectivity matrix between the
$i^\mathrm{th}$ and $(i+1)^\mathrm{th}$ transfer-matrix layers}{sym:Ci}
\addsymbol \hat{C}_{ij}: {generators of SU(4)}{sym:cij}
\addsymbol d: {number of dimensions}{sym:d}
\addsymbol d: {spin/pseudospin flavour $\Downarrow \downarrow$}{sym:dflavor}
\addsymbol d: {dipole matrix element for neutral collective excitations}{sym:dme}
\addsymbol d^\dagger_{n \tau s m}: {creation operator for a hole in graphene (symmetric gauge)}{sym:ddag}
\addsymbol d^\dagger_{n \tau s X}: {creation operator for a hole in graphene (Landau gauge)}{sym:ddagntausX}
\addsymbol D^+: {donor Coulomb impurity}{sym:Dplus}
\addsymbol e: {magnitude of an electron's charge}{sym:e}
\addsymbol \mathbf{e}_j: {real space unit vector in the direction $j$
for $j=x,y,z$}{sym:ej}
\addsymbol E: {band energy of electron}{sym:E}
\addsymbol \mathcal{E}: {related to electric field magnitude via $\mathcal{E}=|\mathbf{E}|/\sqrt{2}$}{sym:mathcalE}
\addsymbol \mathbf{E}: {electric field}{sym:Ebold}
\addsymbol E_0: {energy unit proportional to characteristic energy of Coulomb interaction}{sym:E0}
\addsymbol E_\mathrm{c}: {critical energy or mobility edge}{sym:Ec}
\addsymbol E_e: {single particle energy of an electron in an AB ring}{sym:Ee}
\addsymbol E_\mathrm{F}: {Fermi energy}{sym:EF}
\addsymbol E_\mathrm{int}: {characteristic energy of Coulomb interaction in graphene}{sym:Eint}
\addsymbol E_\mathrm{kin}: {kinetic energy of electron in graphene}{sym:Ekin}
\addsymbol E_n: {energy of the $n^\mathrm{th}$ 2DEG Landau level}{sym:En}
\addsymbol E_{n_2 \tau_2 s_2}^{n_1 \tau_1 s_1}: {total self energy correction for a $\mathcal{N}_2 \to \mathcal{N}_1$ transition}{sym:Entaus}
\addsymbol E_{\mathrm{SE}}\left( n,n'\right): {correction to $\epsilon_n$ due to exchange with electrons in the $n'^{\rm th}$ LL}{sym:ESE}
\addsymbol \tilde{f}_n: {function occuring in expansion of $\Lambda_M$}{sym:fn}
\addsymbol {f_X^\dagger(f_X)}: {raising (lowering) bosonic ladder operators
for $n$ in the Landau gauge}{sym:fX}
\addsymbol F(\Phi): {quantity directly proportional to oscillator strength}{sym:FPhi}
\addsymbol F^{\mathbf{K}(\mathbf{K}')}_{A(B)}: {envelope wave function for electron in $A$($B$) sublattice and $\mathbf{K}(\mathbf{K}')$ valley}{sym:FABKK'}
\addsymbol g: {conductance}{sym:g}
\addsymbol \hbar\omega_{s}: {valley pseudospin splitting energy}{sym:hbaromegas}
\addsymbol \hbar\omega_{v}: {Zeeman spin splitting energy}{sym:hbaromegav}
\addsymbol \hat{H}: {many body Hamiltonian for graphene}{sym:H}
\addsymbol \hat{\bm{\mathrm{H}}}_A: {Anderson Hamiltonian}{sym:HA}
\addsymbol \hat{H}_\mathrm{D}: {Dirac Hamiltonian for a single valley}{sym:HD}
\addsymbol \bm{\mathrm{H}}_i: {2D Hamiltonian for the $i^\mathrm{th}$
transfer-matrix layer}{sym:Hi}
\addsymbol \hat{H}_\mathrm{imp}: {Hamiltonian for $e$-imp and $h$-imp interactions in graphene}{sym:Hint}
\addsymbol \hat{H}_{\mathbf{K}(\mathbf{K}')}^\mathrm{L}:
{Hamiltonian for an electron in the $\mathbf{K}$($\mathbf{K}'$) valley in the
Landau gauge}{sym:HKL}
\addsymbol \hat{H}_{\mathbf{K}(\mathbf{K}')}^\mathrm{s}:
{Hamiltonian for an electron in the $\mathbf{K}$($\mathbf{K}'$) valley in the
symmetric gauge}{sym:HKs}
\addsymbol j_z: {generalised angular momentum projection for one particle}{sym:jz}
\addsymbol \hat{j_z}: {generalised angular momentum projection
operator for one particle}{sym:jzhat}
\addsymbol J_z: {total generalised angular momentum projection}{sym:Jz}
\addsymbol \hat{J}_z: {total generalised angular momentum projection operator}{sym:Jzhat}
\addsymbol J_z^{(0)}: {total generalised angular momentum projection for seed state}{sym:Jz0}
\addsymbol k: {oscillator quantum number for a few particle complex}{sym:k}
\addsymbol\hat{k}_{-}: {raising operator for oscillator quantum number, $k$}{sym:k-}
\addsymbol \hat{\mathbf{k}}: {total operator of magnetic translations}{sym:kbold}
\addsymbol k_\mathrm{F}: {Fermi wave vector}{sym:kF}
\addsymbol \hat{\mathbf{k}_i}: {operator of magnetic translations for one particle}{sym:kboldi}
\addsymbol \mathbf{K},\mathbf{K}': {two distinct valleys in graphene}{sym:KK'}
\addsymbol \ell: {average electron separation}{sym:ell}
\addsymbol \ell_B: {magnetic length}{sym:lB}
\addsymbol \hat{l}_z: {orbital angular momentum projection operator}{sym:lz}
\addsymbol L: {linear size of system}{sym:L}
\addsymbol L_x: {linear system size in the $x$ direction}{sym:Lx}
\addsymbol L_y: {linear system size in the $y$ direction}{sym:Ly}
\addsymbol m: {oscillator quantum number for a single particle}{sym:m}
\addsymbol m_0: {particle mass}{sym:m0}
\addsymbol m_e: {mass of an electron}{sym:me}
\addsymbol m_\mathrm{i}: {order of $\chi_\mathrm{i}$ expansion}{sym:mi}
\addsymbol m_\mathrm{r}: {order of $\chi_\mathrm{r}$ expansion}{sym:mr}
\addsymbol m_z: {orbital angular momentum projection for a single
particle}{sym:mz}
\addsymbol M: {linear size of a transfer-matrix layer}{sym:M}
\addsymbol n: {Landau level index}{sym:n}
\addsymbol n_\mathrm{c}: {critical LL index corresponding to $q_\mathrm{c}$}{sym:nc}
\addsymbol n_e: {carrier density of electrons in graphene}{sym:ne}
\addsymbol n_\mathrm{i}: {order of $\Lambda_M$ expansion}{sym:ni}
\addsymbol n_\mathrm{r}: {order of $\tilde{f}_n$ expansion}{sym:nr}
\addsymbol \mathrm{N}: {collective index equal to $\{nms\}$}{sym:Nrm}
\addsymbol N: {total number of electrons in a 2D system}{sym:N}
\addsymbol N: {number of transfer-matrix layers}{sym:Ntm}
\addsymbol N: {collective index equal to $\{n m \tau\}$}{sym:Ngrap}
\addsymbol \mathcal{N}: {collective index equal to $\{n \tau s \}$}{sym:mathcalN}
\addsymbol N_0: {number of available states in a single Landau level}{sym:N0}
\addsymbol \mathbf{p}: {single particle canonical momentum}{sym:pbold}
\addsymbol \mathbf{P}: {canonical centre of mass momentum for exciton}{sym:Pbold}
\addsymbol \mathcal{P}: {dimensionless magnitude of canonical centre of mass momentum}{sym:mathcalP}
\addsymbol \mathrm{q}: {charge of particle}{sym:q}
\addsymbol \mathbf{q}: {electronic wave vector relative to
$\mathbf{K}$ valley}{sym:qbold}
\addsymbol q_\mathrm{c}: {critical wave vector for deviation from a linear dispersion relation}{sym:qc}
\addsymbol Q_0: {SU(4) vaccuum singlet}{sym:Q0}
\addsymbol Q^{\dag}_{\mathcal{ N}_1  \mathcal{ N}_2 J_z }: {creation operator for neutral collective excitation (symmetric gauge)}{sym:Qdag}
\addsymbol Q^\dagger_{\mathcal{N}_1 \mathcal{N}_2}(\mathbf{P}): {creation operator for neutral collective excitation (Landau gauge)}{sym:QdagP}
\addsymbol \mathbf{r}: {position vector}{sym:boldr}
\addsymbol R^{\dag}_{\mathcal{ N}_1  \mathcal{ N}_2 \mathcal{ N}_3}: {creation operator for an $X^-$ particle}{sym:Rdag}
\addsymbol \mathbf{R}: {position vector relative to guiding centre}{sym:R}
\addsymbol \mathbf{R}: {excitonic centre of mass coordinate}{sym:Rcom}
\addsymbol \mathbf{R}_{A(B)}: {graphene lattices vectors for the $A$($B$) sublattices}{sym:RAB}
\addsymbol s: {spin projection}{sym:sspin}
\addsymbol s: {critical exponent for the conductivity}{sym:s}
\addsymbol s: {spin/pseudospin flavour $\Downarrow \uparrow$}{sym:sflavor}
\addsymbol S: {area of a 2D system}{sym:S}
\addsymbol S: {total spin quantum number for a CE}{sym:Sspin}
\addsymbol \hat{S}_\pm: {spin raising/lowering operators, $\hat{S}_x \pm i\hat{S}_y$}{sym:Spm}
\addsymbol \hat{\mathbf{S}}^2: {total spin operator for CEs}{sym:S2}
\addsymbol S_i: {spin projection for a collective excitation ($i=x,y,z$)}{sym:Sz}
\addsymbol \hat{S}_i: {operator for $i^{\rm th}$ component of spin for CEs ($i=x,y,z$)}{sym:Si}
\addsymbol \mathcal{S}_n: {sign($n$)}{sym:Sn}
\addsymbol S^{\dag}_{\mathcal{ N}_1  \mathcal{ N}_2 \mathcal{ N}_3}: {creation operator for an $X^+$ particle}{sym:Sdag}
\addsymbol t: {hopping energy for nearest neighbours}{sym:t}
\addsymbol t': {hopping energy for next nearest neighbours}{sym:t'}
\addsymbol t_{ij}: {energy to hop between $i^\mathrm{th}$ and
$j^\mathrm{th}$ lattice sites}{sym:tij}
\addsymbol T: {total pseudospin quantum number for a CE}{sym:Tpseudospin}
\addsymbol \mathcal{T}: {transmission amplitude}{sym:mcT}
\addsymbol \hat{\mathbf{T}}^2: {total pseudospin operator for a CE}{sym:T2}
\addsymbol \hat{T}_\pm: {pseudospin raising/lowering operators, $\hat{T}_x \pm i\hat{T}_y$}{sym:Tpm}
\addsymbol \mathbf{T}_i: {transfer-matrix}{sym:Ti}
\addsymbol T_i: {pseudospin projection for a collective excitation ($i=x,y,z$)}{sym:Tz}
\addsymbol T_{n_2 n_1}: {transition from $n_2$ to $n_1$ LL}{sym:Tn2n1}
\addsymbol u: {spin/pseudospin flavour $\Uparrow \downarrow$}{sym:uflavor}
\addsymbol u_0: {dimensionless energy associated with electric field}{sym:u0}
\addsymbol u_\mathrm{c}: {critical value of electric field}{sym:uc}
\addsymbol U_0: {constant electric field} {sym:U0}
\addsymbol U_{n_1 m_1  n_2 m_2}^{n_1' m_1'  n_2' m_2'}: {matrix element for Coulomb interaction in 2DEG (symmetric gauge)}{sym:UME}
\addsymbol \mathcal{U}_{\mathcal{N}_1 m_1   \mathcal{N}_2 m_2}^{\mathcal{N}_1' m_1' 
\mathcal{N}_2' m_2'}: {matrix element for Coulomb interaction in graphene (symmetric gauge)}{sym:mathcalU}
\addsymbol \tilde{\mathcal{U}}_{ \hspace{2mm}\mathcal{N}_1 \mathcal{N}_2}^{d \hspace{1mm} \mathcal{N}_1' \mathcal{N}_2' }: {direct graphene matrix element in Landau gauge}{sym:Utilded}
\addsymbol \tilde{\mathcal{U}}_{ \hspace{2mm}\mathcal{N}_1 \mathcal{N}_2}^{x \hspace{1mm} \mathcal{N}_1' \mathcal{N}_2' }: {exchange graphene matrix element in Landau gauge}{sym:Utildex}
\addsymbol U(r): {Coulomb interaction energy}{sym:Uofr}
\addsymbol v_0: {average electron-hole interaction strength in units of
$\epsilon_0$ for AB problem}{sym:v0}
\addsymbol v_\mathrm{F}: {Fermi velocity}{sym:vF}
\addsymbol \tilde{v}_\mathrm{F}: {Fermi velocity renormalised by exchange self energy correction}{sym:vFtilde}
\addsymbol V\left[ R\right]: {electron-hole interaction potential in AB problem}{sym:VR}
\addsymbol V_0: {constant in $\delta$-function impurity potential in graphene}{sym:V0}
\addsymbol V^{A(B)}: {$e$-imp interaction for a $\delta$-function impurity on an $A$($B$) graphene site}{sym:VAB}
\addsymbol V_\mathrm{C}(r): {Coulomb impurity potential}{sym:VC}
\addsymbol V^\mathrm{C}: {$4 \times 4$ matrix for electron-Coulomb impurity interaction in graphene}{sym:VCup}
\addsymbol {\mathcal{V}_i }_{\mathcal{N} m}^{\mathcal{N'} m'}: {impurity interaction matrix elements in graphene}{sym:mathcalV}
\addsymbol V_{n m}: {Coulomb impurity interaction matrix elements for 2DEG}{sym:2degimpmes}
\addsymbol w: {scaled disorder strength}{sym:w}
\addsymbol W: {disorder parameter}{sym:W}
\addsymbol W: {impurity strength for $\delta$-function impurity in graphene}{sym:Wimp}
\addsymbol W_\mathrm{c}: {critical disorder at centre of band}{sym:Wc}
\addsymbol \mathcal{W}: {see definition in text}{sym:mathcalW}
\addsymbol X: {quantum number giving Landau level degeneracy in Landau
gauge}{sym:X}
\addsymbol \mathbf{X}: {guiding centre coordinate for a cyclotron
orbit}{sym:Xbold}
\addsymbol X^+: {positively charged collective excitation or trion}{sym:X+}
\addsymbol X^-: {negatively charged collective excitation or trion}{sym:X-}
\addsymbol y: {FSS parameter}{sym:y}
\addsymbol \hat{Y}: {hypercharge operator}{sym:Y}
\addsymbol z: {$x+iy$, a complex number formed from a particle's $x$ and
$y$ coordinates}{sym:z}
\addsymbol Z: {coordination number}{sym:Z}
\addsymbol Z: {charge of Coulomb impurity in units of $e$}{sym:Zimp}
\addsymbol \alpha: {phase gained by an electronic wave function
due to a vector potential}{sym:alpha}
\addsymbol \alpha^\mathrm{g}: {fine structure constant for graphene}{sym:alphag}
\addsymbol \beta: {$\frac{d\mathrm{ln}\left(g\right)
}{d\mathrm{ln}\left(L\right)}$}{sym:beta}
\addsymbol \gamma: {Lyapunov exponent}{sym:gamma}
\addsymbol \gamma: {decay length for excitonic wave function}{sym:gamma2}
\addsymbol \gamma: {Euler's constant}{sym:Eulergamma}
\addsymbol \Gamma: {matrix composed of transfer-matrices}{sym:Gamma}
\addsymbol \delta \hat{H}_\pm: {correction to single particle graphene Hamiltonian due to $\sigma^\pm$ light}{sym:delHpm}
\addsymbol \delta \hat{H}_\pm^\mathrm{EG}: {correction to single particle 2DEG Hamiltonian due to $\sigma^\pm$ light}{sym:deltaHpmEG}
\addsymbol \delta \mathcal{H}_\pm: {correction to many body graphene Hamiltonian due to $\sigma^\pm$ light}{sym:delmathcalHpm}
\addsymbol \delta\gamma_\mathrm{min}^N: {standard error of the minimum
$\gamma$ after $N$ transfer-matrix multiplications}{sym:delgam}
\addsymbol \Delta: {excitonic energy in units of $\epsilon_0$}{sym:Delta}
\addsymbol \epsilon: {dielectric constant for $e$-$e$ interactions in graphene}{sym:epsdielectric}
\addsymbol \epsilon_0: {$\hbar^2/2m_e\rho^2$, the energy unit for the AB problem}{sym:eps0}
\addsymbol \epsilon_{\mathrm{imp}}: {dielectric constant for Coulomb impurity}{sym:epsimp}
\addsymbol \epsilon_n: {energy of the $n^\mathrm{th}$ Landau level in
graphene}{sym:epsn}
\addsymbol \tilde{\epsilon}_n: {Landau level energy renormalised by exchange self energy correction}{sym:epstilde}
\addsymbol \varepsilon: {a positive real number small compared to 1}{sym:epsAT}
\addsymbol \varepsilon_i: {onsite potential energy in Anderson model}{sym:epsi}
\addsymbol \zeta_n\left(x\right): {see definition in text}{sym:zetan}
\addsymbol \theta_e: {azimuthal coordinate of an electron in a 1D ring}{sym:thetae}
\addsymbol \theta_h: {azimuthal coordinate of a hole in a 1D ring}{sym:thetah}
\addsymbol \lambda: {localisation length}{sym:lambda}
\addsymbol \lambda^{\left( e\right)}_N: {single particle energy of an electron in an AB ring in units of $\epsilon_0$}{sym:lambdae}
\addsymbol \lambda_\mathrm{F}: {Fermi wavelength}{sym:lambdaF}
\addsymbol \lambda^{\left( h\right)}_{N'}: {single particle energy of a hole in an AB ring in units of $\epsilon_0$}{sym:lambdah}
\addsymbol \lambda_\mathrm{TF}: {Thomas-Fermi wavelength}{sym:lambdaTF}
\addsymbol \Lambda_M: {reduced localisation length}{sym:LambdaM}
\addsymbol \mu: {number of degrees of freedom in FSS calculation}{sym:mu}
\addsymbol \mu: {related to LL filling factor $\nu$, by $\nu=\mu\pm\varepsilon$, where $\varepsilon\ll1$}{sym:mugrap}
\addsymbol \nu: {critical exponent for the correlation length}{sym:nu}
\addsymbol \nu: {Landau level filling factor}{sym:filling}
\addsymbol \xi: {correlation length}{sym:xi}
\addsymbol \mathbf{\Pi}: {single particle kinematic momentum}{sym:Pi}
\addsymbol \rho: {radius of Aharonov-Bohm ring}{sym:rho} 
\addsymbol \rho(E): {density of states for graphene}{sym:rhoE}
\addsymbol \sigma: {conductivity}{sym:sigma}
\addsymbol \bm{\sigma}: {vector consisting of the three Pauli
matrices}{sym:sigmabold}
\addsymbol \sigma^{+(-)}: {right (left) circularly polarised light}{sym:sigmaplus}
\addsymbol \sigma_\pm: {a combination of Pauli matrices, $\sigma_x \pm \sigma_y$}{sym:sigmapmlower}
\addsymbol \hat{\Sigma}_z: {flavour isospin projection}{sym:sigmaz}
\addsymbol \tau: {valley pseudospin projection}{sym:tau}
\addsymbol \tau_N: {product of transfer-matrices
$\mathbf{T}_N\mathbf{T}_{N-1}\ldots\mathbf{T}_{1}$}{sym:tauN}
\addsymbol \phi: {electric potential}{sym:phi}
\addsymbol \phi_{nX}^{(e)}\left( \mathbf{r}\right): {single particle
2DEG wave function for electron in the Landau gauge}{sym:phinXe}
\addsymbol \phi_{nX}^{(h)}\left( \mathbf{r}\right): {single particle
2DEG wave function for hole in the Landau gauge}{sym:phinXh}
\addsymbol \varphi_{\mathcal{N}_1\mathcal{N}_2}: {non-interacting $e$-$h$ wave function for graphene}{sym:varphin1n2}
\addsymbol \varphi_{n_1 n_2}^\mathrm{EG}: {non-interacting $e$-$h$ wave function for 2DEG}{sym:varphin1n2eg}
\addsymbol \Phi: {magnetic flux}{sym:Phi}
\addsymbol \Phi: {magnetic flux in units of $\Phi_0$}{sym:Phi2}
\addsymbol \Phi_0: {universal flux quantum, $hc/e$}{sym:Phi0}
\addsymbol \Phi^{(e)}_{n \tau s X}(\mathbf{r}): {single particle
wave function for an electron in graphene (Landau gauge)}{sym:PhientausX}
\addsymbol \Phi^{(h)}_{n \tau s X}(\mathbf{r}): {single particle
wave function for a hole in graphene (Landau gauge)}{sym:PhihntausX}
\addsymbol \Phi^{(e)}_{n X \tau}(\mathbf{r}): {electronic wave function in graphene (Landau gauge) without $\chi_s$}{sym:PhienXtau}
\addsymbol \Phi^{(h)}_{n X \tau}(\mathbf{r}): {hole wave function in graphene (Landau gauge) without $\chi_s$}{sym:PhihnXtau}
\addsymbol \chi_\mathrm{i}: {irrelevant scaling variable in FSS}{sym:chii}
\addsymbol \chi_\mathrm{r}: {relevant scaling variable in FSS}{sym:chir}
\addsymbol \chi_s: {spin part of wave function}{sym:chis}
\addsymbol \psi_{A(B)}: {electronic wave function amplitudes for $A$($B$) sublattice in graphene}{sym:psiAB}
\addsymbol \psi^{\left( e\right) }_N\left( \theta_e\right): {wave function for an electron in an AB ring with an electric field}{sym:psie}
\addsymbol \psi^{\left( h\right) }_{N'}\left( \theta_h\right): {wave function for a hole in an AB ring with an electric field}{sym:psih}
\addsymbol \psi_{i}: {wave function amplitude on the $i^\mathrm{th}$ sublattice
site}{sym:psii}
\addsymbol \psi_{nm}\left(\mathbf{r} \right): {single particle 2DEG wave
function in the symmetric gauge}{sym:psinm}
\addsymbol \Psi_{\mathbf{q}}^{\Uparrow \pm}\left( \mathbf{r}\right):
{wave function of electron in graphene in $\mathbf{K}$ valley with no magnetic
field}{sym:Psiqup}
\addsymbol \Psi_{\mathbf{q}}^{\Downarrow \pm}\left( \mathbf{r}\right):
{wave function of electron in graphene in $\mathbf{K}'$ valley with no magnetic
field}{sym:Psiqdown}
\addsymbol \Psi_{n m \tau}(\mathbf{r}): {electronic wave function in
graphene (symmetric gauge) without $\chi_s$}{sym:Psinmtau}
\addsymbol \Psi_{n \tau s m}(\mathbf{r}): {electronic wave function in
graphene (symmetric gauge) with $\chi_s$}{sym:Psintausm}
\addsymbol \Psi\left( \theta_e,\theta_h\right): {wave function for an exciton in an AB ring}{sym:Psitheetheh}
\addsymbol \omega: {angular frequency of electric field $\mathbf{E}$}{sym:omega}
\addsymbol \omega: {$\mathrm{exp}\left(2\pi i/3\right)$}{sym:omegaval}
\addsymbol \omega_\mathrm{c}: {the cyclotron
resonance frequency (or energy in units of $\hbar$) for graphene}{sym:omega}
\addsymbol \tilde{\omega}_\mathrm{c}: {the exchange renormalised cyclotron
resonance frequency for graphene}{sym:omegactilde}
\addsymbol \omega_\mathrm{c}^{\mathrm{EG}}: {the cyclotron
resonance frequency (or energy in units of $\hbar$) for the 2DEG}{sym:omegaeg}

\end{tabbing} \clearpage}
\def\addsymbol #1: #2#3{$#1$ \> \parbox{5in}{#2 \dotfill \pageref{#3}}\\}
\def\newnot#1{\label{#1}} 

\newcommand{\eins}{\mbox{$1 \hspace{-1.0mm} {\bf l}$}}
\newcommand{\one}{\mbox{$1$}}
\newcommand{\Andream} {\mbox{\textbf{A.~M.~Fischer}}}

 \leftchapter                       

\renewcommand{\thesisdepartmentname}{Department of Physics}    

\renewcommand{\thesissubmission}{Submitted to the University of Warwick\\
                        for the degree of}

\renewcommand{\thesisauthor}{Andrea Mary Fischer}    

\renewcommand{\thesismonth}{January}     

\renewcommand{\thesisyear}{2011}      

\renewcommand{\thesistitle}{Disorder and Interactions in Graphene and other Quantum Systems}     

\renewcommand{\thesisauthorpreviousdegrees}{MPhys/MP}

\renewcommand{\thesisauthoraddress}{....}

\begin{document}
\def\listofnotations{ \clearpage}
\def\addnotation #1: #2#3{$#1$\> \parbox{5in}{#2 \dotfill  \pageref{#3}}\\}
\def\newnot#1{\label{#1}}

\thesistitlecolourpage           

\pagenumbering{roman} 

\tableofcontents                     

\begin{thesisacknowledgments}        
\begin{singlespace}
I am forever indebted to Prof.\ Rudolf R\"omer and Prof.\ Alexander Dzyubenko, who both taught me so much. Rudo first introduced me to scientific research in 2006, when I undertook a summer project with him. During my Ph.D., he found interesting and challenging problems for me to work on and guided me through them. I also appreciate his encouragement, career advice and enthusiasm for sending me to conferences. I would like to thank Sasha for many enlightening scientific discussions, both in California and on skype. I am inspired by his passion for physics. I am also grateful to him and his wife Galina for making my stays in California enjoyable, particularly our trips to the National Parks.

During my Ph.D.\ studies, I was fortunate enough to receive doctoral training funding from EPSRC. I also received funding from the Physics Theory Group at the University of Warwick, the American Study and Student Exchange Committee, the IOP Research Student Conference Fund and the C R Barber Trust Fund. This additional funding was invaluable, allowing me to attend several international conferences and spend a total of six months at California State University Bakersfield, where much of the work on graphene was done, together with Prof.\ Alexander Dzyubenko. I would like to thank CSUB for hospitality during my visits. I also acknowledge lots of help and support from the staff in the Centre for Scientific Computing at the University of Warwick.

I thank my family for all the support they have given me over the years in many different ways. For keeping me sane, I thank my friends both at Warwick and back home in London. For providing me with a home away from home and the odd margarita, when necessary, I thank Brenda Gooch. For preventing my physics career from being cut short before it had even begun, I thank Mr Martin Post and the physics teachers at Watford Grammar School for Boys. I especially thank Alberto Rodr\'{\i}guez Gonz\'{a}lez for his careful proof reading of the thesis and his love and support throughout.
\end{singlespace}                    


\end{thesisacknowledgments}

\begin{thesisdeclaration}        
I hereby declare that this thesis entitled ``Interactions and Disorder in Graphene and other Quantum Systems'' is an original work and has not been
submitted for a degree or diploma or other qualification at any other University. Chapter 1 provides introductory information  gathered from the literature as referenced. Chapter 2 is based on the paper\\
$\bullet$ {\em Critical parameters for the disorder-induced metal-insulator transition in FCC and BCC lattices},
 A.~Eilmes, \Andream, R.~A.~R\"omer, Phys. Rev. B \textbf{77}, 245117-8 (2008).\\
Chapter 3 is based on the paper\\
$\bullet$ {\em Exciton storage in a nanoscale Aharonov-Bohm ring with electric field tuning},
 \Andream, V.~L.~Campo, Jr.~, M.~E.~Portnoi, R.~A.~R\"omer, Phys. Rev. Lett. \textbf{102}, 096405-4 (2009)\\
Chapters 4-7 detail work published in the following papers\\
$\bullet$ {\em Localized collective excitations in doped graphene in strong magnetic fields},\\
 \Andream, A.~B.~Dzyubenko, R.~A.~R\"omer, Phys. Rev. B \textbf{80}, 165410-5 (2009)\\
 $\bullet$ {\em Localized magneto-optical collective excitations of impure graphene},
 \Andream, A.~B.~Dzyubenko, R.~A.~R\"omer, Ann. Phys. (Berlin) \textbf{18}, 944-948 (2009)\\
 $\bullet$ {\em Symmetry content and observation of charged collective excitations for graphene in strong magnetic fields},
 \Andream, R.~A.~R\"omer, A.~B.~Dzyubenko, EPL \textbf{92} 37003 (2010)\\
$\bullet$ {\em Magnetoplasmons and SU(4) symmetry in graphene},
 \Andream, R.~A.~R\"omer, A.~B.~Dzyubenko, to be published in Journal of Physics: Conference Series\\
All work was done in collaboration with the listed authors.
        
\end{thesisdeclaration}

\begin{thesisabstract}               
 \begin{singlespace}     
This thesis examines the topics of disorder and electron-electron interactions in three distinct quantum systems. Firstly, the Anderson transition is studied for the body centred cubic and face centred cubic lattices. We obtain high precision results for the critical disorder at the band centre and the critical exponent using the transfer-matrix method and finite size scaling. Comparing the critical disorder between the simple cubic, body centred cubic and face centred cubic lattices, an increase in the critical disorder is observed as a function of the coordination number of the lattice. The critical exponent is found to be $\nu\approx 1.5$ in agreement with the value for the simple cubic lattice. Energy-disorder phase diagrams are plotted for both lattice types. 

Next, we consider the Aharonov-Bohm effect for an exciton in a 1D ring geometry. The aim is to determine how the addition of a constant electric field in the plane of the ring effects the Aharonov-Bohm oscillations, which occur as a function of the magnetic flux threading the ring. We develop a self consistent equation for the ground state energy, which is then solved numerically. Oscillations in the ground state energy have an increasing amplitude as a function of electric field strength until a critical electric field value. At this point, oscillations in the oscillator strength become inverted, with the oscillation minimum reaching zero at half a magnetic flux quantum. This suggests a possible process for controlling the formation and recombination of excitons through tuning the applied fields.

The final and largest section of the thesis is concerned with collective excitations of graphene in a strong perpendicular magnetic field. The excitations, which are most strongly mixed are identified and used as a basis to diagonalise the Hamiltonian, which includes the Coulomb interaction between electrons and holes. In this way the oscillator strengths and energies of collective excitations are obtained. The good quantum numbers for collective excitations are identified. In particular, we study those arising from the SU(4) symmetry, which is due to two spin and two valley pseudospin projections. This enables us to determine the multiplet structure of the states. In addition to neutral collective excitations or excitons, we investigate the possible formation of charged collective excitations or trions from nearly full or nearly empty Landau levels. The localisation of neutral collective excitations upon a single Coulomb or $\delta$-function impurity is also examined.

\end{singlespace}
\end{thesisabstract}
\markboth{List of Notation \hfill}{List of Notation \hfill}
\begin{thesisabbreviations}       
\begin{itemize}
\item[$e$] electron
\item[$h$] hole
\item[imp] impurity
\item[2DEG] two dimensional electron gas
\item[AB] Aharonov-Bohm
\item[ABE] Aharonov-Bohm effect
\item[ARPES] angle-resolved photoemission spectroscopy
\item[AT] Anderson transition
\item[BCC] body centred cubic
\item[BEC] Bose-Einstein condensate
\item[BZ] Brillouin zone
\item[CE] collective excitation
\item[CRE] cyclotron resonance energy
\item[D] dimensions or dimensional
\item[FCC] face centred cubic
\item[FSS] finite size scaling
\item[GOE] Gaussian orthogonal ensemble
\item[GS] ground state
\item[LE] Lyapunov exponent
\item[LL] Landau level
\item[MIT] metal to insulator transition
\item[MOSFET] metal oxide semiconductor field effect transistor
\item[OPST] one parameter scaling theory
\item[OS] oscillator strength
\item[QED] quantum electrodynamics
\item[SC] simple cubic
\item[TMM] transfer-matrix method
\item[XABE] excitonic Aharonov-Bohm effect

\end{itemize}

\end{thesisabbreviations}
\newpage
\chapter*{List of Symbols\hfill} \addcontentsline{toc}{chapter}{List of Symbols}
\begin{tabbing}
$\Uparrow,\Downarrow$~~~~~~~~~~~\=\parbox{5in}{valley pseudospin projections\dotfill \pageref{sym:thickarrow}}\\
\addsymbol \uparrow,\downarrow: {spin projections}{sym:thinarrow}
\addsymbol a: {spacing between nearest neighbour carbon atoms in graphene
lattice}{sym:a}
\addsymbol a^\dagger(a): {raising (lowering) bosonic ladder operators
for $n$ in the symmetric gauge}{sym:adag}
\addsymbol a_0: {Bohr radius}{sym:a0}
\addsymbol a_n: {$2^{\frac{1}{2}(\delta_{n,0} -1)}$}{sym:an}
\addsymbol a_{nk}: {coefficients in FSS expansion of $\tilde{f}_n$}{sym:ank}
\addsymbol \mathbf{A}: {magnetic vector potential}{sym:boldA}
\addsymbol A^-: {acceptor Coulomb impurity}{sym:Aminus}
\addsymbol A,B: {two distinct sublattices in graphene}{sym:AB}
\addsymbol \mathbf{A}_\mathrm{L}: {magnetic vector potential in the
Landau gauge}{sym:boldAL}
\addsymbol \mathbf{A}_\mathrm{s}: {magnetic vector potential in the
symmetric gauge}{sym:boldAs}
\addsymbol b^\dagger(b): {raising (lowering) bosonic ladder operators
for the $m$ quantum number}{sym:bdag}
\addsymbol b_m: {coefficients in expansion of $\chi_\mathrm{r}$}{sym:bm}
\addsymbol \mathbf{B}: {magnetic field}{sym:boldB}
\addsymbol c: {spin/pseudospin flavour $\Uparrow \uparrow$}{sym:cflavor}
\addsymbol \hat{c}: {charm operator}{sym:chat}
\addsymbol c^\dagger_i(c_i): {creation (annihilation) operator for an electron
at the $i^\mathrm{th}$ lattice site}{sym:cdag}
\addsymbol c_m: {coefficients in expansion of $\chi_\mathrm{i}$}{sym:cm}
\addsymbol c^\dagger_{n \tau s m}: {creation operator for an electron in graphene (symmetric gauge)}{sym:cdaggergrap}
\addsymbol c^\dagger_{n \tau s X}: {creation operator for an electron in graphene (Landau gauge)}{sym:cdagntausX}
\addsymbol C: {real space path around solenoid}{sym:C}
\addsymbol \bm{\mathrm{C}}_{i}: {connectivity matrix between the
$i^\mathrm{th}$ and $(i+1)^\mathrm{th}$ transfer-matrix layers}{sym:Ci}
\addsymbol \hat{C}_{ij}: {generators of SU(4)}{sym:cij}
\addsymbol d: {number of dimensions}{sym:d}
\addsymbol d: {spin/pseudospin flavour $\Downarrow \downarrow$}{sym:dflavor}
\addsymbol d: {dipole matrix element for neutral collective excitations}{sym:dme}
\addsymbol d^\dagger_{n \tau s m}: {creation operator for a hole in graphene (symmetric gauge)}{sym:ddag}
\addsymbol d^\dagger_{n \tau s X}: {creation operator for a hole in graphene (Landau gauge)}{sym:ddagntausX}
\addsymbol D^+: {donor Coulomb impurity}{sym:Dplus}
\addsymbol e: {magnitude of an electron's charge}{sym:e}
\addsymbol \mathbf{e}_j: {real space unit vector in the direction $j$
for $j=x,y,z$}{sym:ej}
\addsymbol E: {band energy of electron}{sym:E}
\addsymbol \mathcal{E}: {related to electric field magnitude via $\mathcal{E}=|\mathbf{E}|/\sqrt{2}$}{sym:mathcalE}
\addsymbol \mathbf{E}: {electric field}{sym:Ebold}
\addsymbol E_0: {energy unit proportional to characteristic energy of Coulomb interaction}{sym:E0}
\addsymbol E_\mathrm{c}: {critical energy or mobility edge}{sym:Ec}
\addsymbol E_e: {single particle energy of an electron in an AB ring}{sym:Ee}
\addsymbol E_\mathrm{F}: {Fermi energy}{sym:EF}
\addsymbol E_\mathrm{int}: {characteristic energy of Coulomb interaction in graphene}{sym:Eint}
\addsymbol E_\mathrm{kin}: {kinetic energy of electron in graphene}{sym:Ekin}
\addsymbol E_n: {energy of the $n^\mathrm{th}$ 2DEG Landau level}{sym:En}
\addsymbol E_{n_2 \tau_2 s_2}^{n_1 \tau_1 s_1}: {total self energy correction for a $\mathcal{N}_2 \to \mathcal{N}_1$ transition}{sym:Entaus}
\addsymbol E_{\mathrm{SE}}\left( n,n'\right): {correction to $\epsilon_n$ due to exchange with electrons in the $n'^{\rm th}$ LL}{sym:ESE}
\addsymbol \tilde{f}_n: {function occuring in expansion of $\Lambda_M$}{sym:fn}
\addsymbol {f_X^\dagger(f_X)}: {raising (lowering) bosonic ladder operators
for $n$ in the Landau gauge}{sym:fX}
\addsymbol F(\Phi): {quantity directly proportional to oscillator strength}{sym:FPhi}
\addsymbol F^{\mathbf{K}(\mathbf{K}')}_{A(B)}: {envelope wave function for electron in $A$($B$) sublattice and $\mathbf{K}(\mathbf{K}')$ valley}{sym:FABKK'}
\addsymbol g: {conductance}{sym:g}
\addsymbol \hbar\omega_{s}: {valley pseudospin splitting energy}{sym:hbaromegas}
\addsymbol \hbar\omega_{v}: {Zeeman spin splitting energy}{sym:hbaromegav}
\addsymbol \hat{H}: {many body Hamiltonian for graphene}{sym:H}
\addsymbol \hat{\bm{\mathrm{H}}}_A: {Anderson Hamiltonian}{sym:HA}
\addsymbol \hat{H}_\mathrm{D}: {Dirac Hamiltonian for a single valley}{sym:HD}
\addsymbol \bm{\mathrm{H}}_i: {2D Hamiltonian for the $i^\mathrm{th}$
transfer-matrix layer}{sym:Hi}
\addsymbol \hat{H}_\mathrm{imp}: {Hamiltonian for $e$-imp and $h$-imp interactions in graphene}{sym:Hint}
\addsymbol \hat{H}_{\mathbf{K}(\mathbf{K}')}^\mathrm{L}:
{Hamiltonian for an electron in the $\mathbf{K}$($\mathbf{K}'$) valley in the
Landau gauge}{sym:HKL}
\addsymbol \hat{H}_{\mathbf{K}(\mathbf{K}')}^\mathrm{s}:
{Hamiltonian for an electron in the $\mathbf{K}$($\mathbf{K}'$) valley in the
symmetric gauge}{sym:HKs}
\addsymbol j_z: {generalised angular momentum projection for one particle}{sym:jz}
\addsymbol \hat{j_z}: {generalised angular momentum projection
operator for one particle}{sym:jzhat}
\addsymbol J_z: {total generalised angular momentum projection}{sym:Jz}
\addsymbol \hat{J}_z: {total generalised angular momentum projection operator}{sym:Jzhat}
\addsymbol J_z^{(0)}: {total generalised angular momentum projection for seed state}{sym:Jz0}
\addsymbol k: {oscillator quantum number for a few particle complex}{sym:k}
\addsymbol\hat{k}_{-}: {raising operator for oscillator quantum number, $k$}{sym:k-}
\addsymbol \hat{\mathbf{k}}: {total operator of magnetic translations}{sym:kbold}
\addsymbol k_\mathrm{F}: {Fermi wave vector}{sym:kF}
\addsymbol \hat{\mathbf{k}_i}: {operator of magnetic translations for one particle}{sym:kboldi}
\addsymbol \mathbf{K},\mathbf{K}': {two distinct valleys in graphene}{sym:KK'}
\addsymbol \ell: {average electron separation}{sym:ell}
\addsymbol \ell_B: {magnetic length}{sym:lB}
\addsymbol \hat{l}_z: {orbital angular momentum projection operator}{sym:lz}
\addsymbol L: {linear size of system}{sym:L}
\addsymbol L_x: {linear system size in the $x$ direction}{sym:Lx}
\addsymbol L_y: {linear system size in the $y$ direction}{sym:Ly}
\addsymbol m: {oscillator quantum number for a single particle}{sym:m}
\addsymbol m_0: {particle mass}{sym:m0}
\addsymbol m_e: {mass of an electron}{sym:me}
\addsymbol m_\mathrm{i}: {order of $\chi_\mathrm{i}$ expansion}{sym:mi}
\addsymbol m_\mathrm{r}: {order of $\chi_\mathrm{r}$ expansion}{sym:mr}
\addsymbol m_z: {orbital angular momentum projection for a single
particle}{sym:mz}
\addsymbol M: {linear size of a transfer-matrix layer}{sym:M}
\addsymbol n: {Landau level index}{sym:n}
\addsymbol n_\mathrm{c}: {critical LL index corresponding to $q_\mathrm{c}$}{sym:nc}
\addsymbol n_e: {carrier density of electrons in graphene}{sym:ne}
\addsymbol n_\mathrm{i}: {order of $\Lambda_M$ expansion}{sym:ni}
\addsymbol n_\mathrm{r}: {order of $\tilde{f}_n$ expansion}{sym:nr}
\addsymbol \mathrm{N}: {collective index equal to $\{nms\}$}{sym:Nrm}
\addsymbol N: {total number of electrons in a 2D system}{sym:N}
\addsymbol N: {number of transfer-matrix layers}{sym:Ntm}
\addsymbol N: {collective index equal to $\{n m \tau\}$}{sym:Ngrap}
\addsymbol \mathcal{N}: {collective index equal to $\{n \tau s \}$}{sym:mathcalN}
\addsymbol N_0: {number of available states in a single Landau level}{sym:N0}
\addsymbol \mathbf{p}: {single particle canonical momentum}{sym:pbold}
\addsymbol \mathbf{P}: {canonical centre of mass momentum for exciton}{sym:Pbold}
\addsymbol \mathcal{P}: {dimensionless magnitude of canonical centre of mass momentum}{sym:mathcalP}
\addsymbol \mathrm{q}: {charge of particle}{sym:q}
\addsymbol \mathbf{q}: {electronic wave vector relative to
$\mathbf{K}$ valley}{sym:qbold}
\addsymbol q_\mathrm{c}: {critical wave vector for deviation from a linear dispersion relation}{sym:qc}
\addsymbol Q_0: {SU(4) vaccuum singlet}{sym:Q0}
\addsymbol Q^{\dag}_{\mathcal{ N}_1  \mathcal{ N}_2 J_z }: {creation operator for neutral collective excitation (symmetric gauge)}{sym:Qdag}
\addsymbol Q^\dagger_{\mathcal{N}_1 \mathcal{N}_2}(\mathbf{P}): {creation operator for neutral collective excitation (Landau gauge)}{sym:QdagP}
\addsymbol \mathbf{r}: {position vector}{sym:boldr}
\addsymbol R^{\dag}_{\mathcal{ N}_1  \mathcal{ N}_2 \mathcal{ N}_3}: {creation operator for an $X^-$ particle}{sym:Rdag}
\addsymbol \mathbf{R}: {position vector relative to guiding centre}{sym:R}
\addsymbol \mathbf{R}: {excitonic centre of mass coordinate}{sym:Rcom}
\addsymbol \mathbf{R}_{A(B)}: {graphene lattices vectors for the $A$($B$) sublattices}{sym:RAB}
\addsymbol s: {spin projection}{sym:sspin}
\addsymbol s: {critical exponent for the conductivity}{sym:s}
\addsymbol s: {spin/pseudospin flavour $\Downarrow \uparrow$}{sym:sflavor}
\addsymbol S: {area of a 2D system}{sym:S}
\addsymbol S: {total spin quantum number for a CE}{sym:Sspin}
\addsymbol \hat{S}_\pm: {spin raising/lowering operators, $\hat{S}_x \pm i\hat{S}_y$}{sym:Spm}
\addsymbol \hat{\mathbf{S}}^2: {total spin operator for CEs}{sym:S2}
\addsymbol S_i: {spin projection for a collective excitation ($i=x,y,z$)}{sym:Sz}
\addsymbol \hat{S}_i: {operator for $i^{\rm th}$ component of spin for CEs ($i=x,y,z$)}{sym:Si}
\addsymbol \mathcal{S}_n: {sign($n$)}{sym:Sn}
\addsymbol S^{\dag}_{\mathcal{ N}_1  \mathcal{ N}_2 \mathcal{ N}_3}: {creation operator for an $X^+$ particle}{sym:Sdag}
\addsymbol t: {hopping energy for nearest neighbours}{sym:t}
\addsymbol t': {hopping energy for next nearest neighbours}{sym:t'}
\addsymbol t_{ij}: {energy to hop between $i^\mathrm{th}$ and
$j^\mathrm{th}$ lattice sites}{sym:tij}
\addsymbol T: {total pseudospin quantum number for a CE}{sym:Tpseudospin}
\addsymbol \mathcal{T}: {transmission amplitude}{sym:mcT}
\addsymbol \hat{\mathbf{T}}^2: {total pseudospin operator for a CE}{sym:T2}
\addsymbol \hat{T}_\pm: {pseudospin raising/lowering operators, $\hat{T}_x \pm i\hat{T}_y$}{sym:Tpm}
\addsymbol \mathbf{T}_i: {transfer-matrix}{sym:Ti}
\addsymbol T_i: {pseudospin projection for a collective excitation ($i=x,y,z$)}{sym:Tz}
\addsymbol T_{n_2 n_1}: {transition from $n_2$ to $n_1$ LL}{sym:Tn2n1}
\addsymbol u: {spin/pseudospin flavour $\Uparrow \downarrow$}{sym:uflavor}
\addsymbol u_0: {dimensionless energy associated with electric field}{sym:u0}
\addsymbol u_\mathrm{c}: {critical value of electric field}{sym:uc}
\addsymbol U_0: {constant electric field} {sym:U0}
\addsymbol U_{n_1 m_1  n_2 m_2}^{n_1' m_1'  n_2' m_2'}: {matrix element for Coulomb interaction in 2DEG (symmetric gauge)}{sym:UME}
\addsymbol \mathcal{U}_{\mathcal{N}_1 m_1   \mathcal{N}_2 m_2}^{\mathcal{N}_1' m_1' 
\mathcal{N}_2' m_2'}: {matrix element for Coulomb interaction in graphene (symmetric gauge)}{sym:mathcalU}
\addsymbol \tilde{\mathcal{U}}_{ \hspace{2mm}\mathcal{N}_1 \mathcal{N}_2}^{d \hspace{1mm} \mathcal{N}_1' \mathcal{N}_2' }: {direct graphene matrix element in Landau gauge}{sym:Utilded}
\addsymbol \tilde{\mathcal{U}}_{ \hspace{2mm}\mathcal{N}_1 \mathcal{N}_2}^{x \hspace{1mm} \mathcal{N}_1' \mathcal{N}_2' }: {exchange graphene matrix element in Landau gauge}{sym:Utildex}
\addsymbol U(r): {Coulomb interaction energy}{sym:Uofr}
\addsymbol v_0: {average electron-hole interaction strength in units of
$\epsilon_0$ for AB problem}{sym:v0}
\addsymbol v_\mathrm{F}: {Fermi velocity}{sym:vF}
\addsymbol \tilde{v}_\mathrm{F}: {Fermi velocity renormalised by exchange self energy correction}{sym:vFtilde}
\addsymbol V\left[ R\right]: {electron-hole interaction potential in AB problem}{sym:VR}
\addsymbol V_0: {constant in $\delta$-function impurity potential in graphene}{sym:V0}
\addsymbol V^{A(B)}: {$e$-imp interaction for a $\delta$-function impurity on an $A$($B$) graphene site}{sym:VAB}
\addsymbol V_\mathrm{C}(r): {Coulomb impurity potential}{sym:VC}
\addsymbol V^\mathrm{C}: {$4 \times 4$ matrix for electron-Coulomb impurity interaction in graphene}{sym:VCup}
\addsymbol {\mathcal{V}_i }_{\mathcal{N} m}^{\mathcal{N'} m'}: {impurity interaction matrix elements in graphene}{sym:mathcalV}
\addsymbol V_{n m}: {Coulomb impurity interaction matrix elements for 2DEG}{sym:2degimpmes}
\addsymbol w: {scaled disorder strength}{sym:w}
\addsymbol W: {disorder parameter}{sym:W}
\addsymbol W: {impurity strength for $\delta$-function impurity in graphene}{sym:Wimp}
\addsymbol W_\mathrm{c}: {critical disorder at centre of band}{sym:Wc}
\addsymbol \mathcal{W}: {see definition in text}{sym:mathcalW}
\addsymbol X: {quantum number giving Landau level degeneracy in Landau
gauge}{sym:X}
\addsymbol \mathbf{X}: {guiding centre coordinate for a cyclotron
orbit}{sym:Xbold}
\addsymbol X^+: {positively charged collective excitation or trion}{sym:X+}
\addsymbol X^-: {negatively charged collective excitation or trion}{sym:X-}
\addsymbol y: {FSS parameter}{sym:y}
\addsymbol \hat{Y}: {hypercharge operator}{sym:Y}
\addsymbol z: {$x+iy$, a complex number formed from a particle's $x$ and
$y$ coordinates}{sym:z}
\addsymbol Z: {coordination number}{sym:Z}
\addsymbol Z: {charge of Coulomb impurity in units of $e$}{sym:Zimp}
\addsymbol \alpha: {phase gained by an electronic wave function
due to a vector potential}{sym:alpha}
\addsymbol \alpha^\mathrm{g}: {fine structure constant for graphene}{sym:alphag}
\addsymbol \beta: {$\frac{d\mathrm{ln}\left(g\right)
}{d\mathrm{ln}\left(L\right)}$}{sym:beta}
\addsymbol \gamma: {Lyapunov exponent}{sym:gamma}
\addsymbol \gamma: {decay length for excitonic wave function}{sym:gamma2}
\addsymbol \gamma: {Euler's constant}{sym:Eulergamma}
\addsymbol \Gamma: {matrix composed of transfer-matrices}{sym:Gamma}
\addsymbol \delta \hat{H}_\pm: {correction to single particle graphene Hamiltonian due to $\sigma^\pm$ light}{sym:delHpm}
\addsymbol \delta \hat{H}_\pm^\mathrm{EG}: {correction to single particle 2DEG Hamiltonian due to $\sigma^\pm$ light}{sym:deltaHpmEG}
\addsymbol \delta \mathcal{H}_\pm: {correction to many body graphene Hamiltonian due to $\sigma^\pm$ light}{sym:delmathcalHpm}
\addsymbol \delta\gamma_\mathrm{min}^N: {standard error of the minimum
$\gamma$ after $N$ transfer-matrix multiplications}{sym:delgam}
\addsymbol \Delta: {excitonic energy in units of $\epsilon_0$}{sym:Delta}
\addsymbol \epsilon: {dielectric constant for $e$-$e$ interactions in graphene}{sym:epsdielectric}
\addsymbol \epsilon_0: {$\hbar^2/2m_e\rho^2$, the energy unit for the AB problem}{sym:eps0}
\addsymbol \epsilon_{\mathrm{imp}}: {dielectric constant for Coulomb impurity}{sym:epsimp}
\addsymbol \epsilon_n: {energy of the $n^\mathrm{th}$ Landau level in
graphene}{sym:epsn}
\addsymbol \tilde{\epsilon}_n: {Landau level energy renormalised by exchange self energy correction}{sym:epstilde}
\addsymbol \varepsilon: {a positive real number small compared to 1}{sym:epsAT}
\addsymbol \varepsilon_i: {onsite potential energy in Anderson model}{sym:epsi}
\addsymbol \zeta_n\left(x\right): {see definition in text}{sym:zetan}
\addsymbol \theta_e: {azimuthal coordinate of an electron in a 1D ring}{sym:thetae}
\addsymbol \theta_h: {azimuthal coordinate of a hole in a 1D ring}{sym:thetah}
\addsymbol \lambda: {localisation length}{sym:lambda}
\addsymbol \lambda^{\left( e\right)}_N: {single particle energy of an electron in an AB ring in units of $\epsilon_0$}{sym:lambdae}
\addsymbol \lambda_\mathrm{F}: {Fermi wavelength}{sym:lambdaF}
\addsymbol \lambda^{\left( h\right)}_{N'}: {single particle energy of a hole in an AB ring in units of $\epsilon_0$}{sym:lambdah}
\addsymbol \lambda_\mathrm{TF}: {Thomas-Fermi wavelength}{sym:lambdaTF}
\addsymbol \Lambda_M: {reduced localisation length}{sym:LambdaM}
\addsymbol \mu: {number of degrees of freedom in FSS calculation}{sym:mu}
\addsymbol \mu: {related to LL filling factor $\nu$, by $\nu=\mu\pm\varepsilon$, where $\varepsilon\ll1$}{sym:mugrap}
\addsymbol \nu: {critical exponent for the correlation length}{sym:nu}
\addsymbol \nu: {Landau level filling factor}{sym:filling}
\addsymbol \xi: {correlation length}{sym:xi}
\addsymbol \mathbf{\Pi}: {single particle kinematic momentum}{sym:Pi}
\addsymbol \rho: {radius of Aharonov-Bohm ring}{sym:rho} 
\addsymbol \rho(E): {density of states for graphene}{sym:rhoE}
\addsymbol \sigma: {conductivity}{sym:sigma}
\addsymbol \bm{\sigma}: {vector consisting of the three Pauli
matrices}{sym:sigmabold}
\addsymbol \sigma^{+(-)}: {right (left) circularly polarised light}{sym:sigmaplus}
\addsymbol \sigma_\pm: {a combination of Pauli matrices, $\sigma_x \pm \sigma_y$}{sym:sigmapmlower}
\addsymbol \hat{\Sigma}_z: {flavour isospin projection}{sym:sigmaz}
\addsymbol \tau: {valley pseudospin projection}{sym:tau}
\addsymbol \tau_N: {product of transfer-matrices
$\mathbf{T}_N\mathbf{T}_{N-1}\ldots\mathbf{T}_{1}$}{sym:tauN}
\addsymbol \phi: {electric potential}{sym:phi}
\addsymbol \phi_{nX}^{(e)}\left( \mathbf{r}\right): {single particle
2DEG wave function for electron in the Landau gauge}{sym:phinXe}
\addsymbol \phi_{nX}^{(h)}\left( \mathbf{r}\right): {single particle
2DEG wave function for hole in the Landau gauge}{sym:phinXh}
\addsymbol \varphi_{\mathcal{N}_1\mathcal{N}_2}: {non-interacting $e$-$h$ wave function for graphene}{sym:varphin1n2}
\addsymbol \varphi_{n_1 n_2}^\mathrm{EG}: {non-interacting $e$-$h$ wave function for 2DEG}{sym:varphin1n2eg}
\addsymbol \Phi: {magnetic flux}{sym:Phi}
\addsymbol \Phi: {magnetic flux in units of $\Phi_0$}{sym:Phi2}
\addsymbol \Phi_0: {universal flux quantum, $hc/e$}{sym:Phi0}
\addsymbol \Phi^{(e)}_{n \tau s X}(\mathbf{r}): {single particle
wave function for an electron in graphene (Landau gauge)}{sym:PhientausX}
\addsymbol \Phi^{(h)}_{n \tau s X}(\mathbf{r}): {single particle
wave function for a hole in graphene (Landau gauge)}{sym:PhihntausX}
\addsymbol \Phi^{(e)}_{n X \tau}(\mathbf{r}): {electronic wave function in graphene (Landau gauge) without $\chi_s$}{sym:PhienXtau}
\addsymbol \Phi^{(h)}_{n X \tau}(\mathbf{r}): {hole wave function in graphene (Landau gauge) without $\chi_s$}{sym:PhihnXtau}
\addsymbol \chi_\mathrm{i}: {irrelevant scaling variable in FSS}{sym:chii}
\addsymbol \chi_\mathrm{r}: {relevant scaling variable in FSS}{sym:chir}
\addsymbol \chi_s: {spin part of wave function}{sym:chis}
\addsymbol \psi_{A(B)}: {electronic wave function amplitudes for $A$($B$) sublattice in graphene}{sym:psiAB}
\addsymbol \psi^{\left( e\right) }_N\left( \theta_e\right): {wave function for an electron in an AB ring with an electric field}{sym:psie}
\addsymbol \psi^{\left( h\right) }_{N'}\left( \theta_h\right): {wave function for a hole in an AB ring with an electric field}{sym:psih}
\addsymbol \psi_{i}: {wave function amplitude on the $i^\mathrm{th}$ sublattice
site}{sym:psii}
\addsymbol \psi_{nm}\left(\mathbf{r} \right): {single particle 2DEG wave
function in the symmetric gauge}{sym:psinm}
\addsymbol \Psi_{\mathbf{q}}^{\Uparrow \pm}\left( \mathbf{r}\right):
{wave function of electron in graphene in $\mathbf{K}$ valley with no magnetic
field}{sym:Psiqup}
\addsymbol \Psi_{\mathbf{q}}^{\Downarrow \pm}\left( \mathbf{r}\right):
{wave function of electron in graphene in $\mathbf{K}'$ valley with no magnetic
field}{sym:Psiqdown}
\addsymbol \Psi_{n m \tau}(\mathbf{r}): {electronic wave function in
graphene (symmetric gauge) without $\chi_s$}{sym:Psinmtau}
\addsymbol \Psi_{n \tau s m}(\mathbf{r}): {electronic wave function in
graphene (symmetric gauge) with $\chi_s$}{sym:Psintausm}
\addsymbol \Psi\left( \theta_e,\theta_h\right): {wave function for an exciton in an AB ring}{sym:Psitheetheh}
\addsymbol \omega: {angular frequency of electric field $\mathbf{E}$}{sym:omega}
\addsymbol \omega: {$\mathrm{exp}\left(2\pi i/3\right)$}{sym:omegaval}
\addsymbol \omega_\mathrm{c}: {the cyclotron
resonance frequency (or energy in units of $\hbar$) for graphene}{sym:omega}
\addsymbol \tilde{\omega}_\mathrm{c}: {the exchange renormalised cyclotron
resonance frequency for graphene}{sym:omegactilde}
\addsymbol \omega_\mathrm{c}^{\mathrm{EG}}: {the cyclotron
resonance frequency (or energy in units of $\hbar$) for the 2DEG}{sym:omegaeg}

\end{tabbing} \clearpage
\pagenumbering{arabic} 

	
\chapter{Introduction}
\label{chap-INTRO}
The behaviour of electrons in quantum systems depends on the way they interact
with their surroundings, their interactions with other electrons and the
presence of external fields. In this work, we shall explore these themes in
three different systems. Firstly we examine Anderson localisation \cite{And58}
in three-dimensional crystalline solids. This neglects electron-electron
($e$-$e$) interactions, but takes into account the effect of lattice
imperfections, commonly referred to as disorder, which occur in any real
crystalline material. There exists a critical disorder above which the
electronic wave function becomes localised, due to quantum interference and the
material displays insulating behaviour. Secondly, we consider the Aharonov-Bohm
(AB) effect \cite{AhaB59}, which is another quantum interference phenomenon,
exhibited by systems including an external magnetic field in a certain geometry.
It provides a pleasing illustration of how the magnetic vector potential is not
merely a mathematical convenience and can actually influence a system's physical
properties. The excitonic AB effect is of particular interest, due to competing
effects of the magnetic field and electron-hole ($e$-$h$) interactions. Thirdly
we study graphene \cite{CasGPN09}, a two-dimensional carbon system, which has
been the subject of an immense research effort in the past few years. The work
presented here, which comprises the bulk of the thesis, focuses on the
collective behaviour of its electrons in a strong magnetic field in both the
presence and absence of impurities. Much of the theory developed for the
two-dimensional electron gas can be used to describe graphene, although there
are some important differences between the systems. Each of these fields is
introduced more fully in the following.

\section{Anderson localisation}
\label{sec-Aloc}
\subsection{Background and basic ideas}
\label{sec-bg}
\begin{figure}
\centering
\subfloat[]{\includegraphics[width=1.7in]{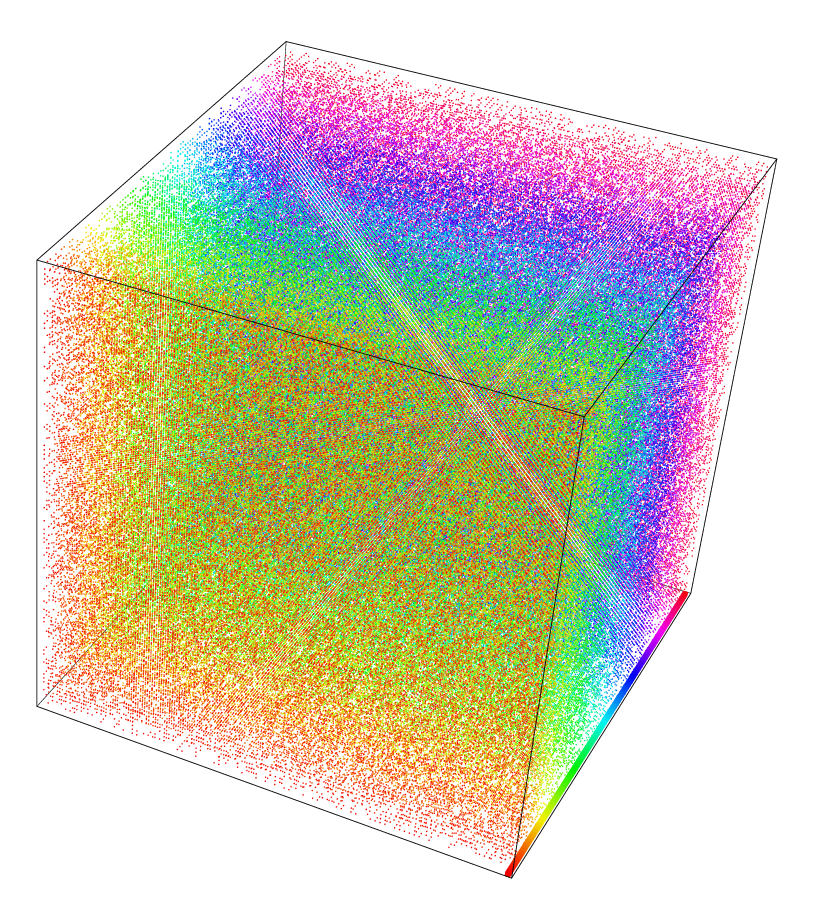}}
\subfloat[]{\includegraphics[width=1.7in]{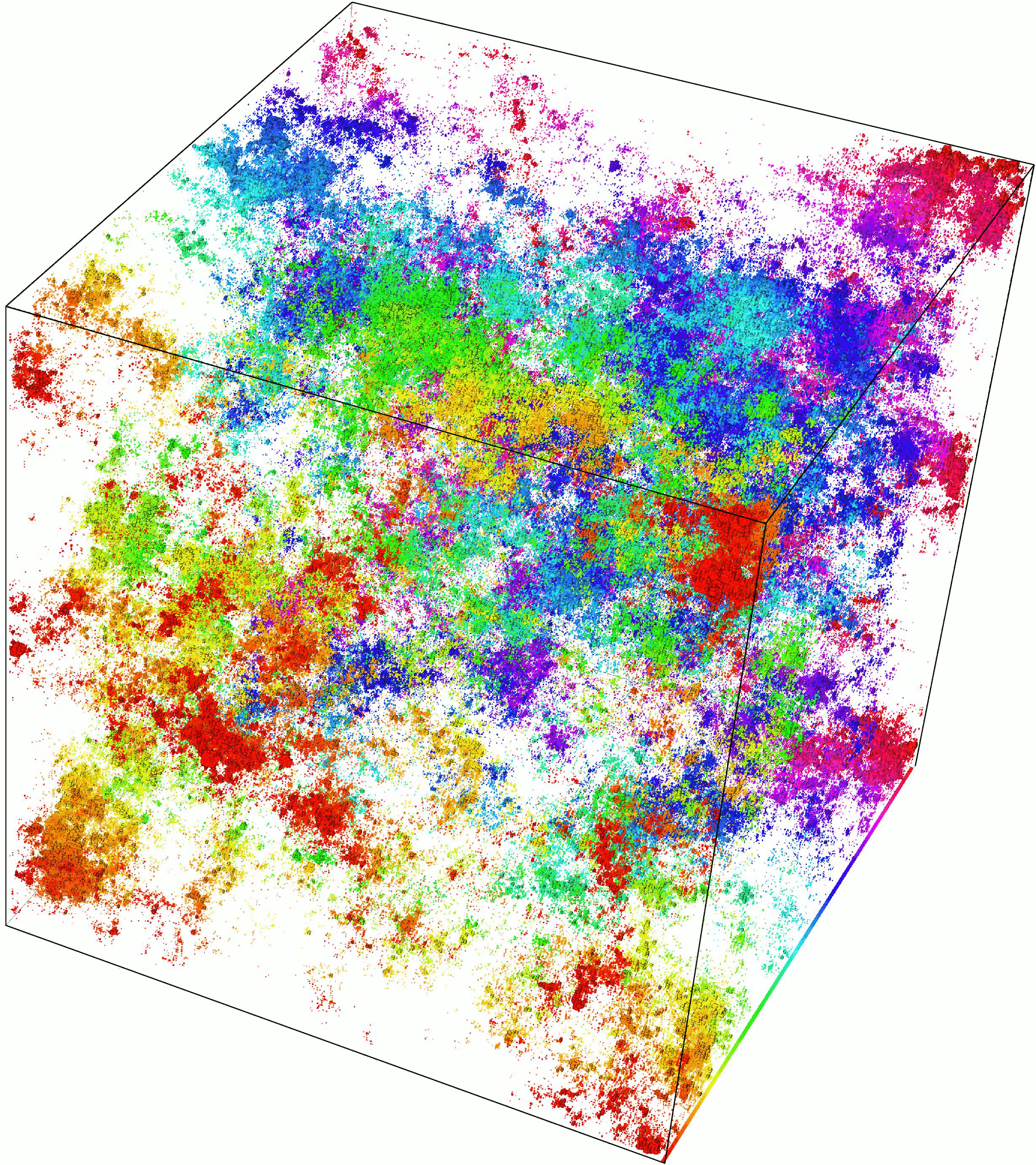}}
\subfloat[]{\includegraphics[width=1.7in]{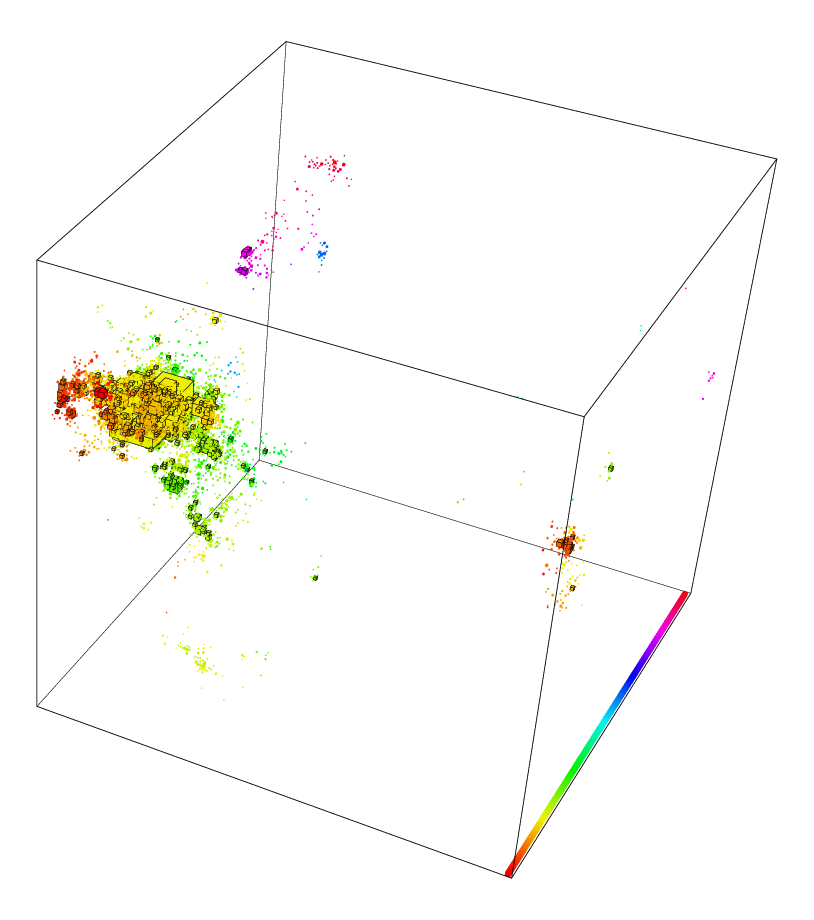}}
   \caption{(Images courtesy of A.\ Rodriguez, R.A.\ R\"omer, K.\ Slevin and
L.J.A.\ Vasquez). Three-dimensional electronic wave function plots for a simple
cubic lattice at the band centre for (a) an extended state ($W=1$), (b) a state
at the critical disorder ($W=W_\mathrm{c}=16.5$) and (c) a localised
state ($W=20$). In (a) and (c), the system has $100^3$
lattice sites and in (b) $240^3$ lattice sites. At a lattice site with index
$j$, the wave function amplitude
is denoted by $\psi_j$ and the size of the box is directly proportional to
$|\psi_j|^2$. The colour of boxes indicates which slice of the system the
lattice site is in i.e.\ it clarifies the third dimension.}
\label{fig-wfplots} 
\end{figure}
Understanding electronic transport in crystalline solids is one of the main goals
of condensed matter theory. Paul Drude was first to seriously tackle the problem in the early
$1900\mathrm{s}$ by using classical kinetic theory. Following this
came various quantum mechanical models such as the free
electron model, Bloch's theorem for electrons in a periodic potential and the
tight binding method. One very fundamental question to ask is: what makes some
materials metallic whilst others are
electrically insulating? Electronic band structure calculations go some way towards answering this question. For example, a material with a completely filled valence band separated from the conduction band by an energy gap large enough to make excitation between the two bands improbable, would be expected to exhibit insulating behaviour. However, such band theories do not include $e$-$e$ interactions and consequently predict some materials, known as Mott insulators, to be conductors when they are in fact insulators \cite{deBV37,MotP37,Mot49}. Another important aspect affecting the itinerancy of electrons is disorder. Here I refer not to the structural disorder of amorphous lattices, rather to the compositional disorder present in crystalline solids. This may be caused by dislocated or missing atoms (vacancies) and impurity atoms.

The framework for such a theory was first laid down by P.W.\ Anderson in 1958
\cite{And58}. In this seminal paper he models electronic transport in a
disordered lattice via quantum jumps between localised states. This model is
only reasonable at temperatures below about $10\mathrm{K}$, where all scattering
is elastic. Both electron-phonon and $e$-$e$ interactions are ignored. Electrons
are treated as spinless particles. The disorder is introduced via a random
potential at the lattice sites with the disorder parameter, $W$\newnot{sym:W},
characterising
the width of the probability distribution of onsite potential energies. In 3D
there is a critical disorder at which there is a phase
transition between extended and exponentially localised states of the electron
-- the disorder-induced metal-insulator transition (MIT) or Anderson transition
(AT) \cite{KraM93,EveM08}. Figure \ref{fig-wfplots} shows some
examples of squared wave function amplitude plots for extended, critical and
localised wave functions at the band centre for different disorder
values. I use $W_\mathrm{c}$\newnot{sym:Wc} to denote the critical disorder
at the band centre.

\begin{figure}
\centering
  \includegraphics[width=3.5in]{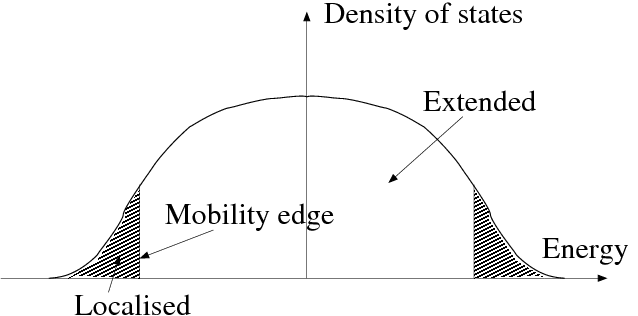}
   \caption{Rough sketch of the density of states in the presence of disorder. The shaded region represents localised states and the unshaded region extended states; these are separated by the mobility edge.}
\label{fig-dos} 
\end{figure}

Let us discuss this AT in more detail. In the absence of disorder, in a
perfectly periodic potential due to the ionic lattice, all electronic states are
Bloch states and extend throughout the system. Hence the system has a finite
conductivity as the absolute zero of temperature is approached and by this
definition behaves as a metal. Upon the introduction of disorder, localised
states form at the band edges. A schematic diagram of the density of states for
a fixed disorder value below $W_\mathrm{c}$ is shown in Fig.\ \ref{fig-dos} from
which it can be seen that localised and extended states exist in
different regions of the band. 
The lines of constant energy separating them are called mobility edges
\cite{Mot68a}. Thus we see that for a fixed disorder below $W_\mathrm{c}$, there
is a localisation-delocalisation transition as a function of energy. As the
degree of disorder is increased, more states become localised and the mobility
edges move towards the band centre. They meet at the band centre for critical disorder,
$W_\mathrm{c}$, at which point the whole band consists of localised states. The
transport properties of the system depend on those states close to the Fermi
energy. If the Fermi energy is well within the extended region, the system will
behave metallically. Conversely if it is within the localised region, the system
acts as an insulator and the conductivity will vanish in the limit of absolute
zero temperature. Keeping the Fermi energy constant and increasing the disorder
from zero will result in a transition from metallic to insulating behaviour at
the critical disorder. 

The mechanism for the localisation of an electron in the presence of disorder is
quantum interference of the electronic wave function. Classically one would
expect an electron to be itinerant if its energy exceeds the average lattice
site potential and localised if it possesses a lower energy. In quantum
mechanics the situation is much more complicated. The electron acts as a wave
packet, so that upon encountering a potential barrier, it splits into a
transmitted and reflected wave. In this way the electronic wave function may
interfere with itself and this results in an increased probability of
backscattering. This has been studied particularly in the limit of weak disorder
\cite{Ber84}, where constructive interference between clockwise and
anticlockwise paths results in a positive correction to the resistivity: the
weak localisation correction. Such behaviour is the precusor to the strong
Anderson localisation discussed here, when the interference is so strong that
the electronic propagation is prohibited altogether.

\subsection{Scaling theory}
\label{sec-scaltheory}
\begin{figure}
  \centering
  \includegraphics[width=3in]{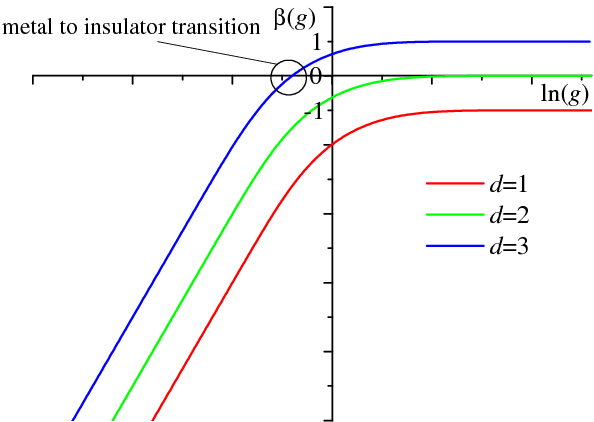}
  \caption{A qualitative plot of the scaling function $\beta\left( g\right)$ against $\mathrm{ln}\left( g\right)$ for different dimensions $d$.}
\label{fig-beta}
\end{figure}

After Anderson's original paper, the problem was re-expressed in the language of
renormalisation group theory and it was shown that the AT can be described as a
second order phase transition \cite{Weg76,Hik81}. In 1979 Abrahams {\it et al.}
developed the one-parameter scaling theory (OPST) of localisation
\cite{AbrALR79}, which examines the scaling of the conductance $g$
\newnot{sym:g} of a
hypercube of volume $L^d$ \newnot{sym:d} with the system size $L$
\newnot{sym:L}. The key assumption of the
theory is that the quantity $\beta=\frac{d\mathrm{ln}\left(g\right)
}{d\mathrm{ln}\left(L\right)}$ \newnot{sym:beta} depends solely on the
conductance $g\left( L\right)$ and not separately on $L$. The qualitative
dependence of $\beta$ upon
$g$ is shown in Fig.\ \ref{fig-beta} and estimated by interpolating between the
known forms in the limits of large and small $g$ \cite{Jan98}. In the limit of
small $g$, the electronic wave function is localised, so $g=g_0e^{-L/\lambda}$.
Here $\lambda$ \newnot{sym:lambda} is the localisation length, which is the
exponential decay length
of the envelope of the wave function. Hence for small $g$,
$\beta\left(g\right)=\mathrm{ln}\left(g/g_0\right)$ (see Fig.\ \ref{fig-beta}).
For large $g$, there is metallic behaviour, so $g=\sigma L^{d-2}$, where
$\sigma$ \newnot{sym:sigma} is the conductivity, yielding
$\beta\left(g\right)=d-2$.
An MIT can take place at the critical disorder when $\beta\left( g\right)=0$.
Thus the main result of OPST is revealed: a transition is only possible in
dimensions $d>2$ and all states are localised for any finite amount of
disorder in lower dimensions. Despite this, many experiments reaching
temperatures as low as $~10^{-1}\rm{K}$ claim to have observed an MIT in
effectively 2D systems such as silicon MOSFETs (metal oxide semiconductor field
effect transisitors), where $e$-$e$ interactions play a role \cite{KraS04}. 
The interaction strength may drive a transition and can be tuned by altering the
electronic density. At higher densities the $e$-$e$ interactions are weaker,
which can result in metallic behaviour \cite{AniKPF07}.
\subsection{Characterising the Anderson transition}
\label{sec-charMIT}
Previously, the temperature dependence of the conductivity, $\sigma$, was
alluded to as a way of distinguishing between metallic and insulating behaviour.
An alternative way, used most commonly in theoretical studies, is to examine the
form of the wave function, in particular the localisation length, $\lambda$. It
tends to infinity in the thermodynamic limit as the critical disorder is
approached from the insulating region. In practice, localisation lengths can
only be calculated for finite system sizes; a material is metallic if its
localisation length as a fraction of the system size increases with increasing
system size and insulating if it decreases with system size. Fortunately both
the conductivity \cite{MelK04} and
the localisation length \cite{KraM93} are self-averaging quantities; the values
obtained by averaging over different impurity configurations at a fixed
disorder, correspond to the value that will be observed with highest probability
in the thermodynamic limit. 

Studies of the AT seek to determine the critical disorder $W_{\mathrm{c}}$ and
the critical exponents for the transition, $s$ \newnot{sym:s} and
$\nu$, \newnot{sym:nu} such that close to the
transition $\sigma \propto | W-W_\mathrm{c} | ^s$ and
$\xi \propto| W-W_\mathrm{c}| ^\mathrm{-\nu}$, where $\xi$ \newnot{sym:xi} is
the
correlation length for an infinite system. Values of these critical exponents
determine the universality class of the system. One may also consider a
transition as a function of energy, $E$, \newnot{sym:E} at constant disorder, as
should be
clear from the explanation of mobility edges. In this case, the above relations
describing the behaviour of $\sigma$ and $\xi$ still hold (with
the same critical exponents) but with $W$ and $W_\mathrm{c}$ replaced by $E$
and $E_\mathrm{c}$ \newnot{sym:Ec} respectively. It should be noted that the
quantities
$E_\mathrm{c}$ and $W_{\mathrm{c}}$ are not universal, depending, for example,
on the distribution of onsite potential energies \cite{SleO99a} and, as shown in
the work presented in Chapter 2, on the lattice type.

Sadly we are far from achieving a generally accepted analytical theory for
Anderson localisation in $d$ dimensions. However, for $d=2+\varepsilon$, where
$\varepsilon<<1$, \newnot{sym:epsAT} it has been shown using perturbation theory
and field theoretic
techniques that $s=\left( d-2\right)\nu$ \cite{Weg76}, so that $s=\nu$ in 3D.
Early attempts to extend this approach to $d$ dimensions solved a self
consistent equation for the diffusion coefficient, resulting in $\nu =
1/\left(d-2 \right)$ \cite{VolW82}, which is equal to $1$ in 3D. This is in
agreement with many experimental studies, which reported $\nu \approx 1.0$
\cite{StuHLM93,WafPL99,ItoWOH99,ItoHBH96,WatOIH98}. The most recent analytical
study \cite{Gar08} finds $\nu=\frac{1}{2}+\frac{1}{d-2}$, agreeing more with
numerical studies, which find $\nu\approx 1.5$ for $d=3$.

A common numerical approach is to obtain localisation lengths via the
transfer-matrix method (TMM) \cite{KraM93}; this is explained in Chapter 2.
Another technique is to diagonalise the Hamiltonian directly and examine energy
level statistics \cite{HofS93,HofS94b,ZhaK97,MilR98}. The distribution of nearest
neighbour level spacings is exponential for localised states, whereas
metallic states obey GOE (Gaussian orthogonal ensemble) statistics. The 3D
eigenstate at criticality is multifractal, so that multifractal analysis
techniques are also commonly employed \cite{Vas10}. Finite size scaling (FSS) is
used to correct for the use of finite system sizes in an attempt to model
physics in the thermodynamic limit \cite{SleO99a}. The exact values of critical
parameters are still not agreed upon. Recent TMM studies have found $\nu=1.57\pm
0.02$, $W_\mathrm{c}=16.54\pm0.02$ \cite{SleO99a} and $\nu=1.515 \pm 0.033$,
$W_\mathrm{c}=16.50\pm 0.05$ \cite{Mac94} for the case of a box distribution of
onsite potentials. See Ref.\ \cite{OhtSK99} for a review. Energy level
statistics \cite{ZhaK95c} yield $\nu=1.45 \pm 0.08$,
$W_\mathrm{c}=16.35$. Multifractal analysis \cite{RodVSR10} finds $\nu=1.58\pm
0.03$, $W_\mathrm{c}=16.57 \pm0.03$. Clearly an accurate determination of the
critical parameters at the transition is still an active area of research.
\subsection{Experiment}
\label{sec-expt}
\begin{figure}
\centering
  \includegraphics[width=4in]{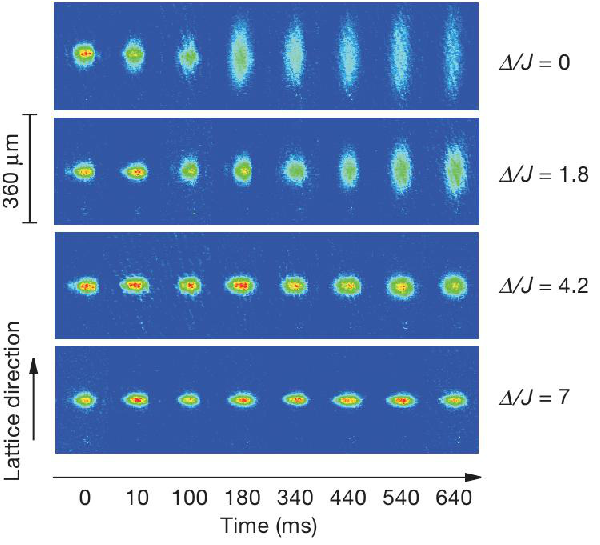}
   \caption{(Taken from Ref.\ \cite{RoaDFF08}). Absorption images of a BEC in a quasiperiodic optical lattice at different times and disorder values. The disorder, $\mathit{\Delta}$, is measured in units of the hopping parameter, $J/h=153\mathrm{Hz}$.}
\label{fig-roadff08} 
\end{figure}
There has been a lot of experimental work in which an MIT has been observed
\cite{StuHLM93,ItoHBH96}. However, a true AT may not have been experimentally
observed in crystalline solids. This is because $e$-$e$ interactions cannot be
sufficiently reduced and phonons are always present, so that a transition
between extended and localised behaviour cannot unambiguously be attributed to
disorder alone. However, it has been seen in a broad range of other physical
systems. Some time ago, light localisation was observed due to scattering from
suspended polystyrene balls \cite{VanL85} and more recently due to scattering
from semiconductor powders \cite{WieBLR97}. The localisation of ultrasound waves
was also seen and the measurement of localisation lengths possible
\cite{HuSPS08}. These systems have the advantage that interparticle interactions
are automatically eliminated. 

For a long time it has been the goal to directly see Anderson localisation of
matter waves. Imaging the local density of states provides a direct way to probe
the electronic probability density. This has been done using scanning tunnelling
microscopy in thin semiconductor samples \cite{RicRMZ10} and also by scanning
tunnelling spectroscopy in Quantum Hall systems \cite{HasSWI08}. However, the
$e$-$e$ interactions have been seen to be strong in these experiments. The last
ten years has seen incredible advances in the ability to control gases of
ultracold atoms \cite{Bou10}; such systems have potential for use as quantum
simulators to experimentally mimic theoretical Hamiltonians. They are
particularly advantageous, since the disorder is tunable and interactions
between atoms can be controlled and made negligible. Using these techniques has
allowed the very recent observation of exponential spatial localisation of
matter waves in 1D \cite{BilJZB08, RoaDFF08}. Quantum kicked rotors, in which
ultracold atoms are exposed to a pulsed laser, have been seen to be analogous to
spatially disordered systems, opening another way to experimentally probe the AT
\cite{ChaLGD08,LemCSG09}. In Ref.\ \cite{RoaDFF08} a transition was seen for a
Bose-Einstein condensate (BEC) in a quasi-periodic 1D optical lattice in an
experimental realisation of the Aubry-Andr\'{e} model \cite{AubA80}. Fig.\
\ref{fig-roadff08} shows absorption images of the BEC as a function of time
after the potential initially confining the atoms is removed. The disorder
increases from top to bottom. At large times, one can see that the condensate
has spread out for low disorders, but remains confined for higher disorders,
clearly indicating localisation.
\section{The Aharonov-Bohm effect}
\label{sec-ABE}
\begin{figure}
\centering
  \includegraphics[width=4in]{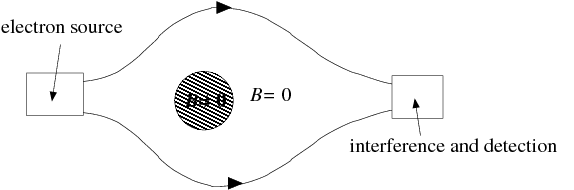}
   \caption{Schematic diagram of the experimental set up required to view the AB effect. The magnetic field is confined to a region from which the electrons are excluded.}
\label{fig-ABscheme} 
\end{figure}
The AB effect \cite{AhaB59} is a purely quantum mechanical phenomenon, which
cannot be understood using classical physics. I shall begin by illustrating it
for a particular example. Let us consider the situation (shown in Fig.\
\ref{fig-ABscheme}) where a beam of electrons is split in two, such that some of
the electrons pass above a region which contains a magnetic field, while some
pass below it and they recombine on the other side. It is important that the
magnetic field is confined to a region from which the electrons are excluded.
This could be achieved by using a tightly wound current carrying solenoid. Each
electron is subject to zero Lorentz force from the magnetic field, so that
classically one might expect it to behave in exactly the same way as it would if
there were no solenoid. However, each electron experiences a finite vector
potential, $\bf{A}(r)$, \newnot{sym:boldA} \newnot{sym:boldB} where $\mathbf{r}$
\newnot{sym:boldr} is the position vector, since for any path $C$
\newnot{sym:C} enclosing the solenoid,
\begin{equation}
\label{eq-Stokes}
\oint_{C}{d\bf{r}.\bf{A}(r)}=\int_{S}{d\bf{S}.\left( \nabla \times \bf{A}\right) }=\int_{S}{d\bf{S}.B }=\Phi,
\end{equation}
where the flux $\Phi$ \newnot{sym:Phi} is finite. Let $\psi_0(\bf{r})$ be the
electron's wave
function and $E$ its energy when there is no solenoid. In the presence of a
solenoid, the Hamiltonian is
\begin{equation}
\label{eq-bham}
\hat{H}= \frac{\left(-i\hbar\nabla+\frac{e}{c}\bf{A} \right)^2}{2 m_e},
\end{equation}
where $e$ \newnot{sym:e} is equal to negative the charge on an electron and
$m_e$ \newnot{sym:me} is the electron's mass. One can show that for
$\psi(\mathbf{r})=e^{i\alpha(\bf{r})}\psi_0(\bf{r})$, with $\nabla
\alpha=-\frac{e\bf{A}}{\hbar c}$,
\begin{equation}
\label{eq-schro}
\hat{H} \psi=e^{i\alpha(\bf{r})}\left( \frac{-\hbar^2}{2m_e}\nabla^2\right)  \psi_0(\mathbf{r})= E\psi.
\end{equation}
Hence the addition of a solenoid alters the wave function by a phase
$\alpha(\bf{r})$.\newnot{sym:alpha}
Let $\psi_1$ be the wave function for a beam that goes on one side of the
solenoid and $\psi_2$ that for a beam on the other. When the two beams
recombine, the intensity of the total beam depends on $| \psi_1+\psi_2|^2$,
which in turn depends on the phase difference
\begin{equation}
\label{eq-delalpha}
\Delta \alpha=\frac{e}{\hbar c}\oint d\mathbf{r}.\mathbf{A(\mathbf{r})}=\frac{e}{\hbar c}\Phi= 2\pi\frac{\Phi}{\Phi_0}.
\end{equation}
Here $\Phi_0=\frac{hc}{e}$ \newnot{sym:Phi0} is the universal flux quantum. Thus
any measurement of the interference pattern will depend upon the flux $\Phi$,
even though the electrons never experience the magnetic field. 

This very striking effect was first predicted by W. Ehrenberg and R.~E. Siday in
1949 \cite{EhrS49}. However, the idea remained largely unnoticed until it was
rediscovered ten years later by Y. Aharonov and D. Bohm \cite{AhaB59}. The
latter paper also describes an electric AB effect, whereby two electron beams
experience different time varying electrical potentials without being in an
electric field and their interference pattern depends on the field. It was later
found by M.~V. Berry that the phase given by  Eq.~(\ref{eq-delalpha}) is a
particular example of a more general geometrical phase, which a particle's wave
function acquires, upon completion of an adiabatic loop in parameter space
\cite{Ber84b}. Indeed the inclusion of such phases is currently revolutionising
semiclassical transport theory \cite{XiaCN10}.

The AB electron interference pattern has been seen in multiple experiments
\cite{Cha60,TonMSF82,OsaMKE86,TonOMK86}, which use a variety of techniques to
confine the magnetic field including whiskers, solenoids, toroidal magnets and
the Meissner effect. The AB effect has also been found to affect the transport
properties in materials with non simply connected geometries. Oscillations in
the magnetoresistance of metal rings with period $\Phi_0$ have been predicted
and experimentally observed \cite{WebWUL85}. The rings must be small enough that
the electron's phase is not randomised by \emph{inelastic} scattering events as
the electron moves about the arm of the ring. However, it was argued that the
thermodynamic properties may still have $\Phi_0$ periodicity for the case when
the \emph{elastic} mean free path is small compared to the size of the ring
\cite{ButIL83}. It is interesting to consider the AB effect in superconducting
quantum rings. It can be shown using Ginzburg-Landau theory that the free energy
oscillates as a function of $\Phi$ with minima at integer multiples of
$\Phi_0/2$ \cite{Ann04}. One can attribute the extra factor of 2 to the
formation of Cooper pairs of electrons. Hence the flux threading a
superconducting quantum ring is quantised in units of $\Phi_0/2$
\cite{DeaF61,TonOMK86}; this is achieved by supercurrents which generate an
additional magnetic field to that applied externally. From this it follows that
all thermodynamic quantities have this flux periodicity \cite{deG66}. For
example, the critical temperature for a normal-superconducting phase transition
has been seen to be $\Phi_0/2$ periodic \cite{LitP62,ParL64}.
\section{Two-dimensional electron gas in a magnetic field}
\label{sec-two-deg}
In the two-dimensional electron gas (2DEG) model, electron motion is confined to
a plane, which is embedded in three dimensions, meaning that any electromagnetic
fields present are not restricted to the plane. During the latter half of the
previous century it was a popular research topic, largely motivated by its
application to heterostructures, where the electrons are trapped in a layer less
than $1\mathrm{nm}$ thick \cite{AndFS82}. In particular, electrons in the
inversion layer of silicon MOSFETs, one of the main electronic components used
in computers, behave in
accordance with the 2DEG model. In addition, effects of fundamental physical
interest such as the integer and fractional quantum Hall effects \cite{Tsu99}
and Wigner crystallisation \cite{Wig34}, have been found to exist in
two-dimensional electron systems. The 2DEG model can also be used to describe
electrons in thin films and to some extent in graphene. Hence it is now
appropriate to examine the single particle problem for the 2DEG in a
perpendicular magnetic field. Also, in Chapter \ref{chap-MPsPristineGraphene},
when the many body graphene problem in a magnetic field is addressed, the
techniques used will be inspired by the interacting 2DEG model. We consider the
2DEG in a portion of the $x-y$ plane of area $S$, \newnot{sym:S} with a
perpendicular magnetic field, $\mathbf{B}=B\mathbf{e}_z$, where $\mathbf{e}_z$
\newnot{sym:ej} is a unit vector in the $z$ direction. For most of the work
presented here it will be convenient to use the symmetric gauge,
$\mathbf{A}_\mathrm{s}\left(\mathbf{r} \right)
=\frac{1}{2}\mathbf{B}\times\mathbf{r}$, \newnot{sym:boldAs} but for some
problems use of the Landau gauge, $\mathbf{A}_\mathrm{L}\left(\mathbf{r}
\right)=Bx\mathbf{e}_y$, \newnot{sym:boldAL} will make for a simpler solution.
We review below the single particle problem in both gauges \cite{Eza08}.
\subsection{Single particle problem in the symmetric gauge}
\label{subsec-symgauge}
The symmetric gauge is the appropriate choice for systems with a disk geometry. The single particle Hamiltonian for an electron can be shown to be
\begin{equation}
\label{eq-2deghamsym}
\hat{H}_{\mathrm{EG}}^\mathrm{s}=\hbar \omega_\mathrm{c}^{\mathrm{EG}}\left( a^\dagger a +\frac{1}{2}\right),
\end{equation}
where $a^\dagger$ and $a$ \newnot{sym:adag} are given (up to a phase) by:
\begin{equation}
\label{eq-ladopsa}
a=\frac{i}{\sqrt{2}}\left( \frac{z^{\ast}}{2\ell_B}+2\ell_B\frac{\partial}{\partial z}\right), \hspace{5mm} \hspace{5mm}
a^\dagger=\frac{-i}{\sqrt{2}}\left( \frac{z}{2\ell_B}-2\ell_B\frac{\partial}{\partial z^{\ast}}\right).
\end{equation}
Here $z=x+iy$,\newnot{sym:z} $\ell_B=\sqrt{\frac{\hbar c}{e B}}$
\newnot{sym:lB} is the magnetic length and
$\omega_\mathrm{c}^{\mathrm{EG}}=\frac{e B}{m_e c}$ \newnot{sym:omegaeg} is the
cyclotron resonance frequency. The eigenkets of
$\hat{H}_\mathrm{EG}^\mathrm{s}$ are
\begin{equation}
\label{eq-eigkets}
|nm\rangle =\frac{1}{\sqrt{ n! m!}}\left(a^\dagger \right)^n \left(b^\dagger \right)^m|00\rangle,
\end{equation}
where
\begin{equation}
\label{eq-ladopsb}
b=\frac{1}{\sqrt{2}}\left( \frac{z}{2\ell_B}+2\ell_B\frac{\partial}{\partial z^{\ast}}\right),\hspace{5mm}
b^\dagger=\frac{1}{\sqrt{2}}\left( \frac{z^{\ast}}{2\ell_B}-2\ell_B\frac{\partial}{\partial z}\right)
\end{equation}
\newnot{sym:bdag}and $n,m$ are non-negative integers. The state, $|00\rangle$,
satisfies $a|00\rangle=b|00\rangle=0$, from which its coordinate form can be
deduced. In the coordinate representation, the eigenstates for general $n,m$ are\footnote{This may be proved using Eq.\ (\ref{eq-eigkets}) and induction.}
\begin{eqnarray}
\label{eq-eigstates}
\langle \mathbf{r}|nm\rangle & = &\psi_{nm}\left(\mathbf{r} \right)\nonumber \\
& = &\left(-1 \right)^{n+m}i^n \left( \frac{m!}{2^{n-m+1}\pi \ell_B^2 n!}\right)^\frac{1}{2} \mathrm{exp}\left( -\frac{r^2}{4 \ell_B^2}\right) \left(\frac{z}{\ell_B} \right)^{n-m} L_{m}^{n-m}\left( \frac{r^2}{2 \ell_B^2}\right) \nonumber \\
& = & i^n  \left( \frac{n!}{2^{m-n+1}\pi \ell_B^2 m!}\right)^\frac{1}{2}\mathrm{exp}\left( -\frac{r^2}{4 \ell_B^2}\right)\left(\frac{z^\ast}{\ell_B} \right)^{m-n} L_{n}^{m-n}\left( \frac{r^2}{2 \ell_B^2}\right),
\end{eqnarray}
\newnot{sym:psinm}where $L_n^\alpha$ is a generalised Laguerre polynomial.
The eigenenergies corresponding to $\psi_{n m}\left(\mathbf{r} \right)$ are 
\begin{equation}
\label{eq-egengs}
E_n=\hbar\omega_\mathrm{c}^{\mathrm{EG}}\left(n +\frac{1}{2} \right).
\end{equation}

\newnot{sym:En}From this we see that electrons occupy discrete energy states
equally separated
by the cyclotron resonance energy (CRE), $\hbar\omega_\mathrm{c}^{\mathrm{EG}}$.
These are Landau levels (LLs) and are labelled by the LL index
$n$\newnot{sym:n}. The LLs are
degenerate with respect to the $m$ \newnot{sym:m} quantum number, which is
called the
oscillator quantum number. The filling factor $\nu$ \newnot{sym:filling} is
defined by $\nu=\frac{N}{N_0}$, where $N$ \newnot{sym:N}is the total
number of electrons in the system
and $N_0=\frac{S}{2 \pi \ell_B^2}=\frac{\Phi}{\Phi_0}$ \newnot{sym:N0} is the LL
degeneracy
associated with $m$. From Eq.\ (\ref{eq-eigkets}) it is clear that
$a^\dagger\left(a \right)$ raises (lowers) an electron by one LL and
$b^\dagger\left(b \right)$ raises (lowers) an electron's $m$ quantum number
within the same LL. The operators $a$, $a^\dagger$ and $b$, $b^\dagger$ thus
form two independent sets of ladder operators. Each set obeys bosonic
commutation relations i.e. $\left[a, a^\dagger\right]= \left[b,
b^\dagger\right]=1$. 

\begin{figure}
\centering
  \includegraphics[width=2.5in]{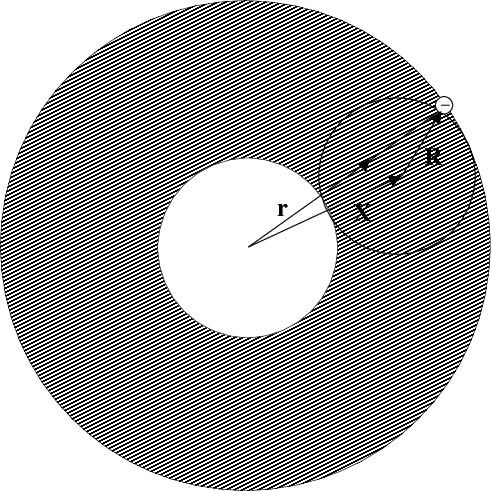}
   \caption{Diagram to explain the decomposition of the electron coordinate into a guiding centre, $\mathbf{X}$, and a relative coordinate, $\mathbf{R}$. The shaded region indicates the area where an electron in a given LL with a given value of the oscillator quantum number, $m$, will exist.}
\label{fig-2degcoords} 
\end{figure}
It is instructive to consider the physical meaning of the $n$ and $m$ quantum
numbers. Fig.\ \ref{fig-2degcoords} indicates how the electron coordinate
$\mathbf{r}$ may be decomposed according to $\mathbf{r}=\mathbf{R}+\mathbf{X}$,
where $\mathbf{X}$ is the guiding centre \newnot{sym:Xbold} and $\mathbf{R}$
\newnot{sym:R} is the relative coordinate. One can show
\begin{equation}
\label{eq-relcoord}
\left\langle\mathbf{R}^2 \right\rangle = \left( 2n+1\right)\ell_B^2, \hspace{5mm} \left\langle\mathbf{X}^2 \right\rangle = \left( 2m+1\right)\ell_B^2,\hspace{5mm} \left\langle\mathbf{r}^2 \right\rangle = \left( n+m+1\right)2\ell_B^2.
\end{equation}
Hence if we consider the electrons classically as particles undergoing circular motion in a magnetic field, the quantum number $n$ determines the radius of the circular motion, whereas the quantum number $m$ determines the distance of the guiding centre, about which the circular motion occurs, from the origin. The averages in Eq.\ (\ref{eq-relcoord}) are taken with respect to the state given in Eq.\ (\ref{eq-eigkets}). By setting $n=0$ in Eq.\ (\ref{eq-relcoord}), we can understand the physical meaning of the magnetic length $\ell_B$; it is the root mean square radius of the cyclotron motion of an electron in the degenerate ground state.

The orbital angular momentum projection is well defined for the states
$|nm\rangle$. It is easily shown that the operator of angular momentum
projection is \newnot{sym:lz}
\begin{equation}
\label{eq-orbangmom}
\hat{l}_z=\mathbf{r}\times-i \hbar \nabla=\hbar\left(a^\dagger a-b^\dagger b \right),
\end{equation}
which clearly commutes with the Hamiltonian in Eq.\ (\ref{eq-2deghamsym}) and
yields an eigenvalue $m_z=\left(n-m \right)\hbar$ \newnot{sym:mz} for the state
$|nm\rangle$.

In the equivalent problem for a hole, the single particle Hamiltonian is the same as that for an electron, given in Eq.\ (\ref{eq-2deghamsym}), but with $a,a^\dagger$ replaced by $b,b^\dagger$. Hence it is $b^\dagger$ ($b$) which raises (lowers) the LL index of a hole and $a,a^\dagger$, which are responsible for the LL degeneracy. It follows that a hole with LL index $n_h$, oscillator quantum number $m_h$ and coordinates $\mathbf{r}_h$, has wave function $\psi_{m_h n_h}(\mathbf{r}_h)$ and orbital angular momentum projection, $(m_h-n_h)\hbar$. From Eq.\ (\ref{eq-relcoord}), it is also clear that the LL index for a hole is associated with the guiding centre of cyclotron motion and the oscillator quantum number for a hole is associated with the radius of the cyclotron orbit. This is the opposite way round compared to an electron.
\subsection{Single particle problem in the Landau gauge}
\label{subsec-lgauge}
The Landau gauge is often used when modelling rectangular systems of area,
\newline $S=L_x L_y$.\newnot{sym:Lx}\newnot{sym:Ly} The non-interacting
Hamiltonian for the electron is
\begin{equation}
\label{eq-2deghamlandau}
\hat{H}_{\mathrm{EG}}^\mathrm{L}=\frac{1}{2m_e}\left(-\hbar^2\frac{\partial^2}{\partial x^2}-\hbar^2\frac{\partial^2}{\partial y^2}-2i\hbar eBx\frac{\partial}{\partial y} + e^2B^2x^2 \right).
\end{equation}
Clearly $[\hat{H}_{\mathrm{EG}}^\mathrm{L},\hat{p_y}]=0$, where $\hat{p_y}$ is
the $y$-component of the canonical momentum, \newnot{sym:pbold}
$\hat{\mathbf{p}}=-i \hbar \nabla$. Since $p_y=\hbar k_y$ is a good quantum
number, we try a solution of the form, $e^{ik_y y}g(x)$, which gives
\begin{equation}
 \label{eq-2deghamlandautwo}
\hat{H}_{\mathrm{EG}}^\mathrm{L} e^{ik_y y}g(x) = e^{ik_y y}\hbar \omega_\mathrm{c}^{\mathrm{EG}}( f^\dagger_X f_X +\frac{1}{2})g(x).
\end{equation}
Here
\begin{equation}
\label{eq-fdagger}
f_X=\frac{-i}{\sqrt{2}}\left(\ell_B \frac{\partial}{\partial x}+ \frac{x-X}{\ell_B} \right), \hspace{5mm} f^\dagger_X=\frac{-i}{\sqrt{2}}\left(\ell_B \frac{\partial}{\partial x}- \frac{x-X}{\ell_B} \right),
\end{equation}
\newnot{sym:fX}are Bose ladder operators, satisfying
$[f_X,f^\dagger_{X'}]=\delta_{X X'}$,
where \newnot{sym:X}$X=-k_y\ell_B^2$. Hence the electronic eigenstate is
\newnot{sym:phinXe}
\begin{equation}
\label{eq-2degwflandau}
\phi_{nX}^{(e)}\left( \mathbf{r}\right)=\frac{1}{\sqrt{L_y}}
\mathrm{exp}\left(-i \frac{Xy}{\ell_B^2}\right)\zeta_n(x-X),
\end{equation}
where \newnot{sym:zetan}
\begin{equation}
\label{eq-chin}
\zeta_n(x)=\langle \mathbf{r} \mid n \hspace{2mm}
X=0\rangle=\frac{1}{\sqrt{n!}}\langle
\mathbf{r}|\left(f^{\dagger}_0\right)^n|00\rangle
=\frac{i^n}{(\sqrt{\pi} \ell_B2^n n!)^\frac{1}{2}}\mathrm{exp}\left(-\frac{x^2}{2\ell_B^2} \right)H_n\left(\frac{x}{\ell_B}\right)
\end{equation}
and $H_n$ is a Hermite polynomial. The eigenstate $\phi_{nX}^{(e)}$ corresponds
to the energy $\hbar \omega_\mathrm{c}^{\mathrm{EG}}\left(n+\frac{1}{2}
\right)$, which is (reassuringly!) the same as that obtained using the symmetric
gauge. The wave function for the hole can be found similarly and is
\newnot{sym:phinXh}
\begin{equation}
\label{eq-2degwfhole}
\phi_{nX}^{(h)}\left( \mathbf{r}\right)=\frac{1}{\sqrt{L_y}} \mathrm{exp}\left(i
\frac{Xy}{\ell_B^2}\right)\zeta_n(x-X).
\end{equation}
Clearly Eqs.\ (\ref{eq-2degwflandau}) and (\ref{eq-2degwfhole}) describe an
extended plane wave in the $y$-direction and a Gaussian peak in the
$x$-direction centred about $x=X$. Since $k_y$ is an integer multiple of
$2\pi/L_y$, $X$ can take the values $X=-\frac{L_x}{N_0},-\frac{2
L_x}{N_0},\ldots,-\frac{(N_0-1) L_x}{N_0},L_x$. 
The $n$ quantum number denotes the LL index as for the symmetric gauge. The
angular momentum projection, $m_z$, is not a good quantum number in the Landau
gauge. The symmetric and Landau gauges are related according to
${\mathbf{A}_\mathrm{s}\left(\mathbf{r}
\right)=\mathbf{A}_\mathrm{L}\left(\mathbf{r} \right)-\nabla
\Lambda\left(\mathbf{r} \right)}$, where $\Lambda\left(\mathbf{r}
\right)=\frac{1}{2}Bxy$. The unitary operator, which transforms a state in the
symmetric gauge into a state in the Landau gauge is
$\mathcal{G}=\mathrm{exp}\left( -\frac{ie}{\hbar c}\Lambda\left(\mathbf{r}
\right)\right)$.
\section{Graphene}
\label{sec-graphene}
\subsection{Background}
\label{subsec-graphbg}
Graphene is the name given to a single layer of graphite \cite{CasGPN09}. Hence it is an atomically two-dimensional structure consisting of carbon atoms, which are sp$^2$-bonded together to form a honeycomb lattice. Historically there has been a lot of interest in carbon crystalline lattices, particularly in those which are comprised predominantly of benzene rings (with the hydrogen atoms removed), such as graphite, fullerenes \cite{And00} and carbon nanotubes \cite{SaiDD98}. Graphene can be viewed as the building block of all these structures and as such has been studied theoretically for a long time \cite{Wal47,Sem84}. Many scientists were of the opinion that it could not be synthesised, as it would be unstable with respect to scrolling up. However, our current understanding is that graphene is most likely created every time someone writes with a graphite pencil. Various early works reported the production and observation of extremely thin layers of graphite \cite{BoeCFH62,VanCT75,LuYHR99,BerSLL04}. However, physicists at the University of Manchester, UK, lead by K.S.\ Novoselov and A.K.\ Geim were the first to isolate true monolayer graphene in 2004 and to clearly demonstrate the number of layers in their samples \cite{NovGMJ04}. They realised the potential of this result and are largely responsible for making graphene one of the most active research areas in condensed matter physics today. For their contribution they have recently been awarded the 2010 Nobel prize in physics \cite{Ger10}.

There are currently several techniques for the fabrication of monolayer graphene. The first samples \cite{NovGMJ04} were created using mechanical exfoliation, whereby graphene layers are removed from graphite flakes using sticky tape. Other commonly used techniques \cite{ChoLSK10,SolMD10} include epitaxial growth, thermal decomposition of the surface of SiC crystals, chemical vapour decomposition, the reduction of graphene oxide and unzipping carbon nanotubes. Monolayer graphene can be identified under a carefully tuned optical microscope, when placed upon an appropriate substrate of the correct thickness to provide a noticeable contrast. Graphene is the first atomically two-dimensional material to have been synthesised and has inspired a search for further examples of atomic membranes \cite{NovJSB05}.

According to OPST described in Section \ref{sec-scaltheory}, the electronic
states in graphene should be localised, provided there is some disorder present,
which is always true for a real sample. Hence one would expect graphene to be an
insulator. Examining the problem from a different perspective, the chemist Linus
Pauling, predicted in the 1960s that graphene, like benzene, has a resonant
valence bond structure and hence should be a Mott insulator \cite{Pau60}. More
recent theoretical studies of graphite find a gap for high enough interaction
strengths \cite{Khv01,GorGMS02}. A study based on Monte Carlo simulations of
graphene \cite{DruL09a,DruL09b}, indicates that a gap should be present when the
surrounding environment has a high enough dielectric constant. However, the size
of the gap is not exactly determined and may not be large enough to induce
insulating behaviour. Indeed, numerous experiments have found that graphene is
in fact an excellent conductor \cite{Per10}. The cause of the apparent
discrepancy between theory and experiment is unclear. It may be due to the
intermediate strength of $e$-$e$ interactions in graphene, which are large
enough to delocalise Anderson electrons, but not strong enough to open a gap.
In addition, localisation by extended defects will not occur due to Klein
tunnelling \cite{Bee08,PerPCF10}.
\subsection{Massless Dirac Fermions}
\label{subsec-masslessfermions}
\begin{figure}
\centering
  \includegraphics[width=3.5in]{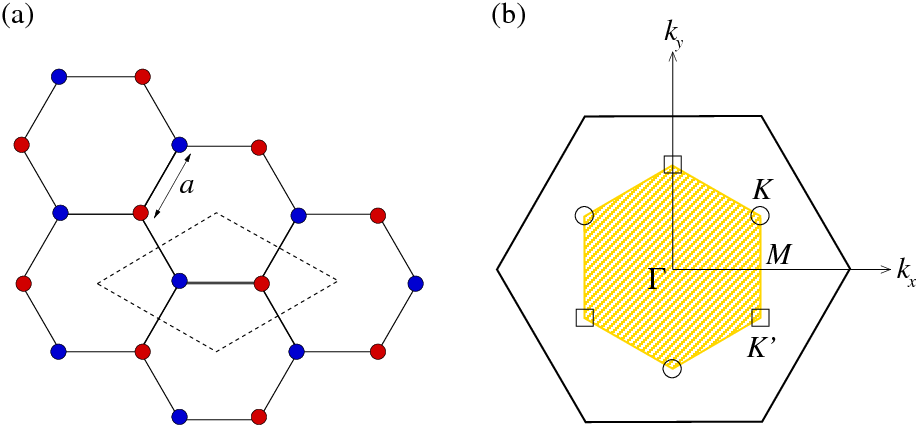}
   \caption{Diagram of the graphene lattice structure (a) and its Brillouin zone
(b). In (a) the red and blue circles are used to indicate carbon atoms on the
$A$ and $B$ sublattices. The dashed rhombus indicates a unit
cell. In (b) the yellow shaded area represents the Brillouin zone with the high
symmetry points $K$,$K'$,$M$,$\Gamma$ indicated. The circles and squares denote
positions equivalent to the $K$ and $K'$ points respectively.}
\label{fig-graphenelattice} 
\end{figure}
The structure of the graphene lattice is indicated in Fig.\
\ref{fig-graphenelattice}(a). It is bipartite and composed of two triangular
sublattices, commonly labelled \newnot{sym:AB} $A$ and $B$, so that there are
two atoms per unit cell. The distance between nearest neighbour atoms is
\newnot{sym:a} $a \approx 1.42 \mathrm{\AA}$. The reciprocal lattice of a
triangular lattice is also triangular, leading to a hexagonal Brillouin zone
(BZ), as shown in Fig.\ \ref{fig-graphenelattice}(b). Only two of the six
vertices of the BZ are inequivalent i.e.\ cannot be joined by a reciprocal
lattice vector. These are typically labelled the $\mathbf{K}$ and $\mathbf{K}'$
\newnot{sym:KK'} valleys and behave like a pseudospin degree of freedom, so I
shall sometimes use the alternative notation \newnot{sym:thickarrow} $\Uparrow$
and $\Downarrow$.

The dispersion relation for graphene was derived by P.R.\ Wallace in 1947
\cite{Wal47} assuming no disorder and no $e$-$e$ interactions, as outlined in
the following. Graphene has only one conduction electron per atom, which is in
the $2p_z$ state (see the following section). Let $\phi(\mathbf{r})$ be the
$2p_z$ orbital wave function for an electron in an isolated atom. Then in the
tight binding approximation, the wave function for an electron in the bulk
material is given by a linear combination of $\Phi_A$ and $\Phi_B$, with
\begin{equation}
\label{eq-phiA}
\Phi_A=\sum_{\mathbf{r}_A} \mathrm{exp}\left( 2\pi i\mathbf{k}.\mathbf{r}_A\right) \phi\left(\mathbf{r}- \mathbf{r}_A \right),
\end{equation}
where the sum over $\mathbf{r}_A$ is over all points in the $A$ sublattice;
$\Phi_B$ is defined analogously. Neglecting the overlap of $2p_z$ wave functions
centred on different lattice sites, it can be shown that the dispersion is
$E\left(\mathbf{k} \right)=H_{AA}\pm|H_{AB}|$, where
\begin{equation}
\label{eq-HAB}
H_{AB}=\frac{1}{N}\int d\mathbf{r}^2 \Phi_A^* \hat{H} \Phi_B
\end{equation}
and $H_{AA}$ is given by replacing $B$ with $A$ in Eq.\ (\ref{eq-HAB}). Here $N$
is the number of unit cells and $\hat{H}$ the Hamiltonian. 
\begin{figure}
\centering
  \includegraphics[width=4in]{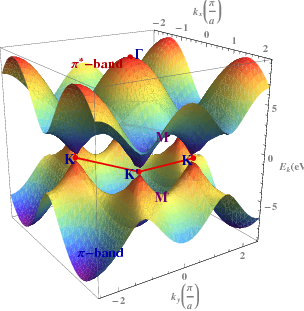}
   \caption{(Produced using source code from \hspace{2mm}the Wolfram
Demonstrations Project). Plot of the electronic dispersion relation,
$E\left(\mathbf{k} \right)$, for the case $t=3\mathrm{eV}$ and $t'=0$ i.e.\ no
next nearest neighbour hopping.}
\label{fig-spdispersion} 
\end{figure}
By substituting the
expressions for $\Phi_A$ and $\Phi_B$ into $H_{AA}$ and $H_{AB}$ and considering
only nearest neighbour \newnot{sym:t} and next nearest neighbour
\newnot{sym:t'} hopping, with energies $t$ and $t'$ respectively, it can be
shown that the dispersion relation measured relative to the energy of a $2p_z$
state in an isolated atom is
\begin{equation}
\label{eq-disp}
E\left(\mathbf{k} \right)=\pm t \left(3+ f\left(\mathbf{k}\right)\right) ^\frac{1}{2}-t'f\left(\mathbf{k} \right) .
\end{equation}
The function $f\left( \mathbf{k} \right)$ is given by
\begin{equation}
\label{eq-fofk}
f\left( \mathbf{k} \right)=2\mathrm{cos}\left(\sqrt{3} k_y a\right)+
4\mathrm{cos}\left(\frac{3}{2} k_x a \right)\mathrm{cos}\left(\frac{\sqrt{3}}{2}
k_y a\right).
\end{equation}
Fig.\ \ref{fig-spdispersion} shows a plot of the dispersion relation for particular values of $t$ and $t'$. The upper band is called the $\pi^*$ band and the lower band the $\pi$ band.
Evaluating $f\left( \mathbf{k} \right)$ at any corner of the BZ, such as
$\mathbf{K}=\frac{2\pi}{3a}\left(1,\frac{1}{\sqrt{3}} \right)$ and
$\mathbf{K'}=\frac{2\pi}{3a}\left(1,-\frac{1}{\sqrt{3}} \right)$, yields
$f\left( \mathbf{k} \right)=-3$. Hence if we include only nearest neighbour
hopping ($t'=0$), which I shall assume from now on, $E\left(\mathbf{k} \right)
=0$ at each corner of the BZ. There are two atoms and hence two electrons per
unit cell, so the $\pi$ band is completely full and the $\pi^*$ band completely
empty. The bands meet at $E=0$, so the Fermi energy is $E_\mathrm{F}=0$
\newnot{sym:EF} and the Fermi surface consists of the six points, which are the
BZ corners. The touching of conduction and valence bands means graphene may be
thought of as a zero-gap semiconductor.

An approximate form of the dispersion relation at low energies, for example
close to the $\mathbf{K}$ point, may be obtained by substituting
$\mathbf{k}=\mathbf{K}+\mathbf{q}$ \newnot{sym:qbold} into Eq.\ (\ref{eq-disp})
and assuming $qa=|\mathbf{q}|a\ll 1$. This gives
\begin{equation}
\label{eq-lowengdisp}
E\left( \mathbf{q}\right) \approx \pm \frac{3}{2} a t q=\pm \hbar v_\mathrm{F} q,
\end{equation}
where $v_\mathrm{F}=\frac{3 a t}{2\hbar}$ \newnot{sym:vF} is the Fermi
velocity; using $t\approx 2.8 \mathrm{eV}$, $v_\mathrm{F} \approx 10^6
\mathrm{m}\mathrm{s}^{-1}$. The linear dispersion relation given by Eq.\
(\ref{eq-lowengdisp}) is that of massless relativistic Dirac particles, but with
the speed of light, $c$, replaced by $v_\mathrm{F}$. Notice that the energy
depends only on the magnitude of $\mathbf{q}$, so that the dispersion relation
at low energies looks like two cones with their apexes touching at the
$\mathbf{K}$-points, as shown in Fig.\ \ref{fig-spdispersion}. These are refered
to as Dirac cones and the $\mathbf{K}$ and $\mathbf{K}'$ points as Dirac points.

One normally assumes that electrons are not scattered between the $\mathbf{K}$ and $\mathbf{K}'$ points, due to the large change in momentum that such a scattering would require. However, it is possible in the presence of $\delta$-function impurities as we shall see in Chapter 7. The single particle Hamiltonian for electrons in the vicinity of the $\mathbf{K}$-point was shown using the continuum approximation to be the 2D Dirac Hamiltonian \cite{Sem84}\newnot{sym:HD}:
\begin{equation}
\label{eq-diracham}
\hat{H}_\mathrm{D}= v_\mathrm{F} \bm{\sigma} \cdot \hat{\mathbf{\Pi}},
\end{equation}
where $\hat{\mathbf{\Pi}}=-i\hbar\nabla-\frac{\mathrm{q}}{c}\hat{\mathbf{A}}$
\newnot{sym:Pi} is
the kinematic momentum with $\mathrm{q}$ \newnot{sym:q} the particle's charge
and the vector
$\bm{\sigma}$ \newnot{sym:sigmabold} contains the Pauli matrices. In the absence
of electromagnetic
fields, the eigenstates \newnot{sym:Psiqup} \newnot{sym:Psiqdown} with energies
$\pm \hbar v_\mathrm{F} q$
are\footnote{Please note the difference between $\mathbf{q}$ the wave vector, its modulus $q=|\mathbf{q}|$ and $\mathrm{q}$ the
particle's charge.} 
\begin{equation}
\label{eq-speigs}
\Psi_{\mathbf{q}}^{\Uparrow \pm}\left( \mathbf{r}\right)=\frac{1}{\sqrt{2S}}e^{i\mathbf{q}.\mathbf{r}}
\left( \begin{array}{c}
\pm 1  \\
e^{i \theta\left(\mathbf{q} \right) }
\end{array}\right),
\end{equation}
where $S$ is the area of the graphene sheet, $\mathrm{tan}\theta\left(\mathbf{q}
\right)=q_y/q_x$ and $\Uparrow$ denotes the $\mathbf{K}$ valley. The appropriate
Hamiltonian for electrons in the vicinity of the $\mathbf{K}'$-point can be
found by changing $\partial_y \to -\partial_y$ in Eq.\ (\ref{eq-diracham}),
which results in eigenstates $\Psi_{\mathbf{q}}^{\Downarrow \pm}\left(
\mathbf{r}\right)$, which can be obtained from $\Psi_{\mathbf{q}}^{\Uparrow
\pm}\left( \mathbf{r}\right)$ by switching the vector components. In both
$\Psi_{\mathbf{q}}^{\Uparrow \pm}\left( \mathbf{r}\right)$ and
$\Psi_{\mathbf{q}}^{\Downarrow \pm}\left( \mathbf{r}\right)$, the first and
second vector components are the amplitudes for finding the electron in
the $A$ and $B$ sublattices respectively. The conical dispersion relation has
been seen explicitly in angle-resolved photoemission spectroscopy (ARPES) measurements
\cite{GruARV09}. In addition, other relativistic behaviour such as Klein
tunnelling has been observed experimentally for electrons in graphene
\cite{Bee08,PerPCF10}.

\begin{figure}
\centering
  \includegraphics[width=3in]{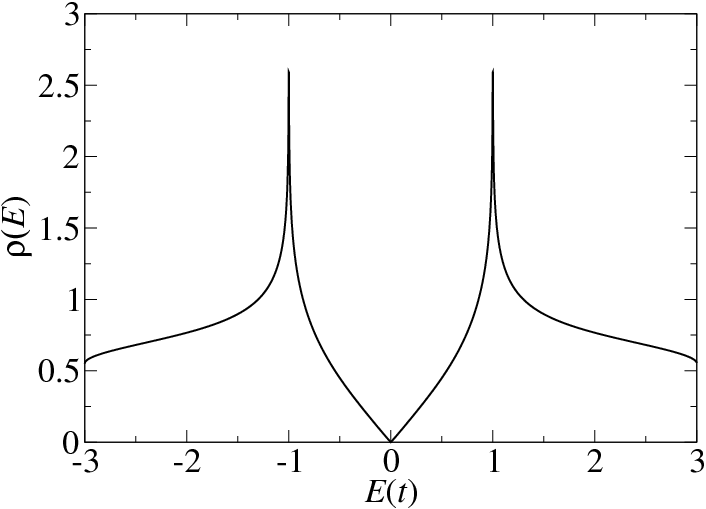}
   \caption{Density of states, $\rho(E)$, where the energy is measured in units of the nearest neighbour hopping parameter, $t$. Next nearest neighbour hopping is neglected ($t'=0$).}
\label{fig-grapdos} 
\end{figure}

Let us now examine the form of the density of states, \newnot{sym:rhoE}
$\rho(E)$. It is shown in Fig.\ \ref{fig-grapdos} for the case $t'=0$, where
there is nearest neighbour hopping only. An analytical solution has been derived
for this case \cite{HobN53}. Close to the Dirac points, it is easy to show
\begin{equation}
\label{eq-dos}
\rho(E)=\frac{2S}{\pi \hbar^2 v_\mathrm{F}^2}|E|.
\end{equation}
The effect of $t'\ne0$ is to distort the density of states and move the minimum away from $E=0$ \cite{CasGPN09}. The effect of an isolated impurity on the density of states has also been studied \cite{BenK05}.
\subsection{Graphene chemistry}
\label{subsec-chem}
A carbon atom has six electrons with orbital configuration $1s^22s^22p^2$. The $2s$, $2p_x$ and $2p_y$ orbitals hybridise to form three $sp^2$ in-plane orbitals each containing one of the four outer shell electrons. Covalent $\sigma$-bonds form between $sp^2$ orbitals in neighbouring carbon atoms and are responsible for graphene's honeycomb lattice structure. The remaining electron exists in the $2p_z$ orbital, which is perpendicular to the lattice plane. As mentioned in the previous section, the $2p_z$ electrons may hop from one atom to another and hence are delocalised. In graphite it is these out of plane orbitals which result in the weak coupling between graphene layers.

The strength of the $\sigma$-bonds between the carbon atoms results in graphene
having some remarkable properties. Its intrinsic strength was measured via
nanoindentation of suspended samples using an atomic force microscope tip and it
was found to be one
of the strongest materials ever tested with a breaking force of $1.8 \mu
\mathrm{N}$ for a tip of radius $16.5 \mathrm{nm}$ \cite{LeeWKH08}. In addition,
such strong bonds make it difficult
for impurity atoms to replace carbon atoms in the honeycomb lattice, so that
graphene is naturally a very clean material. The purity or quality of a sample
is characterised experimentally by the mobility, which is the ratio of the drift
speed to the applied electric field. Graphene samples have been found to have
mobilities 10-100 times higher than semiconductors such as silicon. Suspended
samples, where there are no substrate impurities present, have the highest
mobilities. The low impurity density allows electrons to travel through the
sample in a ballistic manner with a mean free path of up to $1 \mu
\mathrm{m}\approx10^4 a$ \cite{CasGPN09}. These impressive electronic and
mechanical properties make graphene an excellent candidate for a wide range of
applications (see Section \ref{subsec-grapapps}).

\subsection{Graphene in a magnetic field}
\label{subsec-grapmag}
Since its initial isolation, there has been a lot of research on the properties of graphene in a perpendicular magnetic field. Interestingly it was predicted that in the absence of an applied field, a pseudo-magnetic field may be induced, by straining the sample in a particular way, giving rise to gauge fields $B\sim10 \mathrm{T}$ \cite{GuiKV08,GuiGKN10}. Experimentally, pseudo-magnetic fields greater than $300\mathrm{T}$ have been seen in graphene nanobubbles \cite{LevBMP10}. Early on, the anomalous integer quantum Hall effect was observed,
with plateaus detected at filling factors $\nu=\pm2\left(|n|+1\right)$, where $n$ is the Landau level index \cite{NovGMJ05,ZhaTSK05}.
Subsequently, in higher fields, plateaus were seen at $\nu = 0,\pm1, \pm4$, signifying a lift in the fourfold (spin and valley) degeneracy, most likely due to the many-body effects \cite{JiaZSK07}.
Most recently, fabrication of high quality samples has allowed
the fractional quantum Hall effect to be observed at $\nu=\frac{1}{3}$,
indicating again the presence of a strongly correlated electron state \cite{DuSDL09,BolGSS09}.
Clearly many body effects strongly impact the physics of graphene in a high magnetic field. However, before turning to the complex many electron problem addressed in this thesis, we must examine the single particle problem in a magnetic field. 
\subsubsection{Single particle problem for graphene in a magnetic field}
\label{subsec-spgrapmag}
 As in the zero field case, we assume the Hamiltonian is block diagonal in the valley index:
\begin{equation}
\label{eq-blockdirachamB}
\hat{H}_\mathrm{D}=
\left( \begin{array}{cc}
\hat{H}_\mathbf{K}& 0   \\
0 & \hat{H}_\mathbf{K'}
\end{array}\right).
\end{equation}
We examine the problem in the symmetric and Landau gauges separately. For the
symmetric gauge it is convenient to use the points
$\mathbf{K}=\frac{2\pi}{3a}\left(-1,\frac{1}{\sqrt{3}} \right)$ and
${\mathbf{K'}=-\frac{2\pi}{3a}\left(1,\frac{1}{\sqrt{3}} \right)}$. The block
for an electron in the $\mathbf{K}$ valley can be found by replacing $\Pi_x \to
-\Pi_x$ in Eq.\ (\ref{eq-diracham}). This yields \newnot{sym:HKs}
\begin{equation}
\label{eq-dirachamB}
\hat{H}_\mathbf{K}^\mathrm{s}=
\hbar \omega_\mathrm{c}\left( \begin{array}{cc}
0 & a^\dagger   \\
a & 0
\end{array}\right),
\end{equation}
where $a,a^\dagger$ are defined in Eq.\ (\ref{eq-ladopsa}) and $\hbar
\omega_\mathrm{c}=\hbar\frac{\sqrt{2} v_\mathrm{F}}{\ell_B}$
\newnot{sym:omega} is the CRE (between the $n=0$ and $n=1$ LLs) for graphene. The single particle eigenstates of
$\hat{H}_\mathrm{D}$ for an electron in the $\mathbf{K}$ valley
are\newnot{sym:Psintausm} \newnot{sym:Psinmtau}
\begin{equation}
\label{eq-graspwfs}
\langle \mathbf{r} |n\Uparrow s m \rangle = \Psi_{n \Uparrow s m}(\mathbf{r})=\Psi_{n m \Uparrow}(\mathbf{r})  \chi_s =a_n \left( \begin{array}{c}
\psi_{|n| \,  m}(\mathbf{r})  \\
\mathcal{S}_n \psi_{|n|- 1 \,  m}(\mathbf{r}) \\
0\\
0  
\end{array}\right) \chi_s,
\end{equation}
where $\psi_{nm}$ are the 2DEG wave functions defined in Eq.\
(\ref{eq-eigstates}), $\chi_s$ \newnot{sym:chis} is the spin part
\newnot{sym:thinarrow}\newnot{sym:sspin}($s=\uparrow,\downarrow$),
$a_n=2^{\frac{1}{2}(\delta_{n,0} -1)}$ the normalisation constant \newnot{sym:an}and $\mathcal{S}_n={\rm
sign}(n)$ \newnot{sym:Sn} with
$\mathcal{S}_0=0$. The single particle eigenstates, $\Psi_{n \Downarrow s
m}(\mathbf{r})$, for an electron in the $\mathbf{K'}$ valley are given by
swapping the sublattice spinor components in Eq.\ (\ref{eq-graspwfs}). The
eigenenergies, \newnot{sym:epsn} ${\epsilon_n = {\rm sign}(n) \hbar\omega_c
\sqrt{|n|}}$, are
degenerate with respect to spin, valley and oscillator quantum number. However,
the quadruple degeneracy due to spin and valley pseudospin may be lifted in the
presence of high magnetic fields. From Eqs.\ (\ref{eq-orbangmom}) and
(\ref{eq-dirachamB}) it is clear that $\left[\hat{l}_z,
\hat{H}_\mathrm{D}\right]\ne 0$. Instead we define \cite{DivM84} the operator
\newnot{sym:jzhat}
\begin{equation}
\label{eq-jz}
\hat{j_z}=\hat{l_z}-\frac{\hbar}{2} \eins_2 \otimes \sigma_z,
\end{equation}
where $\eins_2$ is the $2\times2$ identity matrix and $\otimes$ denotes
the Kronecker product. Then $\hat{j_z}$ is a generalised angular momentum
projection operator, with the second term introducing a sublattice component and
satisfies $\left[\hat{j}_z, \hat{H}_\mathrm{D}\right]= 0$. For an electron with
wave function $\Psi_{n \tau s m}(\mathbf{r})$, where $\tau=\Uparrow,
\Downarrow$,\newnot{sym:tau} $j_z = |n| - m - \frac{1}{2}$\newnot{sym:jz}, takes
half integer values. Using
Eq.\ (\ref{eq-relcoord}) one can show that for graphene
\begin{equation}
\label{eq-r2}
\langle n \tau s m |  \mathbf{r}^2 |  n \tau s m \rangle = \left( 2(|n|+ m) + 1 + \delta_{n,0} \right) \ell_B^2 ,	
\end{equation}
so as for the 2DEG, large $|n|$,$m$ values mean the particle is far from the origin. The equivalent problem for a hole can be followed through analogously, beginning with the single particle Hamiltonian in the $\mathbf{K}$ valley
\begin{equation}
\label{eq-sphamhole}
\hat{H}_\mathbf{K}^{\mathrm{s} \hspace{1mm} (h)}\propto
\hbar \omega_\mathrm{c}\left( \begin{array}{cc}
0 & b   \\
b^\dagger & 0
\end{array}\right).
\end{equation}

When solving the problem in the Landau gauge, we use the conventional choices of
valley, $\mathbf{K}=\frac{2\pi}{3a}\left(1,\frac{1}{\sqrt{3}} \right)$ and
$\mathbf{K'}=\frac{2\pi}{3a}\left(1,-\frac{1}{\sqrt{3}} \right)$. As for the
analogous problem in the 2DEG, we try a wave function of the form, $e^{ik_y}
\left(\begin{array}{c}
g_A(x)\\
g_B(x)\end{array}\right)$, for an electron in the $\mathbf{K}$
valley:\newnot{sym:HKL}
\begin{equation}
\label{eq-dirachamBlandau}
\hat{H}_\mathbf{K}^\mathrm{L} e^{ik_y} \left(\begin{array}{c}
g_A(x)\\
g_B(x)\end{array}\right) 
= e^{ik_y}\hbar \omega_\mathrm{c}\left( \begin{array}{cc}
0 & f_X   \\
f_X^\dagger & 0
\end{array}\right)
\left(\begin{array}{c}
g_A(x)\\
g_B(x)\end{array}\right).
\end{equation}
Hence the wave function for an electron in the $\mathbf{K}$ valley is
\newnot{sym:PhientausX}\newnot{sym:PhienXtau}
\begin{eqnarray}
\label{eq-graspwfslandau}
\langle \mathbf{r} |n\Uparrow s X \rangle & = & \Phi^{(e)}_{n \Uparrow s
X}(\mathbf{r}) = \Phi^{(e)}_{n X \Uparrow}(\mathbf{r})  \chi_s \\ \nonumber
&=&\frac{a_n}{\sqrt{L_y}}\mathrm{exp}\left(-i \frac{Xy}{\ell_B^2}\right) \left(
\begin{array}{c}
\mathcal{S}_n \zeta_{|n|- 1 }(x-X)  \\
\zeta_{|n| }(x-X) \\
0\\
0  
\end{array}\right) \chi_s,
\end{eqnarray}
where $\zeta_n$ is defined in Eq.\ (\ref{eq-chin}).
It has the same eigenenergy, $\epsilon_n$, which was calculated for the
symmetric gauge. The wave function for a hole in the $\mathbf{K}$ valley
is \newnot{sym:PhihntausX}\newnot{sym:PhihnXtau}
\begin{eqnarray}
\label{eq-graspwfslandauhole}
\langle \mathbf{r} |n\Uparrow s X \rangle &=& \Phi^{(h)}_{n \Uparrow s
X}(\mathbf{r})=\Phi^{(h)}_{n X \Uparrow}(\mathbf{r})  \chi_s\\ \nonumber
&=&\frac{a_n}{\sqrt{L_y}}\mathrm{exp}\left(i \frac{Xy}{\ell_B^2}\right) \left(
\begin{array}{c}
\zeta_{|n| }(x-X)   \\
\mathcal{S}_n \zeta_{|n|- 1 }(x-X) \\
0\\
0  
\end{array}\right) \chi_s.
\end{eqnarray}
The corresponding wave functions in the $\mathbf{K'}$ valley, $\Phi^{(e)}_{n
\Downarrow s X}(\mathbf{r})$ and $\Phi^{(h)}_{n \Downarrow s X}(\mathbf{r})$,
are given by swapping the sublattice spinor components in Eqs.\
(\ref{eq-graspwfslandau}) and (\ref{eq-graspwfslandauhole}) respectively.
\begin{figure}
\centering
  \includegraphics[width=4in]{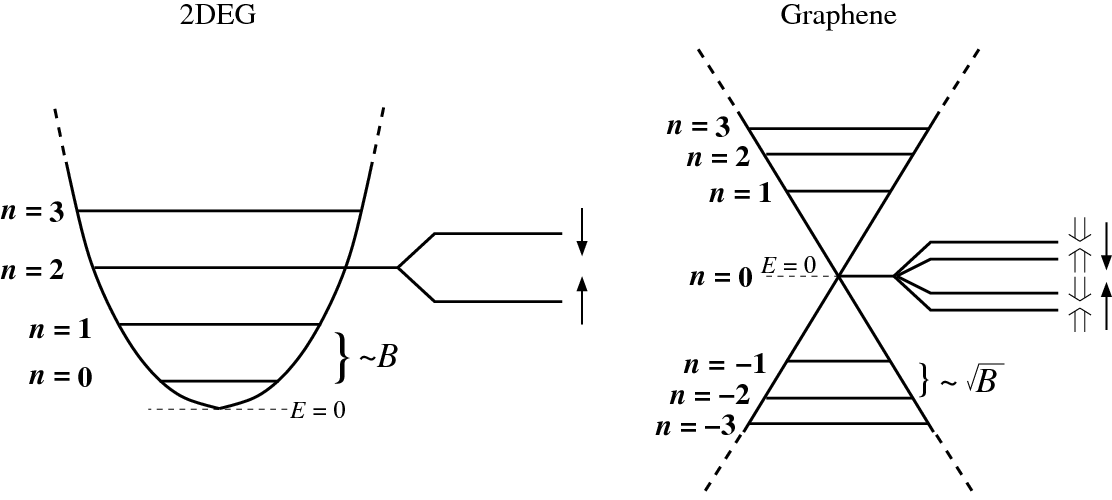}
   \caption{Diagram indicating the structure of LLs for graphene and the 2DEG.}
\label{fig-grapvs2deg} 
\end{figure}
It is instructive to reflect upon the differences between the single electron
problem in the 2DEG and graphene, as emphasised in Fig.\ \ref{fig-grapvs2deg}.
First note that in graphene, unlike for the 2DEG, the LL index $n$ may take
negative integer values and these correspond to negative energies or hole states
in the neutral system. In the 2DEG the ground state energy, $\frac{1}{2}\hbar
\omega_\mathrm{c}$, is positive, whereas in graphene there exists a state with
energy zero. This state has the unique property that the electronic wave
function is confined entirely to one sublattice, which is opposite in the
different valleys. In the 2DEG, the LLs are equally spaced, whereas in graphene,
the spacing between neighbouring levels with indices $n$ and $n+1$ depends on
$\sqrt{n+1}-\sqrt{n}$, which is a decreasing function of $n$. Thus in graphene,
high LLs may become mixed particularly in the presence of disorder broadening.
The LL energies for graphene $ \sim \sqrt{B}$ (with the exception of the $n=0$
LL which
is independent of $B$), so that increasing the magnetic field strength can help
define neighbouring LLs better, assuming that the disorder broadening does not
grow faster with $B$. This is one of several reasons for requiring a strong
magnetic field for the results presented in this thesis.
\subsection{Applications}
\label{subsec-grapapps}
A large proportion of current graphene research is device motivated. Its remarkable electronic and mechanical properties make it an ideal candidate for a huge and varied range of applications. I shall summarise some of the most important ones here; please see Refs.\ \cite{Gei09,GeiN07,Cas10} for a more comprehensive overview. 

One aspect, which has received a lot of media coverage, is the use of graphene
within the electronics industry. Currently, integrated circuits are made with
silicon using complementary metal-oxide-semiconductor technology. However, it is
well known that in approximately ten years, the reasonable limits of silicon
will be reached in terms of minituarising devices, increasing their speed and
reducing the cost. Indeed, there is a physical limit imposed by the fact that
silicon loses its crystalline structure beyond 10 nm due to increasing thermal
fluctuations of the atoms. Thus a new material is required for progress to be
made and both carbon nanotubes and graphene are being seriously investigated for
use in MOSFETs. The
properties recommending graphene for such a purpose are its high mobility
electronic transport, two-dimensional nature, strength and high melting
temperature. Consequently there has been a sustained effort towards the
fabrication of high performance graphene FETs. Most recently a group at IBM
achieved cut-off frequencies of 100 GHz \cite{LinDJF10}, which is markedly
better than the highest performing silicon MOSFETs. These have a high chance of
being used in analog transistors for communications technology and imaging.
However, the fact that graphene has no intrinsic band gap means that it is
difficult to turn off the electrical current, so on-off current ratios are much
too low \cite{XiaFLA10}, ruling out their use in digital circuits for the
moment. Since digital transistors make up the bulk of applications, overcoming
this problem is now one of the main goals of graphene research. Numerous ways of
opening a gap have been suggested. There already exists a gap in bilayer
graphene, which is tunable by a gate voltage \cite{AbeABZ10}. Graphene strongly
interacting with certain substrates may open a gap, particularly if this induces
an asymmetry between the two sublattices \cite{ZhoGFF07}. A gap can be opened by
constricting the path of electrons, for example in nanoribbons \cite{HanOZK07},
quantum dots \cite{SilE07} or nanoconstrictions \cite{TerDVT11}. Chemically
modifying graphene by doping it, with hydrogen \cite{SofCB07,EliNMM09} or
fluorine \cite{RobBJB10}, for example, will also induce a transition to a gapped
insulating state. We have already seen that an excitonic gap should exist in the
regime of strong $e$-$e$ interactions. However, there are problems associated
with each of these approaches and it is still not possible to control the band
gap of graphene to the degree required for competitive digital transistors.
Indeed, the full extent of graphene's future role in the electronics industry
remains unclear. 

Graphene is also used in creating nanoelectromechanical systems (NEMs), due to
its strength and stiffness \cite{BunZAV07}. For example, drum resonators have
been made from graphene oxide with high quality factors \cite{RobZBS08}. It is
hoped that graphene will be used to realise the ultimate goal of
chemical/biological sensors, namely single molecule gas detection
\cite{SchGMH07}. The basic idea is that each adsorbed molecule acts as an
acceptor or donor, causing step-like changes in the resistance. The difficulty
is that such small signals are easily drowned out by the noise. This is reduced
in graphene, since it naturally has a low number of impurities. Other useful
properties in this context are that it remains a conductor even at low carrier
densities \cite{NovGMJ05} and its 2D nature means its entire volume is available
to receive adsorbates.

Another biological application, which has been recently proposed \cite{Pos10},
is using graphene to sequence DNA. How to determine the sequence of
nucleotide bases on a single molecule of DNA at a reasonable cost and speed,
is currently one of the most important and challenging problems in biotechnology
research \cite{PetLA09}. A previous successful technique, known as shotgun
sequencing, involves cutting up the molecule \cite{SanNC77}. In another method,
which does not modify the molecule, the DNA is drawn through a biological
nanopore by an electric field, in an ionic solution \cite{NakAM03}. The basic
idea is that different bases result in different ionic currents. Researchers are
now trying to use holes in graphene instead of biological pores. Graphene
certainly seems to offer several advantages. Its thinness means that only one
base will be in the hole at any given time allowing single base resolution. It
can be used both as a membrane and electrodes. It is also stable and the size
and shape of the hole may be chosen quite precisely.

Graphene is also being investigated for flexible touch screen technology. The material currently used, indium tin oxide, is in short supply.  A graphene sheet supported by a polymer is used to create the screen's transparent electrodes \cite{KimZJL09}. The final aim is to use this in ultrathin computing devices, which can be folded away and easily transported.

With much of the basic physics underlying graphene now understood, there is a strong emphasis on growing larger, cleaner sheets to be used in technological applications. There has been a lot of progress \cite{SolMD10} since the original fabrication of $\mu\rm{m}$ size samples via mechanical exfoliation. The current record is a $30\hspace{0.5mm}\mathrm{inch}$ sample, which was grown on a flexible copper substrate via chemical vapour decomposition using roll-to-roll production, that allows it to be transfered to a target substrate of choice \cite{BaeKLX10}. There is also a lot of interest in the nature of the impurities in graphene, which limit the mobility. This question is important for improving device performance, but also pertains to the fundamental properties of the material. It is also relevant to the work presented in Chapter \ref{chap-nces}, where we examine collective excitations of graphene localised on impurities of different types.
\section{Layout of thesis}
\label{subsec-layout}
The first two Chapters are each self contained. In Chapter \ref{chap-mit}, I
explain the numerical techniques required to study the AT in BCC and FCC
lattices and give the resulting calculated values of the critical parameters. In
Chapter \ref{chap-abe}, a simple model is used to study the excitonic AB effect
in a quantum ring and determine how it is affected by the presence of an
in-plane electric field. Chapters \ref{chap-MPsPristineGraphene}-\ref{chap-nces}
are all concerned with collective excitations of graphene in a strong
perpendicular magnetic field. The single particle problem has already been
treated in Section \ref{subsec-grapmag}. A large amount of the theoretical
framework and background knowledge required to understand the problem is laid
out in Chapter \ref{chap-MPsPristineGraphene}. The focus is on neutral
collective excitations in pristine graphene. In Chapter \ref{chap-su4}, the
dynamical symmetries associated with $e$-$h$ complexes, which have both spin and valley pseudospin quantum numbers, are investigated. Chapters \ref{chap-cces}
and \ref{chap-nces} are mainly concerned with showing and discussing the
results. Chapter \ref{chap-cces} examines charged collective excitations
comprised of three particles, which may form at certain filling factors. Chapter
\ref{chap-nces} deals with the effects of disorder on neutral collective
excitations. A short summary of the most important findings and possible further
avenues of research is given in Chapter \ref{chap-conc}.
\cleardoublepage

\cleardoublepage
\chapter{Critical parameters for the Anderson transition in BCC and FCC lattices}
\label{chap-mit}
The vast majority of Anderson transition (AT) studies use the most elementary crystalline structure, namely the simple cubic (SC) lattice. However, the SC lattice does not commonly exist in nature; the only element known to adopt it is the alpha phase of polonium \cite{AshM76}. Indeed most metals have body centred (BCC) or face centred (FCC) cubic lattices. The aim of this study is to obtain the critical parameters and energy-disorder phase diagrams for the AT in BCC and FCC lattices and to compare the results with the SC lattice. No change in the critical exponent, $\nu \approx 1.5$, is expected, since all the systems belong to the same orthogonal universality class. We use the transfer-matrix method (TMM) to calculate the electronic localisation lengths and finite size scaling (FSS) to analyse the data for different system sizes and acquire the critical parameters, defined in the thermodynamic limit.

Although the localisation properties for an FCC lattice have been studied recently for a vibrational problem \cite{TarLNE01}, to the best of my knowledge no critical parameters were reported for an electronic AT in BCC and FCC lattices prior to this work \cite{EilFR08}. More recently, there has been further interest in modelling more physically relevant lattices \cite{WanZL08,WanZ10}. In Ref.\ \cite{WanZ10} a CsCl-type lattice was studied with site selective disorder and they found $\nu \approx 1.5-1.6$. This models mixed valence semiconductors in which dopants favour the atoms on one of the SC sublattices over the other, so that the disorder is not evenly distributed over the lattice as a whole. The BCC lattice limit was also taken and agreement with our results obtained.
\section{Numerical approach}
\label{sec-num}
\subsection{The transfer-matrix method}
\label{subsec-tmm}
The Anderson Hamiltonian \newnot{sym:HA} for an electron in a lattice with
compositional disorder is
\begin{equation}
  \hat{\bm{\mathrm{H}}}_A = \sum_i \varepsilon_i c^\dagger_i c_i -
  \sum_{i \neq j} t_{ij} c^\dagger_i c_j.
\label{eq-andham}
\end{equation}
Here  $c^\dagger_i$($c_i$) \newnot{sym:cdag} are the creation (annihilation)
operators for
creating (annihilating) an electron in a state localised on an atom at site $i$.
The wave function for such a localised state, $\langle
\mathbf{r}|c^\dagger_i|0\rangle$, is a Wannier function \cite{YuC99}; the
Wannier functions form an orthonormal set. The two terms in Eq.\
(\ref{eq-andham}) describe competing effects. The first term gives the
disordered onsite potential
energy, which has a localising effect on the electron. The onsite energies,
${\varepsilon_i \in [-W/2, W/2]}$\newnot{sym:epsi}, are randomly distributed
according to a
uniform distribution with height $W^{-1}$. The second term is the familiar tight
binding term giving the kinetic contribution, which delocalises the electron.
The $t_{ij}$ \newnot{sym:tij} are the energies required for the electron to hop
from site $i$ to
site $j$. We include nearest neighbour hopping only and do not include disorder
in the hopping term. For a 1D system, for example, this gives
$t_{ij}=t\delta_{|i-j|,1}$. All energies are then measured in units of the
hopping energy, $t$.

In the tight binding approximation, the ket describing an electron's state,
$|\psi\rangle$, can be expressed as a linear combination of the kets for an
electron localised on a particular lattice site:
\begin{equation}
|\psi\rangle=\sum_i \psi_i c^\dagger_i|0\rangle.
\label{eq-elecket}
\end{equation}
Substituting $|\psi\rangle$ into the Schr\"odinger equation,
$\hat{\bm{\mathrm{H}}}_A|\psi\rangle=E|\psi\rangle $, gives in the 1D case, the
lattice Schr\"odinger equation for the \newnot{sym:psii} wave function
amplitudes:
\begin{equation}
\varepsilon_i\psi_i-\psi_{i-1}-\psi_{i+1}=E\psi_i,
\label{eq-lattSchro}
\end{equation}
where I have set $t=1$. Rearranging this yields the matrix equation
\begin{equation}
\left(\begin{array}{c}
\psi _{i+1} \\
\psi _i
\end{array}\right)
= \mathbf{T}_i\left(\begin{array}{c}
\psi _{i} \\
\psi _{i-1}
\end{array}\right),
\label{eq-lattmat}
\end{equation}
where 
\begin{equation}
\mathbf{T}_i=\left(
\begin{array}{cc}
\varepsilon_i-E & -1\\
1 & 0
\end{array}
\right).
\label{eq-ti}
\end{equation}
$\mathbf{T}_i$ is termed the transfer-matrix \newnot{sym:Ti} since it transfers
between the
amplitudes $\psi_i,\psi_{i-1}$ and $\psi_{i+1},\psi_i$ \cite{KraM93,KraS96}.
Eq.\ (\ref{eq-lattmat}) is solved recursively after choosing initial conditions
$\psi_0,\psi_1$. The amplitudes $\psi_N,\psi_{N+1}$ are given by
\begin{equation}
\left(\begin{array}{c}
\psi _{N+1} \\
\psi _{N}
\end{array}\right)
=\mathbf{T}_N\mathbf{T}_{N-1}\ldots\mathbf{T}_{1}\left(\begin{array}{c}
\psi _{1} \\
\psi _{0}
\end{array}\right)
=\tau_N
\left(\begin{array}{c}
\psi _{1} \\
\psi _{0}
\end{array}\right)
\label{eq-tlots}.
\end{equation}
\newnot{sym:tauN} A correct solution is achieved when the iteration is begun
from both ends and the wave function and its derivative can be matched at some
site in the bulk \cite{KraS96}. However, the amplitudes themselves are not needed to determine
the localisation length. The expression for the localisation length is found by
applying Oseledet's multiplicative ergodic theorem \cite{Ose68,Lud91} to the
matrices $\mathbf{T}_N$, $\tau_N$.

The salient points of the theorem (for a rigorous mathematical treatment see
Refs. \cite{Ose68,Lud91}) are: \newline
i.\ The following matrix \newnot{sym:Gamma} exists:
\begin{equation}
\Gamma= \lim_{N \to \infty}\left(\tau_N\tau_N^\mathrm{T}\right)^\frac{1}{2N}.
\label{eq-gamma}
\end{equation}
ii.\ The eigenvalues of $\Gamma$ have the form $e^{\gamma_n}< \infty$. The
$\gamma_n$ \newnot{sym:gamma} are termed the Lyapunov exponents (LEs) of
$\tau_N$. Note that the
matrix $\tau_N$, which is a product of the transfer matrices defined in Eq.\
(\ref{eq-ti}) is symplectic\footnote{A matrix, $A$, is symplectic if it
satisfies
$AJA^\mathrm{T}=J$, where
$J=\tiny{\left(\begin{array}{cc}
0 & -1\\
1 & 0 \end{array}\right)}$.}.
Hence the corresponding eigenvalues of $\Gamma$ are $e^{\pm\gamma}$.\newline
iii.\ For each vector, $\mathbf{x}$, in the eigenspace of $\Gamma$, the
following limit exists:
\begin{equation}
\gamma(\mathbf{x})= \lim_{N \to
\infty}\frac{1}{N}\mathrm{ln}||\tau_N\mathbf{x}||,
\label{eq-lilgamma}
\end{equation}
where $\arrowvert\arrowvert \hspace{0.5mm} \cdotp \arrowvert\arrowvert$ is the
standard Euclidean norm. \newline
iv.\ 
\begin{equation}
\gamma(\mathbf{x})=\mathrm{max}(\{\gamma_i\}),
\label{eq-lilgamma2}
\end{equation}
where $\{\gamma_i\}$ is a subset of the LEs such that the energies
$e^{\gamma_i}$ have eigenvectors, which the vector $\mathbf{x}$ has a non-zero
projection on.

Now consider the propagation of a wave through the system. For a 1D atomic chain
it was shown \cite{KirP84}
\begin{equation}
\gamma=-\lim_{N \to \infty}\frac{\mathrm{ln}\mathcal{T}}{N},
\label{eq-trans}
\end{equation}
where $\mathcal{T}$ is the transmission amplitude \newnot{sym:mcT} at the end of
the system. From
this it follows ${\mathcal{T}_N \sim e^{-N\gamma}}$, where $\mathcal{T}_N$ is
the transmission amplitude at the $N^\mathrm{th}$ site. This expression only
makes physical sense for $\gamma>0$, since more of the wave is reflected and
less transmitted as it propagates through the system. The localisation length is
then defined by
\begin{equation}
\lambda=\frac{1}{\gamma}.
\label{eq-loclen}
\end{equation}
It is the exponential decay length of the transmission and thus also the tails
of the localised wave function.

In the TMM for a 3D system, the lattice is modelled as a quasi-1D bar consisting
of $N$ \newnot{sym:Ntm} 2D layers, each containing $M\times M$ lattice
\newnot{sym:M} sites, where $N>>M$. The localisation length depends on $M$,
the energy $E$ and
disorder $W$. In 3D it is scaled according to the system size (width), defining
the reduced localisation length\newnot{sym:LambdaM}, $\Lambda_M=\lambda/M$. The
length of the system
depends on the number of transfer matrices required to achieve the desired
convergence. The transfer-matrix equation, Eq.\ (\ref{eq-lattmat}), still holds,
but instead of being scalars, the $\psi_i$ are vectors of length $M\times M$
containing the amplitudes for the basis of localised states,
$c^\dagger_j|0\rangle$, for lattice sites in the $i^\mathrm{th}$ layer. The
transfer-matrix in 3D is
\begin{equation}
\mathbf{T}_i=\left(
\begin{array}{cc}
-\bm{\mathrm{C}}_{i+1}^{-1}(E\eins_{M^2}-\bm{\mathrm{H}}_i) &
-\bm{\mathrm{C}}_{i+1}^{-1}\bm{\mathrm{C}}_{i} \\
\eins_{M^2} & \bm{0}_{M^2}
\end{array}
\right).
\label{eq-3d}
\end{equation}
Here $\bm{0}_{M^2}$ and $\eins_{M^2}$ are the $M^2 \times M^2$ zero and identity
matrices respectively, $\bm{\mathrm{H}}_i$ \newnot{sym:Hi} is the 2D Hamiltonian
for the
$i^\mathrm{th}$ layer, containing the diagonal disorder and hopping between
sites in the same layer and $\bm{\mathrm{C}}_{i}$ is the connectivity matrix
\newnot{sym:Ci} describing the connections between the $i^\mathrm{th}$ and
$\left( i-1\right) ^\mathrm{th}$ layers.

Element $c_{jk}$ of the connectivity matrix  equals $1$ if the site $j$ in one
slice is connected to the site $k$ in the other; otherwise $c_{jk}=0$.
In the case of the SC lattice, each site has only one connection to the
succeeding (preceding) layer;
therefore all $\bm{\mathrm{C}}_i$ are unit matrices.
For BCC and FCC lattices, the connectivity matrices take a more complicated
form, but with purely diagonal disorder, they are constant so that the inverse
$\bm{\mathrm{C}}_i^{-1}$ needs to be calculated only once at the beginning of
the TMM calculations for a given size $M$. Nevertheless, the additional need to
multiply all states at each step of the TMM with a dense matrix
$\bm{\mathrm{C}}_i^{-1}$ reduces the speed of the calculation and hence
restricts the attainable system sizes.
The construction of the $\bm{\mathrm{C}}_i^{-1}$ matrices is not necessarily
always possible for a given lattice along all possible lattice vectors. Rather,
only selected directions, boundary conditions and $M$ values will lead to
non-singular $\bm{\mathrm{C}}_i$ matrices. The identification of permissible
directions for the application of the above TMM requires some care.

The numerical technique for obtaining the LEs and thus the localisation lengths
is perhaps different from what one might expect from Oseledet's theorem.
Importantly, one does not diagonalise the matrix $\Gamma$. Instead parts iii.\
and iv.\ of the theorem are utilised. 
The procedure is to begin with a set orthonormal vectors, $\{u_i^1\}$,
and arrange them as the columns of a
matrix $u^1$ - the $M^2 \times M^2$ identity matrix for simplicity. The
matrix $\mathbf{T}_1=\tau_1$ then acts upon
\begin{equation}
\left(\begin{array}{c}
u^1\\
u^0\end{array}\right)
=\left(\begin{array}{c}
\eins_{M^2}\\
\mathbf{0}_{M^2} \end{array}\right),
\label{eq-initialamps}
\end{equation}
where $\eins_{M^2}$ and $\mathbf{0}_{M^2}$ are the $M^2 \times M^2$ identity
and zero matrices respectively. Choosing $u^0=\mathbf{0}_{M^2}$ imposes hard
wall boundary conditions at that end of the system. Such an action yields a new
set of vectors 
\begin{equation}
\left(\begin{array}{c}
u^2\\
u^1\end{array}\right)
=
\mathbf{T}_1\left(\begin{array}{c}
u^1\\
u^0\end{array}\right).
\label{eq-t1action}
\end{equation}
This process
is repeated with the vectors from the previous step being left-multiplied by
$\mathbf{T}_N$ to get the new set of vectors for that set. The idea is that
$u_1^N$ provides the largest LE in the sense of Eq.\ (\ref{eq-lilgamma}),
$u_2^N$ provides the next largest LE etc.\ with $u_{M^2}^N$ giving
the smallest positive LE. The negative LEs are not calculated; they are simply
the negative counterparts to the positive LEs. It is worth noting that the
eigenvalues still come in pairs, $e^{\pm \gamma_i}$, in this case, even though
the matrix in Eq.\ (\ref{eq-3d}) is not generally symplectic. The smallest LE
will give the largest localisation length, so we define
\begin{equation}
\lambda=\frac{1}{\gamma_\mathrm{min}}
\label{eq-loclen3d}
\end{equation}
for the 3D case.

Another important part of the numerical procedure is the reorthonormalisation
of the vectors, $\{u_i^N\}$, at regular intervals. It is the most time consuming
part of the code, but is required for two reasons. Firstly, after a few
transfer-matrix multiplications, the vectors tend to lose their orthogonality.
Orthogonality is essential to ensure that the vectors all yield different LEs,
so that the smallest may definitely be found. Secondly, the size of the vectors
increases exponentially with the number of multiplications, so renormalising
them to one periodically, is needed to avoid the problem of numerical overflow.
The reorthonormalisation is executed after roughly every $10$ steps. However,
the frequency of performing it is varied according to how much the norm of the
vector leading to the smallest LE changes. The reorthogonalisation is achieved
using the Gram-Schmidt process (see, for example Ref.\ \cite{HorJ85}). We have
$\{u_i^N\}$ as the set of vectors, which need orthogonalising. Let $\{v_i^N\}$
be the set of orthogonal vectors constructed from the $\{u_i^N\}$, which
generate the same vector space as the $\{u_i^N\}$. The Gram-Schmidt method sets
$v_1^N=u_1^N$ and for $i>1$
\begin{equation}
v_i^N=u_i^N-\mathrm{proj}_{V_{i-1}}(u_i^N)
=u_i^N-\sum_{j=1}^{i-1}\frac{u_i^N.v_j^N}{||v_j^N||^2}v_j^N,
\label{eq-gramschmidt}
\end{equation}
where $V_i$ is the vector space generated by $\{v_j^N\}_{j=1}^{j=i}$ and
$\mathrm{proj}_V$ the operator which projects a vector onto the vector space
$V$. It is clear that the above definition yields orthogonal vectors, which may
then be normalised to produce an orthonormal set.

After each orthogonalisation step (before normalisation), an approximate
localisation length and its
relative error is calculated. If this error is less than some set value, the
program terminates. After each orthogonalisation step, the vectors are placed in
descending order of their norms, ensuring that it is the final vector,
$v_{M^2}^N$, which approximates the eigenvector corresponding to
$\gamma_\mathrm{min}$. Suppose after $N$ transfer-matrix multiplications, there
have been $n_0$ orthonormalisation steps and that, for the sake of simplicity,
these steps are equally spaced. The estimate for $\gamma_\mathrm{min}$ is then
\begin{equation}
\gamma_\mathrm{min}^N=\frac{1}{N}\sum_{j=1}^{n_0}\mathrm{ln}||v_{M^2}^{jN/n_0}
||.
\label{eq-gamminest}
\end{equation}
The relative error in the estimate of $\gamma_\mathrm{min}$, is the absolute
error\newnot{sym:delgam}, $\delta\gamma_\mathrm{min}^N$, which is the standard
deviation of the $\{\mathrm{ln}||v_{M^2}^{jN/n_0}||/N\}$, divided by
$\gamma_\mathrm{min}^N$:
\begin{equation}
\frac{\delta\gamma_\mathrm{min}^N}{\gamma_\mathrm{min}^N}=\frac{1}{n_0}\sqrt{
\frac{n_0 \sum_{j=1}^{n_0}\left(\mathrm{ln}||v_{M^2}^{jN/n_0}||
\right)^2}{\left(\sum_{j=1}^{n_0}\mathrm{ln}||v_{M^2}^{jN/n_0}||\right)^2}-1}.
\label{eq-error}
\end{equation}
The relative error in the estimate of $\gamma_\mathrm{min}$ is equal in
magnitude to the relative error in the estimate of $\lambda$ and $\Lambda_M$,
due to the relation Eq.\ (\ref{eq-loclen3d}).
\subsection{The lattice structures}
\label{sec-latt}

\begin{figure}
  \centering
  (a)\includegraphics[width=2in]{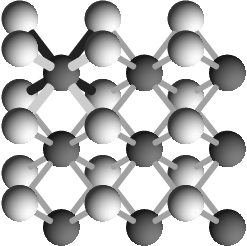}
  (b)\includegraphics[width=2in]{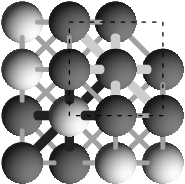}
  \caption{
    (a) Three layers of the 3D BCC lattice along a $\langle 100\rangle$ lattice
vector. Light grey spheres mark the $1^{\rm st}$ and $3^{\rm rd}$ layer, dark
grey ones indicate the central layer. Lines between layers denote the
connections between lattice sites. The connections to the upper-left sphere in
the central layer are emphasised by broad lines, illustrating its 8 neighbours.
The 4 thick light grey lines connect the $1^{\rm st}$ and the central layer, the
black ones
go from the central to the $3^{\rm rd}$ layer.
    (b) Structure of 3 layers of the 3D FCC lattice along a $\langle 111\rangle$
lattice vector. The broken lines mark the cubic unit cell of the lattice. Sites
in the $1^{\rm st}$ and $3^{\rm rd}$ layer are dark grey; sites in the central layer
are light grey. The thick light grey lines represent connections between lattice
sites in the {\em same} layer for one particular site. The thick black lines
represent connections between lattice sites in {\em  neighbouring} layers for
another site. The thin lines indicate connections to other sites. Some sites in
the upper-right corner are removed for clarity.}
\label{fig-bccfcc3D}
\end{figure}

The structure of the BCC lattice is displayed in Fig.\ \ref{fig-bccfcc3D}(a).
The construction of the TMM quasi-1D bar proceeds along a $\langle 100\rangle$
vector. In this case each site within the slice is connected to four sites
in the preceding slice and to four sites in the succeeding one. There are no
connections between sites within the slice, which means that the Hamiltonian
matrix $\bm{\mathrm{H}}_i$ is diagonal.
We use periodic boundary conditions in both transversal directions, which
results in the connectivity matrix for a slice of $M \times M$ sites being
singular for all even $M$, thus restricting the system sizes we can use.  Using
a helical boundary condition \cite{ZhaK99} in one or two directions provides the
same singularities and hence offers no advantage.

Fig.\ \ref{fig-bccfcc3D}(b) shows the structure of the FCC lattice. It proved
convenient to construct the TMM bar along a $\langle 111\rangle$ vector, so the
subsequent layers of the bar are close packed. Within the layer, each site has
six connections to nearest neighbours. In addition there are three connections
to the preceding and three connections to the succeeding layer. The resulting
connectivity matrix can be inverted for each size of the $M \times M$ TMM slice,
but only when we use a mix of periodic boundary conditions in one direction and
helical boundary conditions in the other. We note that it has been shown
previously that critical exponents and transition points are independent of the
boundary conditions \cite{KraM93,SleOK00}. See Appendix \ref{app-con} for
examples of the connectivity matrices for system size $M=3$.
\subsection{Finite size scaling}
\label{sec-fss}
As we saw in Section \ref{sec-charMIT}, the AT is characterised by a divergent
correlation length for the infinite system, $\xi$, so that at fixed energy $E$,
${\xi(W)\propto |W-W_{\rm c}|^{-\nu}}$ and at fixed disorder $W$,
${\xi(E)\propto |E-E_{\rm c}|^{-\nu}}$, where
$\nu$ is the critical exponent and $W_{\rm c}$, $E_{\rm c}$ are the critical
disorder and energy, respectively, at which the AT occurs \cite{KraM93}. In the
following discussion, let's assume the case of fixed energy and varying
disorder; the converse case of fixed disorder and varying energy proceeds
analogously.

In order to extract the critical parameters from the calculated values of
$\Lambda_M(W)$, one applies the FSS procedure outlined in Ref.\ \cite{MacK83}.
The correlation length, $\xi$, may be obtained from the
localisation lengths for finite system sizes $\Lambda_M(W)$ by using
the one-parameter scaling law $\Lambda_M=f(M/\xi)$ \cite{Tho74}.
The FSS can be performed numerically by minimising the deviations of the data
from the common scaling curve; this has two branches for extended and localised
states. The critical parameters are then obtained by fitting the $\xi$ values as
obtained from FSS.
Better numerical accuracy for the FSS procedure can be achieved by fitting
directly the raw data from TMM calculations using the method applied previously
to the TMM data for the 3D SC lattice \cite{SleO99a,MilRSU00}. We introduce a
set of fit functions which include two kinds of corrections to scaling, (i)
nonlinearities of the $W$ dependence of the scaling variables and (ii) an
irrelevant scaling variable which accounts for a shift of the point at which the
$\Lambda_M(W)$ curves cross. We use \cite{MilRSU00}
\begin{equation}
  \label{eq-Slevin}
  \Lambda_M=\tilde{f}(\chi_{\rm r} M^{1/\nu}, \chi_{\rm i} M^{y}),
\end{equation}
where $\chi_{\rm r}$ \newnot{sym:chir} and $\chi_{\rm i}$ \newnot{sym:chii} are
the relevant and irrelevant scaling
variables respectively. Note that $y<0$\newnot{sym:y}, so that the contribution
from the irrelevant scaling variable is reduced for larger system sizes. The
function
$\Lambda_M(W)$ is then Taylor expanded
\newnot{sym:ni}\newnot{sym:nr}\newnot{sym:ank}\newnot{sym:fn}
\begin{eqnarray}
  \label{eq-Slevin2}
  \Lambda_M&= &\sum_{n=0}^{n_{\rm i}} \chi_{\rm i}^n M^{n
    y}\tilde{f}_n(\chi_{\rm r} M^{1/\nu}),\\
%
  \tilde{f}_n&= &\sum_{k=0}^{n_{\rm r}} a_{nk} \chi_{\rm r}^k M^{k/\nu}.
\end{eqnarray}
Nonlinearities are taken into account by expanding $\chi_{\rm r}$ and
$\chi_{\rm i}$ in terms of ${w=(W_{\rm c}-W)/W_{\rm c}}$ \newnot{sym:w}up to
order $m_{\rm r}$ \newnot{sym:mr}and $m_{\rm i}$\newnot{sym:mi}, respectively,
\begin{equation}
  \label{eq-Slevin-Var}
  \chi_{\rm r}(w)=\sum_{m=1}^{m_{\rm r}} b_m w^m , \quad \chi_{\rm
    i}(w)=\sum_{m=0}^{m_{\rm i}} c_m w^m,
\end{equation}
with $b_1=c_0=1$. \newnot{sym:bm} \newnot{sym:cm}
Note that this FSS procedure assures the divergence of $\xi$ and hence it is not
the divergence itself but rather the quality of how the model fits the computed
reduced localisation lengths $\Lambda_M$ which determines the validity of the
scaling hypothesis.
\subsection{Nonlinear fitting}
\label{sec-nlf}
The nonlinear fit was carried out using the built-in \emph{Mathematica}
function \\ \nobreak{``$\mathrm{NonLinearRegress}$''}. It uses the
Levenberg-Marquardt
method \cite{MilRSU00,PreFTV92} to find the  parameters ($a_{ij}$, $b_i$, $c_i$,
$y$, $\nu$, $W_\mathrm{c}$), which provide the best fit of the data
($\{M,W,\Lambda_M\}$) to the model $\tilde{f}$ given in Eqs.\
(\ref{eq-Slevin})-(\ref{eq-Slevin-Var}) with fixed values of $n_\mathrm{r}$,
$n_\mathrm{i}$, $m_\mathrm{r}$, $m_\mathrm{i}$. The data points are weighted
according to $\frac{1}{\Lambda_M^2(\delta\gamma_\mathrm{min}^N
/\gamma_\mathrm{min}^N)^2}$, which is the inverse variance of $\Lambda_M$.
Importantly, it also provides errors for $W_{\rm
c}$ and $\nu$ and statistical measurements of the goodness of fit, which
indicates the validity of the model. This information enables us to choose the
most appropriate values of $n_\mathrm{r}$, $n_\mathrm{i}$, $m_\mathrm{r}$,
$m_\mathrm{i}$. The goodness of fit should be maximised and the errors minimised
whilst simultaneously keeping the number of parameters low. The number of
parameters (not including $W_\mathrm{c}$ and $\nu$) is $(n_\mathrm{i}+1)(n_\mathrm{r}+1)+m_\mathrm{i}+m_\mathrm{r}+2$ for
the case $n_\mathrm{i}>0$ and $n_\mathrm{r}+m_\mathrm{r}+1$ when
$n_\mathrm{i}=0$.

One of the outputs of ``${\mathrm{NonLinearRegress}}$'' is $\chi^2$
\cite{NRC02}. It is related to the issue of how to quantitatively judge how well
a set of parameters fit the data for a particular model. This is done using the
concept of \emph{maximum likelihood estimators}. One calculates the probability
that the data set could have occured, given a particular set of parameters.
Suppose the data is $\{\{x_1,y_1\},\{x_2,y_2\},\ldots,\{x_n,y_n\} \}$ and the
goodness of fit of the model $y(x)$ is to be tested. For data which may take
continuous values, such as our localisation lengths, the aforementioned
probability is zero. Instead we calculate the probability that a data set
contained in ${\{\{x_1,y_1\pm\Delta y\},\{x_2,y_2\pm\Delta
y\},\ldots,\{x_n,y_n\pm\Delta y\} \}}$ could have occured, given the model
$y(x)$, where $\Delta y\ll 1$ is fixed. This is
\begin{equation}
P\propto\prod_{i=1}^{n}\mathrm{exp}\left[-\frac{1}{2}\left(\frac{
y_i-y\left(x_i\right)}{\sigma_i}\right)^2\right]\Delta y.
\label{eq-prob} 
\end{equation}
The expression for $P$ assumes that each data point, $y_i$, is normally
distributed about $y(x_i)$ with standard deviation $\sigma_i$ and that the
errors are independent. The assumption of a normal distribution should hold due
to the central limit theorem \cite{Ara80}. However, one should bare in mind that
this is not always the case in practice. Maximising Eq.\ (\ref{eq-prob}) is
equivalent to minimising
\begin{equation}
\chi^2=\sum_{i=1}^n\left(\frac{y_i-y\left(x_i\right)}{\sigma_i}\right)^2.
\label{eq-chisquared} 
\end{equation}
The method of minimising $\chi^2$ to find the best-fit parameters, is refered to
as a \emph{least-squares fit}. The Levenberg-Marquardt method
\cite{MilRSU00,PreFTV92} minimises $\chi^2$ iteratively. An initial guess for
the parameters is used to calculate an initial value for $\chi^2$. The
parameters are then altered to further lower $\chi^2$. This is done using either
the inverse Hessian method or method of steepest descent, depending on how well
the function $\chi^2(\mathbf{a})$ ($\mathbf{a}$ is a vector containing the
parameters) approximates the behaviour of $\chi^2$ close to its minimum value.

To examine the probability distribution of different values of $\chi^2$, we must
define the incomplete gamma function:
\begin{equation}
\Gamma_\mathrm{p}(a,x)=\frac{1}{\Gamma(a)}\int_0^x dt \hspace{1mm}
t^{a-1}e^{-t},
\label{eq-gammap} 
\end{equation}
where
\begin{equation}
\Gamma(a)=\int_0^\infty dt\hspace{1mm} t^{a-1}e^{-t}
\label{eq-gammafunc} 
\end{equation}
is the ordinary Gamma function. Let $\mu$ \newnot{sym:mu} be the number of
degrees of freedom:
\begin{equation}
\mu=\#\mathrm{data \hspace{2mm} points}-\#\mathrm{parameters}.
\label{eq-mu} 
\end{equation}
Then $\Gamma_\mathrm{p}(\frac{\mu}{2},\frac{\chi^2}{2})$ gives the probability
that the observed chi-squared value for a correct model is less than $\chi^2$.
It is usually the complement of $\Gamma_\mathrm{p}$, i.e.\
$\Gamma_\mathrm{q}=1-\Gamma_\mathrm{p}$, which is considered. The closer
$\Gamma_\mathrm{q}(\frac{\mu}{2},\frac{\chi^2}{2})$ is to 1, the better the
model. However, values $\Gamma_\mathrm{q}(\frac{\mu}{2},\frac{\chi^2}{2})\approx
1$ are suspicious and may mean that the errors have been overestimated. A
reasonable fit is thought to have $\chi^2\approx\mu$.
\section{Calculations and results}
\label{sec-calc}
\subsection{Phase diagrams}
\label{sec-phasediag}

\begin{figure}
  \centering
  \includegraphics[width=3in]{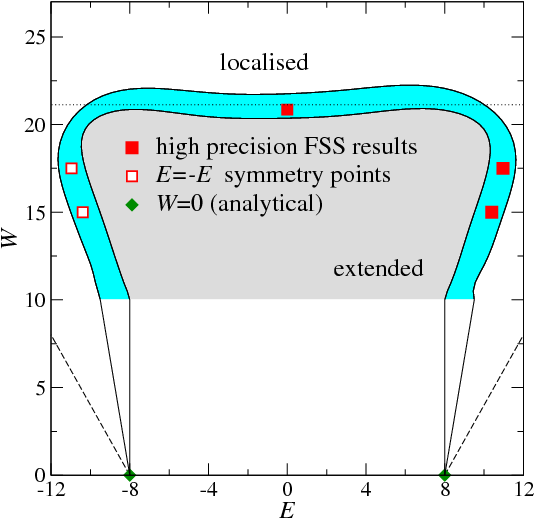}
  \caption{
    Phase diagram for the BCC lattice. The cyan region represents the
approximate location of the phase boundary. Its edges (the solid black lines)
were
    determined by comparing localisation lengths with errors $\le 10 \%$ for
system sizes $M=7$ and $M=9$ in the $(E,W$) plane. The solid red squares
({\color{Red} $\sqbullet$}) are points calculated by performing high-precision
FSS on localisation data with an error ${\le 0.1 \%}$. The hollow squares
({\color{Red} $\square$}) are reflections of the solid squares in the $E=0$
axis. The diamonds ({\color{OliveGreen} $\blackdiamond$}) denote the band edges
at $W=0$. They have been joined to the phase boundary edges calculated for
higher disorders as a guide to the eye. The dashed lines are the theoretical
band edges $\pm (Z + W/2)$, where $Z$ \newnot{sym:Z} is the coordination number.
The horizontal
dotted line is the BCC estimate $21.13$ for $W_{\rm c}$ of Ref.\ \cite{KotS83}.
The light grey, shaded area in the centre contains extended states; states
outside the phase boundary are localised. Error bars are within symbol size for
{\color{Red} $\sqbullet$}, {\color{Red} $\square$}. }

\label{fig-bccphasediag}
\end{figure}

\begin{figure}
  \centering
  \includegraphics[width=3in]{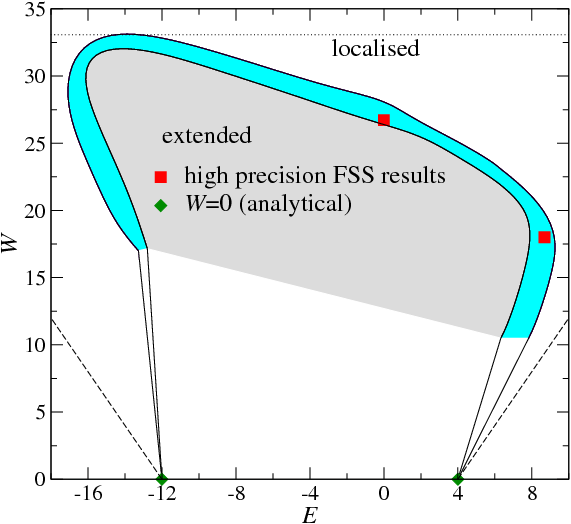}
  \caption{
    Phase diagram for the FCC lattice. Symbols, lines and shaded areas have the
same meaning as in Fig.\ \ref{fig-bccphasediag} with the diamonds
({\color{OliveGreen} $\blackdiamond$}) representing the band edges $-12$ and $4$
at zero disorder \cite{AshM76} and the dotted line corresponding to the
self-consistent estimate $W_\mathrm{c}=33.08$ \cite{KotS83}. Error bars are
within symbol size. }

\label{fig-fccphasediag}
\end{figure}

Fig.\ \ref{fig-bccphasediag}
and Fig.\ \ref{fig-fccphasediag}
show the phase diagrams for the BCC and FCC lattices, respectively. Originally a
grid of $W$ versus $E$ values was created with separation $\Delta E=\Delta W
=0.5t$. At each point the nature of the electronic wave function was determined
by comparing the reduced  localisation lengths, $\Lambda_M$, calculated for
system sizes $M=7$ and $M=9$ with error $\le 10 \%$. If $\Lambda_9 > (<) \Lambda_7$
at the same values of $E$ and $W$ then the point $(E,W)$ in the phase
diagram is identified as extended (localised). The edges of the phase boundary
were obtained by averaging separately over the three extended and localised
points $(E,W)$ nearest to the boundary and then connecting such averages using a
spline fit. Data points are not obtained for lower disorder values, as the
fluctuations in the LEs, due to the small system sizes, become too big; higher
values of disorder smooth out these fluctuations.

A striking difference between the phase diagrams is that for the BCC lattice the phase boundary is symmetric about the line $E=0$, whereas for the FCC lattice it is not. This is due to the bipartiteness of the BCC lattice which consists of two SC sublattices, one displaced half the distance along a body diagonal of the other. Hence for any site in one sublattice, its nearest neighbours are in the other sublattice. Such connections result in states coupled by a bipartite symmetry transformation --- which is exact for the case of no diagonal disorder --- with eigenenergies of the same magnitude but opposite sign having approximately the same localisation lengths; this produces a symmetric phase diagram. The FCC lattice is non-bipartite, so such a symmetry in its phase diagram is not observed.

In Figs.\ \ref{fig-bccphasediag} and \ref{fig-fccphasediag}, we also indicate via horizontal lines previous results for critical disorder strengths \cite{KotS83}, which were based on the self-consistent theory of localisation and obtained for a momentum cut-off of $2 p_{\rm F}$, where $p_{\rm F}$ is the Fermi momentum. These estimates agree well with our results for both BCC and FCC lattices. The SC result $W_{\rm c} = 13.91$ of Ref.\ \cite{KotS83}, however, deviates more strongly from recent $W_{\rm c} = 16.54 \pm 0.02$ \cite{SleO99a,MilRSU00,OhtSK99} estimates.

\subsection{Critical parameters at $E=0$}
\label{sec-critdis}

The TMM calculations were performed for system sizes up to $M = 15$. In order to
examine the localisation properties at the band centre for the BCC lattice and
the barycentre (average energy eigenvalue) for the FCC lattice, we set $E=0$ in
Eq.\ (\ref{eq-3d}). A value of the critical disorder $W_{\rm c}$ was
approximated using the phase diagrams described above and then localisation
lengths $\lambda$ were calculated for a range of $W$ close to this approximate
value with accuracy ranging from $0.1 \%$ for small system sizes $M$ to about
$0.14 \%$ for the largest. Note that the term \emph{critical disorder} is used
to indicate that there are no further extended states at $E=0$ for disorders
$W>W_{\rm c}$; extended states may still exist for $W>W_{\rm c}$ at other
energies $E$, as shown in Fig.\ \ref{fig-fccphasediag}.

\begin{figure}
  \centering
  \includegraphics[width=3in]{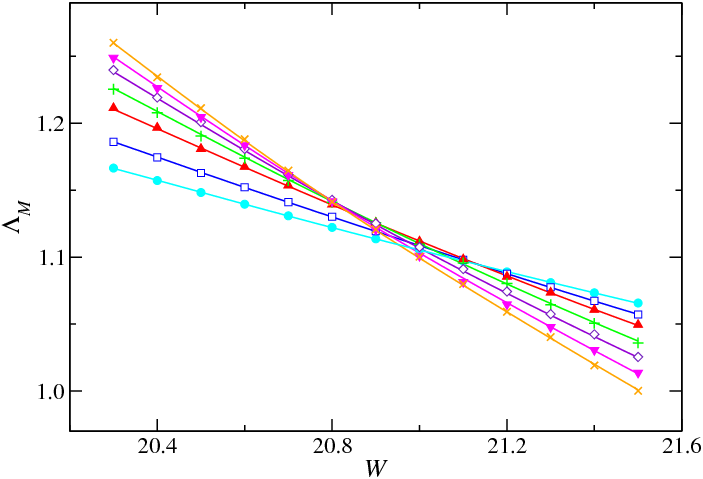}
  \caption{
    Reduced localisation lengths $\Lambda_M$ versus disorder $W$ for BCC lattice.
    System sizes $M$ are $3 ({\color{SkyBlue} \bullet}), 5 ({\color{Blue} \square}), \ldots, 15 ({\color{YellowOrange} \times})$.
    Error bars are within symbol size.
    Lines are fits to the data given by Eqs.\ (\ref{eq-Slevin}) -- (\ref{eq-Slevin-Var})
    with $n_{\rm r}=3, n_{\rm i}=2, m_{\rm r}=3, m_{\rm i}=1$. }
\label{fig-bcc-data}
\end{figure}

\begin{figure}
  \centering
  \includegraphics[width=3in]{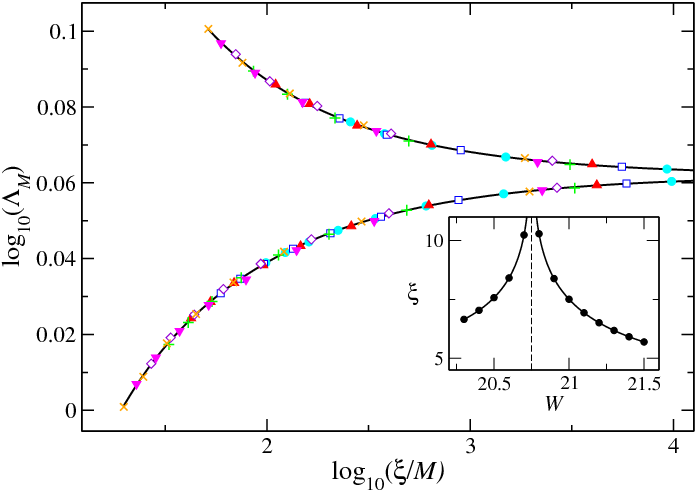}
  \caption{
    Scaling function (solid line) and scaled data points for the BCC lattice and
    $n_{\rm r}=3, n_{\rm i}=2, m_{\rm r}=3, m_{\rm i}=1$. Symbols denote the same values of $M$ as in Fig.\ \ref{fig-bcc-data}. Inset: Dependence of the scaling parameter $\xi$ on the
    disorder strength $W$ for the $13$ $W$ values shown in Fig.\
\ref{fig-bcc-data}. The dashed vertical line indicates the divergence of $\xi$
at $W=20.75$. In all cases error bars are within symbol size. }
\label{fig-bcc-scal}
\end{figure}

The reduced localisation lengths for the BCC lattice are displayed in Fig.\ \ref{fig-bcc-data}.
Since $\Lambda_M$ increases (decreases) as $M$ increases for extended (localised) states, the $\Lambda_M$ curves for different system sizes should cross at $W_\mathrm{c}$. Note how the crossing point of the curves in Fig.\ \ref{fig-bcc-data} shifts with changing $M$. In most cases this indicates the need for an irrelevant scaling variable introduced via non zero values of $n_{\rm i}$ and $m_{\rm i}$ in Eqs.\ (\ref{eq-Slevin2}) and  (\ref{eq-Slevin-Var}). Fig.\ \ref{fig-bcc-scal}
shows the results of the scaling procedure for $n_{\rm r}=3, n_{\rm i}=2, m_{\rm r}=3, m_{\rm i}=1$. The scaling curve exhibits localised (bottom) and extended (top) branches as expected for the MIT. Divergence of the scaling
parameter $\xi$ at ${W \approx 20.75}$ indicates the critical value of the disorder.
Table \ref{tab:bcc}(a)
\begin{table}
  \centering
\caption{Critical parameters for the MIT in the BCC lattice. All errors quoted are standard errors.
(a) 3 examples of FSS results with varying $n_{\rm r}, n_{\rm i}, m_{\rm r}, m_{\rm i}$ at fixed energy $E=0$. We use $91$ data points, equally spaced in the indicated intervals (cp.\ Fig.\ \ref{fig-bcc-data}), for each set of $n_{\rm r}, n_{\rm i}, m_{\rm r}, m_{\rm i}$. Varying $n_{\rm r}, n_{\rm i}, m_{\rm r}, m_{\rm i}$, we obtain $41$ best fit models in order to produce the indicated averages.
(b) Similar FSS results obtained for $3$ out of $14$ best fit models from $82$ non-equally spaced data points at fixed $W=15$ for the indicated energy intervals. Some of the best fit models use irrelevant scaling, although the examples explicitly detailed do not, so that $m_i=n_i=0$ and the parameter $y$ is not used.
(c) Results at $W=17.5$ (cp.\ Fig.\ \ref{fig-bcc-E-scal}) for $3$ out of $8$ best fit models with $108$ non-equally spaced data points used in each FSS procedure. All the best fit models use no irrelevant scaling.
The numerical fitting procedure continued in all cases until convergence was reached or (a,b) $5000$, (c) $1000$ iterations had been completed. When averaging, non-converged results were neglected.}

%
\flushleft{(a)}\\[2ex]
\begin{tabular*}{\hsize}{@{\extracolsep{\fill}}cccccccccc} \hline\hline
    $\Delta M$  & $E$  & $\Delta W$& $n_{\rm r}$ & $n_{\rm i}$ & $m_{\rm r}$ & $m_{\rm i}$  & $W_{\rm c}$ & $\nu$ & $| y |$ \\ \hline
    3 - 15 & 0 & 20.3 - 21.5 & 2 & 0 & 1 & 0 & 20.95(1) & 1.67(5) & - \\
    3 - 15 & 0 & 20.3 - 21.5 & 3 & 1 & 1 & 4 & 20.92(2) & 1.51(9) & 1.7(5) \\
    3 - 15 & 0 & 20.3 - 21.5 & 3 & 2 & 3 & 1 & 20.75(3) & 1.70(9) & 3.0(5) \\
\vdots & \vdots & \vdots & \vdots & \vdots & \vdots & \vdots & \vdots & \vdots & \vdots \\ \hline
\multicolumn{7}{l}{averages:} & 20.81(1) & 1.60(2) & \\ \hline
\end{tabular*}
\\[3ex]
\flushleft{(b)}\\[2ex]
\begin{tabular*}{\hsize}{@{\extracolsep{\fill}}ccccccc}
\hline\hline
    $\Delta M$  & $\Delta E$& $W$  & $n_{\rm r}$  & $m_{\rm r}$  & $E_{\rm c}$ & $\nu$ \\ \hline
9 - 13 & 9.9 - 10.9 & 15 & 2 & 1 & 10.38(1) & 1.32(5) \\
9 - 13 & 9.9 - 10.9 & 15 & 2 & 2 & 10.38(1) & 1.22(5) \\
9 - 13 & 9.9 - 10.9 & 15 & 3 & 4 & 10.40(1) & 1.03(3) \\
\vdots & \vdots & \vdots & \vdots & \vdots & \vdots & \vdots \\ \hline
\multicolumn{5}{l}{averages:} & 10.39(1) & 1.21(2) \\ \hline
%
\end{tabular*}
\\[3ex]
\flushleft{(c)}\\[2ex]
\begin{tabular*}{\hsize}{@{\extracolsep{\fill}}ccccccc}
\hline\hline
    $\Delta M$  & $\Delta E$& $W$  & $n_{\rm r}$  & $m_{\rm r}$ & $E_{\rm c}$ & $\nu$ \\ \hline
7 - 15 & 10.5 - 11.5 & 17.5 & 2 & 1 & 10.98(1) & 1.55(6)\\
7 - 15 & 10.5 - 11.5 & 17.5 & 3 & 2 & 10.99(1) & 1.48(6)\\
7 - 15 & 10.5 - 11.5 & 17.5 & 3 & 4 & 10.99(1) & 1.36(7)\\
\vdots & \vdots & \vdots & \vdots & \vdots & \vdots & \vdots \\ \hline
\multicolumn{5}{l}{averages:} & 10.99(1) & 1.45(3) \\ \hline \hline
\end{tabular*}
\label{tab:bcc}
\end{table}
gives some examples of models providing the best fits and the resulting critical parameters. The values of the critical disorder and critical exponent are obtained by averaging over all the best fit models.

Let me emphasise that given the precision of our data, simply presenting the data with minimal $\chi^2$ value for a given set of $n_{\rm r}, m_{\rm r}, n_{\rm i}, m_{\rm i}$, systematically underestimates the true error. The reason is twofold. Firstly, there is no {\em a priori} justification for which set of $n_{\rm r}, m_{\rm r}, n_{\rm i}, m_{\rm i}$ values to use. Also the variation in results for different $n_{\rm r}, m_{\rm r}, n_{\rm i}, m_{\rm i}$ is typically larger than the error bars suggested for each individual set. Secondly, one can change the range of system sizes and disorder/energy values for the localisation data to be included in the fit. This again leads to changes in the critical parameters which are usually beyond the error bars generated in each individual $n_{\rm r}, m_{\rm r}, n_{\rm i}, m_{\rm i}$ fit. Hence, in the absence of a clear criterion for which of these fits to choose, the strategy is to (i) delete all obviously erroneous fits, i.e.\ those which do not converge or which converge to unphysical values; (ii) to average over the remaining results with a proper estimation of accumulated error, based on the individual errors for each $n_{\rm r}, m_{\rm r}, n_{\rm i}, m_{\rm i}$ choice. Last, all this should be done while keeping the number of parameters --- as determined by $n_{\rm r}, m_{\rm r}, n_{\rm i}, m_{\rm i}$ --- as small as possible. The accumulated standard error \cite{NRC02} is given by
\begin{equation}
\sigma=\frac{1}{n}\left(\sum_{i=1}^n \sigma_i^2 \right)^{\frac{1}{2}},
 \label{eq-accerr}
\end{equation}
where $n$ is the number of models being averaged over and $\sigma_i$ is the standard deviation for the $i^\mathrm{th}$ model.
Note that in ultrahigh precision studies \cite{RodVSR10}, the issues mentioned above are less of a problem, so such an averaging procedure is not used.

\begin{figure}
  \centering
  \includegraphics[width=3in]{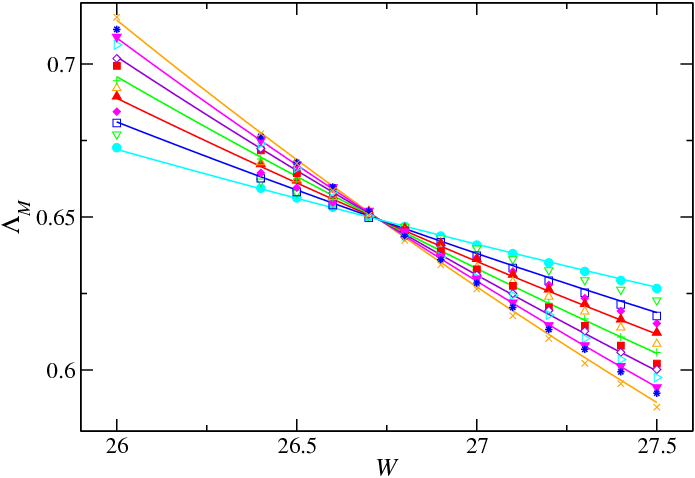}
  \caption{
   Reduced localisation lengths $\Lambda_M$ versus disorder $W$ for FCC lattice.
   Symbol sizes $M$ are $3 ({\color{SkyBlue} \bullet}), 4 ({\color{LimeGreen} \triangledown}), \ldots, 15 ({\color{YellowOrange} \times})$. Error bars are within symbol size. Lines are fits to the data given by Eqs.\ (\ref{eq-Slevin}) -- (\ref{eq-Slevin-Var})
   with $n_{\rm r}=2, m_{\rm r}=2$. Lines for even $M$ have been removed for clarity. }
\label{fig-fcc-data}
\end{figure}

\begin{figure}
  \centering
  \includegraphics[width=3in]{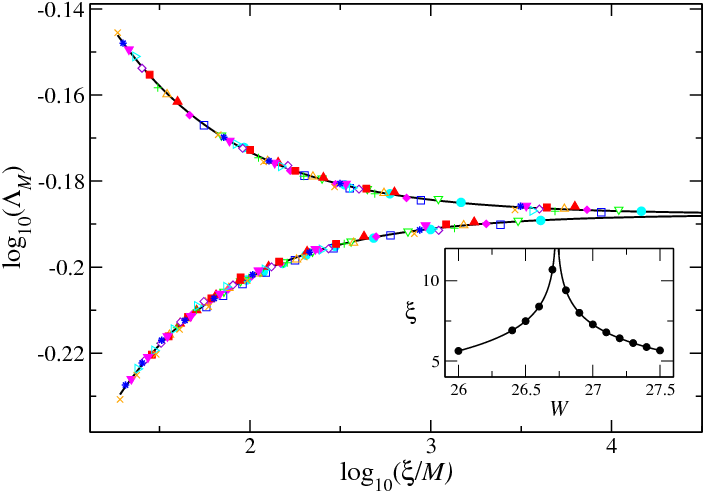}
  \caption{
   Scaling function (solid line) and scaled data points for the FCC lattice and
   $n_{\rm r}=2, m_{\rm r}=2$. Symbols denote the same values of $M$ as in Fig.\ \ref{fig-fcc-data}. Inset: Dependence of the scaling parameter $\xi$ on the
    disorder strength $W$ for the $13$ $W$ values shown in Fig.\ \ref{fig-fcc-data}. In all cases error bars are within symbol size. }
\label{fig-fcc-scal}
\end{figure}

Results of the TMM calculations for the FCC lattice are shown in Fig.\ \ref{fig-fcc-data}.
In this case the lines for constant $M$ cross at the same point --- at least within the accuracy of the calculated $\Lambda_M$ --- indicating that the use of the irrelevant variables in Eqs.\ (\ref{eq-Slevin2}) and  (\ref{eq-Slevin-Var}) is not necessary in most cases and $n_{\rm i}=m_{\rm i}=0$. Results of the fit for $n_{\rm r}=2, m_{\rm r}=2$ are displayed in Fig.\ \ref{fig-fcc-scal}.
The transition at $W \approx 26.73$ is clearly indicated. More examples of best fit models can be found in Table \ref{tab:fccdis} as well as the average values of the critical parameters.
\begin{table}
  \centering
\caption{Critical parameters for the MIT in the FCC lattice at $E=0$ (cp.\ Fig.\ \ref{fig-fcc-data}). We use $91$ data points for each FSS and the obtained $31$ best fit models average as indicated. The fitting procedure was continued until convergence was reached or until $5000$ iterations had been completed, although only models for which convergence was reached were included in the averaging process. No irrelevant scaling was necessary, so $n_{\rm i}=m_{\rm i}=0$.}

  \begin{tabular*}{\hsize}{@{\extracolsep{\fill}}ccccccc} \hline\hline
    $\Delta M$  & $E$  & $\Delta W$& $n_{\rm r}$ & $m_{\rm r}$ & $W_{\rm c}$ & $\nu$  \\ \hline
    3 - 15 & 0 & 26 - 27.5 & 1 & 2 & 26.73(1) & 1.58(2) \\
    3 - 15 & 0 & 26 - 27.5 & 2 & 2 & 26.73(1) & 1.58(3) \\
    3 - 15 & 0 & 26 - 27.5 & 3 & 3 & 26.73(1) & 1.67(5) \\
\vdots & \vdots & \vdots & \vdots & \vdots & \vdots & \vdots \\ \hline
\multicolumn{5}{l}{averages:} & 26.73(1) & 1.60(1) \\ \hline
\end{tabular*}
\label{tab:fccdis}
\end{table}


\subsection{Critical parameters away from the band centre}
\label{sec-criteng}

\begin{figure}
  \centering
  \includegraphics[width=4.5in]{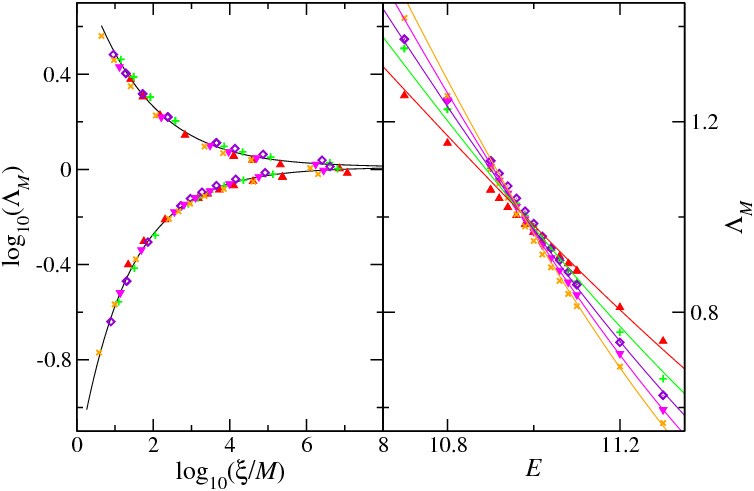}
  \caption{
Localisation data for the BCC lattice with $W=17.5$. System sizes $M$ are $7 ({\color{Red} \blacktriangle}), 9 ({\color{YellowGreen} +}), \ldots, 15 ({\color{YellowOrange} \times})$ as in Fig.\ \ref{fig-bcc-data}. Error bars are within symbol size. Left: Scaling function (solid line) and scaled data points using
   $n_{\rm r}=2, m_{\rm r}=1$. Right: Reduced localisation lengths $\Lambda_M$ versus energy $E$. Lines are fits to the data given by Eqs.\ (\ref{eq-Slevin}) -- (\ref{eq-Slevin-Var})
    with $n_{\rm r}=2, m_{\rm r}=1$.  }
\label{fig-bcc-E-scal}
\end{figure}

We also perform calculations where we fix the disorder and allow the energy to vary across a critical value $E_{\rm c}$ for the transition. It is known that such investigations are numerically more difficult due to the influence of density of states effects \cite{CaiRS99}. Results of the TMM and FSS calculations for the BCC lattice with $W = 17.5$ can be seen in Fig.\ \ref{fig-bcc-E-scal}.
The poorer quality of the fit compared to the calculations where the energy was fixed at zero is evident. This is not due to using lower accuracy data, as the maximum raw-data error remained at $0.1 \%$. Hence we attribute it to complications arising from a varying density of states close to $E_{\rm c}$ at the attainable values of $M$. Results for the critical parameters are shown in Table \ref{tab:bcc} (b) and (c) for $W=15$ and $W=17.5$ respectively. The low value of $\nu$ for the case $W=15$ can be attributed to the use of fewer data points in the FSS and only using three values of $M$. We note that this is consistent with the lower values of $\nu$ obtained in the diagonalisation studies \cite{ZhaK95c} as mentioned in the Section \ref{sec-charMIT}. It appears that the FSS procedure systematically reduces the values of the critical exponent for data from smaller systems or of lower accuracy.

\begin{figure}
  \centering
  \includegraphics[width=4.5in]{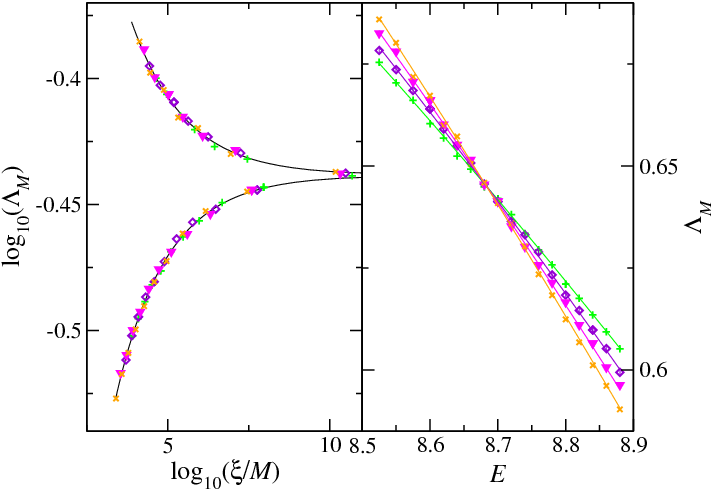}
  \caption{
Data for the FCC lattice with $W=18$. System sizes $M$ are $9 ({\color{YellowGreen} +}), 11 ({\color{Purple}\Diamond}), \ldots, 15 ({\color{YellowOrange} \times})$ as in Fig.\ \ref{fig-fcc-scal}. Error bars are within symbol size. Left: Scaling function (solid line) and scaled data points using
   $n_{\rm r}=1, m_{\rm r}=2$. Right: Reduced localisation lengths $\Lambda_M$ versus energy $E$. Lines are fits to the data given by Eqs.\ (\ref{eq-Slevin}) -- (\ref{eq-Slevin-Var})
    with $n_{\rm r}=1, m_{\rm r}=2$.  }
\label{fig-fcc-E-scal}
\end{figure}

Results for the TMM and FSS calculations for the FCC lattice with $W=18$ can be seen in Fig.\ \ref{fig-fcc-E-scal}.
Table \ref{tab:fcceng} gives examples of the best fit models and shows the resulting average critical parameters. 
\begin{table*}
\centering
\caption{Critical parameters for the MIT in the FCC lattice at $W=18$ (cp.\ Fig.\ \ref{fig-fcc-E-scal}). We use $83$ data points to perform the FSS and use $8$ best fit models to obtain the averages. None of the best fit models use irrelevant scaling. The fitting procedure was continued until convergence was reached or until $1000$ iterations had been completed; only models for which the method converged were used to obtain the averages. $\mu$ is used to denote the number of degrees of freedom.}
   \begin{tabular*}{\hsize}{@{\extracolsep{\fill}}cccccccccc} \hline\hline
    $\Delta M$  & $\Delta E$  & $W$& $n_{\rm r}$ & $m_{\rm r}$ & $E_{\rm c}$ & $\nu$ & $\chi^2$ & $\Gamma_\mathrm{q} $ & $\mu$  \\ \hline
9 - 15 & 8.52 - 8.88 & 18 & 1 & 2  & 8.683(3) & 1.63(5) & 79.0 & 0.44 & 78\\
9 - 15 & 8.52 - 8.88 & 18 & 2 & 2  & 8.687(4) & 1.65(5) & 78.0 & 0.44 & 77\\
9 - 15 & 8.52 - 8.88 & 18 & 3 & 1  & 8.685(3) & 1.62(5) & 77.6 & 0.46 & 77\\
\vdots & \vdots & \vdots & \vdots & \vdots & \vdots & \vdots & \vdots & \vdots & \vdots \\ \hline
\multicolumn{5}{l}{averages:} & 8.684(2) & 1.63(2) & \\ \hline \hline
\end{tabular*}
\label{tab:fcceng}
\end{table*}
Note that both estimates of $\nu$ in Table \ref{tab:fccdis} and Table \ref{tab:fcceng} are consistent with the result $1.57(2)$ for the SC lattice \cite{SleO99a}.
The TMM data for this system has the highest quality out of all the data presented in this study. Thus for this case only, we include the $\chi^2$ values and $\Gamma_\mathrm{q}\left( \mu/2,\chi^2/2\right) $ values to demonstrate the goodness of fit. For all the models $\chi^2 \approx \mu$, which indicates a good fit. 
%
We remark that the results presented in Tables \ref{tab:bcc}, \ref{tab:fccdis} have lower precision, but this is clear since the accuracy of the individual data points is less, due to smaller system sizes and fewer available data points very close to the transition.

\section{Conclusions}
\label{sec-concl}
The transfer-matrix method and FSS were used to determine the critical parameters of the Anderson transition for the BCC and FCC lattices. The values of the critical exponent $\nu$ are in good agreement with the results obtained previously for other systems belonging to the Gaussian orthogonal universality class. The increase of the critical disorder $W_{\rm c}$ from $16.54$ for the SC lattice to $20.81$ for BCC and $26.73$ for FCC lattice may be attributed to an increasing number of nearest neighbours which for the above structures equals $6$, $8$ and $12$, respectively. More nearest neighbours connected to a given site provide more paths for electronic transport, so stronger disorder is needed to localise eigenstates of the system. The universal localisation properties of a 3D system and the presence of an MIT are however not affected in accordance to the scaling theory of localisation \cite{AbrALR79} and are in agreement with results \cite{SchO92} showing that they depend only on the dimensionality of the system, but not on the number of nearest neighbours in the lattice.

Our results and their interpretation are consistent with investigations of classical bond and site percolation models on SC, BCC and FCC lattices. In Ref.\ \cite{GalM96} it was found that the percolation thresholds for these lattices decrease with increasing number of nearest neighbours; more neighbours allow for easier formation of a percolating cluster, or, as in our case, the formation of extended states.
\cleardoublepage

\cleardoublepage
\chapter{Excitonic Aharonov-Bohm effect with external fields}
\label{chap-abe}

\section{Introduction}
\label{sec-abintro}
The manipulation of light and excitons is an area which has sparked much recent
interest \cite{WinHB07,LunSLP99,But04,MitKM00,HauHDB99}. The speed of light has
been slowed down to an amazing $17 \mathrm{ms}^{-1}$ in an ultracold gas of
sodium atoms \cite{HauHDB99} by the use of electromagnetically induced
transparency techniques \cite{HarFI90}. Bose-Einstein condensation of an
excitonic gas is a phenomenon considered theoretically a long time ago
\cite{KelK64}. However, it is only recently that the ability to grow high
quality layered semiconductor structures has allowed for a successful
experimental investigation of excitonic superfluids. Spatial patterns in
photoluminescence measurements of an exciton gas in GaAs/AlGaAs coupled quantum
wells have been observed \cite{ButGC02} and are thought to be a signature of
quantum degeneracy \cite{LevSB05}. Zero Hall voltages measured in bilayer
quantum well systems for particular layer separations and magnitudes of a
magnetic field applied perpendicular to the layers, are also understood to
indicate the presence of an excitonic condensate \cite{KelEPW04}. There has also
been experimental evidence for the formation of a polariton BEC \cite{KasRKB06} and a room temperature polariton laser
\cite{ChrBGL07}. 

 Such discoveries are very important for quantum computing applications \cite{DAmDBP02}. In current computers, electrons are used for information processing and photons for communication. Conversion between the two media places limitations upon the machine's efficiency. The creation of an exciton-based integrated circuit \cite{HigNBH08} could provide an exciting solution to this problem. The possibility of using excitons for data storage has also been investigated using indirect excitons in self-assembled quantum dots \cite{LunSLP99} and coupled quantum wells \cite{WinHB07}. Here the electron and hole are separated using a gate voltage, significantly prolonging their lifetimes.

In this Chapter we study the Aharonov-Bohm effect (ABE) for an exciton in a
nanoring with an in-plane electric field. Oscillations in the physical
properties of the exciton as a function of magnetic field are signatures of the
quantum mechanical interference between the electron and hole wave functions. A
mechanism for controlling the excitonic lifetime by tuning the electric and
magnetic fields is discovered. The excitonic Aharonov-Bohm effect (XABE) is
particularly interesting; as a neutral particle, one would not expect an exciton
to couple to a magnetic vector potential. However, an exciton's finite size and
the internal motion of the electron and hole mean that the XABE is possible. AB
oscillations have been predicted for optically created excitons in the ring
geometry by a variety of different models. Initial treatments used a simple 1D
continuous model that could be solved analytically with a perpendicular
solenoidal magnetic field through the centre of the ring and a contact
electron-hole interaction potential \cite{Cha95,RomR00,MouC04}. AB oscillations
were found both in the excitonic ground state (GS) and in the corresponding
oscillator strength (OS) for rings with diameters smaller than or comparable to
the excitonic Bohr radius, $a_B=\hbar^2\epsilon/\mu e^2$ ($\epsilon$ is
dielectric constant and $\mu$ reduced mass) \cite{RomR00}. 

Other models tried to capture the essence of what can currently be achieved
experimentally. An important property that has been widely treated is the finite
ring width
\cite{HuZLX01,SzaAB02,UllGKWK02,PalDER05,daSUS05,GroGZ06,DaiZ07,GroZ07}. In a 2D
Hubbard model approach with a short range electron-hole interaction, excitonic
AB oscillations in the GS energy were seen to survive for finite
ring widths \cite{PalDER05}. In contrast, a continuous approach introducing the
ring geometry by a confining potential and using a screened Coulombic
interaction found no GS AB oscillations for any finite ring widths
\cite{SonU01,HuZLX01}. However, oscillations in other low lying energy states
were detected \cite{HuZLX01} for rings of small enough diameter and width.
Eccentric ring geometries have also been examined for different confining
potentials \cite{GroZ07}. Although the eccentricity reduces the amplitude of
oscillations, it also causes previously dark states to become optically active,
so a slightly eccentric ring could be a good candidate for the experimental
observation of the XABE. Another approach thought to enhance the effect is the
polarisation of the exciton \cite{GovUKW02,daSUS05,GroGZ06}. This occurs
naturally for particular choices of semiconductor materials. For example, in
self assembled InAs rings on a GaAs substrate, the electron is most likely to be
found towards the centre of the ring, whereas the hole is largely confined to
the outer radius of the ring \cite{LorLGK00}. 

The role of the electron-hole interaction has also been investigated
\cite{MouC04,daSUS05}; the oscillation amplitude was found to decrease as the
Coulombic interaction becomes less shielded, in agreement with intuition. 
The effects of impurities \cite{daSUG04,daSUS05} and more general disorder
\cite{MasMTK01,MeiTK01} have also been studied. The introduction of an isolated
impurity potential breaks the rotational symmetry of the ring, resulting in the
coupling of excitonic states with different angular momenta and anticrossings
appearing in the energy levels. The absorption coefficient is also modulated,
since scattering by impurities causes optical emission from previously optically
inactive states. The introduction of disorder to the onsite potential energies
has been seen to suppress the ABE for certain interaction strengths, when many
realisations of disorder are averaged over \cite{MeiTK01}. However, the shift in
linear absorption peak as a function of magnetic flux is dependent on the
realisation of disorder. For certain realisations and interaction strengths, the
disorder actually enhances the effect. In view of studies conducted on
attractive two particle systems in a disordered ring, which show a non monotonic
dependence of the localisation length upon the interaction strength
\cite{RomP95}, it is probable that disorder enhances the ABE for particular
interaction strengths even when averaging over disorder realisations is
undertaken.

It is only very recently (after the publication \cite{FisCPR09} of the work described in this
Chapter) that the XABE has been observed experimentally for the ring geometry
\cite{TeoCLM10,DinALP10}. In Ref. \cite{TeoCLM10} the energy peak in the
photoluminescence spectrum of an ensemble of type-I InAs/GaAs quantum rings was
seen to oscillate as a function of magnetic field. In Ref.\ \cite{DinALP10}
oscillations in the photoluminescence peak of a single InAs/GaAs quantum ring
were observed. It should be noted that such experiments do not strictly meet the
criteria set out in Section \ref{sec-ABE} for the observation of the ABE, since
the magnetic field penetrates the ring and so the particles are not confined to
a magnetic field free region. However, the observed effect can be modelled as if
the magnetic field was concentrated in the middle of the ring, so is considered
to be an ABE. Prior to this, AB oscillations were observed for negatively charged
excitons or trions (a bound structure of two electrons and a hole)
\cite{BayKHG03}. AB oscillations were also seen for excitons confined to type-II
quantum dots \cite{RibGCM04,SelWKG08}. In this case, only one of the particles
(either the electron or the hole depending on the material) exists in a ring
geometry created by the circumference of the quantum dot. The other is localised
in the centre of the quantum dot. Hence these AB oscillations are really due to
the motion of a single particle.

With quite a large body of theoretical work already existing on the XABE, the
aim of the project described below was to see how the application of an in-plane
electric field might affect the amplitude of AB oscillations in the exciton's GS
energy and OS. It was previously suggested that the electric field would enhance
the XABE \cite{MasC03}; our results suggest this is only partially true.
Interestingly they indicate the vanishing of the OS for particular values of
electric and magnetic field, pointing to the possibility of trapping light. This
result may lead to applications in data storage.
\section{Theoretical approach}
\label{sec-theorapp}
\subsection{Single particle problem}
\label{subsec-spp}
\begin{figure}
  \centering
  \includegraphics[width=2in]{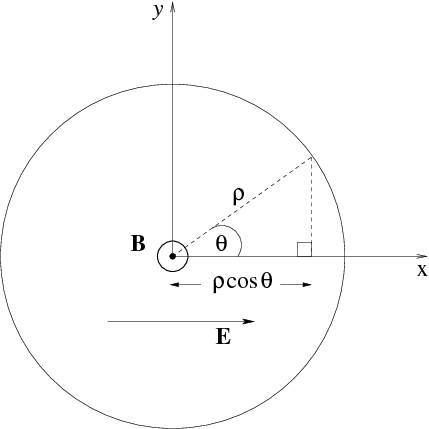}
  \caption {The geometry of a ring in the presence of electric and magnetic
fields.}
\label{fig-ringgeom}
\end{figure}

First consider the case for a single charged particle, since it is the wave
functions of the electron and hole, which provide a basis for the desired
excitonic wave function.  The standard Hamiltonian for a particle with charge
$\mathrm{q}$ and mass $m_0$ \newnot{sym:m0} in the presence of an electric potential $\phi$ \newnot{sym:phi} and a magnetic
vector potential $\bm{\mathrm{A}}$ is
\begin{equation}
 \label{hamil-em}
\hat{H}=\frac{\left( -i\hbar\nabla-\frac{\mathrm{q}}{c}\bm{\mathrm{A}}\right)
^2}{2m_0}+\mathrm{q}\phi.
\end{equation}
The ring geometry is illustrated in Fig.\ \ref{fig-ringgeom}.
The ring is modelled as a 1D circle of radius \newnot{sym:rho} $\rho$. It is threaded by a
magnetic flux perpendicular to the plane of the ring and is in the presence of a
constant in-plane electric field in the $x-$direction of magnitude \newnot{sym:U0} $U_0$. Let
$\theta_e$ \newnot{sym:thetae} and $\psi^{\left( e\right) }$ \newnot{sym:psie} denote the azimuthal coordinate and
wave function of the electron respectively. Then using the Hamiltonian in Eq.\
(\ref{hamil-em}) with $\mathrm{q}=-e$, the Schr\"odinger equation for the electron may be
written in terms of dimensionless quantities as
\begin{equation}
 \label{schro-elec}
\left(\frac{d^2}{d{\theta_e}^2}-2i\Phi\frac{d}{d\theta_e}-\Phi^2+u_0\mathrm{cos}
\left( \theta_e\right) +\lambda^{\left(e\right) }\right) \psi^{\left(
e\right)}\left( \theta_e\right) =0, 
\end{equation}
where \newnot{sym:eps0} $\epsilon_0=\frac{\hbar^2}{2m_e\rho^2}$, \newnot{sym:Phi2} $\Phi=-\frac{{\rho}Ae}{\hbar c}$
(number of flux quanta), \newnot{sym:u0} $u_0=\frac{e{\rho}U_0}{\epsilon_0 c}$ and \newnot{sym:lambdae}
$\lambda^{\left( e\right)} =\frac{E_e}{\epsilon_0}$, where $E_e$ \newnot{sym:Ee} is the energy
of the electron. In the absence of the electric field ($u_0=0$), Eq.\
(\ref{schro-elec}) has translational symmetry and the solutions are plane waves
\cite{RomR00}:
\begin{equation}
\psi^{\left( e\right) }_N\left( \theta_e\right)=\frac{1}{\sqrt{2\pi}}e^{iN
\theta_e}, \hspace{10mm} \lambda^{\left( e\right)}_N=(N-\Phi)^2.
\label{eq-spnoefield}
\end{equation}
For the \newnot{sym:psih} \newnot{sym:thetah} \newnot{sym:lambdah} hole:
\begin{equation}
\psi^{\left( h\right) }_{N'} \left( \theta_h\right)=\frac{1}{\sqrt{2\pi}}e^{iN'
\theta_h}, \hspace{10mm} \lambda^{\left( h\right)}_{N'}=(N'+\Phi)^2,
\label{eq-spnoefieldhole}
\end{equation}
where $N,N'$ are integers and I have assumed the electron and hole masses are the same.

The electric field breaks this symmetry and we use a Fourier Ansatz for the
electronic wave function
\begin{equation}
 \label{fou-ans}
\psi^{\left( e\right)}\left(
\theta_e\right)=\frac{1}{\sqrt{2\pi}}\sum_{m=-\infty}^{\infty}c_{m}^{\left(
e\right) }e^{im{\theta}_e}.
\end{equation}
By substituting this into Eq.\ (\ref{schro-elec}) and using the orthogonality of
the complex exponential functions, one may show that for any integer $m$,
\begin{equation}
 \label{fou-coeffs}
\lambda^{\left( e\right)}c_{m}^{\left( e\right)}={\left(
m-\Phi\right)}^2c_{m}^{\left( e\right)}-\frac{u_0}{2}c_{m-1}^{\left(
e\right)}-\frac{u_0}{2}c_{m+1}^{\left(e\right)}.
\end{equation}
Eq.\ (\ref{fou-coeffs}) can be expressed as an eigenvalue problem:
\begin{equation}
M{\bm{\mathrm{c}}}^{\left( e\right)} 
=\lambda^{\left( e\right)}\bm{\mathrm{c}}^{\left( e\right)},
\label{eq-evp}
\end{equation}
where
\begin{equation}
 \label{eigval-elec}
M=\left( \begin{array}{ccccccccc}
\ddots & & & &  \\
 -\frac{u_0}{2} & \left(-m -\Phi\right) ^2 & -\frac{u_0}{2} & & & & & & \\ 
 & & \ddots & & & & & & \\
 & & -\frac{u_0}{2} & \left(-1 -\Phi\right) ^2 & -\frac{u_0}{2} & & & & \\ 
 & & & -\frac{u_0}{2} &\left(-\Phi\right) ^2 & -\frac{u_0}{2} & & &  \\ 
& & & & -\frac{u_0}{2} & \left(1 -\Phi\right) ^2 & -\frac{u_0}{2} & & \\  
& & & & & & \ddots & &\\
 & & & & & & -\frac{u_0}{2} & \left(m -\Phi\right) ^2 & -\frac{u_0}{2}  \\
 & & & & & & & & \ddots \\
\end{array}\right)
\end{equation}
and $\bm{\mathrm{c}}^{\left( e\right)} =\left(\hdots,\mathrm{c}_{-m}^{\left(
e\right)} ,\hdots,\mathrm{c}_{-1}^{\left( e\right)},\mathrm{c}_{0}^{\left(
e\right)},\mathrm{c}_{1}^{\left( e\right)},\hdots,\mathrm{c}_{m}^{\left(
e\right)},\hdots\right)$. Truncating the summation in Eq.\ (\ref{fou-ans}) at
some finite integer $M_\mathrm{max}$ allows Eq.\ (\ref{eq-evp}) to be solved
numerically to obtain the dimensionless electronic energies, $\lambda_N^{\left(
e\right)}$, and the coefficients $ \bm{\mathrm{c}}^{\left( e\right)}$. 
These can be substituted into Eq.\ (\ref{fou-ans}) to approximate the
electronic wave functions $\psi_N^{\left( e\right)}\left( \theta_e\right) $. 
The hole wave functions $ \psi_{N'}^{\left(h\right)}\left( \theta_h\right)$ and
energies $\lambda_{N'}^{\left( h\right)}$ are obtained by solving an
eigenequation similar to Eq.\ (\ref{eq-evp}) with $\Phi$ and $u_0$ replaced by
$-\Phi$ and $-u_0$ respectively.

\begin{figure}
  \centering
  \includegraphics[width= 3.5in]{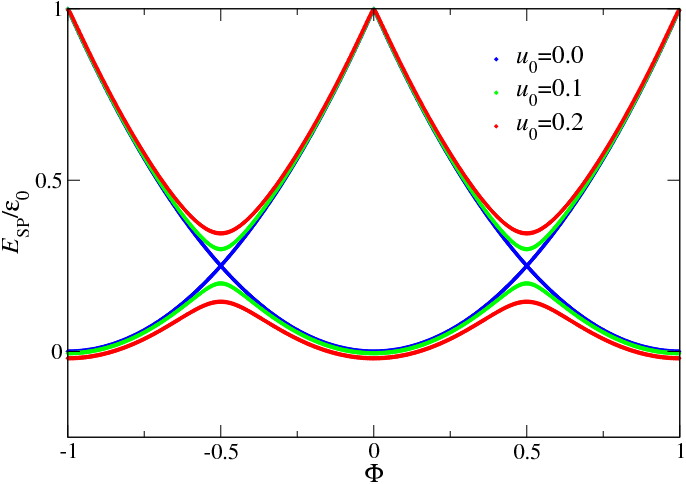}
  \caption{\label{fig-SP-phi} The two lowest single particle energy states as a
function of magnetic flux $\Phi$ at various
    values of electric field $u_0$ with $M_\mathrm{max}=30$.}
\end{figure}

Fig.\ \ref{fig-SP-phi} shows the single particle energies plotted as a function
of magnetic flux $\Phi$ for the case $N,N'=0,1$.
The AB oscillations are clearly visible. The energy values are the same for the
electron and hole. This can be seen by studying the matrix in Eq.\
({\ref{eigval-elec}); it is clearly similar to a matrix with $\Phi$ and $u_0$
replaced by $-\Phi$ and $-u_0$. For the case $u_0=0$ we obtain quadratic
behaviour in agreement with Eqs.\ (\ref{eq-spnoefield}) and
(\ref{eq-spnoefieldhole}).

Fig.\ \ref{fig-SP-u} shows the single particle energies plotted as a function of
electric field $u_0$ for various values of $\Phi$ and $N,N'$ with $0\le N,N'
\le3$. 
\begin{figure}
  \centering
  \includegraphics[width=3.5in]{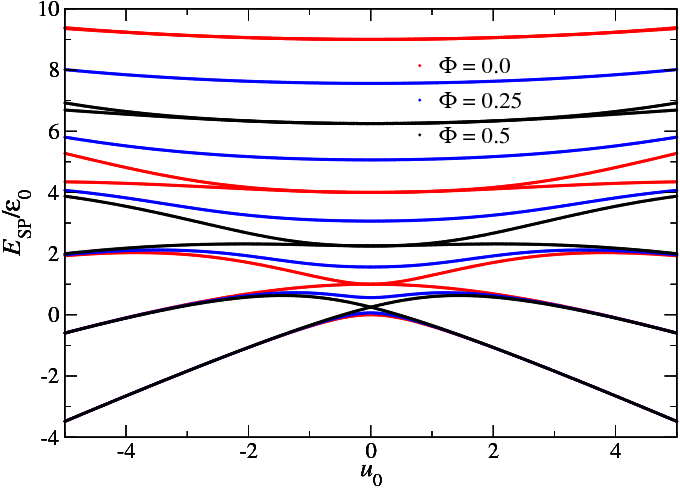}
  \caption{\label{fig-SP-u} Single particle energies as a function of external
field $u_0$ at various
    values of magnetic flux $\Phi$ with $M_\mathrm{max}=30$.}
\end{figure}
The $N,N'$ value associated with a particular curve can be deduced by noting
its value at $\Phi=0, u_0=0$ and comparing this with Eqs.\ (\ref{eq-spnoefield})
and (\ref{eq-spnoefieldhole}). 
All the curves are symmetric about the line $u_0=0$ due to the symmetry of the
problem. This symmetry can also be seen in the matrix in Eq.\
(\ref{eigval-elec}), which is similar to a matrix equal to it, aside from a sign
reversal in the off-diagonal entries. For the case $\Phi=0$, Eq.\
(\ref{schro-elec}) reduces to the Mathieu equation \cite{AbrS72} and the single
particle energies are proportional to the characteristic Mathieu values.

\subsection{Self-consistent solution for the exciton}
\label{sec-exciton-solution}

We are now in a position to solve the problem of an exciton in a nanoring with
external fields. The excitonic wave function \newnot{sym:Psitheetheh} can be expanded in a basis of
electron and hole wave functions:
\begin{equation}
 \label{wf-exciton}
\Psi\left( \theta_e,\theta_h\right) =\sum_{N,N'} A_{NN'}\psi^{\left(
e\right)}_N\left( \theta_e\right)\psi^{\left( h\right)}_{N'}\left(
\theta_h\right).
\end{equation}
The Hamiltonian is given by $\hat{H}=\hat{H}_e+\hat{H}_h+\hat{V}$, where
$\hat{H}_e$ is the Hamiltonian for the electron, $\hat{H}_h$ is the Hamiltonian
for the hole and $\hat{V}$ the interaction term, which depends on the distance
$R$ between the electron and hole. We assume a short range interaction \newnot{sym:VR}
\begin{equation}
\frac{V\left[ R\left( \theta_e-\theta_h\right)\right] }{\epsilon_0}=2\pi
v_0\delta\left( \theta_e-\theta_h\right),
\label{eq-intpot}
\end{equation}
where $v_0$ \newnot{sym:v0} is the average interaction strength in units of
$\epsilon_0$. Values for $v_0$ are chosen as integer multiples of $-1/\pi^2$, to
ensure that the inverse decay length \newnot{sym:gamma2} for the excitonic wave function,
$\gamma=\pi|v_0|\epsilon_0 \sqrt{\frac{\mu}{\hbar^2 \epsilon_0}}$ \cite{RomR00}, is comparable
to the inverse of the ring circumference. The Schr\"odinger equation for the
exciton may now be written as
\begin{equation}
 \label{schro-exciton}
\sum_{N,N'}A_{NN'}\left( \lambda_N^{\left( e\right)}+\lambda_{N'}^{\left(
h\right)}-\Delta\right)\psi^{\left( e\right)}_N\left(
\theta_e\right)\psi^{\left( h\right)}_{N'}\left( \theta_h\right)
 + 2\pi v_0\delta\left(\theta_e-\theta_h \right)\Psi\left(
\theta_e,\theta_h\right) =0,
\end{equation}
where $\Delta$ \newnot{sym:Delta} is the excitonic energy in units of $\epsilon_0$. Multiplying by
$\left[  \psi^{\left( e\right)}_N\left( \theta_e\right)\psi^{\left(
h\right)}_{N'}\left( \theta_h\right)\right]  ^\dagger$ for particular $N,N' \in
\mathbb{Z}$ and integrating over $\theta_{e},\theta_h \in \left[0,2\pi \right]$
allows us to obtain an expression for the coefficients in Eq.\
(\ref{wf-exciton}):
\begin{equation}
 \label{exciton-coeffs}
A_{NN'}=-\frac{2\pi v_0}{\lambda_{N}^{\left( e\right)}+\lambda_{N'}^{\left(
h\right)}-\Delta}
\int_0^{2\pi}d\theta \Psi\left( \theta,\theta \right){\psi^{\left(
h\right)}_{N'}\left( \theta\right)}^\dagger {\psi^{\left( e\right)}_{N}\left(
\theta\right)}^\dagger.
\end{equation}
Using Eq.\ (\ref{wf-exciton}) with $\theta_e=\theta_h=\theta$, we may write
\begin{equation}
 \label{wf-exciton2}
\int_0^{2\pi}d\theta \Psi\left( \theta,\theta \right){\psi^{\left(
h\right)}_{N'_0}\left( \theta\right)}^\dagger {\psi^{\left(
e\right)}_{N_0}\left( \theta\right)}^\dagger =\sum_{N,N'}
A_{NN'}\int_0^{2\pi}d\theta \psi^{\left( e\right)}_N\left( \theta
\right)\psi^{\left( h\right)}_{N'}\left( \theta\right){\psi^{\left(
h\right)}_{N'_0}\left( \theta\right)}^\dagger {\psi^{\left(
e\right)}_{N_0}\left( \theta\right)}^\dagger.
\end{equation}
Let 
\begin{equation}
\label{g}
 G_{NN'}=\int_0^{2\pi}d\theta \Psi\left( \theta,\theta\right)
{\psi_{N'}^{\left(h \right) }}^\dagger \left( \theta
\right){\psi_N^{\left(e\right) }}^\dagger \left( \theta \right) 
\end{equation}
and 
\begin{equation}
\label{p}
P_{NN'N_0N'_0}=-\frac{2\pi v_0}{\lambda_{N}^{\left(
e\right)}+\lambda_{N'}^{\left( h\right)}-\Delta} \int_0^{2\pi} d\theta
\psi_N^{\left(e \right) } \left( \theta \right)\psi_{N'}^{\left(h \right)}
\left( \theta \right){\psi_{N'_0}^{\left(h \right) }}^\dagger \left( \theta
\right){\psi_{N_0}^{\left(e \right) }}^\dagger \left( \theta \right).
\end{equation}
Then Eq.\ (\ref{wf-exciton2}) can be expressed as
\begin{equation}
\label{exciton-eigeneq4d}
 G_{N_0N'_0}=\sum_{N,N'}G_{NN'}P_{NN'N_0N'_0}\left(\Delta \right).
\end{equation}
In order to find approximate solutions to \eqref{exciton-eigeneq4d}, we cut-off
the sums at a maximum value $N_\mathrm{max}$ for $N,N',N_0,N'_0$. We have chosen
$N_\mathrm{max}=15$ for calculation of the excitonic energies and
$N_\mathrm{max}=11$ for the oscillator strengths; we have tested that the
results do not change appreciably for the range of $\Phi$, $u_0$ and $v_0$
considered here. Mapping $N,N'\to K$ and $N_0,N'_0\to K_0$ according to
${K=\left( N-1\right)N_\mathrm{max}+N'}$, allows Eq.\ (\ref{exciton-eigeneq4d})
to be reformulated in 2D:
\begin{equation}
\label{exciton-eigeneq2d}
 G_{K_0}=\sum_K G_{K}P_{KK_0}\left(\Delta \right).
\end{equation}
The excitonic energies may now be calculated numerically by determining the
values of $\Delta$ which result in the matrix $P_{KK_0}^{\rm T}$ having an eigenvalue
equal to 1. Note that it is the transpose of $P_{KK_0}$, which must be diagonalised to obtain the correct eigenvectors, since Eq.\ (\ref{exciton-eigeneq2d}) is an eigenequation from the left.  For fixed values of $u_0$ and $\Phi$, we test a range of $\Delta$
values with separation 0.01 until we find a pair such that for one, the
eigenvalue of $P_{KK_0}$ is less than 1 and for the other, it is greater than 1.
Previous analytical results at $u_0=0$ indicate at which energies to look for
the ground state \cite{RomR00}. A crude approximation for the excitonic energy
is found by linearly interpolating between these two points. The process is
repeated closer to this approximate value using a smaller separation of 0.002 to
produce a set of coordinates $\left\lbrace \Delta, P_{KK_0}\left(\Delta \right) 
\mathrm{eigenvalue}\right\rbrace$. We numerically find a function that
interpolates these points and obtain a final solution for the excitonic energy
by setting this equal to 1 and solving for $\Delta$. The coefficients $A_{NN'}$
may then be found by using the obtained values of $\Delta$ and $G_{NN'}$ in Eq.\
(\ref{exciton-coeffs}).
\section{Numerical results for the energy of the exciton}
\label{sec-abres}
Fig.\ \ref{fig-ex-phi} shows the resulting excitonic ground state energies
plotted as a
function of magnetic flux for different electric field strengths.
\begin{figure}[tb]
  \centering
  \includegraphics[width=0.8\columnwidth]{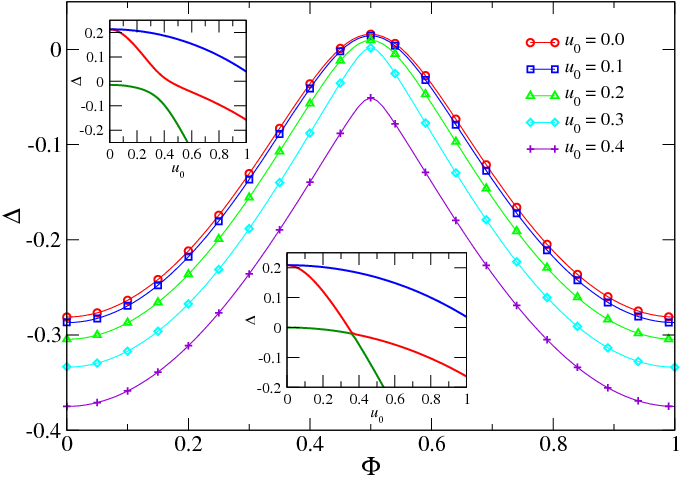}
  \caption{\label{fig-ex-phi} Excitonic energy as a function of
    magnetic flux $\Phi$ at various values of electric field $u_0$ for
    interaction strength $v_0=-2/\pi^2$. Only every $4^{\rm th}$ data point is
shown for
clarity in each curve. Inset top: The energies of the first $3$ states at
$\Phi=0.45$ as a function of $u_0$. Inset bottom: The first $3$ states at
$\Phi=0.5$ as a function of $u_0$. Results displayed in insets were provided by
Vivaldo L. Campo, Jr..}
\end{figure}
For small enough electric fields, the excitonic energy oscillates as a
function of the magnetic field as seen previously for $u_0=0$ \cite{Cha95}. This
is due to the electron and hole, which were created simultaneously at the same
position, having a finite probability to tunnel in opposite directions and meet
each other on the opposite side of the ring. The dependence of the oscillation
amplitude, $\Delta(\Phi=0.5)-\Delta(\Phi=0.0)$, upon electric field strength is
shown in Fig.\ \ref{fig-osc-amp} for
different values of the interaction strength.
\begin{figure}[tb]
  \centering
  \includegraphics[width=0.8\columnwidth]{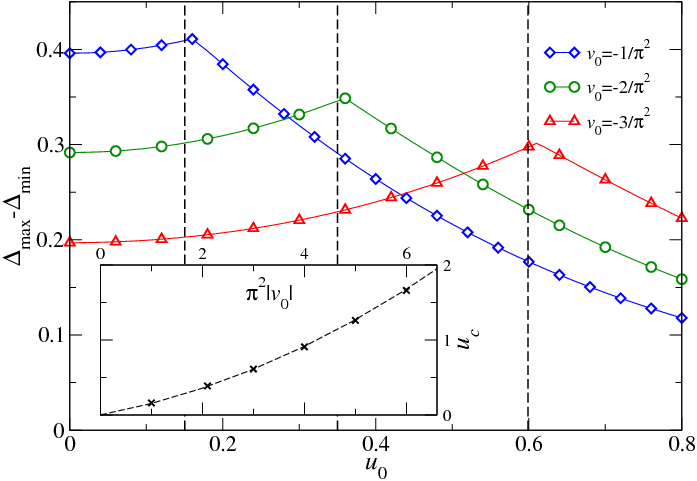}
  \caption{\label{fig-osc-amp} Amplitude of the Aharonov-Bohm
    oscillations as a function of external field $u_0$ at various values of the
interaction strength $v_0$. Only every $6^{\rm th}$ data point is shown in
each curve for clarity.
Inset:  critical external field $u_{\rm c}$ as a function of $v_0$. The dashed
curve, $u_{\rm c} = \pi |v_0| (\pi |v_0|/4 +2/5)$, fits the numerical points
quite
well.
The vertical dashed lines in the main plot indicate $u_{\rm c}(v_0)$.
}
\end{figure}
In all cases, there is initially a small increase in amplitude, as speculated in
\cite{MasC03} and then it
decreases strongly as a function
of electric field strength.
This change happens when the polarisation energy, ${\langle u_0
|\mathrm{cos}(\theta)|\rangle=2u_0/\pi}$, becomes comparable to
the binding energy of an exciton. Using the binding energy for an exciton on a
straight line, $\pi^2 v_0^2/2$ \cite{RomR00}, yields an estimate for the
critical $u_0$ value, \newnot{sym:uc}$u_\mathrm{c}=\pi^3 v_0^2/4$. As seen in the
inset of Fig.~\ref{fig-osc-amp}, a better fit to computed $u_\mathrm{c}$ values
is given by the function $u_{\rm c} = \pi |v_0| (\pi |v_0|/4 +2/5)$.
More precisely, the change of behaviour corresponds to a level crossing between
the ground and first excited state, which only occurs at $\Phi=0.5$, as shown in
the insets of Fig.\ \ref{fig-ex-phi}. The level crossing is
associated
with an exact symmetry at $\Phi=0.5$ upon exchanging $\theta_{
e,h}\rightarrow-\theta_{ e,h}$. It is also evident in Fig.\
\ref{fig-ex-u-v2}, where the excitonic energy is plotted as a function of
the electric field for various values of magnetic flux.
\begin{figure}[tb]
  \centering
  \includegraphics[width=0.8\columnwidth]{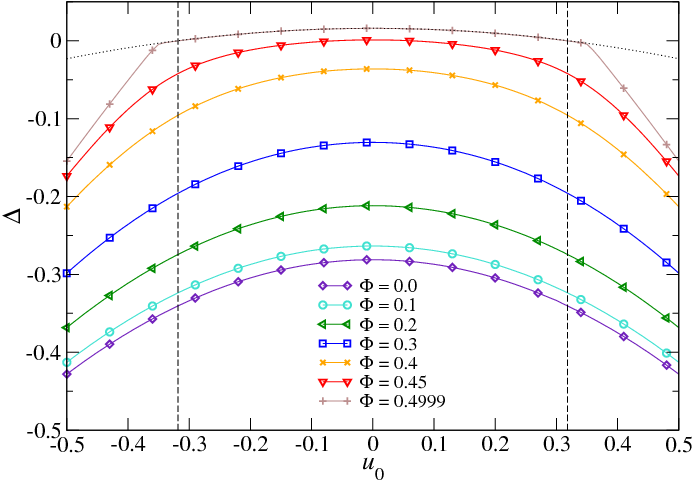}
  \caption{\label{fig-ex-u-v2} Excitonic energy as a function of $u_0$ for
interaction strength $v_0=-2/\pi^2$ and various values of $\Phi$. Only every
$6^{\rm th}$
data point is shown. The dotted lines show the results of second order
perturbation theory
performed on the ground state. The vertical dashed lines indicate $u_{\rm c}$.}
\end{figure}
The quadratic behaviour for small enough magnetic fluxes and the negative
curvature of $\Delta$ are as expected from second order perturbation theory in
$u_0$. For magnetic flux values close to
$\Phi=0.5$, the excitonic energy remains almost constant as a function of
electric field until the critical $u_0$ value, when the level crossing occurs.
\section{Vanishing oscillator strength}
\label{sec-os}
\begin{figure}[tb]
  \centerline{
  \includegraphics[width=0.8\columnwidth]{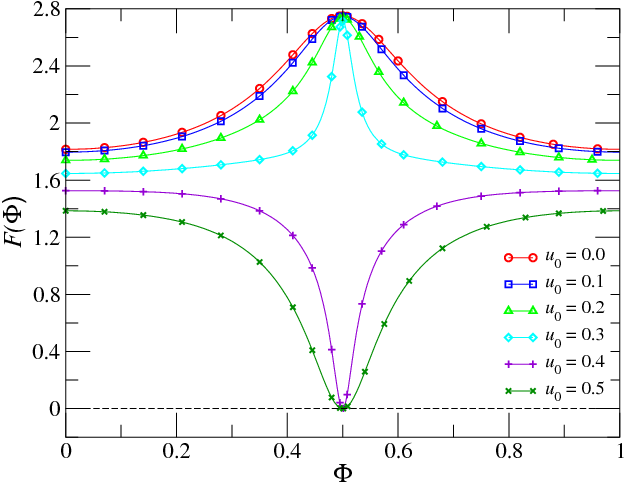}  }
  \caption{\label{fig-OS-v2} Oscillator strength $F$ as a function
    of $\Phi$ for electron-hole interaction strength $v_0=-2/\pi^2$ and various
values of the electric field $u_0$. The horizontal dashed line indicates
$F=0$. Only every $6^{\rm th}$ data point is shown for clarity in each
curve.}
\end{figure}
The OS is the strength of the transition from the excitonic ground state,
$\Psi(\theta_e,\theta_h)$, to the vacuum via photoemission. It can be shown to
be directly proportional to the quantity \newnot{sym:FPhi}
\begin{equation}
F=\left|\int_0^{2\pi} d\theta\Psi\left( \theta,\theta\right)\right|^2,
\label{eq-os}
\end{equation}
where the wave function, $\Psi(\theta_e,\theta_h)$, is correctly normalised
\cite{HenN70}. Figure \ref{fig-OS-v2} shows $F$ plotted as a function of
magnetic flux for different electric field strengths; the AB oscillations are
clearly visible. Interestingly, the oscillations become inverted for large
enough values of $u_0$. Most importantly, for these $u_0$ values, the OS
vanishes at $\Phi=0.5$. This means physically that an exciton created at some
$\Phi,u_0$ with $F\ne 0$ is prevented from recombining once $\Phi,u_0$ are tuned
to values for which $F=0$. In addition, no excitons can be created when $F=0$.
Suppression of ground state exciton emission has already been seen for polarised
excitons in the ring geometry for particular magnetic field strengths
\cite{GovUKW02}. However, this is different to the phenomenon we observe, since
in Ref.\ \cite{GovUKW02}, electron-hole recombination is prevented by confining
the particles to different nanostructures.

The transition from optically active to dark coincides with the change in slope
of the amplitude of the excitonic AB oscillation as in Fig.\ \ref{fig-osc-amp}
and also the reinstatement of the parabolic behavior in Fig.\ \ref{fig-ex-u-v2}.
In order to explain the mechanism for this behaviour, the probability density function for the exciton is shown in Fig.\
\ref{fig-ex-prob} at different
values of $\Phi$, $u_0$ and $v_0$.
\begin{figure}[tb]
  \centerline{
  (a)\includegraphics[width=0.45\columnwidth]{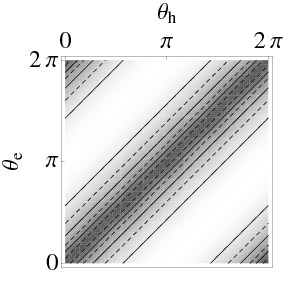} 
  (b)\includegraphics[width=0.45\columnwidth]{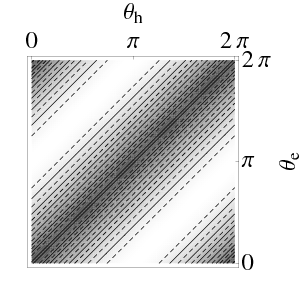}}
  \centerline{
  (c)\includegraphics[width=0.45\columnwidth]{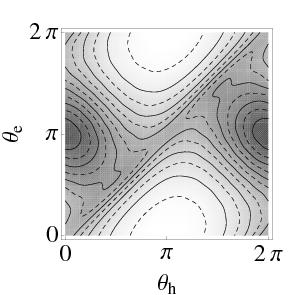}
  (d)\includegraphics[width=0.45\columnwidth]{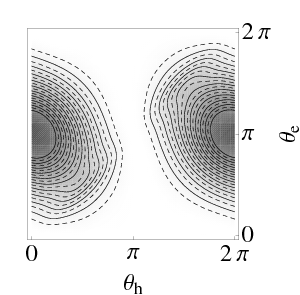}
  }
  \caption{
    Excitonic probability density
    $|\Psi(\theta_e,\theta_h)|^2$ at $v_0=-2/\pi^2$ and
    (a) $u_0=0$,
    $\Phi=0$,
    (b) $u_0=0$,
    $\Phi=0.4999$,
    (c) $u_0=0.5$,
    $\Phi=0$, and
    (d) $u_0=0.5$,
    $\Phi=0.4999$. The solid and dashed lines are contour lines separated in
height by $1/10\pi^2$. }
\label{fig-ex-prob}
\end{figure}
In Fig.\ \ref{fig-ex-prob}(a) the magnetic flux and electric field have been set
to zero. The dark grey diagonal region indicates a peak in probability density
where the electron and hole coordinates are close together, so the exciton is
very small. In Fig.\ \ref{fig-ex-prob}(b) the flux has been increased from its
zero value, whilst keeping the interaction strength and electric field constant.
There is a slight narrowing of the peaked region and the contours are closer
together, showing that the exciton has become smaller. In Fig.\
\ref{fig-ex-prob}(c) the magnetic flux and interaction strength are the same as
in Fig.\ \ref{fig-ex-prob}(a), but the electric field has been increased. We see
the dark grey region has begun to split in two, suggesting the exciton is
beginning to separate. In Fig.\ \ref{fig-ex-prob}(d) the magnetic flux has been
increased from its zero value in Fig.\ \ref{fig-ex-prob}(c). The result is two
dark grey peaks indicating that the exciton has been broken \cite{LunSLP99} into
an electron and hole, which are located on opposite sides of the ring.

Figure \ref{fig-OS-v2} suggests a method for controlling the excitonic lifetime. The procedure is (i) begin with $\Phi=0$ and $u_0=0$ (ii) increase the flux adiabatically to $\Phi=0.5$ so an exciton can easily be created from the vacuum (iii) decrease $\Phi$ (iv) increase $u_0$ until it appreciably exceeds the critical value (v) increase flux to $\Phi=0.5$. The oscillator strength is now zero, so the exciton is trapped and unable to decay until the external fields are further tuned. This has important implications for excitonic data storage.
\section{Experimental work}
\label{sec-abexpt}
After publication of the work presented here, the XABE was observed for the first time in type-I structures in which both the electron and hole move in a ring geometry \cite{TeoCLM10,DinALP10}. In both of these experiments, the effect was enhanced using an electric field, raising it to a measureable level. Experimentally, AB oscillations in the excitonic energy appear on top of a quadratic dependence on the magnetic field, termed the diamagnetic shift \cite{RibGCM04,LinLLF09}. One might also expect to see the linear Zeeman splitting, although this is only observed when single rings are studied, not for ensembles of rings \cite{DinALP10}. 

In Ref. \cite{TeoCLM10}, an array of self assembled InAs/GaAs rings were studied using magneto-photoluminescence techniques. The emission spectra display a weak shoulder at lower energies indicating the presence of two emission bands. Clear oscillations were seen in the integrated intensity as a function of magnetic field for each band, although they have different periods and their maxima and minima occur in the opposite order. The different periods are simply due to different ring sizes. The inversion of the oscillation pattern is caused by an inbuilt electric field for the larger quantum rings, due to piezoelectricity from uniaxial strain fields formed in the growth process. It is known that such strains are particularly strong for InAs/GaAs rings \cite{SchWB07}. It was also found using a theoretical model that inversion of the maxima/minima pattern in oscillations of the oscillator strength can be caused by increasing the temperature, so that excitons are excited out of the bright ground state into dark excited states. However, this was not the correct explanation for the observed behaviour.
\begin{figure}[tb]
  \centerline{
  \includegraphics[width=0.8\columnwidth]{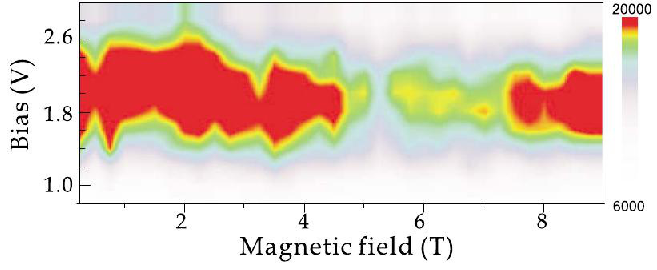}}
  \caption{\label{fig-DinAPL10} (Taken from Ref.\ \cite{DinALP10}). Density plot for the photoluminescence intensity of light emitted by an exciton in a InGaAs/GaAs quantum ring in the presence of external magnetic and electric fields, as a function of the field strengths.}
\end{figure}

In Ref.\ \cite{DinALP10}, InGaAs quantum dots are grown using molecular beam epitaxy and capped with GaAs. Ring structures are then etched from these and the magneto-photoluminescence spectra studied for a single quantum ring. A vertical electrical field is applied, which can be used to alter the electron-hole separation and thus the period of AB oscillations. As suggested in our study, it is found that by tuning the electric and magnetic fields, the photoluminescence intensity is almost quenched. This is shown in Fig.\ \ref{fig-DinAPL10}, which was taken from Ref.\ \cite{DinALP10}. Subsequently, the XABE was studied theoretically in the presence of a perpendicular electric field \cite{LiMP10}. The electric field was used to tune the ground state energy, so that weak AB oscillations were found. The amplitude and period of these oscillations can be tuned by varying the strength of the electric field.
\cleardoublepage
\cleardoublepage
	
\chapter{Neutral collective excitations in pristine graphene}
\label{chap-MPsPristineGraphene}
\section{The importance of interactions in graphene}
\label{sec-grapints}
How to model $e$-$e$ interactions is one of the greatest challenges of condensed matter physics. This is because at any given instant each electron moves in a potential, which depends upon the configuration of all the other electrons. Hence problems involving a few electrons are difficult to solve even using numerical approaches and those treating a macroscopic system are intractable via direct consideration of the forces at a microscopic level. To make progress, it is necessary to ``zoom out'' from these microscopic details. In the mean field approach \cite{AshM76}, for example, the interaction of an electron with other electrons in the system is modelled, by considering the remaining electrons as a smooth charge density representative of the average electronic configuration. In fact in many systems, the addition of $e$-$e$ interactions to the non-interacting problem only results in quantitative changes. Such systems are well described by Landau's Fermi liquid theory \cite{LanP80} in which the problem of interacting electrons is replaced by one of non-interacting quasiparticles, which are elementary excitations of the system. This results in the renormalisation of their non-interacting properties. Systems in which $e$-$e$ interactions induce a qualitative departure from the non-interacting behaviour are termed strongly correlated and are fascinating to study.

The role of $e$-$e$ interactions in graphene \cite{KotUPC10} is not fully understood. Some calculations suggest it is a 2D Fermi liquid \cite{DasHT07} with a Fermi velocity renormalised by interactions, whereas others indicate strong correlations \cite{MulSF09}. An important parameter is the ratio of the characteristic energy of Coulomb interactions, \newnot{sym:Eint}$E_\mathrm{int}$, to the kinetic energy, \newnot{sym:Ekin}$E_\mathrm{kin}$. For the 2DEG, $E_\mathrm{kin} =\frac{\hbar^2 k_\mathrm{F}^2}{2m}\sim \frac{\hbar^2 }{2m \ell^2}$, \newnot{sym:kF} at the Fermi energy, where $\ell$ \newnot{sym:ell} is the average electron separation and $E_\mathrm{int} =\frac{e^2}{\epsilon \ell}$, where $\epsilon$ \newnot{sym:epsdielectric} is the dielectric constant which depends on the surrounding environment. Hence the ratio is
\begin{equation}
\label{eq-engratio2deg}
\left( \frac{E_\mathrm{int}}{E_\mathrm{kin}}\right)_\mathrm{2DEG}\sim \frac{me^2}{\epsilon \hbar^2} \ell=\frac{\ell}{a_\mathrm{0}},
\end{equation}
where $a_\mathrm{0}$ is the \newnot{sym:a0} Bohr radius. Thus we see that for the 2DEG, $e$-$e$ interactions become more important relative to the kinetic energy as the electron separation is increased. Indeed below a critical electron density, the 2DEG will undergo Wigner crystallisation \cite{Wig34}, forming a triangular lattice in order to minimise the potential energy of interactions. For the case of graphene, $E_\mathrm{kin}\sim \hbar v_\mathrm{F} k_\mathrm{F}$, yielding
\begin{equation}
\label{eq-engratiograp}
\frac{E_\mathrm{int}}{E_\mathrm{kin}}\sim \frac{e^2}{\epsilon \hbar v_\mathrm{F} },
\end{equation}
which is independent of $\ell$. In fact $\alpha^\mathrm{g}=\frac{e^2}{\epsilon \hbar v_\mathrm{F} }$ is called the graphene fine structure constant\newnot{sym:alphag}. Depending on the dielectric constant, $\alpha^\mathrm{g} \sim 10^{-1}$, which is large relative to the fine structure constant in QED, $\alpha=1/137$. The fine structure constant also describes the coupling between matter and light and a large value for graphene compared to non-relativistic materials means that it has a high opacity, considering it is only one layer thick, which has been confirmed experimentally \cite{NaiBGN08}.

Another important consideration is to what extent the $e$-$e$ interactions are
screened in graphene \cite{GonGV94}. The Thomas Fermi wavelength\newnot{sym:lambdaTF},
$\lambda_\mathrm{TF}$, gives the length beyond which the potential from a given
electron is screened out. In 2D, \newnot{sym:lambdaF}$\lambda_\mathrm{TF}\simeq \lambda_\mathrm{F}
\frac{E_\mathrm{kin}}{E_\mathrm{int}}$, so for graphene $\lambda_\mathrm{TF} \sim\frac{1}{\sqrt{n_e}}$, where $n_e$ \newnot{sym:ne} is
the carrier density. For neutral graphene $n_e\rightarrow 0$, as the Fermi
energy is approached, since the conduction and valence bands touch at this
point. Hence we conclude that for graphene with little or no doping, the $e$-$e$
interaction potential can be treated as the long-range Coulomb interaction,
\newnot{sym:Uofr}${U\left( r\right) =\frac{e^2}{\epsilon r}}$, where the only source of screening
is from $\epsilon$.


In the presence of a magnetic field, $e$-$e$ interactions play an important role, due to the single particle LL energies (calculated in Section \ref{subsec-spgrapmag}) being highly degenerate.
For the 2DEG
\begin{equation}
\label{eq-engratiomag2deg}
\left( \frac{E_\mathrm{int}}{E_\mathrm{kin}}\right)_\mathrm{2DEG}^{B \neq 0}=\sqrt{\frac{e^3 {m_e}^2 c}{B\epsilon^2 \hbar^3}}\sim \frac{1}{\sqrt{B}}.
\end{equation}
Hence $E_\mathrm{int}/E_\mathrm{kin}$ depends on the magnetic field strength for the 2DEG and interactions can be made less important relative to the kinetic energy, by going to higher field strengths. In contrast, for graphene, 
$E_\mathrm{int}/E_\mathrm{kin}=\alpha^\mathrm{g}$, as in the absence of a magnetic field. The average electronic separation cancels out as before, although for $B \neq 0$ it scales as $\ell_B$. The size of  
$E_\mathrm{int}/E_\mathrm{kin}$ in graphene can be considered medium. It is small enough to minimise spontaneous inter-LL excitations or mixing of LLs in the absence of an external energy source, but large enough not to be ignored altogether. In the absence of LL mixing, in a partially filled LL, the kinetic energy is effectively quenched and the interaction energy is the only remaining energy scale. 

In high magnetic fields, the LLs may be split into four sublevels due to Zeeman splitting \newnot{sym:hbaromegas}($\hbar\omega_s$) and an additional valley pseudospin splitting ($\hbar\omega_{v}$)\newnot{sym:hbaromegav}. This LL splitting is not yet well understood; it is even unclear under what circumstances spin or valley splitting is larger. Lifting of the four-fold degeneracy has been seen in quantum Hall experiments in high magnetic fields ($B>20 \hspace{1mm} \mathrm{T}$) for the $n=0$ LL, with evidence for the $n=\pm1$ LLs observed, but somewhat weaker \cite{ZhaJSP06, JiaZSK07}. Large splittings on the order of the characteristic Coulomb energy, $\frac{e^2}{\epsilon \ell_B}$, have been predicted \cite{YanDM06} and observed \cite{JiaZSK07}. Infrared spectroscopy experiments have found non-monotonic behaviour of the CRE as the filling factor of the zeroth LL is increased, which was interpreted as an indication of LL splitting \cite{HenCJL10}.  Theoretical studies \cite{NomM06,AbaLL06,FerB06,GoeMD06,AliF06,FucL07} state that it is a many body effect, due to some spontaneous symmetry breaking between the two sublattices. 

Another way in which the LL energies are affected by $e$-$e$ interactions is via the exchange self energy correction, which is calculated below (Section \ref{sec-selfeng}). In graphene, energies for different LLs are renormalised differently, so that the energies for the optical excitation of an electron between LLs are altered and this occurs differently for different resonances; such effects have been seen experimentally. In contrast, in the 2DEG, the unique CRE remains unchanged by $e$-$e$ interactions. This is known as Kohn's theorem \cite{Koh61} and is discussed in detail in this Chapter, in particular how it breaks down in graphene. Within the electric dipole approximation, optically induced excitations carry no spin. The case of spin flip transitions within the same LL, often referred to as spin waves, is dealt with for the 2DEG by Larmor's theorem \cite{DobKW88}. It states that the energy of a spin wave excitation at wave vector $\mathbf{k} = 0$ is unchanged by interactions and remains equal to the bare Zeeman splitting, $E\left( \mathbf{k}=0\right)=g \mu_B B$. Larmor's theorem also holds in graphene \cite{RolFG10b}.

The central theme of the work on graphene presented here is collective excitations (CEs) of the monolayer system at zero temperature and in the presence of a strong perpendicular magnetic field. We focus on a system close to neutrality, where the conduction electrons behave as massless Dirac particles. Thus all calculations use ground states, $|\nu \rangle$, with integer filling factors $\nu=-2,-1,0,1,2$, corresponding to the possible sublevel fillings of the $n=0$ LL. Considering only LLs close to neutrality, which have maximum separation for neighbouring levels, should minimise the effects of LL broadening present in real samples due to disorder and phonons; this work requires different LLs to be distinguishable. It should be noted however that the broadening of LLs in graphene is poorly understood at the moment, with qualitative disagreement between theory \cite{PerGC06,YanPX10} and experiment \cite{JiaHTW07,OrlFPN08}. This is hardly surprising, since an accurate theory requires knowledge of the type of disorder causing the broadening, which is difficult to determine and may be different in different experiments. A neutral CE of the ground state occurs when an electron in one of the uppermost filled LLs within the Dirac sea is excited to an empty state in a higher lying LL leaving behind a hole; the bound state of an electron and hole is an exciton. The formation of charged CEs is also possible and will be discussed below. The term ``collective" is used because of the LL degeneracy and also the fact that different excitations are mixed by the Coulomb interaction. 

This Chapter focuses on the clean system and presents some of the theory required for the calculations. This includes identifying the good quantum numbers, which allows classification of CEs according to the geometrical symmetries of the system. Classification of CEs according to their dynamical symmetries is discussed in Chapter \ref{chap-su4}. Optical properties of the CEs are of particular interest, so the selection rules are derived here for circularly polarised light. In addition, I contextualise the work in terms of pre-existing research. 

The majority of work on CEs of graphene in a strong perpendicular magnetic field follows a similar approach to that used to study the analogous problem for the 2DEG \cite{LerL78,KalH84,KalH85}. In the 2DEG, bound electron-hole ($e$-$h$) complexes in a magnetic field have a well defined linear momentum of magnitude $k$. A classical way to visualise this is that at a certain $e$-$h$ separation, the attractive and repulsive forces exerted on them, due to their Coulomb interaction and the Lorentz force respectively, cancel. Hence the excitonic dispersion relation is well defined with the $e$-$e$ interactions causing a deviation from the bare CRE of order \newnot{sym:E0}$E_0=\sqrt{\frac{\pi}{2}}\frac{e^2}{\epsilon \ell_B}$. In Ref.\ \cite{KalH84} the dispersion relation for both filled and half-filled (spin polarised) LLs is calculated using a Hartree-Fock approach to first order in the parameter $E_\mathrm{int}/E_\mathrm{kin}$. For small values of $k$, the dispersion relation looks like that of magnetoplasmons (charge density waves) or spin waves. Corresponding studies of graphene have found excitation spectra qualitatively similar to those for the 2DEG, but with stronger many body corrections \cite{IyeWFB07,BycM08}. Chapter \ref{chap-nces} addresses the addition of impurities and the resulting localised neutral CEs, which appear as discrete states below the band of extended CEs. The identification of these localised states requires knowledge of the width of the band and so for completeness we review in Section \ref{sec-disp} the derivation of the excitonic dispersion relation for graphene. In addition to spin and valley pseudospin preserving CEs, spin flips, spin waves \cite{RolFG10b} and skyrmions \cite{YanDM06, Eza07} (coherent charged topological CEs) have also been studied in the context of graphene in a perpendicular magnetic field.
\section{Neutral and charged collective excitations}
\label{sec-neucharge}
Recall from Section \ref{subsec-grapmag} that in the symmetric gauge, single electron states in graphene are labelled by the quantum numbers $n,\tau,s,m$. Let \newnot{sym:cdaggergrap}$c^\dagger_{n \tau s m}$ be the fermionic creation operator for an electron in graphene, so that $\Psi_{\mathcal{N} m}(\mathbf{r})=\langle \mathbf{r} |c^\dagger_{\mathcal{N} m} |0 \rangle $, where \newnot{sym:mathcalN}$\mathcal{N}=\{n \tau s \}$ is a collective index and $|0\rangle$ represents the vacuum. I shall use the hole representation,\newnot{sym:ddag} $c_{\mathcal{N} m} \rightarrow d^{\dag}_{\mathcal{N} m}$ and $c^{\dag}_{\mathcal{N} m} \rightarrow d_{\mathcal{N} m}$ for all filled levels. The usual anticommutation relations for fermions are obeyed:
\begin{equation}
\label{eq-anticomrel}
\{c_{\mathcal{N}_1 m_1},c^\dagger_{\mathcal{N}_1' m_1'}\}=\delta_{\mathcal{N}_1\mathcal{N}_1'}\delta_{m_1 m'_1}, \hspace{5mm} \{d_{\mathcal{N}_2 m_2},d^\dagger_{\mathcal{N}_2' m_2'}\}=\delta_{\mathcal{N}_2\mathcal{N}_2'}\delta_{m_2 m'_2}.
\end{equation}

Consider a neutral CE, 
corresponding to an electron being promoted from a state $\mathcal{N}_2$ to an empty state $\mathcal{N}_1$ in a higher lying LL. This excitation can be created from the ground state, $|\nu \rangle$, by the creation operator
\begin{equation}
\label{eq-Qdagger} 
 \mbox{} \hspace{-5pt}
      Q^{\dag}_{\mathcal{ N}_1  \mathcal{ N}_2}  =
          \sum_{m_1 , m_2 = 0}^{\infty}
           A_{\mathcal{N}_1 \mathcal{N}_2}(m_1,m_2) \,
          c^{\dag}_{\mathcal{N}_1 m_1} d^{\dag}_{\mathcal{N}_2 m_2}.
\end{equation}
Suppose now that the filling factor is $\nu=\mu+\varepsilon$ with $\varepsilon \ll 1$ and \newnot{sym:mugrap}$\mu=0,\pm1,\pm2$. Then an integer number of sublevels of the $n=0$ LL are filled and there are a few electrons in the next sublevel (see Fig.\ \ref{fig-triondiag}(a)). 
\begin{figure}
\centering
  \includegraphics[width=2in]{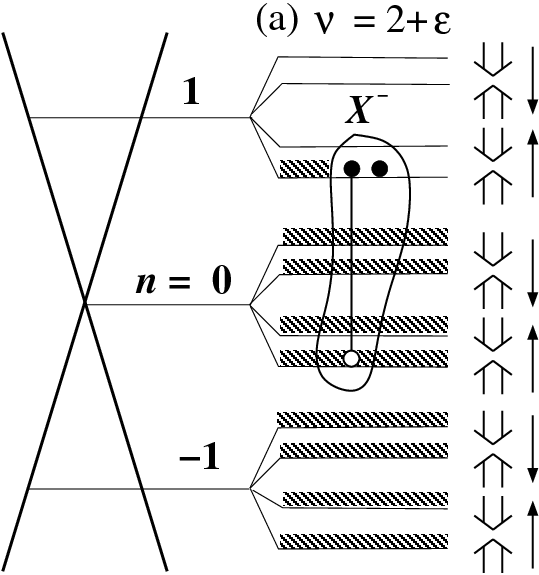}
  \includegraphics[width=2in]{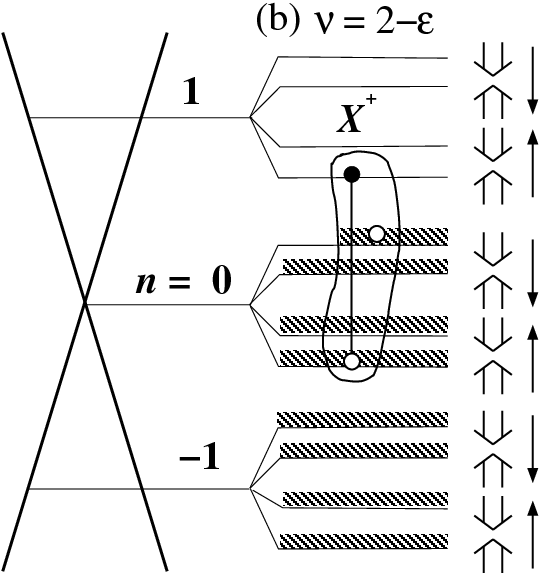}
   \caption{Schematic diagram of the formation of (a) $X^-$ trions at filling factor $\nu=2+\varepsilon$ and (b) $X^+$ trions at filling factor $\nu=2-\varepsilon$.}
\label{fig-triondiag} 
\end{figure}
In this case it is possible for an neutral CE to bind to an additional electron forming a negatively charged CE or trion, \newnot{sym:X-}$X^-$. This excitation can be created by acting on $|\mu \rangle$, by the creation operator
\begin{equation}
\label{eq-Rdagger} 
 \mbox{} \hspace{-5pt}
      R^{\dag}_{\mathcal{ N}_1  \mathcal{ N}_2 \mathcal{ N}_3}  =
          \sum_{m_1 , m_2, m_3 = 0}^{\infty}
           A_{\mathcal{N}_1 \mathcal{N}_2 \mathcal{ N}_3}(m_1,m_2,m_3) \,
          c^{\dag}_{\mathcal{N}_3 m_3} c^{\dag}_{\mathcal{N}_1 m_1} d^{\dag}_{\mathcal{N}_2 m_2}.
\end{equation}
Similarly if the filling factor is $\nu=\mu-\varepsilon$, so that the uppermost filled sublevel is almost completely filled, a positively charged collective excitation, \newnot{sym:X+}$X^+$, may form (see Fig.\ \ref{fig-triondiag}(b)). The creation operator for the $X^+$ can be formed in an analogous way to that for the $X^-$. As we shall see in Chapter \ref{chap-cces}, the charged CEs appear as discrete states below the excitonic continuum. The existence of trions in semiconductors was first proposed in 1958 \cite{Lam58} and observed experimentally during the 1990s \cite{KheCdM93}.
\section{Hamiltonian of collective excitations}
\label{sec-grapham}
The Hamiltonian, $\hat{H}=\hat{H_0}+\hat{U}_{eh}+\hat{U}_{ee}+\hat{U}_{hh}$, for collective excitations is\newnot{sym:H}
\begin{eqnarray}
\label{eq-ham} 
\hat{H} & = & \sum_{\mathcal{N},m} \tilde{\epsilon}_n c^{\dag}_{\mathcal{N} m}c_{\mathcal{N} m}-\sum_{\mathcal{N},m} \tilde{\epsilon}_n d^{\dag}_{\mathcal{N} m}d_{\mathcal{N} m} \nonumber\\ 
&  &+ \sum_{\substack{\mathcal{N}_1, \mathcal{N}_2 \\ \mathcal{N}_1', \mathcal{N}_2'} }\sum_{\substack{m_1, m_2 \\ m_1', m_2'} }\mathcal{W}_{\mathcal{N}_1 m_1   \mathcal{N}_2 m_2}^{\mathcal{N}_1' m_1'  \mathcal{N}_2' m_2'}c^{\dag}_{\mathcal{N}_1' m_1'}d^{\dag}_{\mathcal{N}_2' m_2'}d_{\mathcal{N}_2 m_2}c_{\mathcal{N}_1 m_1}\nonumber\\
 &  &+ \frac{1}{2}\sum_{\substack{\mathcal{N}_1, \mathcal{N}_3 \\ \mathcal{N}_1', \mathcal{N}_3'} }\sum_{\substack{m_1, m_3 \\ m_1', m_3'} }\mathcal{U}_{\mathcal{N}_1 m_1   \mathcal{N}_3 m_3}^{\mathcal{N}_1' m_1'  \mathcal{N}_3' m_3'}c^{\dag}_{\mathcal{N}_3' m_3'}c^{\dag}_{\mathcal{N}_1' m_1'}c_{\mathcal{N}_1 m_1}c_{\mathcal{N}_3 m_3}\nonumber\\
&  & +\frac{1}{2}\sum_{\substack{\mathcal{N}_2, \mathcal{N}_3 \\ \mathcal{N}_2', \mathcal{N}_3'} }\sum_{\substack{m_2, m_3 \\ m_2', m_3'} }\mathcal{U}_{\mathcal{N}_2' m_2'   \mathcal{N}_3' m_3'}^{\mathcal{N}_2 m_2  \mathcal{N}_3 m_3}d^{\dag}_{\mathcal{N}_2' m_2'}d^{\dag}_{\mathcal{N}_3' m_3'}d_{\mathcal{N}_3 m_3}d_{\mathcal{N}_2 m_2}.
\end{eqnarray}
It is important to note that the range over which given $\mathcal{N}$ quantum numbers are summed, depends on whether they index an electron or hole operator. 
The \newnot{sym:epstilde}$\tilde{\epsilon}_n$ are the single particle energies, which have been renormalised by self energy corrections (see Section \ref{sec-selfeng} below). I assume that the previously mentioned spin and valley splitting is negligible compared to the LL energy: $\hbar\omega_{v}<\hbar\omega_s<<\epsilon_1$ and hence set $\hbar\omega_{v}= \hbar\omega_s=0$ in all calculations. The only effect of such splittings is to fix the spin and pseudospin quantum numbers for the different ground states. 

The last three terms give the pairwise particle interactions. We assume that the direct interaction with the Dirac sea is cancelled by a neutral positive background charge from the carbon atoms. The $\mathcal{U}$ are the matrix elements of the Coulomb interaction,
$U\left( |\mathbf{r}_1-\mathbf{r}_2|\right)=e^2/\epsilon |\mathbf{r}_1-\mathbf{r}_2|$, in the basis of single electron
wave functions (see Eq.\ (\ref{eq-graspwfs})) defined by\newnot{sym:mathcalU}
 \begin{equation}
\label{eq-uelt} 
\mathcal{U}_{\mathcal{N}_1 m_1   \mathcal{N}_2 m_2}^{\mathcal{N}_1' m_1' 
\mathcal{N}_2' m_2'}=\delta_{s_1 s_1'}\delta_{s_2 s_2'}\int \int d \mathbf{r}_1
^2 d \mathbf{r}_2 ^2\Psi^\dag_{ N_1'}(\mathbf{r}_1)\otimes\Psi^\dag_{
N_2'}(\mathbf{r}_2)U\left( |\mathbf{r}_1-\mathbf{r}_2|\right)\Psi_{ N_1}
(\mathbf{r}_1)\otimes\Psi_{ N_2 } (\mathbf{r}_2),
\end{equation}
where \newnot{sym:Ngrap}$N=\{n m \tau\}$ is a collective index. Note that
$\mathcal{U}_{\mathcal{N}_1 m_1   \mathcal{N}_2 m_2}^{\mathcal{N}_1' m_1' 
\mathcal{N}_2' m_2'} \propto \delta_{s_1 s_1'}\delta_{s_2 s_2'}\delta_{\tau_1
\tau_1'}\delta_{\tau_2 \tau_2'}$; the long-range Coulomb potential cannot
provide enough momentum to scatter between the valleys \cite{AndN98}.
Eq.\ (\ref{eq-uelt}) can be written in terms of the 2DEG matrix elements\newnot{sym:UME}, 
\begin{equation}
\label{eq-u2deg}
U_{n_1 m_1  n_2 m_2}^{n_1' m_1'  n_2' m_2'}=\int \int d \mathbf{r}_1 ^2 d
\mathbf{r}_2 ^2\psi^\ast_{ n_1'm_1'}(\mathbf{r}_1)\psi^\ast_{
n_2'm_2'}(\mathbf{r}_2)U\left( |\mathbf{r}_1-\mathbf{r}_2|\right)\psi_{ n_2 m_2}
(\mathbf{r}_2)\psi_{ n_1 m_1} (\mathbf{r}_1), 
\end{equation}
according to
\begin{eqnarray}
\label{eq-U2DEG} 
\mathcal{U}_{\mathcal{N}_1 m_1   \mathcal{N}_2 m_2}^{\mathcal{N}_1' m_1' 
\mathcal{N}_2' m_2'} & = & \delta_{s_1 s_1'}\delta_{s_2 s_2'} \delta_{\tau_1
\tau_1'}\delta_{\tau_2 \tau_2'} 
          a_{n_1} a_{n_2} a_{n_1'} a_{n_2'}  \times
\biggl[U_{|n_1|  \, m_1 \,\,  |n_2| \,  m_2}^{|n_1'| \, 
m_1' \,\, |n_2'| \,  m_2'} \nonumber\\ 
	  &  & + \mathcal{S}_{n_1}  \mathcal{S}_{n_1'}   U_{|n_1|-1 \, m_1 \,\, |n_2| \, 
m_2}^{|n_1'|-1 \, m_1' \,\, |n_2'| \, m_2'}  
             +  \mathcal{S} _{n_2} \mathcal{S}_{n_2'}   U_{|n_1| \,  m_1 \,\, |n_2|-1 \,
m_2}^{|n_1'|  \, m_1' \,\, |n_2'|-1 m_2'}\nonumber\\ 
          & & +  \mathcal{S}_{n_1} \mathcal{S}_{n_2} \mathcal{S}_{n_1'} \mathcal{S}_{n_2'}  U_{|n_1|-1 \, m_1 \,\,
|n_2|-1
\, m_2}^{|n_1'|-1 \, m_1' \,\, |n_2'|-1 \, m_2'} \biggr].         
\end{eqnarray}
The first of the two-body terms in Eq.\ (\ref{eq-ham}) gives the $e$-$h$
interaction, $\hat{U}_{eh}$, where\newnot{sym:mathcalW}
\begin{equation}
\label{eq-w} 
\mathcal{W}_{\mathcal{N}_1 m_1   \mathcal{N}_2 m_2}^{\mathcal{N}_1' m_1' 
\mathcal{N}_2' m_2'}=
-\mathcal{U}_{\mathcal{N}_1 m_1   \mathcal{N}_2' m_2'}^{\mathcal{N}_1' m_1' 
\mathcal{N}_2 m_2}
+
\mathcal{U}_{\mathcal{N}_1 m_1   \mathcal{N}_2' m_2'}^{\mathcal{N}_2 m_2 
\mathcal{N}_1' m_1'}.
\end{equation}
The two parts correspond to the direct and exchange interactions; the latter
comprises the random phase approximation. The last two terms, $\hat{U}_{ee}$ and
$\hat{U}_{hh}$, give the $e$-$e$ and $h$-$h$ interactions respectively. These
terms are only needed when considering charged CEs. Techniques for calculating the
matrix elements in Eq.\ (\ref{eq-u2deg}) are given in Ref.\ \cite{Dzyhab}} and Appendix \ref{app-mes}.
\section{Geometrical symmetries of the interacting Hamiltonian}
\label{sec-geom_sym}
We wish to diagonalise the Hamiltonian, $\hat{H}$, in order to obtain the
eigenenergies and amplitudes of CEs. This complex problem can be made easier by
determining the maximal set of mutually commuting operators containing
$\hat{H}$, since these operators will have simultaneous eigenstates. According
to Noether's theorem \cite{Tav71}, corresponding to each symmetry of a physical system is a
conserved quantity and by Heisenberg's equation of motion, the operators of
conserved quantities commute with the Hamiltonian. Thus to find a set of
mutually commuting operators, it makes sense to consider the symmetries of the
system. We examine at this point only the geometrical symmetries; the dynamical
SU(4) symmetry arising from spin and pseudospin degrees of freedom is covered in
Chapter \ref{chap-su4}.

Let us consider a clean monolayer of graphene in the symmetric gauge. Then there
is rotational symmetry and the resulting conserved quantity is the total
generalised angular momentum projection, \newnot{sym:Jz}$J_z$. This corresponds to the operator
\newnot{sym:Jzhat}$\hat{J_z}=\sum_i (\hat{j_z})_i$, where $(\hat{j_z})_i$ is the generalised
angular momentum projection for the $i^\mathrm{th}$ electron defined in Eq.\
(\ref{eq-jz}). One can show explicitly $[\hat{J_z},\hat{H}]=0$, confirming that
$J_z$ is a well defined quantity for CEs. For a neutral CE consisting of an electron in
an $\mathcal{N}_1$ state and a hole in an $\mathcal{N}_2$ state,
$J_z=|n_1|-m_1-|n_2|+m_2$ and the sublattice parts cancel. Thus one of the
summations in Eq.\ (\ref{eq-Qdagger}) can be removed and the operator for neutral CEs
can be re-written more simply as\newnot{sym:Qdag}
\begin{equation}
\label{eq-Q2} 
 \mbox{} \hspace{-5pt}
      Q^{\dag}_{\mathcal{ N}_1  \mathcal{ N}_2 J_z }  =
          \sum_{m_1 = 0}^{\infty}
           A_{\mathcal{N}_1 \mathcal{N}_2 J_z }(m_1) \,
          c^{\dag}_{\mathcal{N}_1 m_1} d^{\dag}_{\mathcal{N}_2 m_2(m_1)},
\end{equation}
where $m_2(m_1)=J_z-|n_1|+m_1+|n_2|$. For the $X^-$ state,
$J_z=|n_1|-m_1-|n_2|+m_2+|n_3|-m_3-\frac{1}{2}$, so that one of the summations
in Eq.\ (\ref{eq-Rdagger}) may also be removed.

The quantum number $J_z$ has a direct geometrical meaning for neutral CEs determining
the average positions of the electron and hole relative to the origin. 
Using Eq.\ (\ref{eq-r2}), one may show
\begin{equation}
\label{eq-r2diff}
\langle \mathcal{ N}_1  \mathcal{N}_2 J_z |  \mathbf{r}_h^2 - \mathbf{r}_e^2
|\mathcal{ N}_1  \mathcal{N}_2 J_z \rangle 
= \left( 2 J_z + 4(|n_2| - |n_1|)  + \delta_{n_2,0} - \delta_{n_1,0} \right)
\ell_B^2,
\end{equation}
where $|\mathcal{ N}_1  \mathcal{N}_2 J_z \rangle=Q^{\dag}_{\mathcal{ N}_1 
\mathcal{ N}_2 J_z }|\nu\rangle$.
Therefore, the sign and magnitude of $J_z$ (together with the LL indices $n_1$
and $n_2$), 
determine whether the electron or hole is closer to the origin. This has
physical relevance for the systems studied in Chapter \ref{chap-nces}, which
have a single impurity located at the origin.
For example, for the excitations involving the initial and final states in the
$n=0$ LLs ($n_1=0$ and $n_2=0$),
we get $\langle \mathbf{r}_h^2 - \mathbf{r}_e^2 \rangle_{0 \rightarrow 0} = 2J_z
\ell_B^2$, so that for $J_z  > 0 $, for example, the excitation is 
a low-energy (high-energy) state on the donor (acceptor) impurity. 
For transitions between the zero and the first LL's in Graphene ($n_1=1$ and
$n_2=0$)
we have 
$\langle \mathbf{r}_h^2 - \mathbf{r}_e^2 \rangle_{0 \rightarrow 1} =
(2J_z-3)\ell_B^2$, while 
for transitions between the LL's  $n_2=-1$ and $n_1=0$,  
$\langle \mathbf{r}_h^2 - \mathbf{r}_e^2 \rangle_{-1 \rightarrow 0} =
(2J_z+3)\ell_B^2$. 

\begin{figure}
\centering
  \includegraphics[width=2in]{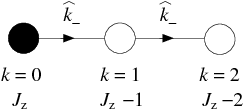}
   \caption{Schematic diagram illustrating how the operator $\hat{k}_-$
simultaneously raises the value of $k$ and lowers the value of $J_z$.}
\label{fig-kdiag} 
\end{figure}

Due to the homogeneity of space, the operator of magnetic translations\newnot{sym:kboldi} is
conserved. For a single particle it is
\begin{equation}
 \label{eq-k}
 \hat{\mathbf{k}}_i=\hat{\mathbf{\Pi}}_i-\frac{\mathrm{q}_i}{c}\hat{\mathbf{r}_i
} \times \mathbf{B}.
\end{equation}
The total operator \newnot{sym:kbold}summed over all electrons is then $\hat{\mathbf{k}}=\sum_i
\hat{\mathbf{k}}_i$. Noting that (in the symmetric gauge only)
$\hat{\mathbf{k}}(\mathbf{B})=\hat{\mathbf{\Pi}}(-\mathbf{B})$ simplifies
calculations. Conservation of $\hat{\mathbf{k}}$ is easily seen from Eq.\
(\ref{eq-k}), since 
\begin{equation}
 \label{eq-kdiff}
 \frac{d\hat{\mathbf{k}}_i}{dt}=\frac{d\hat{\mathbf{\Pi}}_i}{dt}-\frac{\mathrm{q
}_i}{c}\frac{d\hat{\mathbf{r}_i}}{dt}\times \mathbf{B}=0
\end{equation}
by the Lorentz force law; in Eq.\ (\ref{eq-kdiff}) and below $t$ is used to
represent time. This together with the fact that $\hat{\mathbf{k}}$ does not depend
explicitly on time, yields $[\hat{H},\hat{\mathbf{k}}]=0$ by the Heisenberg
equation of motion. However, $[\hat{k}_x,\hat{k}_y] \neq 0$ for charged
complexes and $[\hat{J_z},\hat{\mathbf{k}}] \neq 0$. 
Specifically
\begin{equation}
 \label{eq-kjzcomm}
 [\hat{J_z},\hat{k_x}]=i \hbar\hat{k_y},\hspace{5mm}[\hat{J_z},\hat{k_y}]=-i
\hbar\hat{k_x}.
\end{equation}
Instead one can show $[\hat{J_z},\hat{\mathbf{k}}^2]=0$ and hence $\hat{H}$,
$\hat{J_z}$ and $\hat{\mathbf{k}}^2$ are mutually commuting. The eigenvalues of
$\hat{\mathbf{k}}^2$ are $(2k+1)(\hbar/\ell_B)^2$ with $k = 0, 1, \ldots$
\cite{DzyS00}. $k$ is called the oscillator quantum number \newnot{sym:k} and indeed $k=m$ for
a single particle. Analogously to the single particle case, $k$ determines the
distance from the origin of the guiding centre of a few particle complex.
For a uniform system this leads to the Landau degeneracy in $k$.
Few particle complexes occur in families starting with a seed state, which has
$k=0$ and $J_z=J_z^{(0)}$;\newnot{sym:Jz0} the latter cannot be guessed {\it a priori}.
The offspring $k = 1, 2, \ldots$ are generated by successive action of
the raising ladder operator\newnot{sym:k-} ${\hat{k}_{-}= \frac{l_B}{\sqrt{2}\,\hbar}\left(
\hat{k}_x - i \hat{k}_y\right)}$, since $[\hat{\mathbf{k}}^2,\hat{k}_{-}]=2(\hbar/\ell_B)^2 \hat{k}_{-}$
(see Fig.~\ref{fig-kdiag}).
They have decreasing eigenvalues of $\hat{J}_z$ equal to 
$J_z^{(0)} - k$ because of the relation $\left[ \hat{J}_z,\hat{k}_{-} \right] =
- \hbar \hat{k}_{-}$, which is clear from Eq.\ (\ref{eq-kjzcomm}).
In addition to having the same energies as each other, states in a given family
also exhibit the same optical properties. 

In order to use the $k$ quantum number to remove a degree of freedom, a
Bogoliubov transformation of the basis must be made. This is a challenging
approach, which unfortunately we did not have time to investigate. Translational
symmetry is broken in any numerical calculation, since they can only treat
finite system sizes; the system size is controlled by the value at which
summations over the $m$ quantum number are cut off. Thus even in our
calculations for charged CEs in pristine graphene, $k$ is not strictly speaking an exact quantum number.
However, for large enough system sizes the eigenstates have a large amplitude
for a single $k$ state and in Chapter \ref{chap-cces} we are able to clearly see
the degeneracies described above in the discrete states which appear below the
continuum. All the original work presented here on neutral CEs includes the effects of
a single impurity thus breaking translational symmetry, so $k$ is not a good
quantum number. If we were to remove the impurity, $k$ would in principle be
defined, but then the states would form a continuum hiding the families. In
Section \ref{sec-disp} below, previous work on neutral CEs of clean graphene is
examined, but in this case the Landau gauge is used resulting in different
quantum numbers.
\section{Optical selection rules}
\label{sec-optics}
We begin by determining the perturbation to the single particle Hamiltonian
given in Eq.\ (\ref{eq-blockdirachamB}) by an incoming circularly polarised beam
of light. We work in the electric dipole approximation, ignoring the magnetic
component of the light wave. The electric field for right (left) circularly
polarised light, which I shall refer to as the \newnot{sym:sigmaplus}$\sigma^+$ ($\sigma^-$) 
polarisation, is the real part of the complex field\newnot{sym:Ebold},
$\mathbf{E}=\mathcal{E}e^{-i\omega t}(\mathbf{e}_x \pm i\mathbf{e}_y)$, where
$\mathcal{E}\propto|\mathbf{E}|$\newnot{sym:mathcalE}, $\omega$\newnot{sym:omega} is the angular frequency and $\mathbf{e}_x$
and $\mathbf{e}_y$ are unit vectors in the $x$ and $y$ directions respectively.
Using Weyl's gauge in which the scalar potential vanishes, we obtain
$\mathbf{E}=-\partial (\delta \mathbf{A})/\partial t$, which gives $\delta
\mathbf{A}=-i\mathbf{E}/\omega $, where $\delta \mathbf{A}$ is the correction by
light to the vector potential $\mathbf{A}$ due to the constant magnetic field.
It follows that the correction to the single particle Hamiltonian due to
incoming $\sigma^\pm$ light is\newnot{sym:delHpm}
\begin{eqnarray}
 \label{eq-delh}
\delta \hat{H}_\pm & = & \frac{e v _\mathrm{F}}{c} \left(\begin{array}{cc}
-\sigma_x \delta A_x \pm \sigma_y \delta A_y & 0   \\
0 & -\sigma_x \delta A_x \mp \sigma_y \delta A_y
\end{array}\right) \nonumber\\
& = & \frac{i e v _\mathrm{F} \mathcal{E}}{\omega c} e^{-i\omega t}
\left(\begin{array}{cc}
\sigma_\mp  & 0   \\
0 & \sigma_\pm 
\end{array}\right),
\end{eqnarray}
where \newnot{sym:sigmapmlower}$\sigma_\pm=\sigma_x \pm i \sigma_y$. Here we have assumed that any linear
momentum transfered to the system by a photon is negligible, so the excited
electron cannot change valley, which would require a large momentum boost. In
addition, the interaction with light does not change the spin of an electron, at
least within the electric dipole approximation. Using (un)primed symbols for the
(initial) final state of the electron, these two conservation rules are
expressed as $\tau=\tau'$, $s=s'$.

Using Eqs.\ (\ref{eq-graspwfs}) and (\ref{eq-delh}) the single particle optical
selection rules can be easily found by calculating the dipole matrix elements, since the oscillator strength is directly proportional to the modulus
squared of these matrix elements. The selection rules for the orbital quantum
numbers for an electron, with $\tau=\tau'=\Uparrow$ and illuminated by
$\sigma^+$ light for example, can be found by considering the dipole matrix element
\begin{eqnarray}
\label{eq-dme}
\langle n'\Uparrow s m' \mid \delta \hat{H}_+ \mid n\Uparrow s m \rangle & \propto &
\left( \langle \mid n'|, m'\mid\ \mathcal{S}_{n'}\langle |n'|-1, m'\mid \right)
\left(\begin{array}{cc}
       0 & 0 \\
       1 & 0
      \end{array}
\right)\\
&  & \times
\left( \begin{array}{c}
        \mid |n|, m \rangle\\
         \mathcal{S}_n \mid |n|-1, m\rangle 
       \end{array} \right)	\nonumber\\
& \propto & \mathcal{S}_{n'} \langle |n'|-1, m'\mid |n|, m \rangle \nonumber\\
& \propto & \delta_{|n'|-1,|n|}\delta_{mm'}.
\end{eqnarray}
From this and similar calculations, the single particle optical selection rules
are found to be 
\begin{eqnarray}
\label{eq-spselrules}
\tau=\tau',\hspace{3mm} s=s',\hspace{3mm} m=m', \hspace{3mm} |n'|-|n|=\pm1
\end{eqnarray}
for the $\sigma^\pm$ polarisations, irrespective of the valley index (so long as
it is conserved).

Eq.\ (\ref{eq-spselrules}) is the starting point for calculating the optical
selection rules for CEs. The correction to the many body Hamiltonian in Eq.\
(\ref{eq-ham}) due to incoming $\sigma^\pm$ light is\newnot{sym:delmathcalHpm}
\begin{equation}
\label{eq-colldelH}
\delta \mathcal{H}_\pm=\sum_{\substack{\mathcal{N}, m \\ \mathcal{N'}, m'}
}\langle \mathcal{N}' m'\mid \delta \hat{H}_\pm \mid \mathcal{N} m \rangle
c^\dagger_{\mathcal{N}' m'}d^\dagger_{\mathcal{N} m}.
\end{equation}
It can easily be shown that $\langle \mathcal{ N}_1  \mathcal{
N}_2 J_z  |\delta \hat{\mathcal{H}}_\pm |\nu\rangle \propto \delta_{J_z, \pm1}$, so
that the selection rules for a neutral CE are $J_z=\pm1$ and $S_z=T_z=0$, which are the
spin\newnot{sym:Sz} and pseudospin\newnot{sym:Tz} projections of the CE.
Eq.\ (\ref{eq-colldelH}) will be used to calculate the oscillator strengths for
CEs in later Chapters.
\section{Mixing of excitations}
\label{sec-mixing}
Excitations created by Eq.\ (\ref{eq-Q2}) with the same value of $J_z$, but
different values of $n_1, n_2$ are mixed by the $\hat{U}_{eh}$ interaction term
in the Hamiltonian $\hat{H}$. The true neutral CE is thus given by a superposition of
the different excitations, $|\mathcal{ N}_1  \mathcal{N}_2 J_z \rangle$. This is
best illustrated by an example. Consider illuminating a graphene sample with
filling factor $\nu=-1,\ldots,2$ with $\sigma^+$ light of frequency
$\tilde{\omega}_\mathrm{c}$\newnot{sym:omegactilde} (the tilde denotes renormalisation of the bare
frequency due to $e$-$e$ interactions). Then electrons will be excited from the
$n=0$ LL to the $n=1$ LL ($0\rightarrow1$). These excitations will be mixed to
others with the same $J_z$, for example $-1 \rightarrow 0$ excitations (for
$\nu=-1,0,1$). The latter are in resonance with the $0\rightarrow1$ excitations
and thus significantly mixed. However, excitations that are not in resonance,
$-1\rightarrow2$ for example, are mixed with amplitudes which depend on
$\alpha^\mathrm{g}\sim10^{-1}$. We neglect such virtual transitions and thus
only ever consider the resonant $0\rightarrow1$ and $-1\rightarrow0$
excitations; for the 2DEG this is called the high magnetic field approximation.
Even for the case $\alpha^\mathrm{g}\sim 1$, the mixing in the 2DEG between
non-resonant excitations has been seen to be small \cite{MihF00}. 

\begin{figure}
\centering
  \includegraphics[width=4.5in]{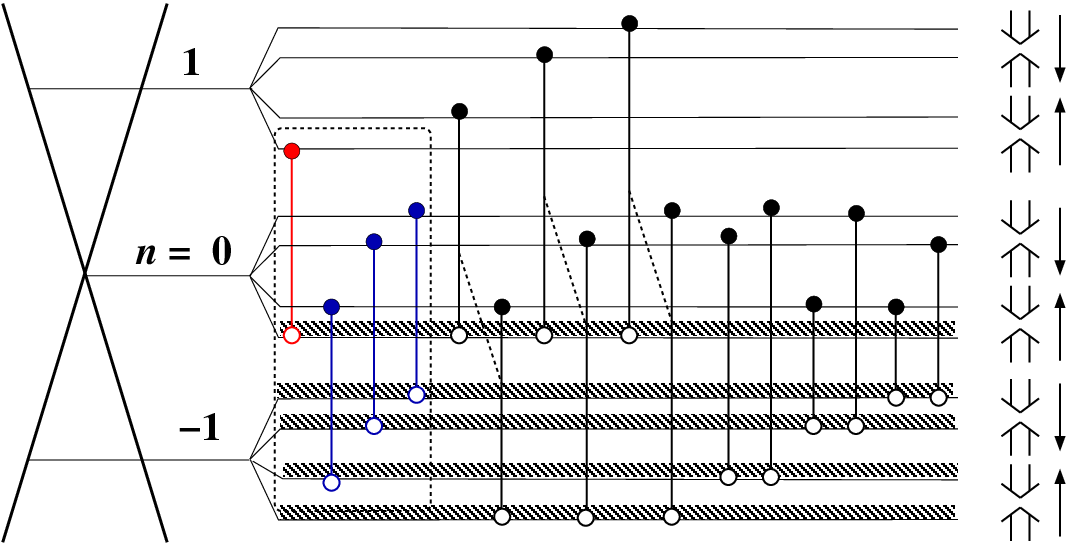}
   \caption{Diagram of lowest energy inter-LL transitions for $\nu=-1$. Shading
indicates the Dirac sea, open circles ($\circ$) holes and closed circles
($\bullet$) electrons. Black transitions are optically dark, whereas coloured
transitions are optically active. Red (blue) indicates transitions which are
active in the $\sigma^+$ ($\sigma^-$) polarisation provided they have $J_z=1$
($J_z=-1$). Excitations in the dashed box are mixed by the exchange Coulomb
interaction, whereas excitations connected by a dashed line are mixed by the
direct interaction.}
\label{fig-allowedtrans} 
\end{figure}

In graphene there are 16 resonant excitations, shown in
Fig.~\ref{fig-allowedtrans} for the particular case $\nu=-1$. The Kronecker's
deltas in Eq.\ (\ref{eq-U2DEG}) together with Eq.\ (\ref{eq-w}) restrict the
possible spin and pseudospin states of mixed transitions. Two excitations, 
$|\mathcal{ N}_1  \mathcal{N}_2 J_z \rangle $ and $|\mathcal{ N}_1' 
\mathcal{N}_2' J_z \rangle$, are mixed by (i) the direct interaction if
$s_1=s_1^{'}$, $\tau_1=\tau_1^{'}$, $s_2=s_2^{'}$ and
$\tau_2=\tau_2^{'}$ and (ii) the exchange interaction if $s_1=s_2$,
$\tau_1=\tau_2$, $s_1^{'}=s_2^{'}$ and $\tau_1^{'}=\tau_2^{'}$. The
energies and eigenstates (the numerical values of $A_{\mathcal{N}_1
\mathcal{N}_2 J_z }$) for CEs are computed numerically by cutting off the
summation over the $m$ quantum numbers in Eq.\ (\ref{eq-ham}) at a finite value
and diagonalising the matrix composed of the matrix elements of the Hamiltonian
\begin{eqnarray}
\label{eq-mes}
\langle \mathcal{N}_1' m_1' \mathcal{N}_2'm_2'|\hat{H}|\mathcal{N}_1 m_1
\mathcal{N}_2 m_2 \rangle & = &
\delta_{m_2, J_z-|n_1|+m_1+|n_2|} \delta_{m_2', J_z-|n_1|'+m_1'+|n_2'|} \\
\nonumber
 & & \times\biggl(\delta_{\mathcal{N}_1 \mathcal{N}_1'}\delta_{m_1
m_1'}\delta_{\mathcal{N}_2 \mathcal{N}_2'}\delta_{m_2
m_2'}(\tilde{\epsilon}_{n_1}-\tilde{\epsilon}_{n_2}
)\\\nonumber
& & + \mathcal{W}_{\mathcal{N}_1 m_1   \mathcal{N}_2 m_2}^{\mathcal{N}_1' m_1' 
\mathcal{N}_2' m_2'}\biggr).
\end{eqnarray}
This truncation of the basis corresponds physically to choosing a finite
system size.

\section{Self energy corrections to single particle energies}
\label{sec-selfeng}
\subsection{Kohn's theorem}
\label{subsec-kohn2deg}
To see how Kohn's theorem breaks down in graphene, let us first examine the
reasons that it holds for the 2DEG. Kohn's theorem \cite{Koh61} states that for
the interacting electron gas in the presence of a magnetic field, the CRE is independent of the interaction strength. The Hamiltonian can
be written as 
\begin{equation}
\label{ham}
\hat{H} =\sum_{i=1}^{N_e}\frac{\hat{\bf{\Pi}}_i^2}{2m_e}+\hat{U},
\end{equation}
where $N_e$ is the number of electrons and $\hat{U}=\sum_{i,j} \hat{u}\left(
\mathrm{\bf{r}}_i-\mathrm{\bf{r}}_j\right)$ is the interaction potential. If the
momentum of the whole system is defined as
$\mathrm{\hat{\bf{\Pi}}}=\sum_i\mathrm{\hat{\bf{\Pi}}}_i$, then it can be shown
that
\begin{equation}
\label{eq-Lorentz}
\frac{d\mathrm{\hat{\bf{\Pi}}}}{dt}=\frac{i}{\hbar}\left[\hat{H},
\hat{\bf{\Pi}}\right] =-\frac{e}{m_e
c}\mathrm{\hat{\bf{\Pi}}}\times\mathrm{\bf{B}},
\end{equation}
which is the many electron version of the Lorentz force law. The derivation of
Eq.\ (\ref{eq-Lorentz}) uses $\left[\hat{U},\hat{\mathbf{\Pi}} \right]=0$, which
holds because $\hat{\mathbf{\Pi}}$ acts only on the centre of mass coordinate,
which $\hat{U}$ is independent of. Defining $\hat{\Pi}_\pm=\hat{\Pi}_x\pm
i\hat{\Pi}_y$ and using Eq.\ (\ref{eq-Lorentz}) gives
\begin{equation}
\label{comm}
\left[\hat{H},\hat{\Pi}_\pm
\right]=\pm\hbar\omega_\mathrm{c}^\mathrm{EG}\hat{\Pi}_\pm.
\end{equation}
Therefore if $\Psi_0$ is an eigenstate of $\hat{H}$ with energy $E^0$,
$\Psi_1=\hat{\Pi}_+\Psi_0$ satisfies
\begin{equation}
\label{eq-psi1}
\hat{H}\Psi_1=\hat{H}\hat{\Pi}_+\Psi_0=(\hbar\omega_\mathrm{c}^\mathrm{EG}\hat{
\Pi}_+ +\hat{\Pi}_+\hat{H})\Psi_0
=(E^0+\hbar\omega_\mathrm{c}^\mathrm{EG})\Psi_1
\end{equation}
and so is also an eigenstate with energy
$E^1=E^0+\hbar\omega_\mathrm{c}^\mathrm{EG}$. The perturbation $\delta
\hat{H}_\pm^\mathrm{EG}$\newnot{sym:deltaHpmEG} to the Hamiltonian $\hat{H}$ due to incoming circularly
polarised light in the $\sigma^\pm$ polarisation can be derived within the
electric dipole approximation in a similar way to that for graphene given by
Eq.\ (\ref{eq-delh}). Assuming light of a long enough wavelength (wavelength in
infra red regime or higher)
\begin{equation}
\label{deltaHEG}
\delta \hat{H}_\pm^\mathrm{EG}=-\frac{ie\mathcal{E}}{m_e \omega c}\hat{\Pi}_\pm
e^{-i\omega t}.
\end{equation}
Eq.\ (\ref{deltaHEG}) indicates that an electron in state $\Psi_0$, which is
excited by a photon can only be promoted to $\Psi_1$, so that there is a sharp
absorption peak at $\omega=\omega_c$. This shows that the CRE is
independent of the interaction strength, proving Kohn's theorem for the 2DEG.

To summarise, the CRE for the 2DEG is unaffected by interactions due to Eqs.\
(\ref{eq-psi1}) and (\ref{deltaHEG}). For the case of graphene, we see from Eq.\
(\ref{eq-delh}), that the Hamiltonian, $\delta \hat{H}_\pm$, describing the
coupling between light and electrons, is directly proportional to $\sigma_+$ or
$\sigma_-$, depending on the valley and polarisation direction. However,
$[\hat{H}_{\rm D},\sigma_\pm]$ is not directly proportional to
$\sigma_\pm$, where $\hat{H}_{\rm D}$ is the graphene Hamiltonian defined in Eq.\ (\ref{eq-diracham}), so there is no graphene analogy to Eq.\ (\ref{comm}). Thus we
conclude that optical excitation energies in graphene can be renormalised by
$e$-$e$ interactions and proceed to calculate these self energy corrections.
This topic has been addressed in several theoretical papers
\cite{IyeWFB07,BycM08,BarLNM09,Shi10}.
\subsection{Renormalisation of single particle energies}
\label{subsec-secorrecs}
Consider the ground state of graphene in a magnetic field with an integer filling
factor. An electron in a given LL, $n$, will interact via the exchange
interaction with each other electron in the system with the same spin and
pseudospin, irrespective of what LL they are in. The two-electron state remains
unchanged by such an interaction. The overall effect is that the energy of this
electron is lowered below the bare single particle energy, $\epsilon_n$. Let's
derive this self energy correction using the symmetric gauge. We have seen that
the exchange interaction energy for an electron in state $\mathcal{N}_1m_1$ and
a hole in state $\mathcal{N}_2m_2$ is given by the matrix element
$\mathcal{U}_{\mathcal{N}_1 m_1   \mathcal{N}_2 m_2}^{\mathcal{N}_2 m_2 
\mathcal{N}_1 m_1}$. Analogously, the interaction energy due to exchange between
two electrons in states, $\mathcal{N} m$ and $\mathcal{N}'m'$ is given by
$-\mathcal{U}_{\mathcal{N} m   \mathcal{N}' m'}^{\mathcal{N}' m'  \mathcal{N}
m}$, where the sign change is due to switching between a positively charged hole
and a negatively charged electron. Indeed, self energy corrections are easier to
visualise by thinking in the electron representation as opposed to the usual
$e$-$h$ representation. Since $\mathcal{U}_{\mathcal{N} m   \mathcal{N}'
m'}^{\mathcal{N}' m'  \mathcal{N} m}>0$, the exchange interaction between two
electrons is attractive. One may assign the energy
$-\frac{1}{2}\mathcal{U}_{\mathcal{N} m   \mathcal{N}' m'}^{\mathcal{N}' m' 
\mathcal{N} m}$ to each electron. From Eq.\ (\ref{eq-uelt}),
$\mathcal{U}_{\mathcal{N} m   \mathcal{N}' m'}^{\mathcal{N}' m'  \mathcal{N}
m}\propto \delta_{ss'} \delta_{\tau \tau'}$ and $\mathcal{U}_{n \tau s
m\hspace{2mm}   n'\tau s m'}^{n' \tau s m' \hspace{1mm}   n \tau s m}$ depends
only on $n,m,n',m'$, not on the specific value of $\tau,s$. Let
$E_{\mathrm{SE}}\left( n,n'\right)$\newnot{sym:ESE} denote the correction to the energy of an
electron in the $n^{\rm th}$ LL due to exchange with electrons in the $n'^{\rm th}$ LL,
so that
\begin{equation}
\label{eq-se}
E_{\mathrm{SE}}\left( n,n'\right)=-\frac{1}{2}\sum_{m'}\mathcal{U}_{n \tau s
m\hspace{2mm}   n'\tau s m'}^{n' \tau s m' \hspace{1mm}   n \tau s m}.
\end{equation}
Clearly $E_{\mathrm{SE}}\left( n,n'\right)=E_{\mathrm{SE}}\left( n',n\right)$.
As usual, the graphene matrix elements can be expressed in terms of 2DEG matrix
elements according to Eq.\ (\ref{eq-U2DEG}). Details of how to calculate
$E_{\mathrm{SE}}\left( n,n'\right)$ are given in Appendix \ref{app-mes}, where it is shown
to be independent of $m$. The complete self energy correction for an electron in
the $n^{\rm th}$ LL is of course given by summing $E_{\mathrm{SE}}\left(
n,n'\right)$ over all filled LLs $n'$. We shall see that the bare single
particle energies are renormalised differently for different $n$, leading to a
renormalisation of the optical excitation energies. 

Let us now focus on the particular example of $\nu=-2,\ldots2$ and calculate the
self energies for different inter-LL transitions. The possible transitions for
${\nu=-1}$ are shown in Fig.\ \ref{fig-allowedtrans}; they fall into two different
classes with the members of each class having the same self energy correction.
All self energy corrections for transitions occuring at other partial fillings
of the $n=0$ LL are identical to the self energies of one of these classes. To
proceed we need the following results, which hold for $n\ge1$:
\begin{equation}
\label{eq-se00}
E_{\mathrm{SE}}\left( 0,0\right)=-E_0,
\end{equation}
\begin{equation}
\label{eq-se0n}
E_{\mathrm{SE}}\left(
0,-n\right)=-\frac{E_0}{2\sqrt{\pi}}\frac{\Gamma\left(n+\frac{1}{2}\right)}{n!},
\end{equation}
\begin{equation}
\label{eq-se10}
E_{\mathrm{SE}}\left( 1,0\right)=-\frac{E_0}{4},
\end{equation}
\begin{equation}
\label{eq-se1n}
E_{\mathrm{SE}}\left(
1,-n\right)=\frac{E_0}{8\sqrt{\pi}}\frac{\Gamma\left(n+\frac{1}{2}\right)}{
\Gamma\left(n+1\right)}\frac{4\sqrt{n}-8n+1}{2n-1},
\end{equation}
\begin{equation}
\label{eq-sneg1n}
E_{\mathrm{SE}}\left(
-1,-n\right)=-\frac{E_0}{8\sqrt{\pi}}\frac{\Gamma\left(n+\frac{1}{2}\right)}{n!}
\frac{8n+4\sqrt{n}-1}{2n-1},
\end{equation}
where the Gamma function is as defined in Eq.\ (\ref{eq-gammafunc}).
In general, higher lying LLs are lowered less by the exchange compared to lower
lying LLs, for example $|E_{\mathrm{SE}}\left(
1,0\right)|<|E_{\mathrm{SE}}\left( 0,0\right)|$. This means that although the
self energy correction to each bare LL energy is negative, the overall self
energy correction to the optical excitation energy is positive.

Let's first consider the transitions with no spin or pseudospin flips, enclosed
by the dashed box in Fig. \ref{fig-allowedtrans}, beginning with the $0 \to 1$
transition. The total self energy correction to the transition energy is\newnot{sym:Entaus}
\begin{equation}
\label{eq-ee01}
E_{0 \tau s}^{1 \tau
s}=\delta\epsilon_1-\delta\epsilon_0=\sum_{n=0}^{n_\mathrm{c}}\left(E_{\mathrm{
SE}}\left( 1,-n\right)
-E_{\mathrm{SE}}\left( 0,-n\right)\right),
\end{equation}
where $n_\mathrm{c}$ is the value of $n$ at which the sum is cut off. This
should correspond to the region in the zero magnetic field case, where the
dispersion relation deviates significantly from being linear (see Fig.\
\ref{fig-spdispersion}). According to the derivation of Eq.\
(\ref{eq-lowengdisp}), this occurs at the critical wave vector\newnot{sym:qc},
$q_\mathrm{c}\sim\frac{1}{a}$. 
Hence an appropriate value of $n_\mathrm{c}$ can be determined from
\begin{equation}
\label{eq-nc1}
\hbar v_\mathrm{F} \frac{1}{a}=\sqrt{2}\frac{\hbar
v_\mathrm{F}}{\ell_B}\sqrt{n_\mathrm{c}},
\end{equation}
which yields\newnot{sym:nc}
\begin{equation}
\label{eq-nc2}
n_\mathrm{c}\approx \frac{16300}{B},
\end{equation}
where it is understood here that $B$ is the magnetic field in Tesla. This
approach is similar to that used in Ref.\ \cite{BarLNM09} to estimate
$n_\mathrm{c}$, although there the authors used
$q_\mathrm{c}\sim\frac{1}{\sqrt{3}a}$, where $\sqrt{3}a$ corresponds to the
lattice parameter of one of the triangular graphene sublattices. The estimates
for $n_\mathrm{c}$ derived here and in Ref.\ \cite{BarLNM09} are both lower than
that given in Ref.\ \cite{IyeWFB07}. Using Eqs.\ (\ref{eq-se00})-(\ref{eq-se1n})
one can show
\begin{equation}
\label{eq-ee012}
E_{0 \tau s}^{1 \tau
s}=\frac{3}{4}E_0+\frac{E_0}{8\sqrt{\pi}}\sum_{n=1}^{n_\mathrm{c}}\frac{
\Gamma\left(n+\frac{1}{2}\right)}{n!}\frac{4\sqrt{n}-3}{2n-1}.
\end{equation}
At large values of $n_\mathrm{c}$
\begin{equation}
\label{eq-secorrhighnc}
E_{0 \tau s}^{1 \tau s} \approx \dfrac{E_0}{4\sqrt{\pi}}\left\lbrace
\sum_{n=1}^\infty\left[\dfrac{\sqrt{n}\Gamma\left( n-\frac{1}{2}\right)
}{n!}-\dfrac{1}{n}
\right]+\gamma+\dfrac{3\sqrt{\pi}}{2}+\mathrm{ln}n_\mathrm{c}\right\rbrace,
\end{equation}
where $\gamma$\newnot{sym:Eulergamma} is Euler's constant. Hence $E_{0 \tau s}^{1 \tau s}$ diverges
logarithmically with $n_\mathrm{c}$ and for very large values of $n_\mathrm{c}$
is hardly altered by small changes in $n_\mathrm{c}$, which is a pleasing
result. Reassuringly Eq.\ (\ref{eq-secorrhighnc}) agrees with the result in
Ref.\ \cite{IyeWFB07}, where it was calculated using the Landau gauge. For
$B=20T$, $n_\mathrm{c}\approx 810$ and $E_{0 \tau s}^{1 \tau s}\approx 1.56E_0$.

One can show $E_{0 \tau s}^{1 \tau s}=E_{-1 \tau s}^{0 \tau s}$, which is due to
the particle-hole symmetry in graphene. Hence the energies of collective
excitations comprised of the four excitations without spin or pseudospin flips
are only shifted by a constant due to self energy corrections. It is clear by
studying Fig.\ \ref{fig-allowedtrans} (see page \pageref{fig-allowedtrans}) that all the transitions which are not
mixed have the same self energy correction, $E_{0 \tau s}^{1 \tau s}$, as the
four transitions which are mixed by the exchange interaction. Further
calculations show
\begin{equation}
\label{eq-semixedpairs}
E_{0 \Uparrow \uparrow}^{1 \Downarrow \uparrow }=E_{0 \Uparrow \uparrow}^{1
\Uparrow \downarrow }=E_{0 \Uparrow \uparrow}^{1 \Downarrow \downarrow }
=E_{-1 \Uparrow \uparrow}^{0 \Downarrow \uparrow }=E_{-1 \Uparrow \uparrow}^{0
\Uparrow \downarrow }=E_{-1 \Uparrow \uparrow}^{0 \Downarrow \downarrow }
=E_{0 \tau s}^{1 \tau s}+\frac{E_0}{4},
\end{equation}
so that all transitions, which are mixed to exactly one other transition by the
direct interaction, have the same self energy as each other. 

Let's consider the self energy correction for another transition with a
different bare energy for comparison. For simplicity I choose the filling factor
$\nu=-2$, so the Fermi energy lies between the $n=-1$ and $n=0$ LLs. All sixteen
possible $-1 \to 0$ transitions will then have the same self energy correction
$E_{-1 \tau s}^{0 \tau s}$. The $-1 \to 2$ transition has the lowest bare
transition energy, $\hbar \omega_\mathrm{c}(\sqrt{2}+1)$, following the $-1 \to
0$ transition. The self energy correction for the $-1 \to 2$ transition is
\begin{eqnarray}
\label{eq-senegonetwo}
E_{-1 \tau_2 s_2}^{2 \tau_1 s_1}&=&\delta\epsilon_2-\delta\epsilon_{-1}\\
\nonumber
&=&\sum_{n=1}^{n_\mathrm{c}}\left(E_{\mathrm{SE}}\left(
2,-n\right)-E_{\mathrm{SE}}\left( -1,-n\right)\right)\\ \nonumber
&=&\frac{E_0}{32\sqrt{\pi}}\Biggl\{2\sqrt{2}\sqrt{\pi}+\frac{29}{2}\sqrt{\pi}\\
\nonumber
& & + \sum_{n=2}^{n_\mathrm{c}}\frac{\Gamma\left(n+\frac{1}{2}\right)}{
\Gamma\left(n+1\right)}\left[\frac{32(\sqrt{2}+1)n^{\frac{3}{2}}-16n-(36\sqrt{2}
+48)n^{\frac{1}{2}}+3}{\left(2n-1\right)\left(2n-3\right)}\right]\Biggr\}.
\end{eqnarray}
For $B=20T$, $n_\mathrm{c}\approx 810$ and $E_{-1 \tau_2 s_2}^{2 \tau_1
s_1}\approx 2.69E_0$. Note that the self energy corrections do not scale in the
same way as the single particle energies.
%
%
\section{Dispersion relation for magnetoexcitons}
\label{sec-disp}
Below I present the derivation of the magnetoexcitonic dispersion relation in
graphene following in part the approach outlined in \cite{IyeWFB07}. It is
similar to that used for the 2DEG \cite{KalH84}. The non-interacting $e$-$h$
wave function, which depends on momentum, is found by manipulating the
Hamiltonian through coordinate transformations and a gauge transformation until
it can be formulated in terms of Bose ladder operators. This is used to find the
second quantised operator, which creates such an $e$-$h$ pair out of the
Dirac sea. The dispersion relation is found by diagonalising the Coulomb interaction matrix in the
basis of these non-interacting two-particle wave functions.
\subsection{Derivation of non-interacting wave function}
\label{subsec-niwf}
We begin by finding the non-interacting $e$-$h$ wave function using the Landau
gauge. Let's assume both the electron and hole are in the $\mathbf{K}$ valley.
As previously remarked, the long-range Coulomb interaction does not scatter
between the valleys. The valley index of either particle can be changed by
interchanging the $A$ and $B$ sublattice components as usual. The two particle
Hamiltonian is then
\begin{eqnarray}
 \label{eq-niham}
\hat{H}_0 & = & \hat{H}_{0e} \otimes \eins_2 +\eins_2 \otimes \hat{H}_{0h} \\
\nonumber
& = & \left(\begin{array}{cccc}
            0 & \hat{H}_{0h}^{12} & \hat{H}_{0e}^{12} & 0 \\
	    \hat{H}_{0h}^{21} & 0 & 0 & \hat{H}_{0e}^{12} \\
	    \hat{H}_{0e}^{21} & 0 & 0 & \hat{H}_{0h}^{12} \\
	    0 & \hat{H}_{0e}^{21} & \hat{H}_{0h}^{21} & 0
            \end{array}
 \right),
\end{eqnarray}
where $\hat{H}_{0e}(\mathbf{r}_1)$ and $\hat{H}_{0h}(\mathbf{r}_2)$ are the
single particle electron and hole Hamiltonians respectively, as given by Eq.\
(\ref{eq-diracham}). After making the coordinate transformation\newnot{sym:Rcom}
\begin{equation}
 \label{eq-coordtransform}
\mathbf{R}=R_x\mathbf{e}_x+R_y\mathbf{e}_y=\frac{\mathbf{r}_1+\mathbf{r}_2}{2},
\hspace{3mm}
\mathbf{r}=r_x\mathbf{e}_x+r_y\mathbf{e}_y=\mathbf{r}_1-\mathbf{r}_2
\end{equation}
and the gauge transformation $U=\mathrm{exp}\left(-\frac{iR_xr_y}{\ell_B^2}
\right)$, we obtain $\hat{\tilde{H}}_0=U^\dagger\hat{H}_0 U$. Now
$\hat{\tilde{H}}_0$ satisfies $[\hat{\tilde{H}}_0,-i\hbar \nabla_\mathbf{R}]=0$,
so that the canonical centre of mass momentum, $\mathbf{P}$\newnot{sym:Pbold}, is a good quantum
number. Since we wish to find simultaneous eigenstates of $\hat{\tilde{H}}_0$
and $-i\hbar \nabla_\mathbf{R}$, we replace the operator $-i\hbar
\nabla_\mathbf{R}$ in $\hat{\tilde{H}}_0$ by the quantity $\mathbf{P}$. The
$\mathbf{R}$ dependence of eigenstates of $\hat{\tilde{H}}_0$ is then included
by the factor $\mathrm{exp}(i \mathbf{P}.\mathbf{R}/\hbar)$. The dependence of
$\hat{\tilde{H}}_0$ on $\mathbf{P}$ can be removed by making another coordinate
transformation 
\begin{equation}
\label{eq-rtilde}
\tilde{\mathbf{r}}=\mathbf{r}-\ell_B^2 (\mathbf{e}_z \times \mathbf{P})/ \hbar.
\end{equation}
This gives
\begin{equation}
 \label{eq-nihamops}
\hat{\tilde{H}}_0=\hbar \omega_\mathrm{c} \left(\begin{array}{cccc}
                       0 & \tilde{b}^\dagger & \tilde{a} & 0 \\
                       \tilde{b} & 0 & 0 & \tilde{a} \\
                       \tilde{a}^\dagger & 0 & 0 & \tilde{b}^\dagger \\
                       0 & \tilde{a}^\dagger & \tilde{b} & 0
                        \end{array}
\right),
\end{equation}
where
\begin{equation}
 \label{eq-adaggertilde}
\tilde{a}^\dagger=\frac{i}{\sqrt{2}}\left(\frac{\tilde{z}}{2 \ell_B}-2\ell_B
\frac{\partial}{\partial \tilde{z}^\ast}  \right), \hspace{3mm}
\tilde{b}^\dagger=-\frac{i}{\sqrt{2}}\left(\frac{\tilde{z}^\ast}{2
\ell_B}-2\ell_B \frac{\partial}{\partial \tilde{z}}  \right)
\end{equation}
and $\tilde{z}=\tilde{r}_x+i\tilde{r}_y$. Notice that the operators
$\tilde{a}^\dagger$, $\tilde{b}^\dagger$ are the same up to a phase as those
defined in Eqs.\ (\ref{eq-ladopsa}) and (\ref{eq-ladopsb}):
$\tilde{a}^\dagger=-a^\dagger$, $\tilde{b}^\dagger=-ib^\dagger$. Hence
\begin{eqnarray}
\label{eq-phitilde}
\tilde{\psi}_{nm}\left(\tilde{\mathbf{r}} \right) & = &\langle
\tilde{\mathbf{r}}|\frac{1}{\sqrt{ n! m!}}\left(\tilde{a}^\dagger \right)^n
\left(\tilde{{b}}^\dagger \right)^m|00\rangle \nonumber \\
& = & (-1)^{n+m}i^m \psi_{nm}\left(\tilde{\mathbf{r}} \right).
\end{eqnarray}
We can now construct the normalised $e$-$h$ eigenstates of $\hat{H}_0$ as
\begin{equation}
\label{eq-eheigstate}
\varphi_{nm}\left( \mathbf{r},\mathbf{R},\mathbf{P}
\right)=\frac{1}{\sqrt{S}}a_n a_m\mathrm{exp}\left(-\frac{iR_x r_y}{\ell_B^2}
\right)\mathrm{exp}(i \mathbf{P}.\mathbf{R}/ \hbar)
\left(\begin{array}{c}
 \mathcal{S}_n \tilde{\psi}_{|n|-1,|m|} \left(\tilde{\mathbf{r}} \right) \\
 \mathcal{S}_n \mathcal{S}_m \tilde{\psi}_{|n|-1,|m|-1} \left( \tilde{\mathbf{r}} \right) \\
\tilde{\psi}_{|n|,|m|} \left(\tilde{\mathbf{r}} \right) \\
\mathcal{S}_m \tilde{\psi}_{|n|,|m|-1} \left( \tilde{\mathbf{r}} \right)
      \end{array}
\right).
\end{equation}
These have corresponding eigenenergies $(\mathcal{S}_n \sqrt{|n|}+\mathcal{S}_m\sqrt{|m|})\hbar
\omega_\mathrm{c}$, which is the sum of the electron and hole single particle
energies, as one would expect. Note that $m$ here represents the LL index of the
hole, not an oscillator quantum number as in the single particle case. Thus it
is more appropriate to use the notation $n_1,n_2$ instead of $n,m$. In addition,
the order of the components of the coordinate part of the wave function depend
on the pseudospins of the electron and hole and of course both particles also
have spin. Hence we switch to the notation \newnot{sym:varphin1n2}$\varphi_{\mathcal{N}_1\mathcal{N}_2
}\left( \mathbf{r},\mathbf{R},\mathbf{P} \right)$, which is connected to the old
notation via $\varphi_{n_1 \Uparrow s_1 n_2 \Uparrow s_2}\left(
\mathbf{r},\mathbf{R},\mathbf{P} \right)=\varphi_{n_1 n_2 }\left(
\mathbf{r},\mathbf{R},\mathbf{P} \right)\chi_{s_1}\chi_{s_2}$.
The analogous $e$-$h$ wave function for the 2DEG\newnot{sym:varphin1n2eg} is very similar to Eq.\
(\ref{eq-eheigstate}):
\begin{equation}
\label{eq-exwf2deg}
\varphi_{n_1 n_2}^\mathrm{EG}\left( \mathbf{r},\mathbf{R},\mathbf{P}
\right)=\frac{1}{\sqrt{S}}\mathrm{exp}\left(-\frac{iR_x r_y}{\ell_B^2}
\right)\mathrm{exp}(i \mathbf{P}.\mathbf{R}/ \hbar)\tilde{\psi}_{n_1 n_2}
\left(\tilde{\mathbf{r}} \right).
\end{equation}
\subsection{Determining the dispersion relation}
\label{subsec-disp}
The next step is to express the state $\varphi_{\mathcal{N}_1
\mathcal{N}_2}\left( \mathbf{r},\mathbf{R},\mathbf{P} \right)$ in second
quantised form. In other words we would like to find the operator\newnot{sym:QdagP}\footnote{Note that $J_z$ is not a good quantum number for these CEs, since we are using the Landau gauge.}
\begin{equation}
\label{eq-qdaggerlandau}
Q^\dagger_{\mathcal{N}_1 \mathcal{N}_2}(\mathbf{P})=\sum_{X_1,X_2}
A_{\mathcal{N}_1 X_1,\mathcal{N}_2 X_2}(\mathbf{P})c^\dagger_{\mathcal{N}_1 
X_1}d^\dagger_{\mathcal{N}_2 X_2},
\end{equation}
which satisfies 
\begin{eqnarray}
\label{eq-aa}
\langle \mathbf{r},\mathbf{R} |Q^\dagger_{\mathcal{N}_1
\mathcal{N}_2}(\mathbf{P})|\nu \rangle&=&
\sum_{X_1,X_2} A_{\mathcal{N}_1 X_1,\mathcal{N}_2 X_2}(\mathbf{P})
\Phi^{(e)}_{n_1 X_1 \tau_1}(\mathbf{r}_1) \otimes\Phi^{(h)}_{n_2 X_2
\tau_2}(\mathbf{r}_2)\chi_{s_1} \chi_{s_2} \nonumber \\&=&\varphi_{\mathcal{N}_1
\mathcal{N}_2}\left( \mathbf{r},\mathbf{R},\mathbf{P} \right).
\end{eqnarray}
The creation operators for electron and hole, \newnot{sym:cdagntausX}$c^\dagger_{n_1 \tau_1 s_1 X_1}$
and \newnot{sym:ddagntausX}$d^\dagger_{n_2 \tau_2 s_2 X_2}$ respectively, are the Landau gauge analogy
of those defined for the symmetric gauge. The $A_{\mathcal{N}_1
X_1,\mathcal{N}_2 X_2}(\mathbf{P})$ are found by using the orthonormality of the
$\phi_{nX}^{(e)}\left( \mathbf{r}\right)$ and $\phi_{nX}^{(h)}\left(
\mathbf{r}\right)$, giving
\begin{equation}
\label{eq-qdaggersimple}
Q^\dagger_{\mathcal{N}_1 \mathcal{N}_2}(\mathbf{P})=\frac{1}{\sqrt{N_0}}\sum_{X}
\mathrm{exp}\left(i \frac{X P_x}{\hbar}\right)c^\dagger_{\mathcal{N}_1  X-P_y
\ell_B^2/2\hbar}d^\dagger_{\mathcal{N}_2 X+P_y \ell_B^2/2\hbar}.
\end{equation}
This is useful, since we already have the second quantised form of the
interacting Hamiltonian. It is given in Eq.\ (\ref{eq-ham}) for the symmetric
gauge, but the corresponding operator for the Landau gauge is simply obtained by
changing all $m \to X$ in Eqs.\ (\ref{eq-ham})-(\ref{eq-w}), all $\Psi \to \Phi$
in Eq.\ (\ref{eq-uelt}) and all $\psi \to \phi$ in Eq.\ (\ref{eq-u2deg}). Let
$|\mathcal{N}_1 \mathcal{N}_2 \mathbf{P} \rangle=Q^\dagger_{\mathcal{N}_1
\mathcal{N}_2}(\mathbf{P})|\nu\rangle$. As discussed in Section
\ref{sec-mixing}, several states $|\mathcal{N}_1 \mathcal{N}_2 \mathbf{P}
\rangle$ are mixed together by the Coulomb interaction to form the CE. The
dispersion relation is then given by diagonalising the matrix with elements
\begin{eqnarray}
\label{eq-melandau}
\langle \mathcal{N}_1'\mathcal{N}_2' \mathbf{P'}|\hat{H}|\mathcal{N}_1
\mathcal{N}_2 \mathbf{P}\rangle & = &
\frac{\delta_{\mathbf{P}\mathbf{P'}}}{N_0}\sum_{X X'}
e^{\frac{i}{\hbar}(X-X')P_x} \biggl[\delta_{\mathcal{N}_1
\mathcal{N}_1'}\delta_{\mathcal{N}_2 \mathcal{N}_2'}\delta_{X X'}\\ \nonumber
& & \times(\tilde{\epsilon}_{n_1}-\tilde{\epsilon}_{n_2})
  +  \mathcal{W}_{\mathcal{N}_1 X-P_y \ell_B^2/2\hbar\hspace{1mm}  \mathcal{N}_2
 X+P_y \ell_B^2/2\hbar}^{\mathcal{N}_1' X'-P_y \ell_B^2/2\hbar \hspace{1mm} 
\mathcal{N}_2' X'+P_y \ell_B^2/2\hbar}\biggr]\\ \nonumber
& = & \delta_{\mathbf{P}\mathbf{P'}} \biggl[\delta_{\mathcal{N}_1
\mathcal{N}_1'}\delta_{\mathcal{N}_2
\mathcal{N}_2'}(\tilde{\epsilon}_{n_1}-\tilde{\epsilon}_{n_2})
 +  \tilde{\mathcal{U}}_{ \hspace{2mm}\mathcal{N}_1 \mathcal{N}_2}^{d
\hspace{1mm} \mathcal{N}_1' \mathcal{N}_2' }
+  \tilde{\mathcal{U}}_{ \hspace{2mm}\mathcal{N}_1 \mathcal{N}_2}^{x
\hspace{1mm} \mathcal{N}_1' \mathcal{N}_2' }
\biggr]
\end{eqnarray}
The first term in Eq.\ (\ref{eq-melandau}) represents the exchange renormalised
single particle energies and is independent of $\mathbf{P}$. The latter two
terms are not and give the direct and exchange contributions respectively. They
are defined as\newnot{sym:Utilded}
\begin{eqnarray}
\label{eq-dispdir}
\tilde{\mathcal{U}}_{ \hspace{2mm}\mathcal{N}_1 \mathcal{N}_2}^{d \hspace{1mm}
\mathcal{N}_1' \mathcal{N}_2' }
=-\frac{1}{N_0}\sum_{X X'} e^{\frac{i}{\hbar}(X-X')P_x}  
\mathcal{U}_{\mathcal{N}_1 X-P_y \ell_B^2/2\hbar\hspace{1mm}  \mathcal{N}_2' 
X'+P_y \ell_B^2/2\hbar}
^{\mathcal{N}_1' X'-P_y \ell_B^2/2\hbar \hspace{1mm}  \mathcal{N}_2 X+P_y
\ell_B^2/2\hbar}
\end{eqnarray}
and\newnot{sym:Utildex}
\begin{eqnarray}
\label{eq-dispex}
\tilde{\mathcal{U}}_{ \hspace{2mm}\mathcal{N}_1 \mathcal{N}_2}^{x \hspace{1mm}
\mathcal{N}_1' \mathcal{N}_2' }
=\frac{1}{N_0}\sum_{X X'} e^{\frac{i}{\hbar}(X-X')P_x}  
\mathcal{U}_{\mathcal{N}_1 X-P_y \ell_B^2/2\hbar\hspace{1mm}  \mathcal{N}_2' 
X'+P_y \ell_B^2/2\hbar}
^{ \mathcal{N}_2 X+P_y \ell_B^2/2\hbar \hspace{1mm} \mathcal{N}_1' X'-P_y
\ell_B^2/2\hbar}.
\end{eqnarray}
The graphene matrix elements, $\mathcal{U}$, can be expressed as a combination
of 2DEG matrix elements, $U$, as we saw for the symmetric gauge in Eq.\
(\ref{eq-U2DEG}). Hence Eqs.\ (\ref{eq-dispdir}) and (\ref{eq-dispex}) can be
written in terms of $\tilde{U}_{ \hspace{2mm}n_1 n_2}^{d \hspace{1mm} n_1' n_2'
}$ and $\tilde{U}_{ \hspace{2mm}n_1 n_2}^{x \hspace{1mm} n_1' n_2' }$, which are
defined by changing $\mathcal{N} \to n$ and $\mathcal{U} \to U$ in Eqs.\
(\ref{eq-dispdir}) and (\ref{eq-dispex}) respectively. One can show
\begin{equation}
\label{eq-dispdir2deg}
\tilde{U}_{ \hspace{2mm}n_1 n_2}^{d \hspace{1mm} n_1' n_2'
}=(-1)^{n_2+n_2'+1}\int d^2\mathbf{r}
\tilde{\psi}_{n_1' n_2'}^\ast \left(\mathbf{r}
\right)U\left(|\mathbf{r}+\ell_B^2 (\mathbf{e}_z \times \mathbf{P})/ \hbar |
\right)\tilde{\psi}_{n_1 n_2}\left(\mathbf{r} \right),
\end{equation}
which is derived using
\begin{equation}
\label{eq-ehwf2degsp}
\varphi_{n_1 n_2}^\mathrm{EG}\left( \mathbf{r},\mathbf{R},\mathbf{P}
\right)=\frac{1}{\sqrt{N_0}}\sum_X \mathrm{exp}\left(i \frac{X
P_x}{\hbar}\right)
\phi_{n_1 X-P_y \ell_B^2/2\hbar}^{(e)}\left( \mathbf{r}_1 \right)\phi_{n_2 X+P_y
\ell_B^2/2\hbar}^{(h)}\left( \mathbf{r}_2 \right)
\end{equation}
and Eq.\ (\ref{eq-exwf2deg}). The exchange interaction can also be simplified as
\begin{equation}
\label{eq-dispex2deg}
\tilde{U}_{ \hspace{2mm}n_1 n_2}^{x \hspace{1mm} n_1' n_2'
}=(-1)^{n_1+n_1'}\tilde{U}(P/\hbar)\tilde{\psi}_{n_1 n_2}^\ast
\left(\mathbf{e}_z \times\mathbf{P} \right)\tilde{\psi}_{n_1' n_2'}
\left(\mathbf{e}_z \times\mathbf{P} \right).
\end{equation}
From Eq.\ (\ref{eq-dispdir2deg}) one can see that taking the matrix element of
the Coulomb interaction with respect to the two-body wave function,
$\varphi_{\mathcal{N}_1\mathcal{N}_2 }\left( \mathbf{r},\mathbf{R},\mathbf{P}
\right)$, yields the direct interaction only. To obtain both the exchange and
direct terms, the many body approach described here is necessary.
\subsection{Worked examples}
\label{subsec-dispeg}
\subsubsection{Intra-LL excitations}
\label{subsubsec-dispintra}
\begin{figure}
\centering
 \includegraphics[width=3.5in]{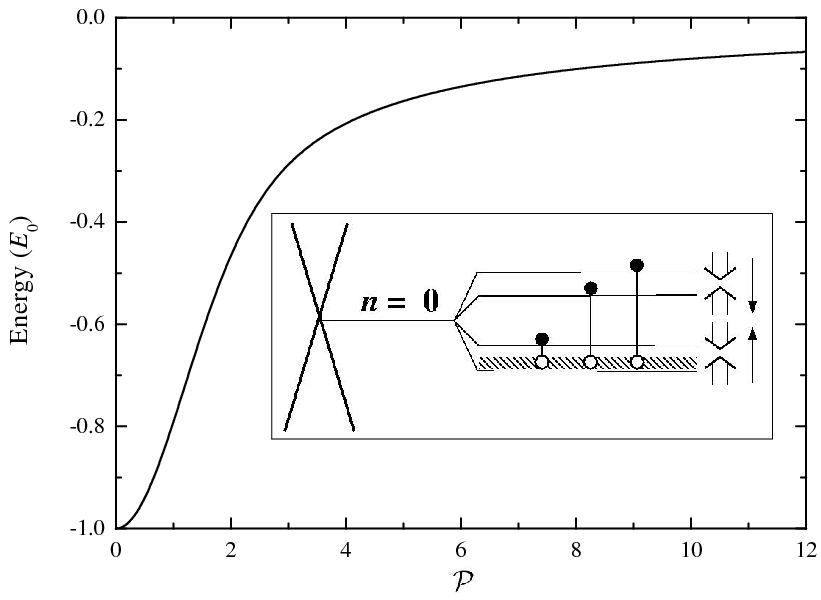}
   \caption{Dispersion relation for transitions within the $n=0$ LL, given by the expression $\tilde{U}_{ \hspace{2mm}0 0}^{d \hspace{1mm} 0 0 }$ (see text). Inset: Diagram of the possible intra-LL transitions for filling factor $\nu=-1$.}
\label{fig-dispintrall} 
\end{figure}

We begin with a simple example of the dispersion relation for the case when both the electron and hole are in the $n=0$ LL. The possible intra-LL neutral CEs are shown in Fig.\ \ref{fig-dispintrall} (inset) for $\nu=-1$; they are not mixed. Since we neglect the energy separation between sublevels, there is no single particle contribution. Also each transition involves a spin or pseudospin flip, so there is no exchange contribution. The direct matrix element is given by
\begin{equation}
\label{eq-u0000}
\tilde{\mathcal{U}}_{ \hspace{2mm}0 \tau_1 s_1\hspace{0.5mm} 0 \tau_2 s_2}^{d \hspace{0.8mm}0 \tau_1 s_1\hspace{0.5mm} 0 \tau_2 s_2}
=\tilde{U}_{ \hspace{2mm}0 0}^{d \hspace{1mm} 0 0 }=-E_0e^{-\frac{\mathcal{P}^2 }{4 }}I_0\left(\frac{\mathcal{P}^2}{4} \right),
\end{equation}
where $\mathcal{P}=P\ell_B/\hbar$\newnot{sym:mathcalP} and $I_n$ are modified Bessel functions of the first kind. Notice that the matrix element depends only on the modulus of $\mathbf{P}$ and not its direction, which is to be expected due to the symmetry of the system. Fig.\ \ref{fig-dispintrall} shows the dispersion relation. Since the self energy is independent of $\mathcal{P}$, we set $E_{0 \tau_1 s_1}^{0 \tau_2 s_2}=0$. By taking the limits $\mathcal{P}\to 0$ and $\mathcal{P}\to \infty$ in Eq.\ (\ref{eq-u0000}), the band width for the case when both electron and hole are in the $n=0$ LL is seen to be $E_0$.

\subsubsection{Inter-LL excitations}
\label{subsubsec-dispinter}
Here I present the dispersion relation for inter-LL magnetoexcitons for the case when an integer number of sublevels of the $n=0$ LL is filled (see Fig.\ \ref{fig-allowedtrans}, page \pageref{fig-allowedtrans}). As described in Section \ref{sec-mixing}, we only take the $0 \to 1$ and $-1 \to 0$ excitations into account. We require the following matrix elements:
\begin{eqnarray}
\label{eq-u0101}
d=\tilde{\mathcal{U}}_{ \hspace{2mm}1 \tau_1 s_1\hspace{0.5mm} 0 \tau_2 s_2}^{d \hspace{0.8mm}1 \tau_1 s_1\hspace{0.5mm} 0 \tau_2 s_2}& =&
\tilde{\mathcal{U}}_{ \hspace{2mm}0 \tau_1 s_1\hspace{0.5mm} -1 \tau_2 s_2}^{d \hspace{0.8mm}0 \tau_1 s_1\hspace{0.5mm} -1 \tau_2 s_2} \\ \nonumber
& = &\frac{1}{2}\left( \tilde{U}_{ \hspace{2mm}1 0}^{d \hspace{1mm} 1 0 }+\tilde{U}_{ \hspace{2mm}0 0}^{d \hspace{1mm} 0 0 }\right)\\ \nonumber
& = & -\frac{E_0}{8}e^{-\frac{\mathcal{P}^2 }{4 }}\left[\left(6 +  \mathcal{P}^2    \right)I_0\left(\frac{\mathcal{P}^2}{4} \right)- \mathcal{P}^2 I_1\left(\frac{\mathcal{P}^2}{4} \right)\right],
\end{eqnarray}
\begin{eqnarray}
\label{eq-u0110}
d'=\tilde{\mathcal{U}}_{ \hspace{2mm}0 \tau_1 s_1\hspace{0.5mm} -1 \tau_2 s_2}^{d \hspace{1mm}1 \tau_1 s_1\hspace{2.5mm} 0 \tau_2 s_2}& =&
\frac{1}{2} \tilde{U}_{ \hspace{2mm}0 0}^{d \hspace{1mm} 1 1 }\\ \nonumber
& = & \frac{E_0}{8}e^{-\frac{\mathcal{P}^2 }{4 }}\left[ \mathcal{P}^2 I_0\left(\frac{\mathcal{P}^2}{4} \right)-\left(2+\mathcal{P}^2 \right)I_1\left(\frac{\mathcal{P}^2}{4} \right)\right],
\end{eqnarray}
\begin{equation}
\label{eq-u01ex}
ex=\tilde{\mathcal{U}}_{ \hspace{2mm}1 \tau' s'\hspace{0.5mm} 0 \tau' s'}^{x \hspace{1mm}1 \tau s\hspace{2mm} 0 \tau s} 
=\tilde{\mathcal{U}}_{ \hspace{2mm}0 \tau' s'\hspace{0.5mm} -1 \tau' s'}^{x \hspace{1mm}0 \tau s\hspace{1.5mm} -1 \tau s} 
=-\tilde{\mathcal{U}}_{ \hspace{2mm}0 \tau' s'\hspace{0.5mm} -1 \tau' s'}^{x \hspace{1mm}1 \tau s\hspace{3mm} 0 \tau s}
=\frac{1}{2}\tilde{U}_{ \hspace{2mm}1 0}^{x \hspace{1mm} 1 0 }
=\frac{E_0}{4}\mathcal{P}e^{-\frac{\mathcal{P}^2 }{2}}.
\end{equation}
Since the single particle energies are independent of $\mathbf{P}$, they only contribute a constant term to the dispersion energy. We renormalise the dispersion energies by setting $\epsilon_1+E_{0 \tau s}^{1 \tau s}=0$.

\begin{figure}
\centering
  \includegraphics[width=3.5in]{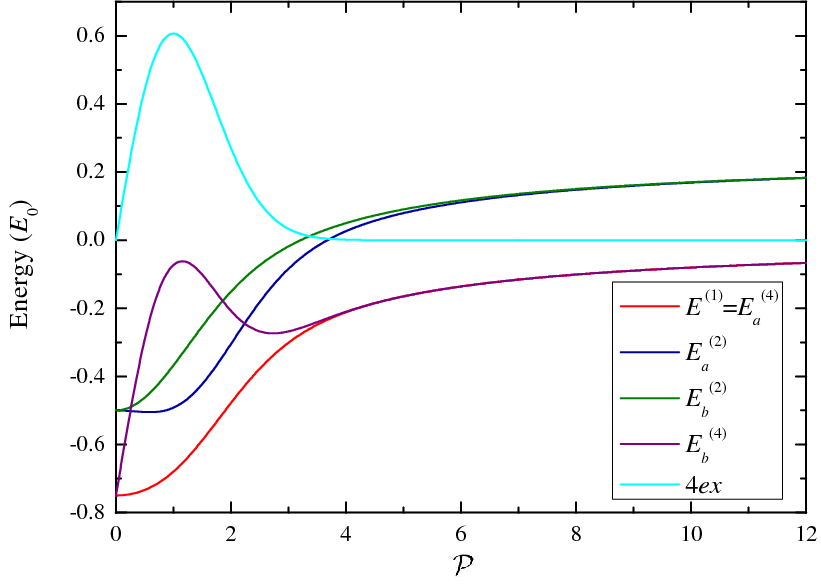}
   \caption{Plots of the various inter-LL dispersion relations described in the text corresponding to partial filling of the $n=0$ LL.}
 \label{fig-dispinterll} 
\end{figure}

We have already seen that there are three types of transitions: those which are not mixed to any other transition, those which exist in a pair and those comprised of four mixed transitions. I denote the (renormalised) dispersions for these three types as $E^{(1)},E^{(2)}$ and $E^{(4)}$ respectively. The resulting dispersion relations are shown in Fig.\ \ref{fig-dispinterll}. 
The degeneracy of each dispersion relation depends upon the filling factor; for example there are no mixed pairs for the case $\nu=2$. It is clear that
\begin{equation}
\label{eq-e1}
E^{(1)}=d,
\end{equation}
which is given explicitly in Eq.\ (\ref{eq-u0101}). The mixed pairs of excitations are found by diagonalising the matrix
\begin{equation}
\label{eq-e2}
\left( \begin{array}{cc}
d+\frac{E_0}{4} & d'\\
d' &  d+\frac{E_0}{4}
\end{array}
\right),
\end{equation}
where the inclusion of $\frac{E_0}{4}$ on the diagonal is due to the additional self energy possessed by mixed pairs of transitions, as seen in Section \ref{subsec-secorrecs}. This yields
\begin{equation}
\label{eq-e2ans}
E^{(2)}_a=d+\frac{E_0}{4}-d', \hspace{5mm}
E^{(2)}_b=d+\frac{E_0}{4}+d',
\end{equation}
where the different letters signify different branches of the dispersion. For the case $\nu=-1$, the dispersions corresponding to the four exchange-mixed transitions are found by diagonalising
\begin{equation}
\label{eq-e4}
\left( \begin{array}{cccc}
d+ex & ex & ex & -ex\\
ex & d+ex & ex & -ex\\
ex & ex & d+ex & -ex\\
-ex & -ex & -ex & d+ex
\end{array}
\right).
\end{equation}
This gives a triply degenerate eigenvalue which equals $E^{(1)}$ and a non-degenerate one:
\begin{equation}
\label{eq-e4eigvals}
E^{(4)}_a=d, \hspace{5mm}
E^{(4)}_b=4ex+d.
\end{equation}
Although the matrices for different sublevel filling factors of the $n=0$ LL may look different to that in Eq.\ (\ref{eq-e4}), they yield the same eigenvalues and degeneracies. In Fig.\ \ref{fig-dispinterll}, in addition to plotting $E^{(4)}_b$, its direct ($d=E^{(1)}$) and exchange ($4ex$) contributions are plotted separately.

We may now determine the excitonic band width for excitations with bare energies $\epsilon_1$. For all the dispersions calculated above, their lowest value occurs at $P=0$ and they are bounded above by the $P\to\infty$ limit. Hence, if only transitions with no spin or pseudospin flip are considered, the band has width $0.75E_0$. If all sixteen transitions are considered however, it has width $E_0$.
\subsection{Experimental evidence}
\label{subsec-seexptevid}
\begin{figure}
\centering
  \includegraphics[width=4in]{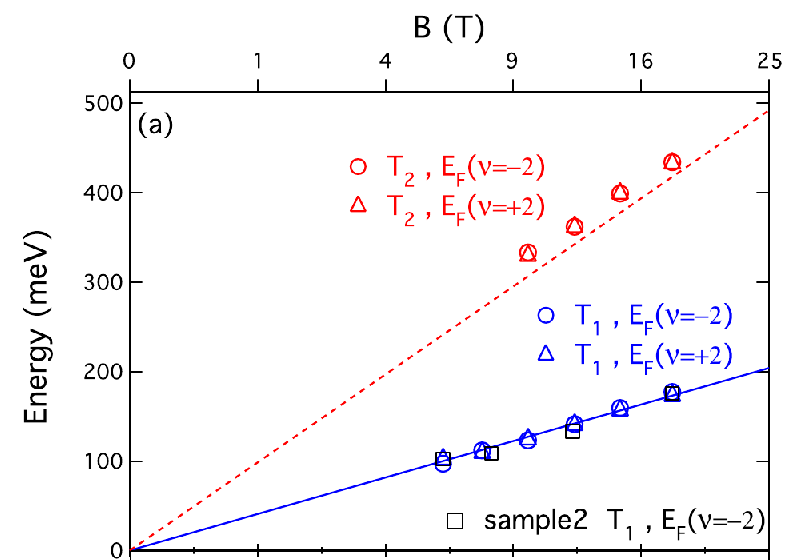}
   \caption{(Taken from Ref.\ \cite{JiaHTW07}). Cyclotron resonance energies plotted as a function of magnetic field, $B$. The transition $\mathrm{T_1}$ corresponds to the $-1 \to 0$ transition ($\nu=-2$), which is equal to the $0 \to 1$ transition ($\nu=2$). The transition $\mathrm{T_2}$ corresponds to the $-1 \to 2$ transition ($\nu=\pm2$). The solid blue line is a best $\sqrt{B}$ fit to the $\mathrm{T_1}$ transition data, yielding $\tilde{v}_\mathrm{F}=(1.12\pm0.02)\times10^6\mathrm{ms^{-1}}$. The dashed red line is a scaling of the solid blue line by a factor of $1+\sqrt{2}$.}
\label{fig-JiaHTW07} 
\end{figure}
There have been numerous experimental studies investigating the CREs in graphene \cite{SadMPB06,JiaHTW07,DeaCNN07,LiHJH08,HenCJL10}. Ref.\ \cite{DeaCNN07} describes a photoconductance study in which modulations of the samples' conductivity upon radiation are used to determine the CRE. An asymmetry between the electron and hole bands was observed, suggesting a difference in the electron and hole velocities. This is interesting, since it is widely thought that graphene has perfect particle-hole symmetry. To the best of my knowledge, such an asymmetry has not been reproduced experimentally, though this may well be due to the high precision of measurements required to observe it. A popular direct probe of the CRE is spectroscopy. Initial far infrared spectroscopy of ultrathin graphite \cite{SadMPB06} reported good agreement with the simple non-interacting Dirac picture. However, many experimental papers on monolayer graphene claim that their results indicate the presence of many body effects \cite{JiaHTW07,LiHJH08,HenCJL10}. The bare transition energy and the many body correction both have approximately the same dependence on the magnetic field, so this cannot be used to measure the correction. The presence of many body effects cannot be demonstrated by examining a single resonance energy, since the effect of $e$-$e$ interactions is just to renormalise $v_\mathrm{F}$ and one cannot turn off the interactions and experimentally determine a bare $v_\mathrm{F}$ for comparison. However, as we have seen in Section \ref{subsec-secorrecs}, the self energy correction for excitations with different bare energies, does not scale according to the ratio of these bare energies. Hence by studying more than one resonance it should be possible to see a signature of many body effects. 

This approach was used in Ref.\ \cite{JiaHTW07}, where the energies of the $-1\to0$ and $-1 \to 2$ transitions were compared at filling $\nu=-2$. The results are shown in Fig.\ \ref{fig-JiaHTW07}, which is taken from Ref.\ \cite{JiaHTW07}.  A renormalised Fermi velocity \newnot{sym:vFtilde}$\tilde{v}_\mathrm{F}$ was calculated, by setting the experimentally determined energy of the $-1\to0$ transition equal to $\sqrt{2}\hbar\tilde{v}_\mathrm{F}/\ell_B$. It was then noted that the experimental data for the $-1 \to 2$ transition energy lies slightly above the line $\sqrt{2}\hbar\tilde{v}_\mathrm{F}(1+\sqrt{2})/\ell_B$ (red dashed line in Fig\ \ref{fig-JiaHTW07}) and that this shift is outside the error bars, which are contained within the symbol size. These results are consistent with the theory discussed in Sections \ref{sec-selfeng} and \ref{sec-disp}, as I shall now explain. We need to calculate the contribution to the transition energy from interactions, assuming no spin or pseudospin flips and $\mathcal{P}=0$ (by conservation of momentum, since the photon does not supply any momentum). Let's denote these interaction energies by $F_{-10}$ and $F_{-12}$ for the $-1 \to 0$ and $-1 \to 2$ transitions respectively. They have direct and self energy components, since the exchange energy vanishes at $\mathcal{P}=0$ for both transitions. For $-1 \to 0$, we have $F_{-10}=d(0)+E_{-1 \tau s}^{0 \tau s}\approx 0.81E_0$, where $d$ is defined in Eq.\ (\ref{eq-u0101}). Using
\begin{eqnarray}
\label{eq-uneg12}
\tilde{\mathcal{U}}_{ \hspace{2mm}2 \tau_1 s_1\hspace{0.5mm} -1 \tau_2 s_2}^{d \hspace{0.8mm}2 \tau_1 s_1\hspace{0.5mm} -1 \tau_2 s_2}& =&
\frac{1}{4}\left( \tilde{U}_{ \hspace{2mm}2 1}^{d \hspace{1mm} 2 1 }+\tilde{U}_{ \hspace{2mm}1 1}^{d \hspace{1mm} 1 1 }+\tilde{U}_{ \hspace{2mm}2 0}^{d \hspace{1mm} 2 0 }+\tilde{U}_{ \hspace{2mm}1 0}^{d \hspace{1mm} 1 0 }\right)\\ \nonumber
& = & -\frac{E_0}{128}e^{-\frac{\mathcal{P}^2 }{4 }}\biggl[\left(66 +  13\mathcal{P}^2 +2\mathcal{P}^4 + \mathcal{P}^6 \right)I_0\left(\frac{\mathcal{P}^2}{4} \right)\\ \nonumber
&   &-\left(23 +4\mathcal{P}^2+\mathcal{P}^4\right) I_1\left(\frac{\mathcal{P}^2}{4} \right)\biggr],
\end{eqnarray}
we have for $-1 \to 2$, $F_{-12}=\tilde{\mathcal{U}}_{ \hspace{2mm}2 \tau_1 s_1\hspace{0.5mm} -1 \tau_2 s_2}^{d \hspace{0.8mm}2 \tau_1 s_1\hspace{0.5mm} -1 \tau_2 s_2}\left(0\right)+E_{-1 \tau_2 s_2}^{2 \tau_1 s_1}\approx2.17E_0$. Hence $F_{-12}/F_{-10}\approx 2.7 > 1+\sqrt{2}\approx 2.4$ in qualitative agreement with the experimental data for the $-1 \to 2$ transition lying above the red dashed line in Fig.\ \ref{fig-JiaHTW07}.
\cleardoublepage

\cleardoublepage
	
\chapter{SU(4) symmetry in graphene}
\label{chap-su4}

Symmetry is ubiquitous in physics. The wave function of a set of
bosons/fermions is symmetric/antisymmetric under the exchange of quantum numbers
for any pair of particles. In this Chapter we focus on the spin and pseudospin
degrees of freedom of CEs, considering them separately from the other quantum
numbers. The two possible spin projections and two possible pseudospin
projections, yield four equivalent possibilities for the spin/pseudospin state
of an electron in graphene, which leads to the SU(4) symmetry. We first classify
the neutral CEs according to their total spin and pseudospin and then explore the SU(4)
symmetry, by applying techniques used in elementary particle physics to classify
hadrons.
\section{Classification of excitons by spin and pseudospin}
\label{sec-spins}
Consider a system of two electrons and suppose each one has spin and no other quantum numbers associated with it. Then it is well known that a basis for the possible states of this two electron system may be constructed in such a way that the basis states are eigenstates of the total spin operator and total spin projection operator. For CEs it is clear that the total spin and pseudospin projections, $S_z$ and $T_z$, are always well defined quantum numbers. Explicitly for the CEs $Q^{\dag}_{\mathcal{ N}_1  \mathcal{ N}_2}|\nu\rangle$, $R^{\dag}_{\mathcal{ N}_1  \mathcal{ N}_2 \mathcal{ N}_3} |\mu\rangle$, $S_z=s_1-s_2$ and $T_z=\tau_1-\tau_2$. We would like to see whether it is also possible to assign a total spin, \newnot{sym:Sspin}$S$ and total pseudospin, $T$\newnot{sym:Tpseudospin}, to CEs. To do this it is necessary to express the operators for these quantities in second quantised form. The operator for total spin\newnot{sym:S2}, $\hat{\mathbf{S}}^2$, can be derived using $\hat{\mathbf{S}}^2=\hat{S}_x^2+\hat{S}_y^2+\hat{S}_z^2$ and\newnot{sym:Si}
\begin{equation}
\label{eq-spini}
\hat{S}_i=\sum_{N,s,s'}\langle Ns|\hat{S}_i|N s'\rangle c^\dagger_{Ns}c_{Ns'},
\end{equation}
where $i=x,y,z$ and $N=\{n m \tau \}$. We stick to the electron representation to simplify the formulae; it is also the natural choice to determine the total spin of the ground state. After rearranging to normal ordered form, we find
\begin{eqnarray}
\label{eq-stot}
\hat{\mathbf{S}}^2 & = & \frac{\hbar^2}{4}\sum_{N,N'}\Bigl(4c^\dagger_{N\uparrow}c^\dagger_{N'\downarrow}c_{N'\uparrow}c_{N\downarrow}
-2c^\dagger_{N\uparrow}c^\dagger_{N'\downarrow}c_{N'\downarrow}c_{N\uparrow} \\ \nonumber
&  & +c^\dagger_{N\uparrow}c^\dagger_{N'\uparrow}c_{N'\uparrow}c_{N\uparrow}
+c^\dagger_{N\downarrow}c^\dagger_{N'\downarrow}c_{N'\downarrow}c_{N\downarrow}\Bigr) \\ \nonumber
&   &+\frac{3}{4}\hbar^2\sum_{N}\left(c^\dagger_{N\uparrow}c_{N\uparrow}+c^\dagger_{N\downarrow}c_{N\downarrow}\right).
\end{eqnarray}
Note that $\hat{\mathbf{S}}^2$ is symmetric under flipping all spins, as expected and that it gives the correct value for the total spin of a single electron.
The operator for total pseudospin, $\hat{\mathbf{T}}^2$\newnot{sym:T2}, is found analogously. It can be obtained from the expression for $\hat{\mathbf{S}}^2$, by changing $\uparrow, \downarrow \to \Uparrow, \Downarrow$ and $N \to \mathrm{N}=\{nms\}$\newnot{sym:Nrm}.

The total spin (pseudospin) of a CE is only well defined if the ground state, $|\nu\rangle$, from which it is excited has total spin (pseudospin) zero. The total spin (pseudospin) of a ground state is found by expressing it in second quantised form and then acting on it by $\hat{\mathbf{S}}^2$ ($\hat{\mathbf{T}}^2$). For example,
\begin{equation}
\label{eq-nu2}
|\nu=2\rangle=\prod_{n=-n_\mathrm{c}}^0\prod_{m}\prod_{s=\uparrow,\downarrow}\prod_{\tau=\Uparrow,\Downarrow}c^\dagger_{n \tau s m}|0\rangle.
\end{equation}
Hence one can show that $S=0$ only for $\nu=\pm2$ (or indeed any completely filled LL) and $T=0$ only for $\nu=0,\pm2$ (or any half or completely filled LL).

We concentrate on neutral CEs from ground states with $S=T=0$. In this case, $\hat{\mathbf{S}}^2 Q^{\dag}_{\mathcal{ N}_1  \mathcal{ N}_2}|\nu\rangle=[\hat{\mathbf{S}}^2,Q^{\dag}_{\mathcal{ N}_1  \mathcal{ N}_2}]|\nu\rangle$ and $\hat{\mathbf{T}}^2 Q^{\dag}_{\mathcal{ N}_1  \mathcal{ N}_2}|\nu\rangle=[\hat{\mathbf{T}}^2,Q^{\dag}_{\mathcal{ N}_1  \mathcal{ N}_2}]|\nu\rangle$. Choosing $\nu=2$ (without loss of generality), one may show
\begin{eqnarray}
\label{eq-ssquaredqdaggercomm}
[\hat{\mathbf{S}}^2,Q^{\dag}_{1 \tau \uparrow \hspace{1mm} 0 \tau \uparrow}]
& = &\hbar^2\sum_{m_1,m_2} A_{1 \tau \uparrow \hspace{1mm} 0 \tau \uparrow}(m_1,m_2) \\ \nonumber
& & \times \sum_N\left( c^\dagger_{N \downarrow} c^\dagger_{1 m_1 \tau \uparrow}c_{N \uparrow} c_{0 m_2 \tau \downarrow}-
c^\dagger_{N \uparrow} c^\dagger_{1 m_1 \tau \downarrow}c_{N \downarrow} c_{0 m_2 \tau \uparrow} \right)
\end{eqnarray}
and hence
\begin{equation}
\label{eq-ssquaredqdagger}
\hat{\mathbf{S}}^2Q^{\dag}_{1 \tau \uparrow \hspace{1mm} 0 \tau \uparrow}|\nu=2 \rangle
= \hbar^2\sum_{m_1,m_2} A_{1 \tau \uparrow \hspace{1mm} 0 \tau \uparrow}(m_1,m_2) \left(c^\dagger_{1 m_1 \tau \uparrow}c_{0 m_2 \tau \uparrow}- c^\dagger_{1 m_1 \tau \downarrow}c_{0 m_2 \tau \downarrow}\right)|\nu=2 \rangle.
\end{equation}
A similar expression for $\hat{\mathbf{S}}^2Q^{\dag}_{1 \tau \downarrow \hspace{1mm} 0 \tau \downarrow}|\nu=2 \rangle$ can be obtained by flipping all spins in Eq.\ (\ref{eq-ssquaredqdagger}). Let's set $A_{1 \tau \uparrow \hspace{1mm} 0 \tau \uparrow}(m_1,m_2)=A_{1 \tau \downarrow \hspace{1mm} 0 \tau \downarrow}(m_1,m_2)$ for all $m_1$, $m_2$ and $\tau$, and let $|\tau_1 s_1 \hspace{1mm} \tau_2 s_2 \rangle=Q^{\dag}_{1 \tau_1 s_1 \hspace{1mm} 0 \tau_2 s_2}|\nu=2 \rangle$. Then
\begin{equation}
\label{eq-spinsinglet}
\hat{\mathbf{S}}^2\left(|\Uparrow \uparrow \hspace{1mm} \Uparrow \uparrow \rangle +|\Downarrow \uparrow \hspace{1mm} \Downarrow \uparrow \rangle
+ |\Uparrow \downarrow \hspace{1mm} \Uparrow \downarrow \rangle + |\Downarrow \downarrow \hspace{1mm} \Downarrow \downarrow \rangle
\right)=0
\end{equation}
and we have shown that the CE, 
$\frac{1}{2}\left(|\Uparrow \uparrow \hspace{1mm} \Uparrow \uparrow \rangle +|\Downarrow \uparrow \hspace{1mm} \Downarrow \uparrow \rangle
+ |\Uparrow \downarrow \hspace{1mm} \Uparrow \downarrow \rangle + |\Downarrow \downarrow \hspace{1mm} \Downarrow \downarrow \rangle
\right)$
is a spin singlet ($S=0$). By similar reasoning it is also a pseudospin singlet ($T=0$), so we define
\begin{equation}
\label{eq-s0t0}
|S=0, T=0; S_z=0, T_z=0\rangle =\frac{1}{2}\left(|\Uparrow \uparrow \hspace{1mm} \Uparrow \uparrow \rangle +|\Downarrow \uparrow \hspace{1mm} \Downarrow \uparrow \rangle
+ |\Uparrow \downarrow \hspace{1mm} \Uparrow \downarrow \rangle + |\Downarrow \downarrow \hspace{1mm} \Downarrow \downarrow \rangle
\right).
\end{equation}
Further calculations find
\begin{equation}
\label{eq-s0t1}
|S=0, T=1; S_z=0, T_z=0\rangle =\frac{1}{2}\left(|\Uparrow \uparrow \hspace{1mm} \Uparrow \uparrow \rangle -|\Downarrow \uparrow \hspace{1mm} \Downarrow \uparrow \rangle
+ |\Uparrow \downarrow \hspace{1mm} \Uparrow \downarrow \rangle - |\Downarrow \downarrow \hspace{1mm} \Downarrow \downarrow \rangle
\right),
\end{equation}
\begin{equation}
\label{eq-s1t0}
|S=1, T=0; S_z=0, T_z=0\rangle =\frac{1}{2}\left(|\Uparrow \uparrow \hspace{1mm} \Uparrow \uparrow \rangle +|\Downarrow \uparrow \hspace{1mm} \Downarrow \uparrow \rangle
- |\Uparrow \downarrow \hspace{1mm} \Uparrow \downarrow \rangle - |\Downarrow \downarrow \hspace{1mm} \Downarrow \downarrow \rangle
\right),
\end{equation}
\begin{equation}
\label{eq-s1t1}
|S=1, T=1; S_z=0, T_z=0\rangle =\frac{1}{2}\left(|\Uparrow \uparrow \hspace{1mm} \Uparrow \uparrow \rangle -|\Downarrow \uparrow \hspace{1mm} \Downarrow \uparrow \rangle
- |\Uparrow \downarrow \hspace{1mm} \Uparrow \downarrow \rangle + |\Downarrow \downarrow \hspace{1mm} \Downarrow \downarrow \rangle
\right).
\end{equation}
Notice that the four states above all have $S_z=0, T_z=0$. The states corresponding to different spin/pseudospin projections can be found by acting on these states by the spin/pseudospin raising and lowering operators. In second quantised form these are\newnot{sym:Spm}
\begin{equation}
\label{eq-splus}
\hat{S}_+=\hat{S}_x+i\hat{S}_y=\hbar \sum_N c^\dagger_{N \uparrow}c_{N \downarrow},
\end{equation}
\begin{equation}
\label{eq-sminus}
\hat{S}_-=\hat{S}_x-i\hat{S}_y=\hbar \sum_N c^\dagger_{N \downarrow}c_{N \uparrow},
\end{equation}
with similar definitions for $\hat{T}_+,\hat{T}_-$\newnot{sym:Tpm}. In total there are 16 states; they are listed in full in Appendix \ref{app-states}. An alternative way of classifying the states is according to the quantum numbers, $S,T,I,I_z$. The quantity $I$ is the total (spin plus pseudospin) angular momentum projection:
\begin{equation}
\label{eq-j}
\hat{I}^2|S, T, I; I_z\rangle=(\hat{S}+\hat{T})^2|S, T, I; I_z\rangle=I(I+1)|S, T, I; I_z\rangle
\end{equation}
and $I_z=S_z+T_z$.

A similar analysis may be done for trions, where the states can be classified as spin/pseudospin doublets or quadruplets. However, this is considerably more complicated than for the two-particle case, since it involves 64 states. In addition, it doesn't explain the degeneracy in the energies of CEs. Hence instead of pursuing this spin/pseudospin classification, we follow another route, which is able to predict the degeneracies of CE energies. To do this, I must first introduce some mathematical concepts.
\section{Lie algebras}
\label{sec-lie}
In this Section, I assume that the reader is familiar with the concept of a discrete group (see for example Ref.\ \cite{Rot95}). An $n$-parameter continuous group consists of the elements $A(\bm{\alpha})=A(\alpha_1,\alpha_2,\ldots,\alpha_n)$ obtained by varying the parameters $\alpha_1,\alpha_2,\ldots,\alpha_n$ continuously, together with a binary operation. By the closure requirement for a group, $A(\bm{\alpha})A(\bm{\beta})=A(\bm{\gamma})$. Thus $\bm{\gamma}=f(\bm{\alpha},\bm{\beta})$, where $f$ is a continuous function. If $f$ is also analytic (it has a convergent Taylor series expansion everywhere in its domain), then the group is a Lie group \cite{Gil08}. Let's define
\begin{equation}
\label{eq-infop}
\mathbf{B}=\bm{\nabla}_{\bm{\alpha}}A(\bm{\alpha})\mid_{\bm{\alpha}=\bm{0}},
\end{equation}
from which it follows
\begin{equation}
\label{eq-a}
A(\bm{\alpha})=e^{\mathbf{B}.\bm{\alpha}}=\mathop {\lim }\limits_{n \to \infty } \left(1+\frac{\mathbf{B}.\bm{\alpha}}{n}\right)^n.
\end{equation}
From Eq.\ (\ref{eq-a}) it is clear that the components of $\mathbf{B}$ generate the Lie group. They form the basis of a vector space, $V$, over some field, $F$. Moreover, one can prove that $V$ together with the binary operation, $[\cdotp,\cdotp]: V \times V \to V$ form a Lie algebra, since they satisfy the axioms for all $a,b\in F$ and for all $x,y,z\in V$:

1. Bilinearity
\begin{equation}
\label{eq-bilin}
[ax+by,z]=a[x,z]+b[y,z],\hspace{5mm} [z,ax+by]=a[z,x]+b[z,y]
\end{equation}

2. Jacobi Identity
\begin{equation}
\label{eq-jacobi}
[[x,y],z]+[[z,x],y]+[[y,z],x]=0
\end{equation}

3. Antisymmetry
\begin{equation}
\label{eq-antisym}
[x,y]=-[y,x].
\end{equation}
We have seen how a Lie algebra emerges from a Lie group. It is also possible to reverse this process and specify a Lie group from a Lie algebra; this is Lie's theorem.

Let us illustrate these ideas on a simple example. $\mathrm{SU}(2)$ is the group of $2\times2$ unitary matrices with determinant one. Each element has the form
\begin{equation}
\label{eq-unimat}
U(a,b)=\left(\begin{array}{cc}
              a & b  \\
	      -b^\ast & a^\ast
             \end{array}
\right),
\end{equation}
where $a,b$ are complex numbers satisfying $|a|^2+|b|^2=1$. Hence $\mathrm{SU}(2)$ is a Lie group with dimension 3. If we consider a spin $\frac{1}{2}$ system, whose state is given in the Pauli two-component formalism by $\left(\begin{array}{c}
a^\uparrow\\
a^\downarrow\end{array}\right)$, a physical application of $\mathrm{SU}(2)$ is that it consists of the matrices which when applied to 
$\left(\begin{array}{c}
a^\uparrow\\
a^\downarrow\end{array}\right)$
yield the new state resulting from a particular rotation of space. The three independent operators $\hat{S}_x$, $\hat{S}_y$ and $\hat{S}_z$ form the basis of a vector space for the Lie algebra. Notice that
\begin{equation}
\label{eq-scomm}
[\hat{S}_i,\hat{S}_j]=\epsilon_{ijk}\hat{S}_k,
\end{equation}
where $i,j,k \in \{x,y,z\}$ and $\epsilon_{ijk}$ is the Levi-Civita symbol. Hence the vector space is closed under commutation, as specified in the definition of a Lie algebra given above. The Lie group is generated by $\hat{S}_x, \hat{S}_y, \hat{S}_z$ or strictly speaking by the corresponding Pauli matrices $\sigma_x, \sigma_y, \sigma_z$, in the sense that each $U(a,b)=\mathrm{exp}\left(\frac{-i \bm{\sigma}.\hat{\bm{n}}\theta}{2}\right)$, where $\hat{\bm{n}}$ is the axis of rotation in 3D space and $\theta$ is the angle of rotation. In fact, for any finite $n$, $\mathrm{SU}(n)$ is a Lie group of rank $n-1$ with $n^2-1$ independent generators.
\section{SU(4) symmetry in graphene}
\label{sec-su4}
Graphene is $\mathrm{SU}(4)$ symmetric in the sense that the ordinary (non-interacting) Hamiltonian commutes with the elements of $\mathrm{SU}(4)$. In addition, the Hamiltonian of the Coulomb interactions is also $\mathrm{SU}(4)$ symmetric up to small symmetry-breaking terms \cite{GoeMD06}. This is due to the equivalence of the two valleys and the two spin states, including the continuous interchange between spins and valleys. We assume here, as above, that the ground state corresponds to the $n=0$ LL being either completely full or completely empty. The four basis states that form a fundamental representation of $\mathrm{SU}(4)$ are $|\Uparrow \uparrow \rangle, |\Downarrow \uparrow \rangle, |\Uparrow \downarrow \rangle, |\Downarrow \downarrow \rangle$. I shall label them in analogy with the four flavours of down, up, strange, and charm quarks\newnot{sym:uflavor}\newnot{sym:dflavor}\newnot{sym:cflavor}\newnot{sym:sflavor}, by $ \{ \Downarrow\downarrow, \Uparrow\downarrow, \Downarrow\uparrow, \Uparrow\uparrow \} \equiv \{ d, u, s, c \}$.

The generators of $\mathrm{SU}(4)$ can be constructed using bilinear combinations of the fermionic creation and annihilation operators conserving
the number of electrons and holes. We have\newnot{sym:cij}
\begin{equation}
\label{eq-cij}
\hat{C}_{ij} =\sum_{nm} c^{\dagger}_{nmi} c_{nmj}
       - \sum_{nm}d^{\dagger}_{nmj} d_{nmi},
\end{equation}
where $i,j\in\left\lbrace d,u,s,c \right\rbrace$. The generators satisfy the commutation relations
\begin{equation}
\label{eq-cijcomm}
[\hat{C}_{ij}, \hat{C}_{kl} ] = \delta_{jk} \hat{C}_{il} - \delta_{il} \hat{C}_{kj},
\end{equation}
so are closed under commutation.
The operators of spin and pseudospin can be expressed in terms of the $\hat{C}_{ij}$.
For example,
$\hat{T}_z =  \frac{1}{2}\left( \hat{C}_{cc} + \hat{C}_{uu} - \hat{C}_{ss} - \hat{C}_{dd}\right)$
and
$\hat{T}_+ = \hat{T}_x + i \hat{T}_y= \hat{C}_{ud} + \hat{C}_{cs}$. The elements of $\mathrm{SU}(4)$ may then be expressed as $\hat{U}=\exp \left( i \sum_{ij }\Theta_{ij} \hat{C}_{ij} \right)$, with $\Theta_{ij}$ the transformation ``angles'' \cite{GreM97,GoeMD06}. Since $\hat{U}$ commutes with the Hamiltonian, applying it to an eigenstate of the Hamiltonian, yields another eigenstate of the Hamiltonian with the same energy. This means that the eigenstates of the Hamiltonian exist in multiplets such that applications of the $\mathrm{SU}(4)$ operators to any state always result in another state in the same multiplet. In addition, all states within the same multiplet are degenerate. It is possible to calculate the multiplicities of our states (CEs) and also to learn something about their symmetry, upon interchanging the quantum numbers of pairs of particles, before recourse to numerical methods. This is done using Young diagrams. 
\subsection{Young diagrams}
\label{subsec-young}
Young diagrams are a useful technique for determining the $\mathrm{SU}(n)$ multiplet structure of a composite system of particles. In the following brief description, I treat them as a useful tool, stating several results without proof. 
For a more rigorous development of the theory, see Ref.\ \cite{Lic10}. 

\begin{figure}
\centering
  \includegraphics[width=2.5in]{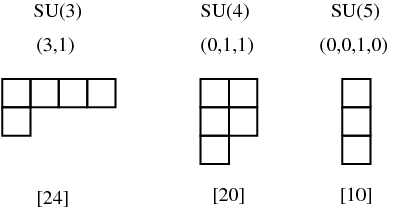}
   \caption{Examples of Young diagrams for different $\mathrm{SU}$ multiplets together with their unique multiplet label and the corresponding multiplicity.}
\label{fig-young} 
\end{figure}

Young diagrams consist of $n$ or fewer left-justified rows of boxes, such that each row is as least as long as the row beneath it. Roughly speaking, each box represents a particle. Each $\mathrm{SU}(n)$ multiplet is uniquely labelled by a list of $n-1$ non-negative integers, $(a_1,a_2,\ldots,a_{n-1})$. The integer in the $i^{\rm th}$ position corresponds to the number of boxes in the $i^{\rm th}$ row minus the number of boxes in the $(i+1)^{\rm th}$ row i.e. the overhang of the $i^{\rm th}$ row. Some examples of Young diagrams for different SU multiplets and their labels are given in Fig.\ \ref{fig-young}.
By the definition of a multiplet label, adding columns of length $n$ to the left of a Young diagram is redundant. Young diagrams can be used to calculate the multiplicity of a multiplet. The multiplicity, $[N(n)]$ of an $\mathrm{SU}(n)$ multiplet is
\begin{equation}
\label{eq-nofn}
[N(n)]=\prod_{i=1}^{n-1}\prod_{j=i}^{n-1}\frac{j-i+1+\sum_{k=i}^ja_k}{j-i+1}.
\end{equation}
Thus for the $n=4$ case,
\begin{equation}
\label{eq-nof4}
[N(4)]=\frac{(a_1+1)}{1}\frac{(a_2+1)}{1}\frac{(a_3+1)}{1}\frac{(a_1+a_2+2)}{2}\frac{(a_2+a_3+2)}{2}\frac{(a_1+a_2+a_3+3)}{3}.
\end{equation}

The boxes in Young diagrams can be labelled by positive integers to represent the quantum state of a particle, according to certain rules. These are i.\ in each row the labels must not decrease, as read from left to right, ii.\ in each column the labels must strictly increase as labelled from top to bottom. The Young diagram then represents a state, which is first symmetrised with respect to exchanging pairs of quantum numbers for boxes in the same row and then antisymmetrised with respect to exchanging pairs of quantum numbers for boxes in the same column. Of course, for bosons the state should always be completely symmetric and the Young diagram a single row and for fermions, the state should always be completely antisymmetric and the Young diagram a single column. However, this holds if all quantum numbers that can label a particle are used. Often this is not the case, for example, spin quantum numbers or in the case of graphene, spin and pseudospin quantum numbers, are treated separately to the orbital quantum numbers. Under these circumstances other symmetric patterns may emerge.

Let's illustrate this on the familiar example of a two-particle system, where each particle has spin $\frac{1}{2}$ and two possible spin projections, which I shall call $1$ and $2$. The first step is to create all allowed Young diagrams consisting of two boxes. In fact there are only two possibilities, $\yng(2)$ and $\yng(1,1)$. As we saw in Section \ref{sec-lie}, these are $\mathrm{SU}(2)$ multiplets. The former has label $(2)$ and multiplicity $[3]$; the latter has label $(0)$ and multiplicity $[1]$. Following the labelling rules, the Young diagram, $\yng(2)$, can be labelled as $\young(11)$, $\young(12)$ or $\young(22)$. Since the state should be symmetric with respect to permuting the quantum numbers of the two particles, we deduce that the states are $|11\rangle$, $\frac{1}{\sqrt{2}}\left(|12\rangle+|21\rangle\right)$ and $|22\rangle$, where $|ij\rangle$ means particle one is described by the quantum number $i$ and particle two is described by the quantum number $j$. We recognise these as the symmetric spin triplet states. The only way of labelling $\yng(1,1)$ is $\young(1,2)$. Since the two boxes are now in the same column, the state must be antisymmetric with respect to interchanging the two particles' quantum numbers, which is satisfied by $\frac{1}{\sqrt{2}}\left(|12\rangle-|21\rangle\right)$. This is the familiar spin singlet.

There are two fundamental representations of $\mathrm{SU}(n)$: $[n]$, with Young diagram $\yng(1)$ and label $(1,0,\ldots0)$ and $[\bar{n}]$ with Young diagram composed of $n-1$ vertically stacked boxes and label $(0\ldots,0,1)$. In general, whenever two multiplets have the same entries in their labels, but in reverse order, they are said to be conjugate and one of their multiplicities is denoted with a bar. The latter represents an antiparticle; this makes sense if we identify $n$ vertically stacked boxes with the vacuum. I shall also use the symbol, $\underline{1}$ to denote the vacuum. For the current purpose, the hole in a CE behaves as an antiparticle. In order to create $\mathrm{SU}(n)$ multiplets corresponding to $e$-$h$ and $e$-$e$-$h$ complexes, we need to be able to combine two Young diagrams correctly. There is a standard way for doing this, which I shall now explain. The general idea is to take the first Young diagram as being the upper left portion of the emerging combined diagrams and label the second diagram such that all boxes in the first row are labelled $1$, all boxes in the second row $2$ and so on. The boxes from the second diagram should then be added to the ends of rows of the first diagram or the bottom of the first diagram in all the different possible ways subject to the following rules: \newline
i.\ The resulting diagram takes the form of an ``allowed'' Young diagram as previously described with a maximum of $n$ rows for $\mathrm{SU}(n)$ multiplets.
\newline
ii.\ Each number should appear no more than once in any column.\newline
iii.\ The sequence of numbers formed by reading from right to left in the first row, then the second row etc. is admissible. I define a sequence as being admissible if at any point in the sequence for any integer $j$, $N_i\ge N_j$ holds for all $i\le j$, where $N_i$ is the number of times the integer $i$ has appeared in the sequence up to that point. This is best illustrated by an example. Two $\mathrm{SU}(3)$ octets may be combined in the following way:
\begin{eqnarray}
\label{eq-combine_young_diags}
\yng(2,1)\otimes\young(11,2) & = & \left(\young(\hfil \hfil 11,\hfil)\oplus\young(\hfil \hfil 1,\hfil1)\oplus\young(\hfil \hfil 1,\hfil,1)
\oplus\young(\hfil \hfil ,\hfil 1,1)\right)\otimes \young(2)\\ \nonumber
& = &  \young(\hfil \hfil 11,\hfil 2) \oplus \young(\hfil \hfil \one \one,\hfil,2)\oplus\young(\hfil \hfil 1,\hfil 12)
\oplus\young(\hfil \hfil 1,\hfil 1,2) \\ \nonumber
& & \oplus \young(\hfil \hfil 1,\hfil 2,1)\oplus \young(\hfil \hfil ,\hfil \one,\one 2)\\ \nonumber
& = & \yng(4,2) \oplus \yng(3) \oplus \yng(3,3) \oplus \yng(2,1) \oplus \yng(2,1) \oplus \underline{1}.
\end{eqnarray}
To obtain the final line of Eq.\ (\ref{eq-combine_young_diags}), I removed $\yng(1,1,1)$ from the left of Young diagrams where possible. Eq.\ (\ref{eq-combine_young_diags}) can also be written in terms of multiplet labels as
\begin{equation}
 \label{eq-younglabels}
(1,1) \otimes (1,1)= (2,2) \oplus(3,0) \oplus(0,3) \oplus(1,1) \oplus(1,1) \oplus(0,0)
\end{equation}
and multiplicities as
\begin{equation}
 \label{eq-youngmults}
[8] \otimes [8] = [27] \oplus [10] \oplus [\bar{10}] \oplus[8] \oplus[8] \oplus[1].
\end{equation}
We are now in a position to work out the multiplet structure for CEs.
\subsection{$\mathrm{SU}(4)$ multiplet structure for excitons}
\label{subsec-exmults}
Our neutral CEs can be thought of as a particle-antiparticle ($e$-$h$) complex where particles can have one of four flavours, as described previously. They are thus in complete analogy with mesons, which are elementary particles composed of a quark-antiquark ($q\bar{q}$) pair, provided the quarks are either first ($u,d$) or second ($c,s$) generation. We find the multiplet structure for excitons by combining the Young diagrams for an electron and hole as follows:
\begin{equation}
\label{eq-youngmultsmeson}
\yng(1) \otimes \yng(1,1,1) = \yng(2,1,1) \oplus \underline{1}.
\end{equation}
Eq.\ (\ref{eq-youngmultsmeson}) is equivalent to $(1,0,0) \otimes (0,0,1) = (1,0,1) \oplus (0,0,0)$ or $[4]\otimes[\bar{4}]=[15]\oplus[1]$. In terms of the classification of excitons described in Section \ref{sec-spins}, the vacuum singlet corresponds to the spin and pseudospin singlet in Eq.\ (\ref{eq-s0t0}). It can be expressed diagrammatically as $\yng(1,1,1,1)$ and symbolically as \newnot{sym:Q0}$Q_0 = \frac{1}{2} \left( d\bar{d} + u\bar{u} +  s\bar{s} + c\bar{c} \right)$.
From Eq.\ (\ref{eq-youngmultsmeson}) we can deduce that all exciton energies will have degeneracy $1$ or $15$. This is useful, since it reduces the numerical effort. For the case of a completely filled LL, we need only calculate the energies of states where there is no spin or pseudospin flip. We have in fact already seen these degeneracies in the excitonic dispersion relations calculated in Section \ref{subsubsec-dispinter}. Note that for $\nu=2$ there are 12 unmixed $0 \to 1$ transitions and 4 with no spin or pseudospin flips, which are mixed by the exchange Coulomb interaction. There are of course no pairs of transitions (one $0 \to 1$ and the other $-1 \to 0$ ), which are mixed by the direct Coulomb interaction for this filling. The possible dispersion relations are then $E^{(1)}$, which has degeneracy $12+3=15$ and $E^{(4)}_b$, which is non-degenerate.

\begin{figure}
\centering
  \includegraphics[width=3in]{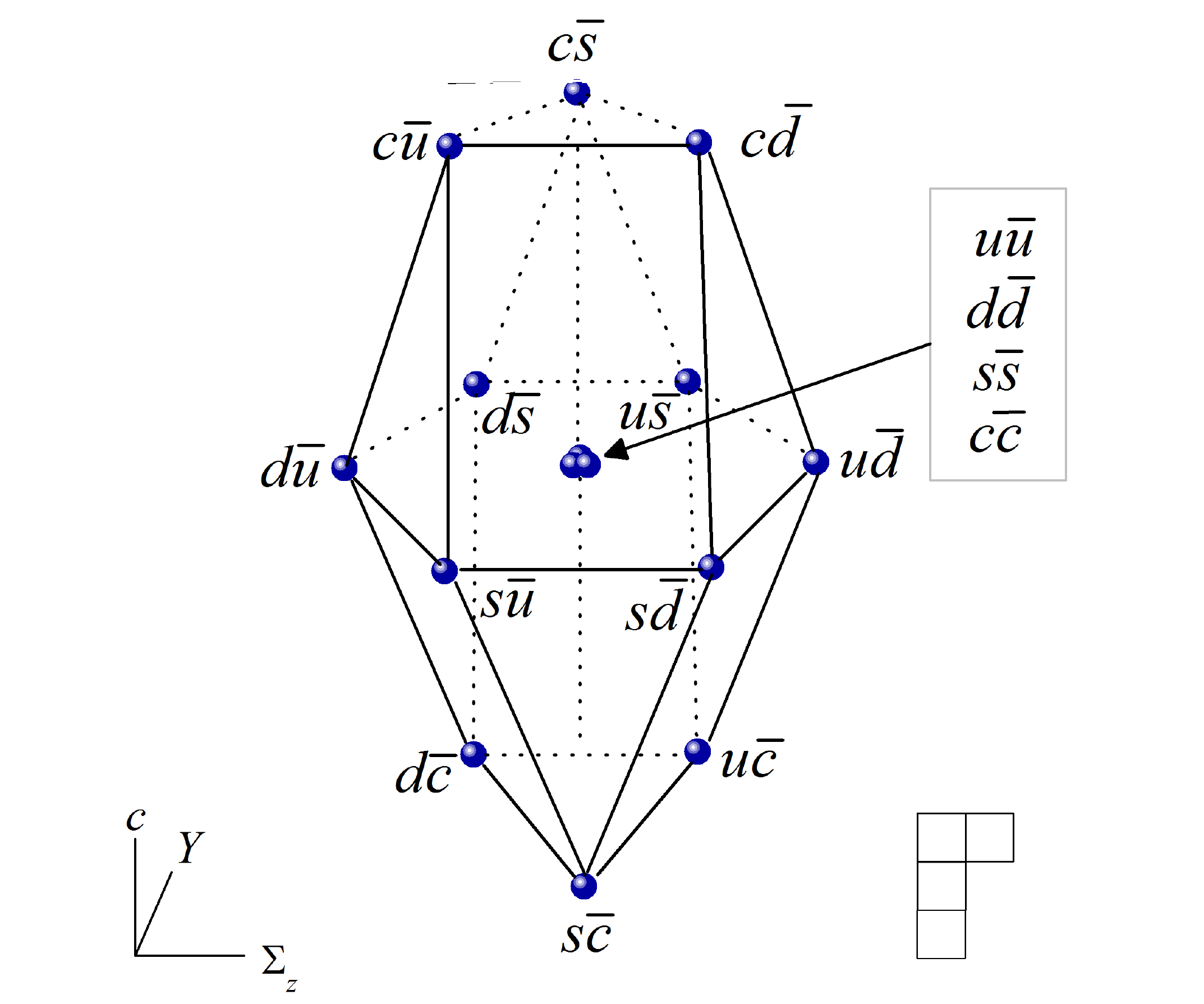} 
   \caption{SU(4) $\left[15 \right]$-plet describing the excitonic states in terms of flavour isospin ($\Sigma_z$), hypercharge ($Y$) and charm ($c$). }
\label{fig-excitonmult} 
\end{figure}

A way to visualise the $[15]$ multiplet is using the standard elementary particle physics $\mathrm{SU}(4)$ representation \cite{GreM97}, as shown in Fig.\ \ref{fig-excitonmult}. 
\begin{figure}[t]
\hspace*{-15mm}
  \includegraphics[width=6in]{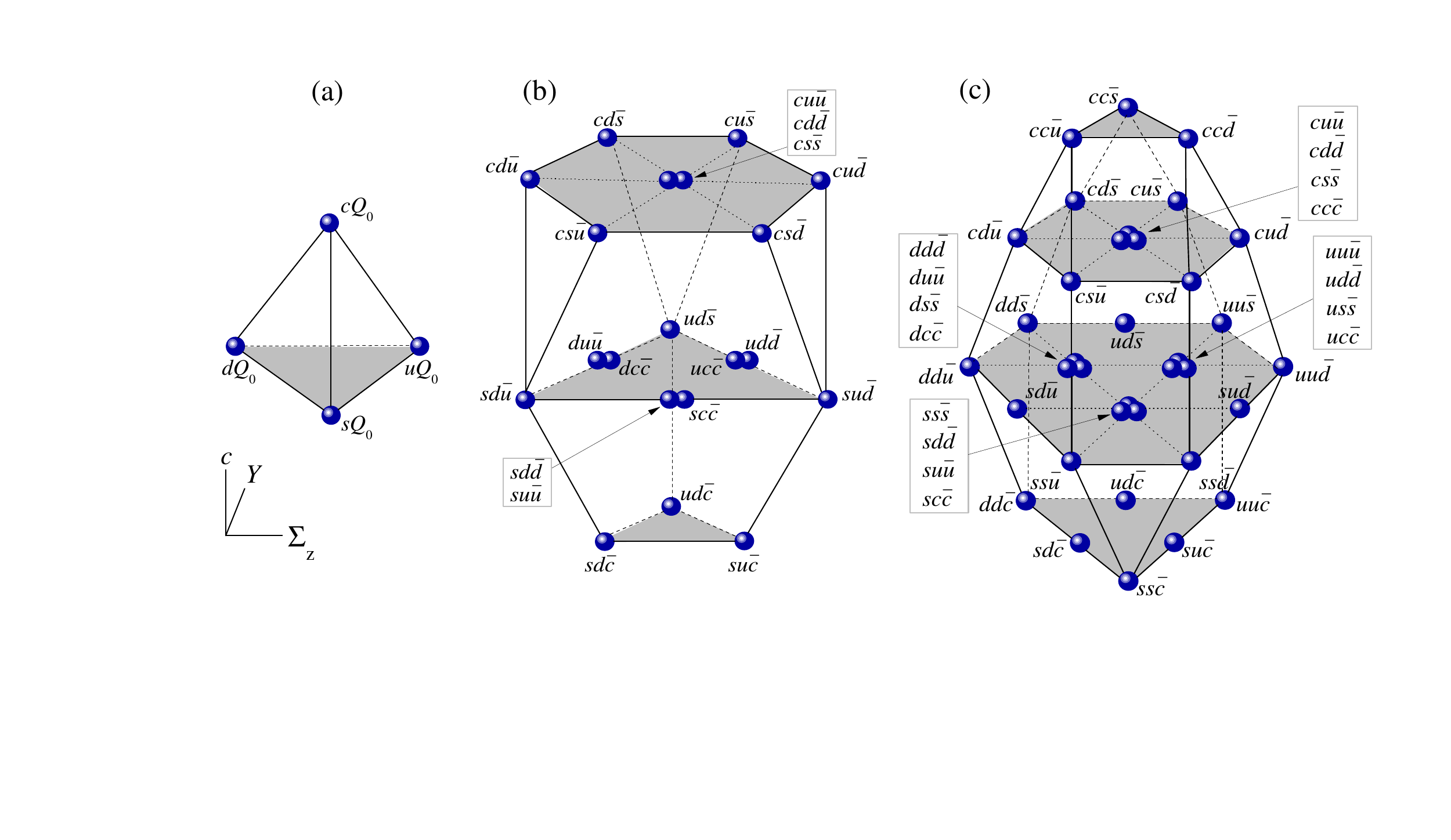}
   \caption{
    (a) Quartet ($\left[4 \right]$), (b) $\left[ \smash{\overline{20}} \right]$
  and (c) $\left[ 36\right]$ SU(4) multiplets
  describing the trion $X^-$ states. States are plotted in terms of flavour isospin ($\Sigma_z$), hypercharge ($Y$) and charm ($c$). Shading denotes the different levels of charm.
  }
\label{fig-trionmult} 
\end{figure}
Namely, the three orthogonal axes show:
the flavour isospin projection \newnot{sym:sigmaz}$\hat{\Sigma}_z = \frac{1}{2}\left( \hat{C}_{uu} - \hat{C}_{dd}\right) $,
the hypercharge \newnot{sym:Y}$\hat{Y} = \frac{1}{3}\left(\hat{C}_{uu}+\hat{C}_{dd}+\hat{C}_{cc} \right)-\frac{2}{3}\hat{C}_{ss} $
and the charm \newnot{sym:chat}$\hat{c} = \hat{C}_{cc}$.
For example, the exciton $d\bar{c}$ (exciton where the electron has flavour $d$ and the hole flavour $c$) has $\Sigma_z=-\frac{1}{2}$, $Y=0$ and $C=-1$.
\subsection{$\mathrm{SU}(4)$ multiplet structure for trions}
\label{subsec-trionmults}
In contrast to the exciton case, the charged CEs or trions have no direct particle physics analogy, since bound structures of the form $qq\bar{q}$ or $q\bar{q}\bar{q}$ have colour and so do not exist. The closest thing is a baryon, which is a bound structure of three quarks. However, we may still use the same techniques. 
For the $X^-$ trion, the multiplet structure is:
\begin{eqnarray}
\label{eq-youngmultstrion}
\yng(1) \otimes \yng(1) \otimes \yng(1,1,1)& = &\yng(1) \otimes \left( \yng(2,1,1) \oplus \underline{1}\right)\\ \nonumber
&=& \yng(3,1,1)\oplus \yng(2,2,1)\oplus \yng(1)\oplus \yng(1)
\end{eqnarray}
or $[4]\otimes[4]\otimes[\bar{4}]=[36]\oplus[\bar{20}]\oplus[4]\oplus[4]$. The multiplet plots are shown in Fig.\ \ref{fig-trionmult}. It is not clear what the physical meaning is behind the presence of two identical Young diagrams in Eq.\ (\ref{eq-youngmultstrion}).

The multiplet structure for the $X^+$ trion is
\begin{eqnarray}
\label{eq-youngmultstrionplus}
\yng(1,1,1) \otimes \yng(1,1,1)\otimes \yng(1)& = &\yng(1,1,1) \otimes \left( \yng(2,1,1) \oplus \underline{1}\right)\\ \nonumber
&=& \yng(3,2,2)\oplus \yng(2,1)\oplus \yng(1,1,1)\oplus \yng(1,1,1)
\end{eqnarray}
or $[\bar{4}]\otimes[\bar{4}]\otimes[4]=[36]\oplus[20]\oplus[\bar{4}]\oplus[\bar{4}]$. Although the Young diagrams in Eqs.\ (\ref{eq-youngmultstrion}) and (\ref{eq-youngmultstrionplus}) look different, the multiplicities are the same. The multiplet plots for the $X^+$ trion look similar to those for the $X^-$ shown in Fig.\ \ref{fig-trionmult}, except that all electrons are replaced by holes and all holes by electrons, which results in each coordinate of the three-particle complex changing sign. Indeed we shall see in the following Chapter that a symmetry exists between positively and negatively charged trions.
\cleardoublepage

\cleardoublepage
\chapter{Charged collective excitations in pristine graphene}
\label{chap-cces}
\begin{figure}
 \centering
 \includegraphics[width=0.29\textwidth]{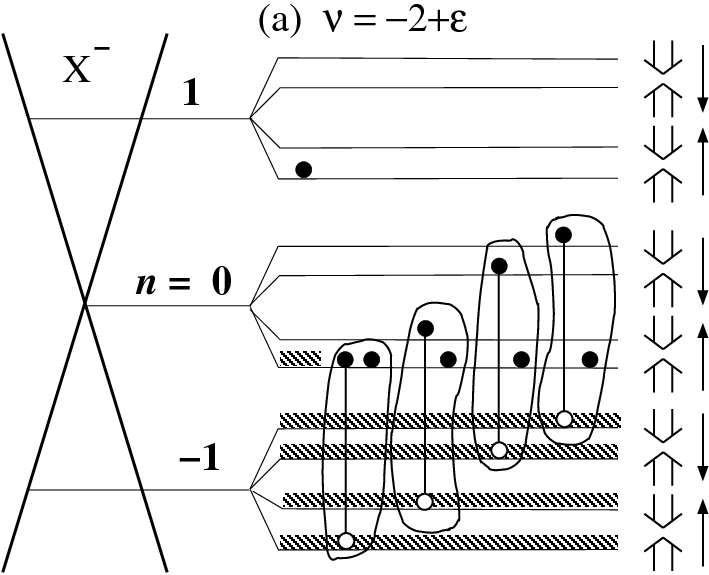}
 \includegraphics[width=0.29\textwidth]{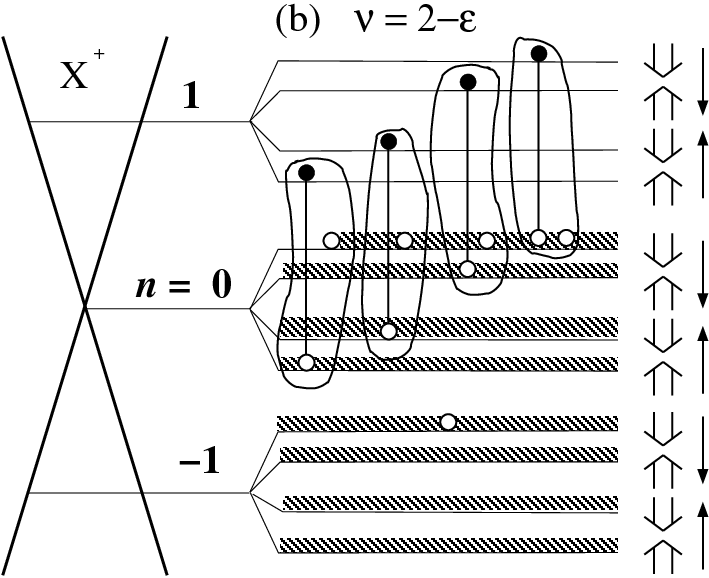}
 \includegraphics[width=0.29\textwidth]{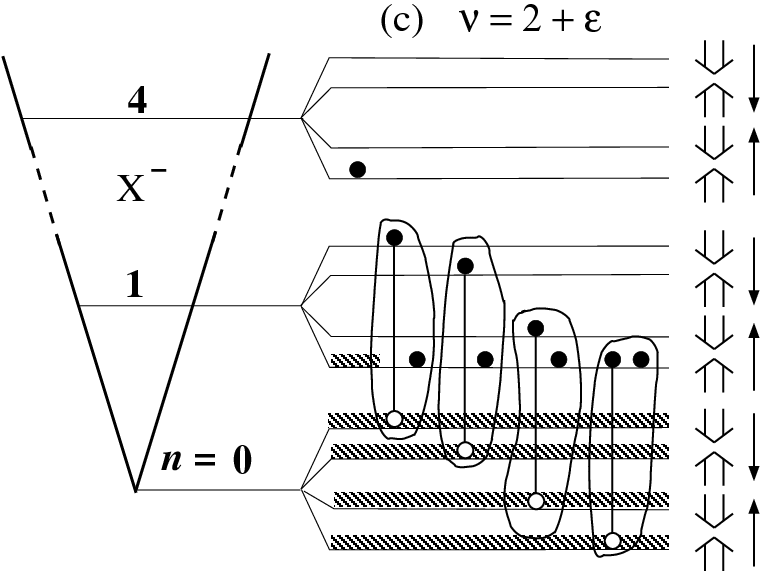}
  \caption {States resonantly mixed by the Coulomb interaction for various filling factors, $\nu$. Filled circles ($\bullet$) denote electrons, empty circles holes ($\circ$) and shaded regions, the Dirac sea. The ringed particle structures are bound trions. (a) $\nu=-2+\varepsilon$. The electron in the $n=1$ LL is only mixed for angular momenta $J_z\le \frac{1}{2}$. (b) $\nu=2-\varepsilon$. The hole in the $n=-1$ LL is only mixed for angular momenta $J_z\ge -\frac{1}{2}$. (c) $\nu=2+\varepsilon$. The electron in the $n=4$ LL is only mixed for angular momenta $J_z\le \frac{7}{2}$. 
}
\label{fig-trionmixing}
\end{figure}
The possibility of the formation of charged collective excitations (charged CEs) or trions in pristine graphene has already been mentioned in Section \ref{sec-neucharge}. The present Chapter expands upon this by applying the techniques detailed in Chapters \ref{chap-MPsPristineGraphene} and \ref{chap-su4} and examines the results. 
\section{Techniques for charged collective excitations}
\label{sec-tech}
\subsection{Creation operators}
\label{subsec-creationops}
We saw in Section \ref{sec-geom_sym} that charged CEs can be labelled by the quantum numbers $J_z$ and $k$. Eq.\ (\ref{eq-Rdagger}) can thus be rewritten to give the creation operator for a negatively charged CE or $X^-$ particle as\newnot{sym:Rdag}
\begin{equation}
\label{eq-Rdaggerminus} 
 \mbox{} \hspace{-5pt}
      R^{\dag}_{\mathcal{ N}_1  \mathcal{ N}_2 \mathcal{ N}_3}  =
          \sum_{m_1 , m_2}
           A_{\mathcal{N}_1 \mathcal{N}_2 \mathcal{ N}_3}^{J_z k}(m_1,m_2) \,
          c^{\dag}_{\mathcal{N}_3 m_3} c^{\dag}_{\mathcal{N}_1 m_1} d^{\dag}_{\mathcal{N}_2 m_2}.
\end{equation}
Notice that the sum over $m_3$ has been removed, since it is now fixed by $m_1,m_2$ and $J_z$ according to $J_z=|n_1|-m_1-|n_2|+m_2+|n_3|-m_3-\frac{1}{2}$. Similarly the creation operator for a positively charged CE or $X^+$ particle is\newnot{sym:Sdag}
\begin{equation}
\label{eq-Sdaggerplus} 
 \mbox{} \hspace{-5pt}
      S^{\dag}_{\mathcal{ N}_1  \mathcal{ N}_2 \mathcal{ N}_3}  =
          \sum_{m_1 , m_2}
           B_{\mathcal{N}_1 \mathcal{N}_2 \mathcal{ N}_3}^{J_z k}(m_1,m_2) \,
           c^{\dag}_{\mathcal{N}_1 m_1} d^{\dag}_{\mathcal{N}_2 m_2}d^{\dag}_{\mathcal{N}_3 m_3}.
\end{equation}
 For the $X^+$, $J_z=|n_1|-m_1-|n_2|+m_2-|n_3|+m_3+\frac{1}{2}$.
\subsection{Mixing of charged collective excitations}
\label{sec-ccemixing}
\begin{figure}
 \centering
 \includegraphics[width=3in]{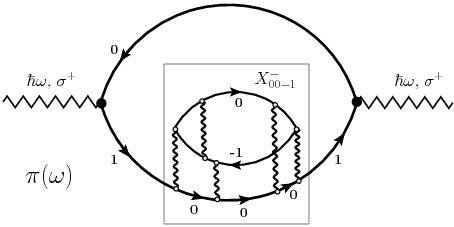}
  \caption { Photon polarisation operator $\pi(\omega)$.
  The box shows the resonant excitation of the \mbox{$e$-$h$}
  pair with the de-excitation of the electron to $n=0$ LL.
  Because of the multiple Coulomb interactions (wiggly vertical lines) the intermediate state
  becomes a bound trion $X^-_{00-1}$, with two electrons in the $n=0$ LL and a hole in the $n=-1$ LL,
  leading to a singularity in $\pi(\omega)$.
}
\label{fig-polarisation}
\end{figure}

We focus on charged CEs where there is no spin or pseudospin flip in the initially excited $e$-$h$ pair, since these are most likely to be optically active. As for the neutral CEs, we only take into account mixing between resonant excitations. Fig.\ \ref{fig-trionmixing} shows examples of which trions are mixed for different filling factors. Surprisingly, a single particle state may be mixed to the three particle complexes. This is an electron for filling factors of the form $\nu=\mu+\varepsilon$ and a hole for filling factors of the form $\nu=\mu-\varepsilon$. Looking at Fig.\ \ref{fig-trionmixing}(a), one can think of this as being due to the recombination of the electron in the $n=0$ LL and the hole in the $n=-1$ LL, which allows the second electron to be promoted from the $n=0$ to $n=1$ LL, since the energies are the same. Such a process is illustrated in Fig.\ \ref{fig-polarisation}.
For the case $\nu=2+\varepsilon$ shown in Fig.\ \ref{fig-trionmixing}(c), an electron from the $n=4$ LL may become mixed, since $\epsilon_4-\epsilon_1=\epsilon_1-\epsilon_0$. It is important to note that a charged CE will only have a non-zero amplitude for the single particle state, provided the generalised angular momentum projection for the single particle, $j_z$, is the same as that for the three particle complexes, $J_z$. Suppose the single particle has orbital quantum numbers $n,m$. Then for the $X^-$, an electron will only be mixed to the three particle states when $j_z\equiv|n|-m-\frac{1}{2}=J_z$. Hence it is only mixed for $J_z\le|n|-\frac{1}{2}$. Similarly for the $X^+$, a hole will only have a non-zero amplitude when $J_z\ge \frac{1}{2}-|n|$.

Once it is understood which states are mixed (significantly) by the Coulomb interaction, the energies and coefficients, $A_{\mathcal{N}_1 \mathcal{N}_2 \mathcal{ N}_3}^{J_z k}(m_1,m_2)$ or $B_{\mathcal{N}_1 \mathcal{N}_2 \mathcal{ N}_3}^{J_z k}(m_1,m_2)$, may be found by diagonalising the matrix for the Hamiltonian, given by Eq.\ (\ref{eq-ham}) with some additional terms to describe the scattering to a single particle state. In our calculations, we typically used $0\le m_1,m_2 \le 15$ which involves diagonalising square matrices with between $600$ and $900$ rows. 
\subsection{Self energy corrections}
\label{subsec-cceselfeng}
Mostly, the self energy correction is easy to handle, since it is the same as that for neutral CEs. Let's first consider the case $\nu=-2+\varepsilon$ shown in Fig.\ \ref{fig-trionmixing}(a). The excitation is a mixture of four states of the form $R^{\dag}_{0 \tau s \hspace{1mm} -1 \tau s\hspace{1mm} 0 \Uparrow \uparrow}|-2\rangle$ and for the case, $J_z\le\frac{1}{2}$, an additional single electron state, $c^\dagger_{1 \Uparrow \uparrow \frac{1}{2}-J_z}|-2\rangle$. Note that the ground state is $c^\dagger_{0 \Uparrow \uparrow m}|-2\rangle$ for some general $m$, not simply $|-2\rangle$. Then, considering the single particle state, the self energy contribution to 
$\langle-2|c_{1 \Uparrow \uparrow \frac{1}{2}-J_z }\hat{H}c^\dagger_{1 \Uparrow \uparrow \frac{1}{2}-J_z }|-2\rangle-\langle-2|c_{0 \Uparrow \uparrow m}\hat{H}c^\dagger_{0 \Uparrow \uparrow m}|-2\rangle$ is $E_{0 \tau s}^{1 \tau s}-0.75E_0$. The self energy correction to \\ $\langle-2|R_{0 \tau s \hspace{1mm} -1 \tau s\hspace{1mm} 0 \Uparrow \uparrow}\hat{H}
R^{\dag}_{0 \tau s \hspace{1mm} -1 \tau s\hspace{1mm} 0 \Uparrow \uparrow}|-2\rangle-\langle-2|c_{0 \Uparrow \uparrow m}\hat{H}c^\dagger_{0 \Uparrow \uparrow m}|-2\rangle$ is $E_{0 \tau s}^{1 \tau s}=E_{-1 \tau s}^{0 \tau s}$, as for neutral CEs. 

There are a couple of cases, where calculating the self energy corrections is more involved. These are $\nu=2+\varepsilon$ for the $X^-$ and $\nu=-2-\varepsilon$ for the $X^+$. As we shall see in the following section, these two cases are dual, so it is enough to consider only the former, which is shown in Fig.\ \ref{fig-trionmixing}(c). The $X^-$ excitation is a mixture of four states, $R^{\dag}_{1 \tau s \hspace{1mm} 0 \tau s\hspace{1mm} 1 \Uparrow \uparrow}|2\rangle$, and for the case, $J_z\le\frac{7}{2}$, an additional single electron state, $c^\dagger_{4 \Uparrow \uparrow\frac{7}{2}-J_z }|-2\rangle$. The ground state is $c^\dagger_{1 \Uparrow \uparrow m}|2\rangle$. The self energy correction to 
$\langle 2|R_{1 \tau s \hspace{1mm} 0 \tau s\hspace{1mm} 1 \Uparrow \uparrow}\hat{H}R^{\dag}_{1 \tau s \hspace{1mm} 0 \tau s\hspace{1mm} 1 \Uparrow \uparrow}|2\rangle
-\langle2|c_{1 \Uparrow \uparrow m}\hat{H}c^\dagger_{1 \Uparrow \uparrow m}|2\rangle$ is $E_{0 \tau s}^{1 \tau s}$, as before. However, the self energy correction to 
$\langle 2|c_{4 \Uparrow \uparrow\frac{7}{2}-J_z }\hat{H}c^\dagger_{4 \Uparrow \uparrow\frac{7}{2}-J_z }|2\rangle
-\langle2|c_{1 \Uparrow \uparrow m}\hat{H}c^\dagger_{1 \Uparrow \uparrow m}|2\rangle$ is $\sum_{n=0}^{n_\mathrm{c}}[E_{\mathrm{SE}}\left( 4,-n\right)-E_{\mathrm{SE}}\left( 1,-n\right)]$. Using $n_\mathrm{c}\approx810$ at $B=20\mathrm{T}$ as before, $\sum_{n=0}^{810}[E_{\mathrm{SE}}\left( 4,-n\right)-E_{\mathrm{SE}}\left( 1,-n\right)]\approx 0.89E_0$.

As previously mentioned, the charged CEs appear as discrete states below the neutral CE continuum discussed in Section \ref{sec-disp}. The binding energy of a charged CE is the difference between the energy of the lower continuum edge and the energy of the charged CE. Thus in the graphical presentation of numerical results for the energies (see Section \ref{sec-resultscces}), the energy of the lower continuum edge is set equal to zero. For the cases $\mu=0,\pm1$, it is possible for an exciton to be created with both electron and hole in the $n=0$ LL and for the second electron to be in the $n=1$ LL. Using results from Section \ref{subsec-dispeg}, this gives a lower continuum edge at the energy $\epsilon_1+E_{0 \tau s}^{1 \tau s}-E_0$. For the cases $\mu=\pm2$, the lower continuum edge is at $\epsilon_1+E_{0 \tau s}^{1 \tau s}-0.75E_0$, since only excitons with one composite particle in the $n=0$ LL and one with $|n|=1$ can be formed. 
\subsection{Symmetry between $X^+$ and $X^-$}
\label{sec-symmetry}
It is only necessary to perform calculations for either positively or negatively charged CEs, since they are linked by a symmetry relation. The energies and expansion coefficients are identical for an $X^+$ state formed at filling factor $\nu=\mu-\varepsilon$ with quantum numbers $J_z$, $k$ and an $X^-$ state formed at filling factor $\nu=-\mu+\varepsilon$ with quantum numbers $-J_z$, $k$. The duality is an immediate consequence of the inherent particle-hole symmetry in pristine graphene. An example of this can be seen in Figs.~\ref{fig-trionmixing}(a) and (b). From now on, the discussion will focus on the $X^-$ particle.
\section{Optical selection rules for charged collective excitations}
\label{sec-optselcces}
Optical selection rules for charged CEs can be derived from the commutation relations:
\begin{equation}
 \label{eq-delHcommk}
[\delta \hat{\mathcal{H}}_\pm,\hat{\mathbf{k}}^2]=0,
\end{equation}
\begin{equation}
 \label{eq-delHcommjz}
[\delta \hat{\mathcal{H}}_\pm,\hat{J}_z]=\pm\delta \hat{\mathcal{H}}_\pm,
\end{equation}
where $\delta \hat{\mathcal{H}}_\pm$ is defined in Eq.\ (\ref{eq-colldelH}). Suppose the ground state is $|e\rangle=c^\dagger_{n \tau s m}|\mu\rangle$ with $j_z=|n|-m-\frac{1}{2}$ and the photocreated trion state, $|eeh\rangle$, has geometric quantum numbers $k$, $J_z$. The intensity of the resulting absorption peak is then directly proportional to the square of the dipole matrix element, $\langle eeh|\delta \hat{\mathcal{H}}_\pm|e\rangle$. Thus from Eq.\ (\ref{eq-delHcommk}), we may derive the optical selection rule, $k-m=0$. This reflects the negligible linear momentum of a photon in the dipole approximation
and the fact that $k$ is associated with translations; it is the discrete analog of linear momentum for charged states in a magnetic field. From Eq.\ (\ref{eq-delHcommjz}), $J_z-j_z=\pm1$ for the $\sigma^{\pm}$ circular polarisations. Combining the selection rules for $k$ and $J_z$ yields $J_z+k=|n|-\frac{1}{2}\pm1$. In Section \ref{sec-geom_sym}, we saw $J_z+k$ is equal to the generalised angular momentum projection of the seed state, $J_z^{(0)}$. Hence if a family of states is optically active in the $\sigma^+$ polarisation, $J_z^{(0)}=|n|+\frac{1}{2}$; if it is active in the $\sigma^-$ polarisation, $J_z^{(0)}=|n|-\frac{3}{2}$.

These selection rules are a necessary, but not sufficient condition for a family of states to be optically active. There is another selection rule associated with the electron flavour discussed in Chapter \ref{chap-su4}. It arises from the fact that $\delta \hat{\mathcal{H}}_\pm \propto Q_0$ (see Eq.\ (\ref{eq-colldelH})), so the photon is flavourless in the $\mathrm{SU}(4)$ sense. From this and considerations of the dipole transition matrix elements, $\langle eeh| Q_0 |e\rangle$, it immediately follows that the states in the $\left[ \smash{\overline{20}} \right]$  and $[36]$ multiplets are all dark, while the states in the $[4]$-multiplet may be bright, provided they have the proper orbital quantum numbers just described.
\section{Numerical results}
\label{sec-resultscces}
We performed calculations for trions with no spin or pseudospin flips for the filling factors $\nu=\mu\pm\varepsilon$ with $\mu=0,\pm1,\pm2$ and charged CEs were only found below the band for $\mu=\pm2$. This can be attributed to the fact that the lower continuum edge for the $\mu=\pm2$ cases is $0.25 E_0$ higher than for the remaining filling factors, as seen in Section \ref{subsec-cceselfeng}. Hence less $e$-$h$ attraction is required to push the energy of the trion below the band. For the $\mathrm{SU}(4)$ symmetric case, $\mu=\pm2$, the trion states with spin or pseudospin flips have the same energies as those with no flips. This is not the case for $\mu=0,\pm1$, but it is unlikely that charged CEs will be found even if spin and pseudospin flips are taken into account, due to the continuum reaching low energies.

\begin{figure}
 \centering
 \includegraphics[width=2.7 in]{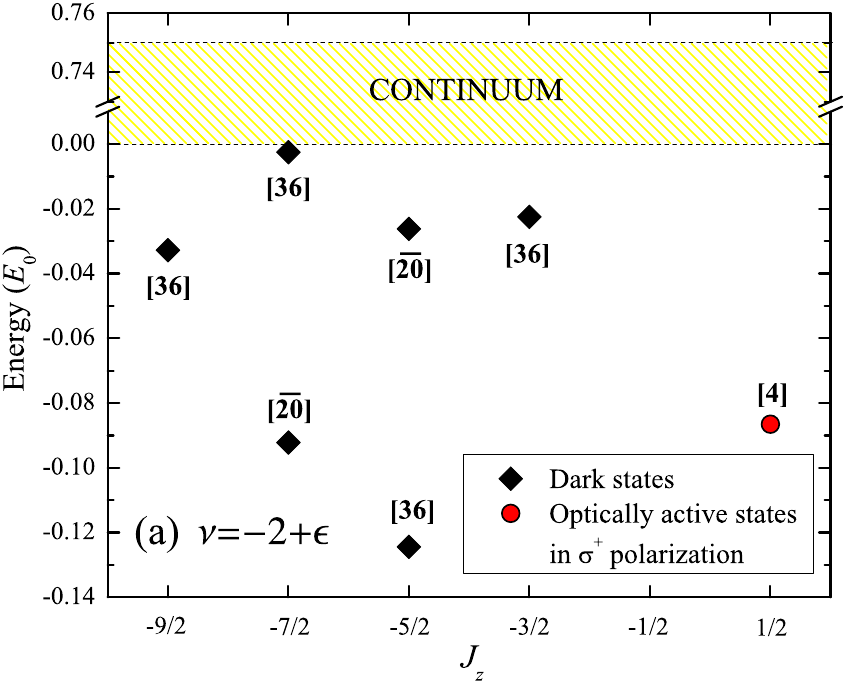}
 \includegraphics[width=2.7 in]{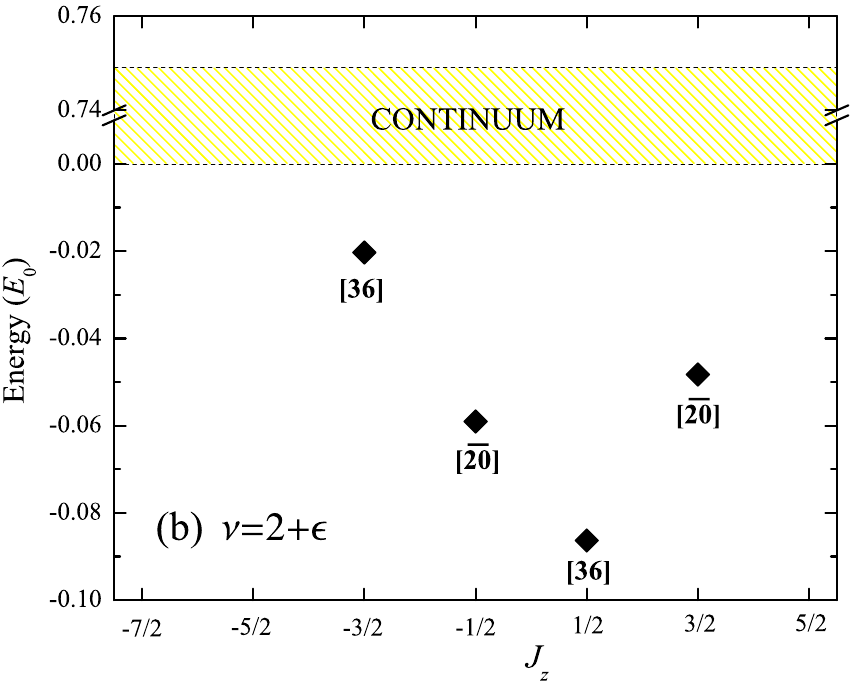}
  \caption { Bound $X^-$ states and their multiplicities as a function of $J_z$ for (a) $\nu= -2+\varepsilon$ and (b) $\nu= 2+\varepsilon$.
  Only the seed states ($k=0$) are shown. Energy zero corresponds to the CRE at $\epsilon_1+E_{0 \tau s}^{1 \tau s}-0.75E_0$.
}
\label{fig-trionres}
\end{figure}
Figure \ref{fig-trionres} shows the energies of $X^-$ states relative to the lower continuum edge for (a) $\nu= -2+\varepsilon$ and (b) $\nu= 2+\varepsilon$. The relevant mixed states for these filling factors are shown in Figs.\ \ref{fig-trionmixing}(a) and (c) respectively. The multiplicities have been marked on, indicating which $\mathrm{SU}(4)$ multiplet the states belong to. Calculations were made assuming no spin/pseudospin flips in the exciton and that the additional bound electron had flavour $c$. Hence the numerical diagonalisation found the state $Q_0c$ from the $[4]$-plet, two states combining $cu\bar{u}$, $cd\bar{d}$ and $cs\bar{s}$ from the $[20]$-plet and three states combining $cu\bar{u}$, $cd\bar{d}$, $cs\bar{s}$ and $cc\bar{c}$ from the $[36]$-plet. The remaining states in a multiplet can be found by repeatedly acting on a single state by the generators. The structure of the Young diagrams (see Eq.\ (\ref{eq-youngmultstrion})) should tell us about the symmetry of the states they represent, although this is less clear when a hole or ``antiparticle'' is involved, so that a single box no longer corresponds to a single particle. The diagrams for the $[20]$-plet, $\yng(2,2,1)$ and the $[36]$-plet, $\yng(3,1,1)$, both have a hole to the left and two electrons to the right. For the $[20]$-plet the electrons are in the same column and states within this multiplet are antisymmetric with respect to interchanging their flavours. For the $[36]$-plet the electrons are in the same row and states within this multiplet are symmetric with respect to interchanging their flavours.

Out of all the states formed at $\mu=\pm2$, only those at one energy, indicated by the red circle in Fig.\ \ref{fig-trionres}(a), are optically active. Notice that these states obey both the flavour selection rule (they are in the $[4]$-plet) and the geometrical selection rules, $J_z^{(0)}=\frac{1}{2}$ for the $\sigma^+$ polarisation. The state in Fig.\ \ref{fig-trionres}(b) with $J_z^{(0)}=\frac{3}{2}$ satisfies the geometrical selection rules, but not the flavour selection rule, so is not optically active.

\section{Experiment}
\label{sec-ccesexperiment}
Charged CEs of the 2DEG created by confining electrons to a semiconductor quantum well have been observed in many optical experiments using absorption and luminescence techniques \cite{KheCdM93,FinSB95,BuhMWB95,ShiOSP95,TisBGP02,AstYRC05}. Both $X^+$ \cite{ShiOSP95} and $X^-$ \cite{KheCdM93,FinSB95,BuhMWB95,TisBGP02,AstYRC05} particles have been detected in the presence and absence of magnetic fields. Signatures of charged CEs have also been seen in semiconductor quantum dots \cite{TisBGP02}. The binding energy is of the order of $\mathrm{meV}$. Theoretical predictions \cite{DzyS00} suggest it should be $\sim 0.1E_0$, which agrees with experiment, since $E_0 \sim 10\mathrm{meV}$. The binding energies we predict for charged CES in graphene are also $\sim 0.1E_0$, so they are similar to those for the 2DEG in units of $E_0$. However, $E_0$ depends inversely on the dielectric constant, $\epsilon$. For GaAs quantum wells, $\epsilon \approx 12$. It is difficult to assign a value to $\epsilon$ for the case of graphene, since this will depend on its environment. At least for suspended graphene, $\epsilon<12$, meaning that trion binding energies should be higher in suspended graphene than in the 2DEG in GaAs quantum wells. Thus trions should be easier to observe in graphene, at least in principle.
\begin{figure}
 \centering
 \includegraphics[width=2.7 in]{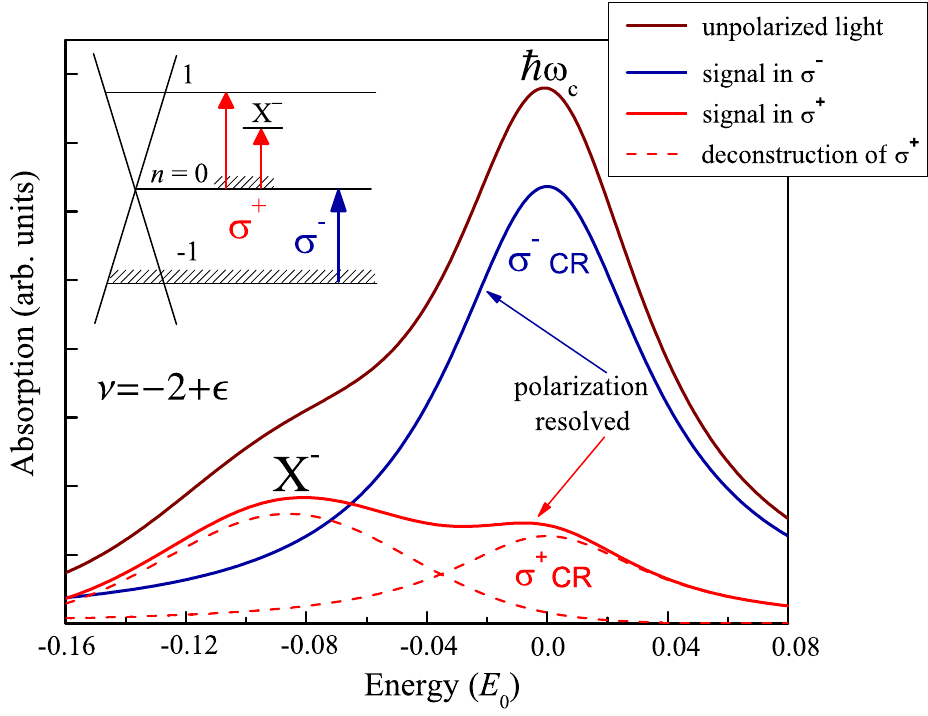}
  \caption {
Predicted absorption spectra in pristine graphene for the bright state at energy \mbox{$-0.09 E_0$} 
  with \mbox{$\nu = -2 + \varepsilon$}. 
  The inset schematically shows the relevant optical transitions. 
  Energies are given in units of $E_0$ with zero energy corresponding to the CR energy, $\epsilon_1+E_{0 \tau s}^{1 \tau s}-0.75E_0$.
}
\label{fig-trionoptics}
\end{figure}

A recent ARPES study of potassium doped graphene in the absence of a magnetic field, indicates that the simple picture of two conical bands touching at the Dirac point is not the full story \cite{BosSSH10}. In addition to these two charge bands, a pair of plasmaron bands are clearly visible in their high resolution data, so that the crossing at the Dirac point is in fact resolved into three crossing points. A plasmaron is formed when a hole becomes bound to a plasmon. Hence it is the zero-field analogue of our charged CEs.

To the best of my knowledge, charged CEs in graphene in high magnetic fields have not yet been observed, which is perhaps surprising since they are now routinely seen in semiconductor quantum wells. Our study found one bright $X^-$ state at $\nu = -2 + \varepsilon$ with binding energy $\approx 0.09 E_0$  ($J_z = \frac{1}{2}$, active in $\sigma^+$, see Fig.~\ref{fig-trionres}(a)). Note that the $eeh$ complexes involved are created when an exciton with the hole in the $n=-1$ LL and the electron in the $n=0$ LL become bound to an additional electron. As such they are not optically active in the $\sigma^+$ polarisation. However, as indicated in Fig.\ \ref{fig-polarisation}, they ``dress'' the optical response of the resonant process by which an electron is promoted from the $n=0$ to $n=1$ LL, which is active in the $\sigma^+$ polarisation. This should be observable in absorption experiments, provided the noise is low enough. It is expected to give rise to a separate absorption peak below the CRE or may show up as a low-energy CRE shoulder, depending on the broadening. This is shown in Fig.\ \ref{fig-trionoptics}, where the spectra  are broadened $\sim 0.05 E_0$, which amounts to 2\,meV at $B=10$\,T. This broadening is slightly below that found in absorption spectra of the cleanest graphene samples currently available \cite{HenCJL10}, which may explain why signatures of charged CEs are not seen in this data. The intensity of both the $X^-$ peak and the main CRE peak scales according to $\varepsilon$ in the $\sigma^+$ polarisation. Notice that because of the $e$-$h$ symmetry, there exists the $X^+$
counterpart at $\nu =  2 - \varepsilon$, which is bright in $\sigma^-$  polarisation.

The dark bound $X^-$ and $X^+$ states may in principle be observed by tunneling spectroscopy,
which is a sensitive tool for probing discrete states in the spectrum. This can be pursued, e.g.,
in tunneling experiments involving gate-tunable graphene quantum dots \cite{GueFSI10}.
Also, the various symmetry-breaking terms such as external fields, disorder and
lattice defects, ripples, and deformations \cite{CasGPN09,AbeABZ10}
may partially lift the limitations following from the
orbital and $\mathrm{SU}(4)$ selection rules. This can make some of the dark states ``grey'' and
detectable in the absorption spectra as well as possibly in photoluminescence \cite{SchBSH03}.
\cleardoublepage

\cleardoublepage
\chapter{Localised neutral collective excitations in dirty graphene}
\label{chap-nces}
The previous three chapters have studied CEs of pristine graphene. Although graphene is relatively clean, when compared to other materials, disorder is still always present. For the purpose of applications, there is a lot of interest in determining the dominant scattering mechanisms, which limit the DC conductivity of graphene. One would expect the type of disorder in a graphene sample to depend on its method of fabrication \cite{ChoLSK10,SolMD10}. Although mechanical exfoliation produces high quality samples, useful for investigating fundamental properties of the material, scalable growth methods such as chemical vapour decomposition and epitaxial growth are the techniques likely to produce large enough samples for industrial applications. Whenever graphene is grown on a substrate, it will experience strain due to the mismatch between the crystal lattices. This may result in ripples or blisters forming in the sheet, or alternatively the breaking of $\sigma$-bonds and the formation of structural defects such as pentagons and heptagons. In samples of graphene grown on nickel, lines of structural defects composed of pentagons and heptagons were observed \cite{LahLBO10}. They act as dopants resulting in a line of charge. Other sources of disorder may be point-like impurities either in the substrate or on top of the graphene layer. In this Chapter we study the localisation of neutral CEs on a single such impurity, which models the physical situation of a low impurity density. Two different impurity types are considered for comparison: i.\ a short range ($\delta$-function) scatterer due to a vacancy or neutral foreign impurity at one of the graphene lattice sites, ii.\ a long-range charged impurity on or close to the graphene layer. The position of a Coulomb impurity within a graphene unit cell is irrelevant due to its long-ranged potential, which barely changes over such length scales. A similar study was carried out for the 2DEG for the case of the Coulomb impurity \cite{DzyL93}.
\section{Impurity Hamiltonian}
\label{sec-impham}
\subsection{Single particle impurity Hamiltonian}
\label{subsec-spimpham}
The impurity is assumed to be at the origin and the symmetric gauge is chosen.
We use the single particle Hamiltonian derived in Ref.\ \cite{AndN98} to incorporate the impurity potential.
The amplitudes of the electronic wave function at positions $\mathbf{R}_A$\newnot{sym:RAB} in
sublattice $A$ and $\mathbf{R}_B$ in sublattice $B$ are written as\newnot{sym:psiAB}\newnot{sym:FABKK'}
\begin{equation}
 \label{eq-wfampa}
\psi_A(\mathbf{R}_A)=e^{i\mathbf{K}.\mathbf{R}_A}F^\mathbf{K}_A(\mathbf{R}_A)+e^
{i\mathbf{K}'.\mathbf{R}_A}F^{\mathbf{K}'}_A(\mathbf{R}_A),
\end{equation}
\begin{equation}
 \label{eq-wfampb}
\psi_B(\mathbf{R}_B)=-\omega
e^{i\mathbf{K}.\mathbf{R}_B}F^\mathbf{K}_B(\mathbf{R}_B)+e^{i\mathbf{K}'.\mathbf
{R}_B}F^{\mathbf{K}'}_B(\mathbf{R}_B),
\end{equation}
where \newnot{sym:omegaval}$\omega=\mathrm{exp}\left(2\pi i/3\right)$. Here the $F$s are envelope
wave functions which change slowly on the scale of the lattice constant, $a$.
Upon substituting these into the tight binding equation and considering nearest
neighbour hopping only, the effective-mass equation is found to be:
\begin{equation}
\label{eq-effmass}
\left[\left(\begin{array}{cc}
      v_{\mathrm{F}}\bm{\sigma}.\hat{\bm{\Pi}} & 0 \\
0 & v_{\mathrm{F}}\bm{\sigma}^\ast.\hat{\bm{\Pi}}
      \end{array}\right)+V\right]\left(\begin{array}{c}
                      F^\mathbf{K}_A\\
		      F^\mathbf{K}_B\\
		      F^{\mathbf{K}'}_A\\
		      F^{\mathbf{K}'}_B
                      \end{array}
\right)
=E
\left(\begin{array}{c}
                      F^\mathbf{K}_A\\
		      F^\mathbf{K}_B\\
		      F^{\mathbf{K}'}_A\\
		      F^{\mathbf{K}'}_B
                      \end{array}
\right),
\end{equation}
where 
\begin{equation}
 \label{eq-impmat}
V=\left(\begin{array}{cccc}
       v_A(\mathbf{r}) &0 & \tilde{v}_A(\mathbf{r}) & 0\\
0 & v_B(\mathbf{r}) & 0 & -\omega^\ast \tilde{v}_B(\mathbf{r}) \\
\tilde{v}_A^\ast(\mathbf{r}) & 0 & v_A(\mathbf{r}) & 0 \\
0 & -\omega \tilde{v}_B^\ast(\mathbf{r}) & 0 & v_B(\mathbf{r})
      \end{array}
\right)
\end{equation}
and $E$ is the single particle energy.
In Eq.\ (\ref{eq-impmat}), 
\begin{equation}
 \label{eq-vdef}
v_A(\mathbf{r})=\sum_{\mathbf{R}_A}g(\mathbf{r}-\mathbf{R}_A)v_A'(\mathbf{r})
\end{equation}
and
\begin{equation}
 \label{eq-vtdef}
\tilde{v}_A(\mathbf{r})=\sum_{\mathbf{R}_A}g(\mathbf{r}-\mathbf{R}_A)e^{
i(\mathbf{K}'-\mathbf{K}).\mathbf{R}_A}v_A'(\mathbf{r}),
\end{equation}
where $v_A'(\mathbf{r})$ is the onsite potential energy and $g(\mathbf{r})$ is a
normalised real function, which only has an appreciable amplitude for
$|\mathbf{r}|$ less than a few times $a$. Similar expressions for
$v_B(\mathbf{r})$ and $\tilde{v}_B(\mathbf{r})$ are obtained by replacing $A\to
B$ in Eqs.\ (\ref{eq-vdef}) and (\ref{eq-vtdef}).

Let us now examine the form of the impurity matrix, $V$, for the Coulomb and
$\delta$-function impurity potentials. The Coulomb impurity potential is
\newnot{sym:VC}$V_\mathrm{C}(r)=\frac{Ze^2}{\epsilon_{\mathrm{imp}}r}$, where $Z$\newnot{sym:Zimp} is the charge
of the impurity in units of $e$ and $\epsilon_{\mathrm{imp}}$\newnot{sym:epsimp} is the dielectric
constant associated with the electron-impurity interaction. 
Screening of Coulomb impurities has been seen to be relevant in graphene
\cite{TerMKS08},
although the situation is not yet fully understood, particularly for a strong
magnetic field.
Coulomb impurities with charge $|Z|=1$ belong to the subcritical regime and
hence screening effects
due to the substrate and the electron system in graphene can be modelled
via an effective charge $|Z_{\rm eff}| <|Z|$ \cite{TerMKS08}, which is
equivalent to assuming an effective dielectric constant,
$\epsilon_{\mathrm{imp}}$.
Since energies are always measured in units of $E_0$ here, which contains
$\epsilon$, it is only the ratio $\epsilon/\epsilon_{\rm imp}$, which is
important. Its value for graphene is not known
\cite{IyeWFB07,BycM08}, so for most results presented below
$\epsilon/\epsilon_{\rm imp}=1$ and $Z=\pm1$ is assumed.

Based on the previous discussion, the Coulomb impurity is long-ranged, so that
$v_A(\mathbf{r})=v_B(\mathbf{r})=V_\mathrm{C}(r)$. In addition, for any
long-range impurity, $\tilde{v}_A(\mathbf{r})$ and $\tilde{v}_B(\mathbf{r})$ are
negligible, since they are reduced to a sum over oscillating functions (see Eq.\
(\ref{eq-vtdef})). Hence for the Coulomb impurity,
$V^\mathrm{C}=V_\mathrm{C}(r)\eins_4$\newnot{sym:VCup}. For a $\delta$-function impurity on an
$A$ sublattice site at the origin,
$v_A(\mathbf{r})=\tilde{v}_A(\mathbf{r})=V_0\delta(\mathbf{r})$ and
$v_B(\mathbf{r})=\tilde{v}_B(\mathbf{r})=0$ for some constant \newnot{sym:V0}$V_0=Wa^2$, where
$W$\newnot{sym:Wimp} is the impurity potential. This gives\newnot{sym:VAB}
\begin{equation}
 \label{eq-himpa}
V^A=V_0\left(\begin{array}{cccc}
\delta(\mathbf{r}) & 0 & \delta(\mathbf{r}) & 0\\
0 & 0 & 0 & 0 \\
\delta(\mathbf{r}) & 0 & \delta(\mathbf{r}) & 0\\
 0 & 0 & 0 & 0
\end{array}
\right).
\end{equation}
Conversely for a $\delta$-function impurity on a $B$ sublattice site at the
origin, $v_B(\mathbf{r})=\tilde{v}_B(\mathbf{r})=V_0\delta(\mathbf{r})$ and
$v_A(\mathbf{r})=\tilde{v}_A(\mathbf{r})=0$, yielding
\begin{equation}
 \label{eq-himpb}
V^B=V_0\left(\begin{array}{cccc}
0 & 0 & 0 & 0\\
0 & \delta(\mathbf{r}) & 0 & -\omega^\ast\delta(\mathbf{r})\\
0 & 0 & 0 & 0 \\
0 & -\omega \delta(\mathbf{r}) & 0 & \delta(\mathbf{r}) 
\end{array}
\right).
\end{equation}
Notice that the Coulomb impurity does not scatter between the valleys or
sublattices, whereas the $\delta$-function impurity does scatter between the
valleys, whilst preserving the sublattice.
\subsection{Second quantised impurity Hamiltonian}
\label{subsec-mbimpham}
The full second quantised Hamiltonian describing the single particle
energies, the $e$-$h$ interactions and the interaction with the impurity is
$\hat{H}+\hat{H}_\mathrm{imp}$, where $\hat{H}$ is given by Eq.\ (\ref{eq-ham}) on page \pageref{eq-ham}
and\newnot{sym:Hint}
\begin{equation}
 \label{eq-himpop}
\hat{H}_\mathrm{imp} = \sum_{\mathcal{N}, \mathcal{N'}} \sum_{m,
m'}{\mathcal{V}_i}_{\mathcal{N}m}^{\mathcal{N'}m'}  c^{\dag}_{\mathcal{N'}
m'}c_{\mathcal{N} m}- \sum_{\mathcal{N}, \mathcal{N'}} \sum_{m,
m'}{\mathcal{V}_i}_{\mathcal{N}m}^{\mathcal{N'}m'} d^{\dag}_{\mathcal{N'}
m'}d_{\mathcal{N} m}.
\end{equation}
The impurity matrix elements are given by \newnot{sym:mathcalV}${\mathcal{V}_i }_{\mathcal{N}
m}^{\mathcal{N'} m'}=\delta_{s s'}\int \int d \mathbf{r} ^2 \Psi^\dag_{
N'}(\mathbf{r})V^i\Psi_{ N}(\mathbf{r})$, where $i \in \left\lbrace A, B,
\mathrm{C}\right\rbrace$ denotes the impurity type and $N=\{n m \tau\}$, $\mathcal{N}=\{n \tau s\}$ as before. For the Coulomb impurity, 
\begin{equation}
\label{eq-impmec}
{\mathcal{V}_\mathrm{C} }_{\mathcal{N} m}^{\mathcal{N'}
m'}=\delta_{\mathcal{N}\mathcal{N}'}\delta_{m m'}a_{n}^2\left( \mathcal{S}_n^2
V_{|n|-1 \,  m} + V_{|n| \, m} \right),
\end{equation}
where $V_{n \, m}=\int \int d \mathbf{r}
^2\psi_{nm}^\ast(\mathbf{r})V_\mathrm{C}(r)\psi_{nm}(\mathbf{r})$\newnot{sym:2degimpmes} are the 2DEG
matrix elements calculated in Appendix \ref{app-mes}. We ignore off-diagonal terms with
$n\ne n'$ or $m \ne m'$, since these give the second order contributions to the
energy, which are small compared to the first order contribution.

The general form of a matrix element for a $\delta$-function impurity is 
\begin{equation}
\label{eq-impmeab}
{\mathcal{V}_{A\left(B \right)} }_{\mathcal{N} m}^{\mathcal{N'} m'}\propto
\delta_{s s'} V_0 \psi_{k  m'}^*\left( \bf{0}\right)\psi_{l  m}\left(
\bf{0}\right),
\end{equation}
where $l \epsilon \left\lbrace|n|,|n|-1 \right\rbrace$ and $k \epsilon
\left\lbrace|n'|,|n'|-1 \right\rbrace$, with exact values set by $\tau,\tau'$
and whether the impurity is on the $A$ or $B$ sublattice. This imposes selection
rules determining which states are connected by the delta impurity, since
$\psi_{n m}\left(\bf{0} \right)  \neq 0$ if and only if $n=m$. For completeness
$\psi_{n n}\left(\bf{r} \right)  = \frac{i^n}{\sqrt{2 \pi \ell _B^2}}L_n\left(
\frac{r^2}{2\ell _B^2}\right)\mathrm{exp}\left(\frac{-r^2}{4\ell_B^2} \right) 
$, where the $L_n$ are Laguerre polynomials, so that $\psi_{n n}\left(\bf{0}
\right)  = \frac{i^n}{\sqrt{2 \pi \ell _B^2}}  $. Exact expressions for
${\mathcal{V}_{A\left(B \right)} }_{\mathcal{N} m}^{\mathcal{N'} m'}$ are:
\begin{equation}
\label{eq-impmeuu}
{\mathcal{V}_A}_{n \Uparrow s m}^{n' \Uparrow s' m'}={\mathcal{V}_B}_{n
\Downarrow s m}^{n' \Downarrow s' m'}=\mathcal{S}_n\mathcal{S}_{n'}
2^{-\frac{1}{2}\left(2-\delta_{0 n}-\delta_{0 n'} \right)}\delta_{s
s'}\delta_{|n'|-1 m'}\delta_{|n|-1 m}V_0\frac{i^{m+m'}}{2\pi \ell_B^2},
\end{equation}
\begin{equation}
\label{eq-impmeud}
{\mathcal{V}_A}_{n \Uparrow s m}^{n' \Downarrow s' m'}=-\omega{\mathcal{V}_B}_{n
\Downarrow s m}^{n' \Uparrow s' m'}=\mathcal{S}_n
2^{-\frac{1}{2}\left(2-\delta_{0 n}-\delta_{0 n'} \right)}\delta_{s
s'}\delta_{|n'| m'}\delta_{|n|-1 m}V_0\frac{i^{m+m'}}{2\pi \ell_B^2},
\end{equation}
\begin{equation}
\label{eq-impmedu}
{\mathcal{V}_A}_{n \Downarrow s m}^{n' \Uparrow s'
m'}=-\omega^\ast{\mathcal{V}_B}_{n \Uparrow s m}^{n' \Downarrow s'
m'}=\mathcal{S}_{n'} 2^{-\frac{1}{2}\left(2-\delta_{0 n}-\delta_{0 n'}
\right)}\delta_{s s'}\delta_{|n'|-1 m'}\delta_{|n| m}V_0\frac{i^{m+m'}}{2\pi
\ell_B^2},
\end{equation}
\begin{equation}
\label{eq-impmedd}
{\mathcal{V}_A}_{n \Downarrow s m}^{n' \Downarrow s' m'}={\mathcal{V}_B}_{n
\Uparrow s m}^{n' \Uparrow s' m'}=2^{-\frac{1}{2}\left(2-\delta_{0 n}-\delta_{0
n'} \right)}\delta_{s s'}\delta_{|n'|m'}\delta_{|n|m}V_0\frac{i^{m+m'}}{2\pi
\ell_B^2}.
\end{equation}

We treat both impurity types as a perturbation, using the first order term only.
For the Coulomb impurity, the ratio of the second order energy correction,
$E^{(2)}$, to the first order energy correction, $E^{(1)}$, is
$E^{(2)}/E^{(1)}\sim\alpha^\mathrm{g}$. For the case of the $\delta$-function
impurity, let us illustrate the details on a particular example, which works
similarly to other scenarios. We shall consider an impurity on the $B$
sublattice and calculate the ratio of the first and second order corrections to
the single particle energy of the $|0\Uparrow \downarrow 
0\rangle=c^\dagger_{0\Uparrow \downarrow  0}|0\rangle$ state. The first order
correction is 
$E^{(1)}=\langle 0\Uparrow \downarrow 0|\hat{H}_\mathrm{imp}|0 \Uparrow
\downarrow  0\rangle=\frac{V_0}{2 \pi \ell_B^2}.$ 
The second order correction is 
\begin{equation}
\label{appBe2}
E^{(2)}={\sum_{n=1}^{n_\mathrm{c}} \sum_{m=0}^\infty}\dfrac{|\langle n\Uparrow
\downarrow m|\hat{H}_\mathrm{imp}|0\Uparrow \downarrow 0\rangle|^2
+|\langle n \Downarrow\downarrow  m|\hat{H}_\mathrm{imp}|0  \Uparrow \downarrow
0\rangle|^2}{\epsilon_0-\epsilon_n}.
\end{equation}
 This represents virtual processes where an electron spontaneously hops to a
higher LL and then decays back to its original state. Using $\langle n \Uparrow
\downarrow  m|\hat{H}_\mathrm{imp}|0 \Uparrow\downarrow
0\rangle={\mathcal{V}_B}_{0  \Uparrow \downarrow  0}^{n \Uparrow\downarrow 
m}=\delta_{|n|m}\frac{i^m V_0}{2 \sqrt{2} \pi \ell_B^2}$ and ${\langle
n\Downarrow \downarrow  m|\hat{H}_\mathrm{imp}|0 \Uparrow \downarrow 
0\rangle=-\omega\delta_{|n|-1 m}\frac{i^m V_0}{2 \sqrt{2} \pi \ell_B^2}}$, we
obtain for Eq.\ (\ref{appBe2})\footnote{Note that $\sum_{n=1}^{n_\mathrm{c}} \dfrac{1}{\sqrt{n}}$ is the generalised Harmonic number of order $n_\mathrm{c}$ of 1/2.}
\begin{equation}
\label{appB_e2}
E^{(2)}=-\dfrac{V_0^2}{4 \sqrt{2} \pi^2 \hbar v_\mathrm{F}
\ell_B^3}\sum_{n=1}^{n_\mathrm{c}} \dfrac{1}{\sqrt{n}}.
\end{equation}
This gives the ratio of the first and second order terms as
\begin{equation}
 \label{appBratio}
\dfrac{E^{(2)}}{E^{(1)}}=\dfrac{-V_0}{2 \sqrt{2} \pi \hbar v_\mathrm{F}
\ell_B}\sum_{n=1}^ {n_\mathrm{c}}\dfrac{1}{\sqrt{n}}.
\end{equation}
It may seem disappointing that $\frac{E^{(2)}}{E^{(1)}}\sim \sqrt{B}$, since we
are in the high magnetic field regime, but the ratio is still small. In fact the
ratio for the $n^{\mathrm{th}}$ order correction behaves as
$\frac{E^{(n)}}{E^{(1)}}\sim \left( \frac{V_0}{\hbar v_\mathrm{F} \ell_B}\right)
^{n-1}$ with $\frac{V_0}{\hbar v_\mathrm{F} \ell_B}<<1$. Using $B=20\mathrm{T}$,
$n_\mathrm{c}=810$ and $V_0=1\mathrm{eV} \times a^2$, we find
$\frac{|E^{(2)}|}{|E^{(1)}|}\approx 0.03$. This suggests that second and higher
order terms in the perturbation expansion can be safely ignored.
\section{Mixing of excitations by the $\delta$-function impurity Hamiltonian}
\label{sec-impmix}
The Coulomb impurity matrix is diagonal, so it does not mix different neutral CEs,
$Q^\dagger_{\mathcal{ N}_1  \mathcal{ N}_2 J_z } |\nu\rangle$. Hence the mixing
of excitations is determined by the $e$-$h$ interaction, which was described in
Section \ref{sec-mixing}. The $\delta$-function impurity behaves quite
differently in this respect, scattering between single particle states in
different valleys. This results in mixing between excitons with and without a
pseudospin flip. We seek optically active states, so the procedure is
to begin with an optically active transition and then determine which other
transitions it is mixed to. Recall for the Coulomb impurity, the mixed
transitions were those four with no spin and no pseudospin flips and with
$J_z=\pm1$. For the $\delta$-function impurity, either six or eight transitions
are mixed depending on the sublattice the impurity is located on, the filling
factor and the light polarisation. Notably the $\delta$-function impurity mixes
states with different values of $J_z$, so $J_z$ is not a good quantum number in
this case.

\begin{figure}
 \centering
 \includegraphics[width=0.47\textwidth]{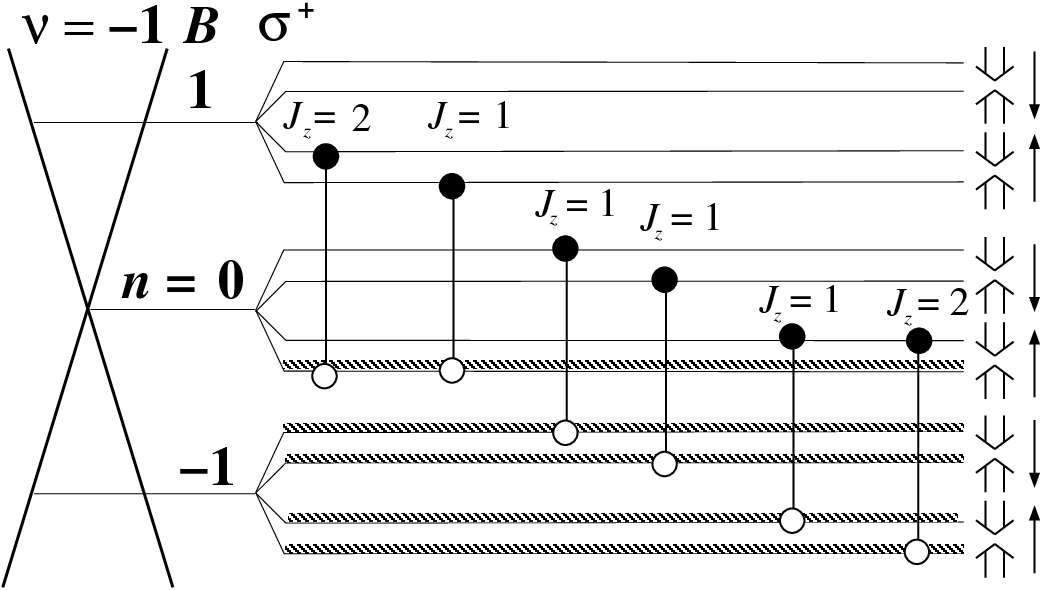}
  \caption {Set of mixed transitions for impurity on $B$ sublattice, $\sigma^+$
polarisation and $\nu=-1$.
}
\label{fig-delmix}
\end{figure}

To make the approach clearer, we now consider a specific example, where the
impurity is on a $B$ sublattice site (at the origin), the filling factor is
$\nu=-1$ and the graphene sheet is illuminated by $\sigma^+$ circularly
polarised light. As can be seen in Fig.\ \ref{fig-delmix}, we have taken the
$\bf{K}$ valley ($\Uparrow$) to be lower than the $\bf{K}'$ valley
($\Downarrow$) in energy. 
This valley splitting leads to a possible inequivalence of the $A$ and $B$
sublattices, so that results may be different for an impurity located on the $A$
sublattice and one located on the $B$ sublattice. However, the same results are
always obtained when moving the impurity from the $A$ to $B$ sublattice, whilst
simultaneously switching the energies of the $\bf{K}$ and $\bf{K'}$ valleys. Let
${|\mathcal{ N}_1 m_1, \mathcal{N}_2 m_2 \rangle=c^{\dag}_{\mathcal{N}_1 m_1}
d^{\dag}_{\mathcal{N}_2 m_2}|\nu=-1 \rangle}$. The least energetic optically
active inter LL transition takes the form $|1 \Uparrow \uparrow m, 0\Uparrow
\uparrow m \rangle$ for $m=0,1,\hdots$, which has $J_z=1$. The only transitions,
which are in resonance are other transitions from LLs $n=0$ to $n=1$ and those
from LLs $n=-1$ to $n=0$. We do not correct the CRE for the valley splitting,
setting $\hbar \omega_v=0$, since $\hbar \omega_v\ll\hbar \omega_\mathrm{c}$.
Thus transitions with pseudospin flips may be considered in resonance with those
without. Spin flip transitions are not involved in the CE, since neither the
$e$-$h$ Coulomb interaction nor the interaction with the impurity flips the
spin. The transition ${|1 \Uparrow \uparrow m, 0 \Uparrow \uparrow m \rangle}$
is mixed via the Coulomb exchange interaction to the optically dark transitions
${|0  \Downarrow \uparrow m, -1 \Downarrow \uparrow m+2 \rangle}$, ${|0 \Uparrow
\downarrow m, -1 \Uparrow \downarrow  m+2 \rangle}$ and ${|0 \Downarrow
\downarrow m, -1 \Downarrow \downarrow m+2 \rangle}$. Here the oscillator
quantum numbers are determined by the conservation of $J_z$. 

According to the selection rules given in Eqs.\
(\ref{eq-impmeuu})-(\ref{eq-impmedd}), a particle can only be scattered to or
from a state with $\tau=\Uparrow$ ($\tau=\Downarrow$) by an impurity on the $B$
sublattice, when $|n|=m$ ($|n|-1=m$). Hence the latter three transitions are not
mixed with any others by the impurity interaction. The hole in ${|1 \Uparrow
\uparrow m, 0 \Uparrow \uparrow m \rangle}$ only interacts with the impurity
when $m=0$, in which case its quantum state is unaltered. The electron only
interacts with the impurity when $m=1$ and the corresponding excitation is ${|1
\Uparrow \uparrow 1, 0 \Uparrow \uparrow 1 \rangle}$. This is scattered to
either ${|1 \Uparrow \uparrow 1, 0 \Uparrow \uparrow 1 \rangle}$ (unchanged) or
to ${|1 \Downarrow \uparrow 0, 0 \Uparrow \uparrow 1 \rangle}$ by the
electron-impurity ($e$-$\mathrm{imp}$) interaction. Note that $J_z=2$ for ${|1
\Downarrow \uparrow 0, 0 \Uparrow \uparrow 1 \rangle}$, so excitations of the
form ${|1 \Downarrow \uparrow m, 0 \Uparrow \uparrow m+1 \rangle}$ must be
included. Such transitions are mixed by the direct $e$-$h$ Coulomb interaction
to those of the form ${|0 \Downarrow \uparrow m, -1 \Uparrow \uparrow m+3
\rangle}$, which are not mixed with other types of transition by the impurity.
We have now obtained the set of six excitations illustrated in Fig.\
\ref{fig-delmix} , ${|1 \Uparrow \uparrow m, 0 \Uparrow \uparrow m \rangle}$,
${|0 \Downarrow \uparrow m, -1 \Downarrow \uparrow m+2 \rangle}$, ${|0 \Uparrow
\downarrow m, -1 \Uparrow \downarrow  m+2\rangle}$, ${|0 \Downarrow \downarrow
m, -1 \Downarrow \downarrow m+2 \rangle}$, ${|1 \Downarrow \uparrow m, 0
\Uparrow \uparrow  m+1 \rangle}$, ${|0 \Downarrow \uparrow m, -1 \Uparrow
\uparrow m+3 \rangle}$, such that each member excitation is only mixed to other
members of the set and which contains the only excitation optically active in
the $\sigma^+$ polarisation for this filling factor. One could of course
construct different sets of mutually mixed excitations, but these would be dark
and the aim of this work is to focus on the optical problem.

It is now straightforward to construct the Hamiltonian matrix and to diagonalise
it numerically. This requires cutting off the sum over $m$ at a finite value.
This is entirely justified physically, since we seek excitations, which are
localised and only have large amplitudes for low $m$ values. We include the
first $30$ $m$-values in our calculations, so that the total matrix size is
$180$ for the example detailed above. Changing the cut off value for $m$ close to $m=30$ does not change the energies appreciably, so that the errors are within symbol size for all results presented below.
\begin{figure}
  \centering
 \includegraphics[width=0.47\textwidth]{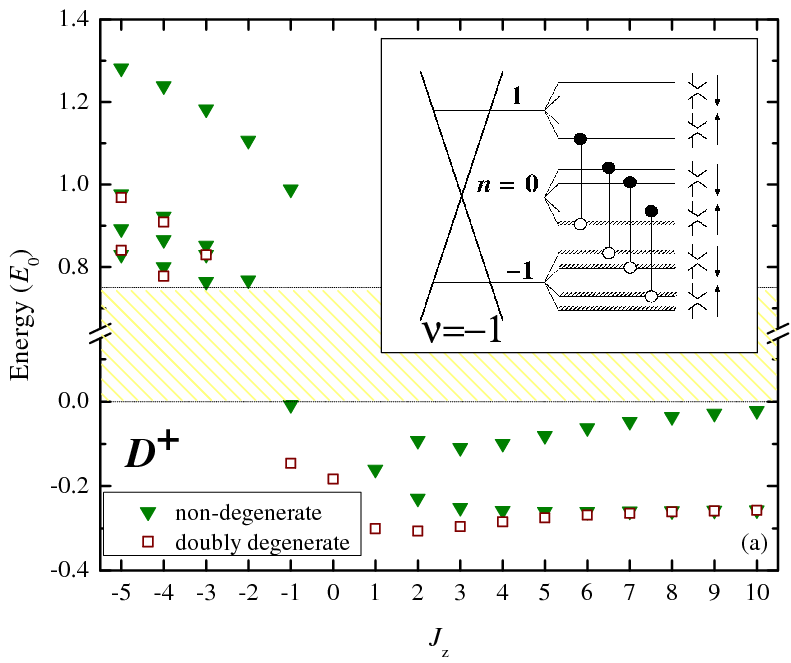}
  \includegraphics[width=0.47\textwidth]{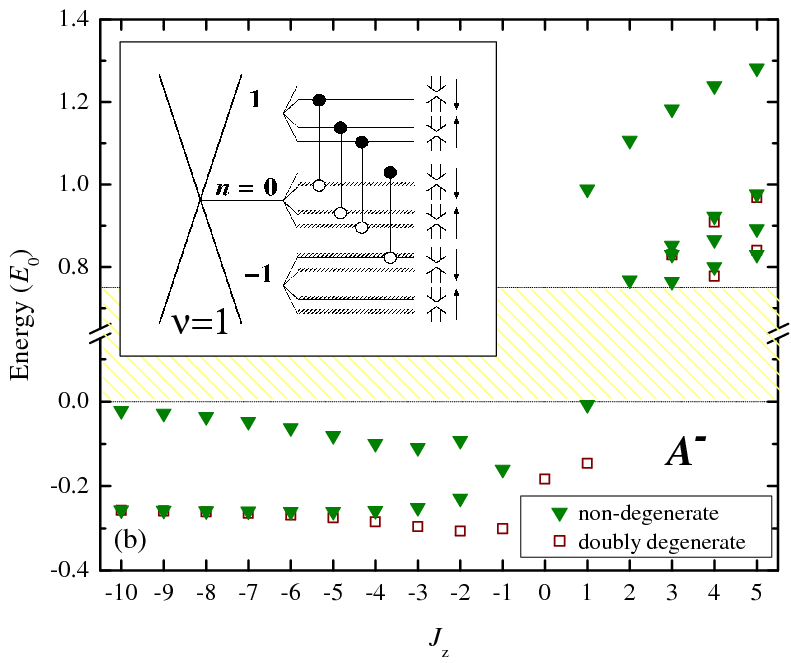}
  \caption {neutral CEs localised on (a) a donor impurity at $\nu=-1$ and
(b) an acceptor impurity at $\nu=1$.
Energies are given relative to $\epsilon_1+E_{0 \tau s}^{1 \tau s}-0.75E_0$
in units of $E_0$ as a function of the generalised angular momentum projection,
$J_z$. The hatched area of width $0.75 E_0$ represents the continuum of extended
neutral CEs.
Quasibound states within the continuum are not shown.
Insets show four branches of resonantly mixed inter-LL transitions conserving
spin and pseudospin.
}
\label{fig-emz}
\end{figure}
\section{Results}
\label{sec-impres}
\subsection{Coulomb impurity}
\label{subsec-Couimp}
A certain duality exists in our results as a consequence of particle-hole
symmetry in graphene. The energies and absorption strengths are the same under
simultaneously changing the sign of the filling factor
($\nu\leftrightarrow-\nu$), the direction of light polarisation
($\sigma^+\leftrightarrow \sigma^-$), the sign of the angular momentum
projection ($J_z \leftrightarrow -J_z$) and the sign of the impurity's charge
(donor, $D^+$\newnot{sym:Dplus} $\leftrightarrow$ acceptor, $A^-$\newnot{sym:Aminus}). An example of this is shown in
Fig.\ \ref{fig-emz}. The neutral CEs which are localised on the impurity appear as
discrete states outside the continuum of delocalised neutral CEs. All the presented
results are for $\epsilon/\epsilon_\mathrm{imp}=1$ unless otherwise stated.

Figure~\ref{fig-emz}(a) shows for $\nu = -1$ four low-energy branches of neutral CEs
localised on the $D^+$ 
for $J_z > 1$; two of these branches are degenerate. Their nature is explained
as follows.
For large positive $J_z$, the hole is on average much farther away from the
impurity than the electron (see Eq.\ (\ref{eq-r2diff})). 
Therefore, the $e$-$D^+$ attraction dominates over the $h$-$D^+$ and $e$-$h$
interactions. 
Hence we find branches with asymptotic $J_z \gg 1$ energies equal to 
$\epsilon_1+E_{0 \tau s}^{1 \tau s}+{\mathcal{V}_\mathrm{C} }_{\mathcal{N}
m}^{\mathcal{N} m}$ ($m=0, 1, \hdots$). 
As an example, notice the three branches approaching energy $-0.25 E_0$ 
and the single branch approaching 
zero energy 
in Fig.~\ref{fig-emz}(a).
These originate, respectively, from the three $n=-1 \rightarrow n=0$ transitions
(denoted hereafter as $T_{-10}$\newnot{sym:Tn2n1})   
and from the single $T_{01}$ transition for $\nu=-1$, since
${\mathcal{V}_\mathrm{C} }_{0 \tau s 0}^{0 \tau s 0}=-E_0$ and
${\mathcal{V}_\mathrm{C} }_{1 \tau s 0}^{1 \tau s 0}=-0.75E_0$.
\begin{figure}
  \centerline{
 \includegraphics[width=0.98\columnwidth]{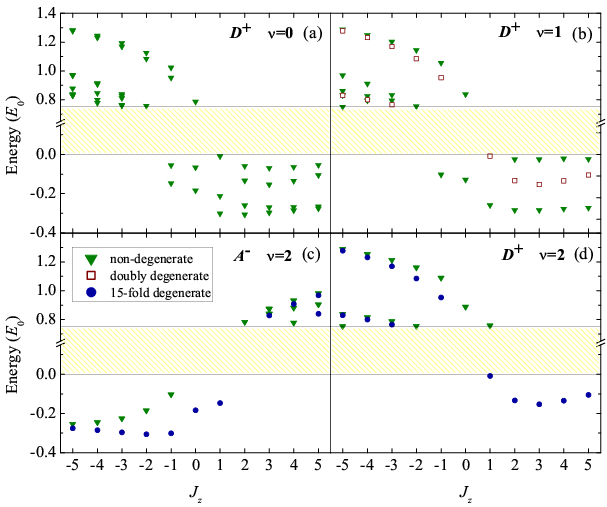} 
  }
  \caption{neutral CEs localised on the $D^+$ at (a) $\nu=0$, (b) $\nu=1$, (d)
$\nu=2$, 
and (c) localised on the $A^-$ at $\nu=2$. 
     }
\label{fig-emzmult}
\end{figure}

Similar asymptotic behaviour can be seen for other filling factors in
Fig.~\ref{fig-emzmult}.
The neutral CEs above the band develop for $J_z < 0$, 
when the hole is closer to the $D^+$ than the electron (or the electron is
closer to the $A^-$ than the hole). 
Such unusual excited states are bound in 2D because of the confining effect of
the magnetic field.
Due to the duality, results for the $A^-$ at $\nu=-1,0$ (not shown) can be
obtained from those for the $D^+$ 
by changing $J_z \rightarrow - J_z$ and $\nu \rightarrow - \nu$.

For the case $\nu=2$, we have the full $\mathrm{SU(4)}$ symmetry, so I have
marked on the true degeneracies of the states in Figs.\ \ref{fig-emzmult}(c) and
(d), as determined in Section \ref{subsec-exmults}. For the other filling
factors, the symmetry between different single particle spin/pseudospin flavours
is broken, but only partially. Hence if the energies of excitons with spin and
or pseudospin flips were calculated, I would expect some of them to be
degenerate with the energies shown. However, this is left to further work and
the degeneracies given are due to excitons with no spin or pseudospin flips
only.
\begin{figure}
  
 \includegraphics[width=0.47\columnwidth]{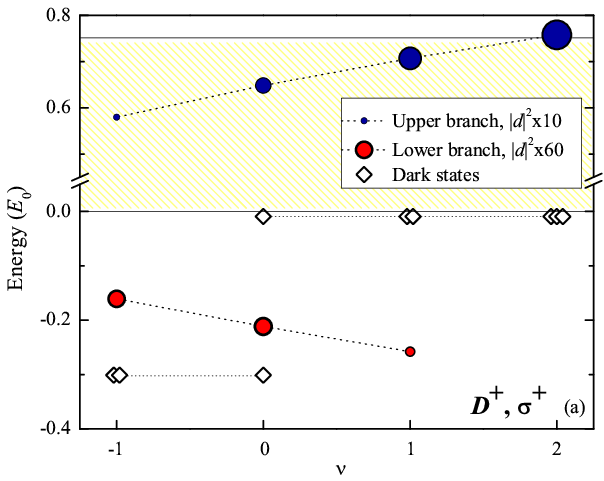} 
  \includegraphics[width=0.47\columnwidth]{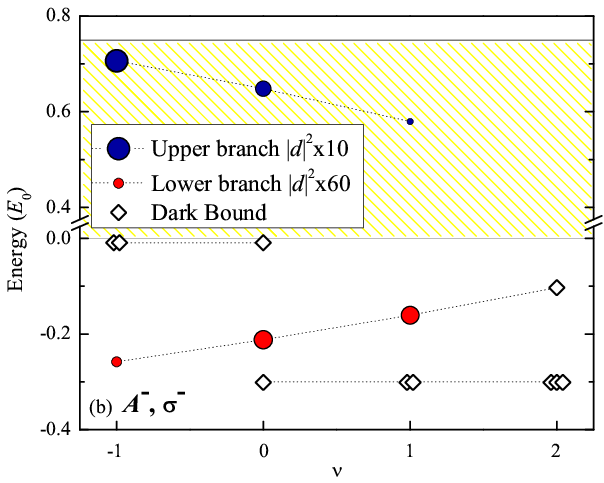}\newline
\includegraphics[width=0.47\columnwidth]{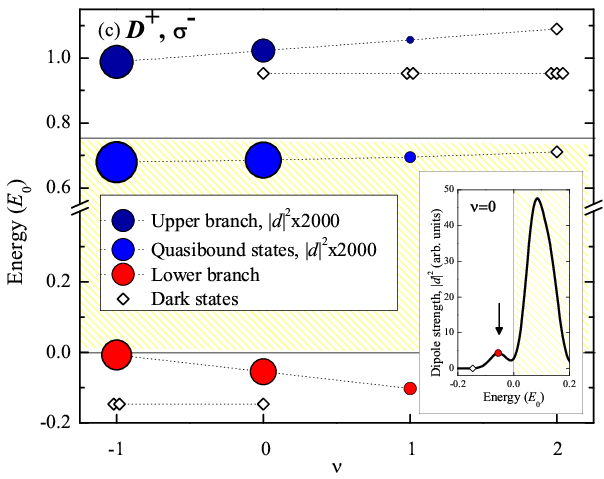}
\includegraphics[width=0.47\columnwidth]{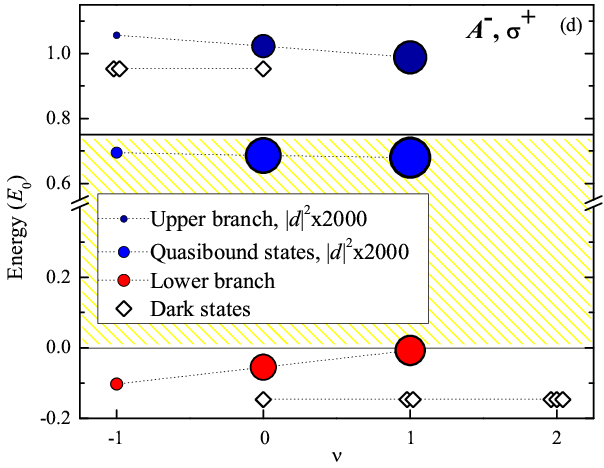}

  \caption{Evolution with filling factor $\nu$ of energies and optical strengths
of neutral CEs localised on (a) the $D^+$
with $J_z=1$ active in the $\sigma^+$ polarisation, (b) the $A^-$ with $J_z=-1$
active in the $\sigma^-$ polarisation, (c) the $D^+$
with $J_z=-1$ active in the $\sigma^-$ polarisation and (d) the $A^-$ with
$J_z=1$ active in the $\sigma^+$ polarisation.
The optically active states are indicated by circles with sizes $\sim |d|^2$.
Their degeneracies are not given for clarity; this information is contained
within Figs.\ \ref{fig-emz} and \ref{fig-emzmult}.
The diamonds represent optically dark states. The degeneracies implied by the
close overlapping symbols are due to excitons with no spin or pseudospin flips
only.
The dotted lines are guides to the eye. Inset of (c): Dipole strength $|d|^2$
versus energy for $\nu=0$.
The spectra were convoluted with a Gaussian of width $0.03E_0$.
The arrow indicates an impurity-related feature below the CRE. 
     }
\label{fig-optics}
\end{figure}
\begin{figure}
  \centering
 \includegraphics[width=3in]{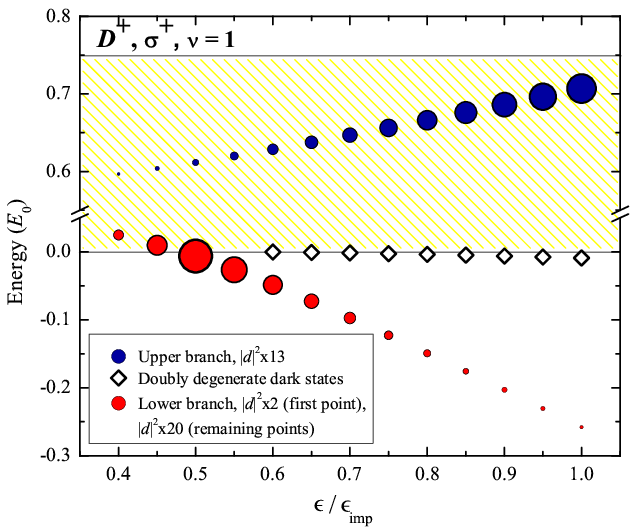} 
  \caption{Energies and absorption strengths of neutral CEs localised on the $D^+$ for
filling factor $\nu=1$ and light polarisation $\sigma^+$ as a function of
$\epsilon/\epsilon_\mathrm{imp}$. 
     }
\label{fig-screening}
\end{figure}

Figure~\ref{fig-optics} shows the optical properties of the localised states for
various different parameters. The duality between states in Figs.\
\ref{fig-optics}(a) and (b) and also between states in Figs.\
\ref{fig-optics}(c) and (d) is evident. For this reason, we restrict the
discussion of results to the $D^+$.
Note that for each polarisation, there are two branches of dark states indicated
by diamonds. Optically active localised states are shown using circles, with
size which scales as the modulus squared of the dipole matrix element,\newnot{sym:dme}
$d=\langle \mathcal{N}_1 \mathcal{N}_2 J_z|\hat{\mathcal{H}}_\pm|\nu\rangle$.
Two types of localised states can be optically observed: 
(i) truly bound states, which are split off the continuum and have normalisable
wave functions with exponentially decaying tails,
(ii) quasibound states within the continuum, which have high probability amplitudes on the impurity and long-range oscillating tails.
The latter may exhibit asymmetric Fano-type optical signatures \cite{Fan61}, which is beyond the scope of the present work.

For both polarisations (Figs.\ \ref{fig-optics}(a) and (c)), the upper branch originates mostly from the $T_{01}$ transitions with some small 
(zero at $\nu=2$) admixture of the $T_{-10}$.
With increasing $\nu$, the number of (strongly mixed by the $e$-$h$ exchange) $T_{01}$ transitions increase (see Fig.~\ref{fig-emz} insets), 
which leads to the enhanced contribution of the repulsive $e$-$h$ exchange interactions. 
This explains the blue shift of the upper branch to higher energies with increasing $\nu$.  
Also, its optical strength, $|d|^2$,  increases for the $\sigma^+$ polarisation (Fig.~\ref{fig-optics}(a)) 
and decreases for the $\sigma^-$ polarisation (Fig.~\ref{fig-optics}(c)). This is explained by the fact that 
the $T_{01}$ transitions 
are optically active in the $\sigma^+$ polarisation 
while $T_{-10}$ transitions are dark. 
Conversely, the strength of the upper branch in the $\sigma^-$ polarisation originates solely from the $T_{-10}$. 
There are fewer of them with increasing $\nu$, and eventually the upper branch becomes 
completely dark in the $\sigma^-$ at $\nu=2$.
Similarly, the lower-energy branch mainly originates from the $T_{-10}$ transitions 
with some small admixture of the $T_{01}$. 
Its red shift to lower energies with increasing $\nu$ is explained by the decreasing number of the $T_{-10}$ transitions 
leading to the decrease of the repulsive $e$-$h$ exchange contribution. 

Figure.\ \ref{fig-screening} addresses the effect of impurity screening, as discussed in Section \ref{subsec-spimpham}. The energies of states localised on a $D^+$ impurity for $\nu=1$ and the $\sigma^+$ polarisation are plotted as a function of $\epsilon/\epsilon_\mathrm{imp}$. As $\epsilon/\epsilon_\mathrm{imp}$ increases, the impurity screening is reduced and the states are pushed away from the band.
\subsection{$\delta$-function impurity}
\label{subsec-delimp}
As for the case of a Coulomb impurity, states localised on a $\delta$-function impurity possess a high level of symmetry. Their energies and optical properties are the same under simultaneously changing the sign of the filling factor ($\nu\leftrightarrow-\nu$), the direction of light polarisation ($\sigma^+\leftrightarrow \sigma^-$), the sign of the impurity potential ($V_0 \leftrightarrow -V_0$) and the sublattice on which the impurity is located ($A \leftrightarrow B$). In addition, changing only the impurity from one sublattice to the other, yields identical results when the filling factor is even, but qualitatively different results when it's odd. This is due to the valley splitting, which allows the sublattices to become inequivalent if the valleys are not equally filled.

\begin{figure}
\includegraphics[width=0.47\columnwidth]{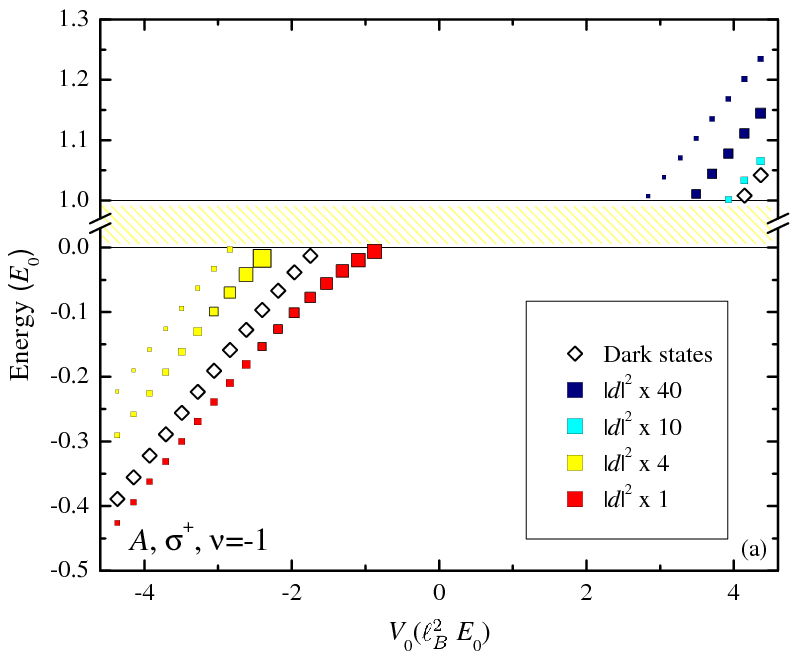}  
\includegraphics[width=0.47\columnwidth]{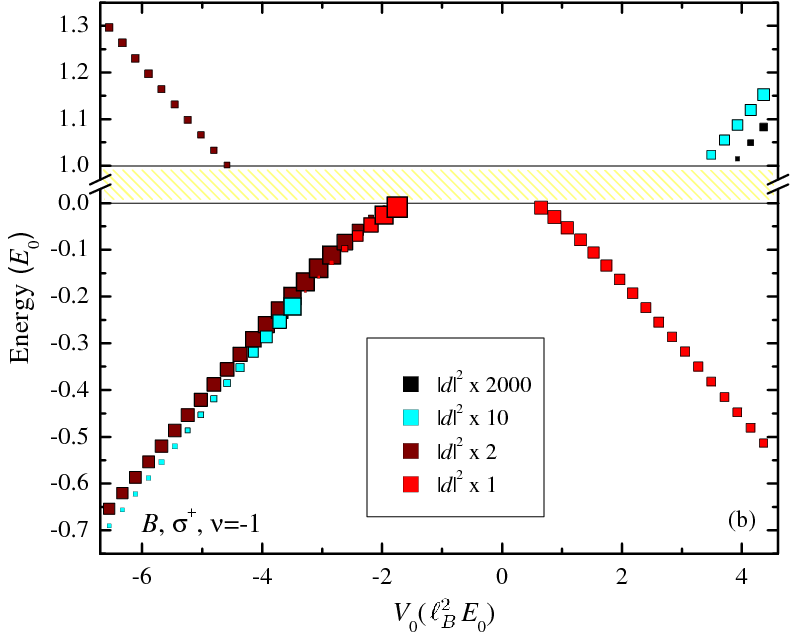}\newline
\includegraphics[width=0.47\columnwidth]{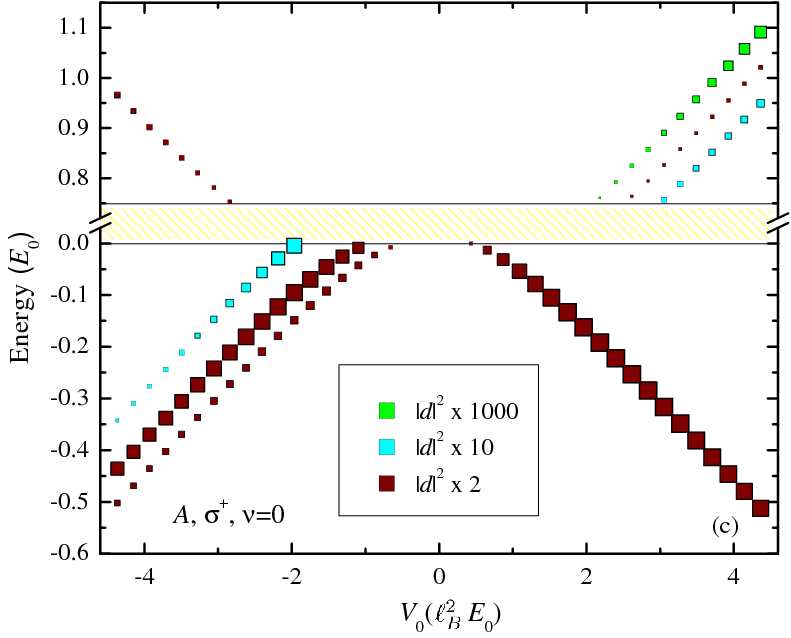}
\includegraphics[width=0.47\columnwidth]{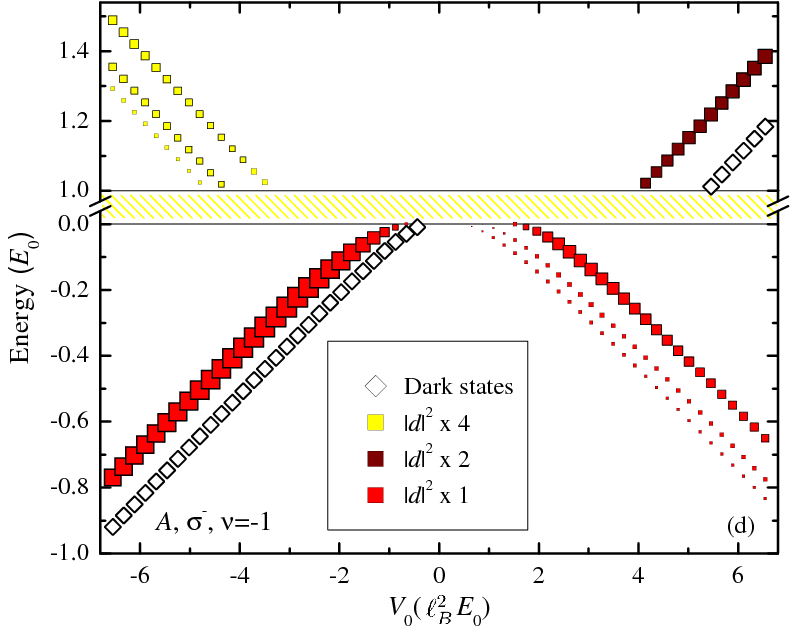}
  \caption{Evolution of energies and optical strengths of neutral CEs localised on a $\delta$-function impurity located on either an $A$ or $B$ sublattice site as a function of the impurity strength, $V_0$, for various filling factors and light polarisations (see labels on plots). The parameters $B=100\mathrm{T}$ and $\epsilon=5$ are used. The yellow shaded area represents the continuum of neutral CEs. The energies are renormalised, so that the CRE, which equals $\epsilon_1+E_{0 \tau s}^{1 \tau s}-0.75E_0$, is at zero energy. Optically dark states are represented by diamonds and optically active states (in the specified polarisation) by squares with sizes $\sim |d|^2$. 
     }
\label{fig-engvsv}
\end{figure}

Figure \ref{fig-engvsv} shows the energies and absorption strengths of localised states for a selection of values of filling factor, light polarisation and sublattice plotted as a function of the impurity strength, $V_0$. In each plot only one of the parameters differs from those in Fig.\ \ref{fig-engvsv}(a) to see how this affects the states. As for the Coulomb impurity case, the localised neutral CEs occupy discrete energy levels outside the continuum of extended neutral CEs. In contrast to the Coulomb impurity, the energies of states localised on the $\delta$-function impurity are non-degenerate. In particular there is no evidence of the SU(4) [15]-plet seen in clean graphene and the Coulomb impurity case for $\nu=2$. This is because the $\delta$-function impurity breaks the symmetry between the sublattices and hence the valleys. In contrast, the long-ranged Coulomb interaction varies slowly on the scale of the lattice spacing and preserves the sublattice symmetry.

\begin{figure} 
\centering
\includegraphics[width=0.47\columnwidth]{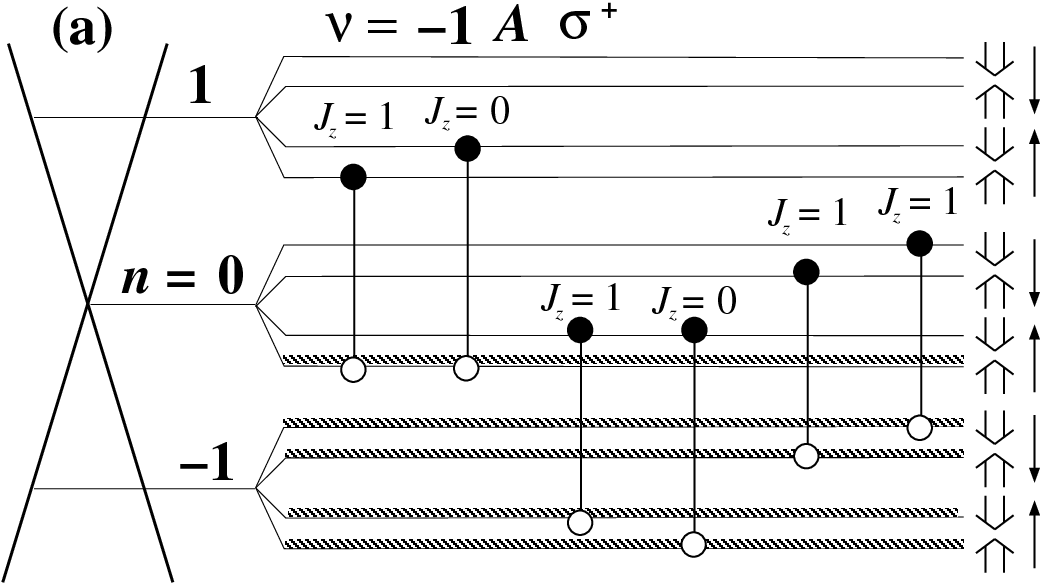}\newline
\includegraphics[width=0.47\columnwidth]{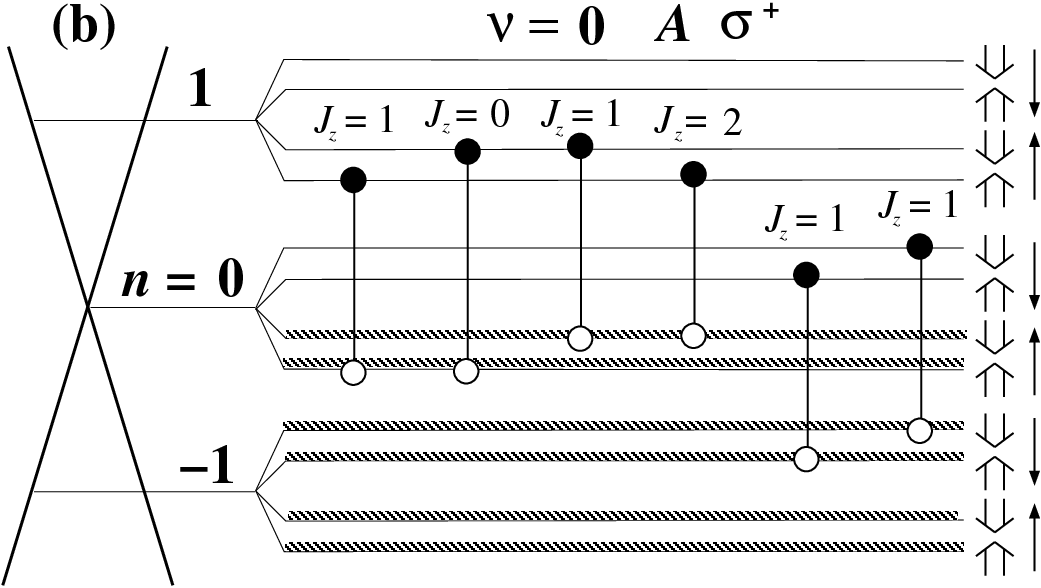}\newline
\includegraphics[width=0.47\columnwidth]{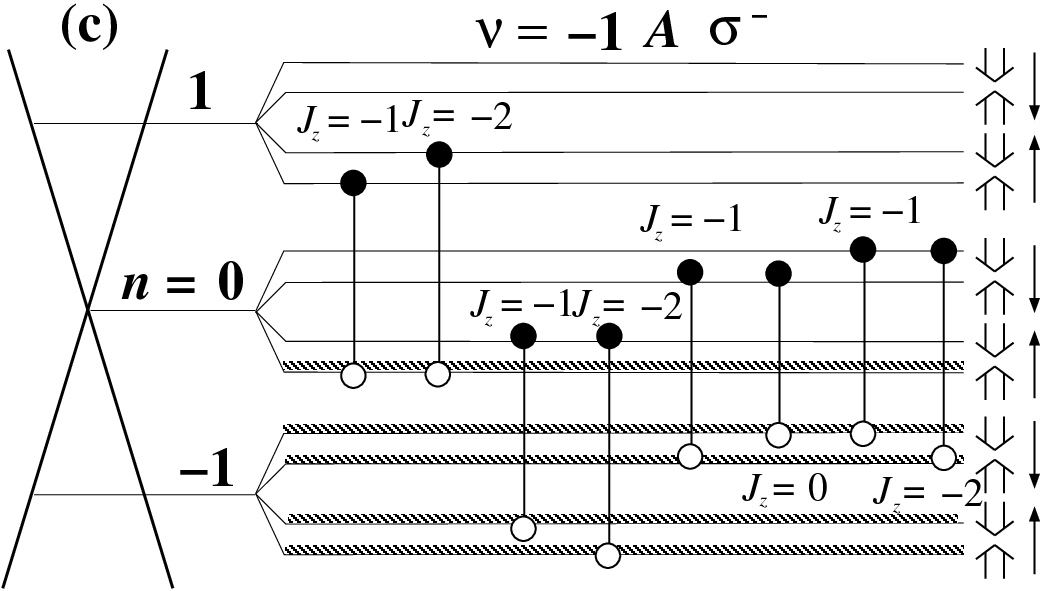}\newline
  \caption{Diagrams indicating which neutral CEs are mixed when the $\delta$-function impurity is on the $A$ sublattice and the sample is illuminated by (a) $\sigma^+$ polarised light with $\nu=-1$, (b) $\sigma^+$ polarised light with $\nu=0$ and (c) $\sigma^-$ polarised light with $\nu=-1$.
     }
\label{fig-deltamixdiag}
\end{figure}

Note that the continuum has width $E_0$ in Figs.\ \ref{fig-engvsv}(a),(b),(d) which have $\nu=-1$ and width $0.75E_0$ in Fig.\ \ref{fig-engvsv}(c) which has $\nu=0$. The band width is determined by which transitions are mixed; this is shown in Figs.\ \ref{fig-delmix} and \ref{fig-deltamixdiag}. Notice that in all cases, there are the four transitions with no spin or pseudospin flips, which are mixed by the exchange $e$-$h$ Coulomb interaction. If these were not mixed to any other transitions, which is the case for the Coulomb impurity, the continuum would have width $0.75E_0$, as seen in Section \ref{subsec-dispeg}. In some cases (specifically those shown in Figs.\ \ref{fig-deltamixdiag}(b),(c)) the $\delta$-function impurity admixes transitions with a pseudospin flip, which are not otherwise mixed by the Coulomb $e$-$h$ interaction to any other transitions; these do not alter the width of the band. For filling factors $\nu=-1$ and $\nu=1$ (latter not shown) however, the $\delta$-function impurity also admixes pseudospin flip transitions, which are each mixed to a single other transition by the direct $e$-$h$ Coulomb interaction. This broadens the band from $0.75E_0$ to $E_0$ (see Section \ref{subsec-dispeg}).\newline
Our results indicate critical $V_0$ values for the formation of localised excitations, although within our approach, it can be difficult to resolve what may be a low energy bound state from the lower continuum edge. The critical value for which bound states appear varies slightly depending on the filling factor, light polarisation and sublattice of the impurity. For example, for a system with filling factor $\nu=-1$ illuminated by $\sigma^+$ circularly polarised light with an impurity on an $A$ sublattice (such a system will henceforth be denoted by $A$, $\sigma^+$, $\nu=-1$), there are no bound states for $-30\mathrm{eV}<W<120\mathrm{eV}$ (for $B=100\mathrm{T}$). Thus it takes a very strong $\delta$-function impurity potential to localise excitations. This is due to the fact that the $\delta$-function impurity only couples to a small subset of the basis states of a neutral CE, which have $m=0,\pm1$. Strictly speaking, such large energies are greater than the width of the $\pi$ band, where the electrons can be reasonably treated as massless Dirac fermions and so should not be treated as a small perturbation.

For both light polarisations, impurity positions and for all four filling factors, larger magnitudes of $V_0$ are required to push states above the band than to pull them below. States below the band form when either the electron or hole is nearer to the impurity and attracted to it; in this case the additional $e$-$h$ attraction also lowers the energy. States above the band form when one particle is held nearer to the impurity due to magnetic confinement, but is repelled by it, whilst the other particle is further away. In this case the $e$-$h$ attraction works against this, trying to lower the energy, so that a larger impurity strength is required to overcome this.

The bound states move further from the band as the impurity strength increases, in accordance with intuition. An interesting question is what happens to the bound states when they approach the band. One possibility is they continue to exist within the band as quasibound states, with a high probability to exist on the impurity in addition to long-range oscillating tails. Although we detect possible signatures of such states, our method is not accurate enough to claim their existence. The hypothesis may be supported however by the observation that in all cases the number of branches below the band for positive (negative) $V_0$ equals that of branches above the band for negative (positive) $V_0$.
\begin{figure} 
\centering
\includegraphics[width=0.47\columnwidth]{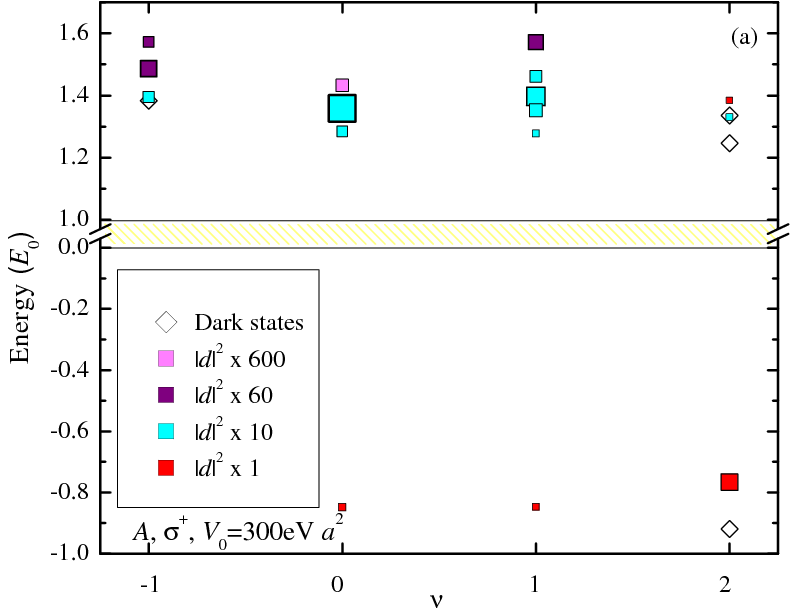}
\includegraphics[width=0.47\columnwidth]{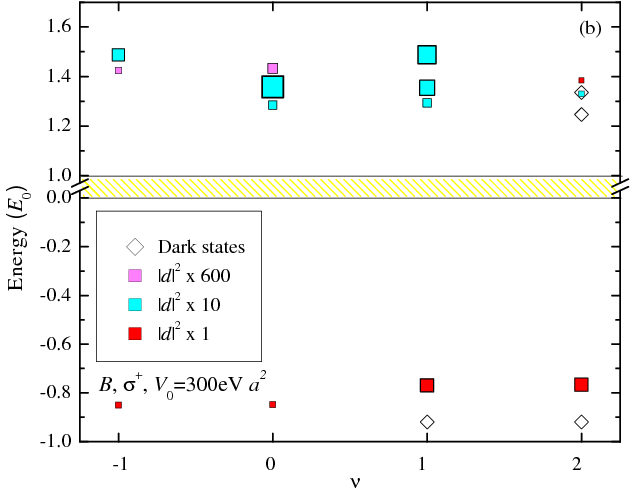}
  \caption{Evolution of energies and optical strengths of neutral CEs localised on a $\delta$-function impurity located at a carbon site on (a) the $A$ sublattice and (b) the $B$ sublattice site as a function of the filling factor $\nu$. The yellow shaded area represents the continuum of neutral CEs. The energies are renormalised, so that the CRE, which equals $\epsilon_1+E_{0 \tau s}^{1 \tau s}-0.75E_0$, is at zero energy. Optically dark states are represented by diamonds and states active in the $\sigma^+$ polarisation by squares with sizes $\sim |d|^2$. 
     }
\label{fig-nuplot}
\end{figure}

In most cases, for large enough $|V_0|$ values, bound states may be found both above and below the band for both $V_0>0$ and $V_0<0$. One exception (see Fig.\ \ref{fig-engvsv}(a)) is the $A$, $\sigma^+$, $\nu=-1$ system. In this case, the impurity selection rules forbid the hole to interact with the impurity, so that all diagonal matrix elements have the same sign as $V_0$ and states above the band are only seen for large enough positive $V_0$ and states below the band are only for large enough negative $V_0$. This behaviour is of course mimicked by the symmetric system $B$, $\sigma^-$, $\nu=1$ except that the electron is forbidden to interact with the impurity, so that states above the band only appear for large enough negative values of $V_0$ and states below the band only for large enough positive $V_0$ values. The other exception is the pair of symmetric systems $A, \sigma^-, \nu=2$ and $B, \sigma^-, \nu=2$, where in both cases only the hole interacts with the impurity so that states above the band only exist for large enough negative values of $V_0$ and states below the band only for large enough positive $V_0$ values. The number of branches due to the $e$-imp and the number of branches due to the $h$-imp interaction can be understood by studying the transitions involved and the impurity selection rules.

Figure \ref{fig-nuplot} shows the energies and absorption strengths of states localised on a $\delta$-function impurity with $V_0=300\mathrm{eV}\hspace{0.5mm}a^2$, as function of filling factor $\nu$, when the sample is illuminated by $\sigma^+$ polarised light. The cases when the impurity is located on the $A$ and $B$ sublattice are compared. One can see that the energies and absorption strengths are identical for filling factors $\nu=0,2$, but differ for $\nu=\pm1$, as previously mentioned.
\section{Discussion}
\label{sec-dis}
It is interesting to compare Figs.\ \ref{fig-optics}(a) and \ref{fig-nuplot} and see how the results differ for the Coulomb and $\delta$-function impurities. For the $\delta$-function impurity, there are more localised states and they do not seem to evolve smoothly as a function of filling factor in contrast to the case of the Coulomb impurity. This is because for the Coulomb impurity there are always four transitions with the same $J_z$ for any filling factor, whereas for the short-range impurity, where the intervalley scattering is significant, changing the filling factor may change the number of transitions that are mixed and their $J_z$ values, which influences the number of branches of bound states that will be seen. The energies may be degenerate for the Coulomb impurity, but not the $\delta$-function impurity. This can again be attributed to the greater complexity of the transitions involved for the $\delta$-function impurity. 

One may also compare the results for the 2DEG with those for graphene in the case of a Coulomb impurity \cite{DzyL93}. They are qualitatively similar with branches of neutral CEs appearing both above and below the continuum with energies $\sim E_0$. However, there are more branches in graphene than in the 2DEG, due to the the additional LL pseudospin splitting.

A possible way of detecting the localised neutral CEs predicted above is via polarisation resolved spectroscopy experiments, in much the same way as was described for charged CEs. The localised neutral CEs below the band may be visible as additional lower intensity peaks in the absorption spectrum below the main CRE peak. This is shown schematically in Fig.\ \ref{fig-optics}(c), inset. Our single-impurity theory is applicable to samples with finite impurity density $n_{\rm imp} < 1/\pi l_{B}^2$, i.e.\ when the mean separation between impurities exceeds the size of localised neutral CEs. The intensity of impurity peaks will then be $I \sim n_{\rm imp} |d|^2$. In current samples, $n_{\rm imp}\sim 10^{12}\mathrm{cm^{-2}}\approx 1/\pi l_{B}^2$ \cite{NovGMJ05,HwaAD07}. Hence it is reasonable to expect that with the rapid advances being made in sample fabrication, clean enough samples will be available in the near future. Experimental verification of our results would enable the presence of short-range and long-range (Coulomb) scatterers to be distinguished, thus contributing towards the ongoing debate regarding the nature of disorder in graphene.
\cleardoublepage

\cleardoublepage
\chapter{Conclusions and outlook}
\label{chap-conc}
This work has focused on several aspects of disorder and many-body interactions
in quantum systems. The analytical treatment of either of these issues is fraught with
difficulty and much of the progress is made using numerical techniques. An even
more challenging problem is the interplay between the two effects, especially
when neither one is clearly dominant. Many theories focus on one or the other,
but usually both will be present. In this thesis, three systems were examined:
i. the Anderson model of BCC and FCC lattices in which disorder is modelled
using random matrices and electron-electron interactions are ignored, ii. an
exciton in a nanoscopic Aharonov Bohm ring, which is disorder-free, but where
the electron and hole interact via a $\delta$-function potential and iii. a
monolayer of graphene in a strong perpendicular magnetic field, where
excitations are collective, due to the screened Coulomb electron-electron
interactions. There are a few other themes that connect the projects. For
example, impurity (disorder) induced localisation of electrons is considered in
i. and of collective excitations in iii. In projects i. and ii. quantum
interference of the particle's wave function plays an important role. In this
final Chapter, I shall summarise the main results arrived at in the previous
Chapters, discuss the remaining open questions and suggest directions in which
the work could be taken forward.

In the Anderson transition study, we implemented the transfer-matrix method to
calculate localisation lengths for electrons in the 3D BCC and FCC lattices with
onsite disorder. Finite size scaling was performed on these results for a range
of different disorders, energies and system sizes in order to obtain critical
disorders (at a constant energy), critical energies (at a constant disorder) and
the critical exponent, which governs the behaviour of the correlation length as
criticality is approached. At energy $E=0$, we found an increase in the critical
disorder, as the coordination number of the lattice increases, in accordance
with intuition: $W_{\rm c}=16.54$ \cite{Mac94,SleO99a,MilRSU00,OhtSK99} for the
SC lattice (coordination number $6$), $20.81$ for the BCC lattice (coordination
number $8$) and $26.73$ for the FCC lattice (coordination number 12). All the
calculated critical exponents for both lattice types, however, are $\nu \approx
1.5$, which is also in agreement with previously calculated results for the SC
lattice. This is because it is a universal localisation property of a 3D system in the Gaussian orthogonal universality class. 
and hence should not change by the scaling theory of localisation
\cite{AbrALR79}. In addition to high precision calculations, we used some of our
slightly lower precision data to obtain the energy-disorder phase diagrams for
BCC and FCC lattices. The diagram for the BCC lattice has a similar shape to
that previously calculated for the SC lattice, but the diagram for the FCC
lattice is assymetric about $E=0$, due to its non-biparticity. 

It would be interesting to apply the techniques outlined in this project to
other lattices. Graphene would perhaps be a good candidate. According to scaling
theory, as a 2D system, the electronic wave function should be completely
localised in the presence of any finite disorder, leading to insulating
behaviour. Indeed this was found to be the case in numerical calculations of the
localisations lengths for the honeycomb lattice \cite{SchO92}. However, the
majority of experiments on graphene indicate that it is in fact a good
conductor. In addition, other numerical studies claim to have found an MIT
\cite{AmiJS09,Hil10}. This prompts the question of how large the system size
should be before localisation is observed and how this compares with the typical
sample sizes used in experiments.

In the Aharonov-Bohm study, we developed a simple 1D model for an exciton in a
nanoring with an in-plane electric field. We were able to express the problem in
terms of an eigenequation, where the matrix depends on the excitonic energy.
Correct energy and wave function amplitudes are provided, when the eigenvalue is
close to 1. We observed clear AB oscillations with period equal to the magnetic
flux quantum. The amplitude of these oscillations first increased as a function
of electric field and then rapidly decreased once the exciton had been pulled
apart. Most importantly we also observed oscillations in the oscillator strength
as a function of magnetic flux. Beyond the critical electric field value, these
became inverted i.e. the maximum became a minimum. Interestingly the minimum
reached zero, indicating a probability zero for the recombination of the
electron and hole and the emission of a photon. Thus by tuning the electric and
magnetic fields, we found a mechanism whereby the exciton could be manipulated
between being light and dark. This could have potential applications in
excitonic data storage and the trapping of light. Two recent experiments have
observed excitonic AB oscillations in the presence of an electric field
\cite{TeoCLM10,DinALP10}. Both an inversion in the oscillation pattern at high
enough electric fields \cite{TeoCLM10} and a dramatic reduction in the emission
intensity for particular values of the external fields \cite{DinALP10} were
observed, in accordance with our predictions.

An obvious extension to this work would be to create a more realistic ring model by introducing a width and/or height to the ring in order to see how this might affect the oscillations. Another interesting avenue of research would be to consider electrons (Cooper pairs) in superconducting quantum rings instead of excitons in semiconducting devices. In the thermodynamic limit, the free energy \cite{Ann04} and all thermodynamic quantities \cite{deG66} should have periodicity $\Phi_0/2$. However, this limit is not reached in mesoscopic systems. The excitonic Bohr radius will be replaced by the coherence length (roughly, the distance between electrons in a Cooper pair) and the penetration depth of the magnetic field may also play a role. In $s$-wave superconducting rings much smaller than the coherence length, oscillations in the supercurrent with period $\Phi_0$ are now well established \cite{TzuG08}. In $d$-wave superconductors a $\Phi_0$ current component has also been found \cite{LodKKM09}. It would be interesting to look at the problem for $p$-wave superconductors. These are particularly interesting in this context, since $\Phi_0/4$-quantised vortices exist there \cite{Vol99}. One topic of investigation could be how the periodicities vary with the ring size close to the coherence length.

In the graphene project we made a comprehensive study of collective excitations of the monolayer system in the presence of a magnetic field. I first reviewed the work that has been done by others on neutral collective excitations of pristine graphene. These can be thought of as excitons, but the term ``collective'' is used to emphasise that several excitations are mixed i.e. there is a superposition, due to the two-body Coulomb interaction. We then considered the localisation of such neutral collective excitations on a single impurity. Both long range (screened Coulomb) and short range ($\delta$-function) impurities were treated. The localised excitations occur in discrete states outside the band of extended collective excitations. They form branches for the case of a Coulomb impurity, although for a $\delta$-function impurity things are somewhat (and maybe counterintuitively) more complicated, as different excitations are mixed for different filling factors. The main difference between the two impurity types is that the Coulomb impurity cannot scatter between the two valleys, $\mathbf{K}$ and $\mathbf{K}'$, in graphene, whereas the $\delta$-function impurity is able to cause this large change in an electron's momentum. Certainly the Coulomb impurity is much more effective at localising the collective excitations. We observed a symmetry between the localised states. Their energies and optical properties are the same under simultaneously changing the sign of the filling factor ($\nu\leftrightarrow-\nu$), the direction of light polarisation ($\sigma^+\leftrightarrow \sigma^-$), the sign of the impurity potential ($V_0 \leftrightarrow -V_0$) and the sublattice on which the impurity is located for the case of the $\delta$-function impurity ($A \leftrightarrow B$). We hope that our work, particularly if it is experimentally verified, may contribute towards the ongoing debate regarding the nature of disorder in graphene.

We also looked at charged collective excitations, which may be thought of as electron-electron-hole ($X^-$) or electron-hole-hole ($X^+$) complexes. Such discrete states may exist below the continuum for very nearly full ($X^+$) or very nearly empty ($X^-$) Landau levels, making them energetically favourable; these two types of excitation are connected by a symmetry relation similar to that for localised neutral collective excitations. The very recent observation of plasmarons in graphene \cite{BosSSH10}, which are the zero magnetic field equivalent of our $X^+$ states, makes us optimistic about future experimental observation of the states we predict. We found that a few are optically active and as such may be visible in optical spectroscopy experiments as a low energy shoulder to the cyclotron resonance or additional peak below it.

In our calculations we made use of both the geometrical and dynamical symmetries of our system. There is always rotational symmetry about a perpendicular axis through the origin, where the impurity is located. This enables us to label the collective excitations by a generalised angular momentum, $J_z$, which has an orbital and sublattice component. For the pristine system, translational symmetry meant that the charged collective excitations could be labelled by a $k$ quantum number associated with the operator of magnetic translations. By the dynamical symmetry, I mean the SU(4) symmetry associated with the presence of two spin and two valley pseudospin degrees of freedom, yielding four possible flavours in total. For a fully or approximately fully filled Landau level, a symmetry arises due to the possible exchange of these four equivalent flavours. (Note that we do not have the $\delta$-function impurity in mind here, since in this case the impurity introduces an inequivalency between the two pseudospin projections). With this in mind, we introduced an analogy between neutral collective excitations and mesons and also between charged collective excitations and baryons, where only the first and second generation quarks are considered. Using this, we could employ Young diagram techniques to understand and correctly predict the degeneracies of our states. We also produced multiplet plots similar to the style of the SU(4) multiplet polygons seen in particle physics, as a way of visualising our states and their dynamical symmetries.

There are still several unanswered questions within the work presented here. Most importantly, we would like to fully understand the dualities that exist between our results for different filling factors, light polarisations and impurity potentials. There should of course be a similarity transformation connecting any two Hamiltonians describing symmetric cases. Another way of taking the graphene work further, would be to examine collective magneto-excitations of the bilayer system, which has already attracted a lot of attention \cite{AbeABZ10}. Its neutral collective excitations (magnetoplasmons) have already been studied for a clean system \cite{TahS08,BerGL08}, but not, so far as I'm aware, for a system containing impurities.
\cleardoublepage

%

 \appendix                            
\chapter{Connectivity matrices}
\label{app-con}
Given below are explicit expressions for the connectivity matrices for BCC and FCC lattices with $M=3$. Recall that $\bm{\mathrm{C}}_{i}$ is the connectivity matrix describing the connections of the $i^\mathrm{th}$ slice to the $(i-1)^\mathrm{th}$ slice. Element $c_{jk}$ of the connectivity matrix  equals $1$ if site $j$ in the $i^\mathrm{th}$ slice is connected to site $k$ in the $(i-1)^\mathrm{th}$ slice; otherwise $c_{jk}=0$. The boundary terms are indicated in italics.
For the BCC lattice for odd layers,
\begin{equation}
\label{eq-cibccodd}
\boldsymbol{\mathrm{C}}_{2i-1} =
\left( \begin{array}{ccccccccc}
1 & 0 & {\it 1} & 0 & 0 & 0 & {\it 1} & 0 & {\it 1} \\
1 & 1 & 0 & 0 & 0 & 0 & {\it 1} & {\it 1} & 0 \\
0 & 1 & 1 & 0 & 0 & 0 & 0 & {\it 1} & {\it 1} \\
1 & 0 & {\it 1} & 1 & 0 & {\it 1} & 0 & 0 & 0 \\
1 & 1 & 0 & 1 & 1 & 0 & 0 & 0 & 0 \\
0 & 1 & 1 & 0 & 1 & 1 & 0 & 0 & 0 \\
0 & 0 & 0 & 1 & 0 & {\it 1} & 1 & 0 & {\it 1} \\
0 & 0 & 0 & 1 & 1 & 0 & 1 & 1 & 0 \\
0 & 0 & 0 & 0 & 1 & 1 & 0 & 1 & 1 \end{array} \right).
\end{equation}
For even layers
\begin{equation}
\label{eq-cibcceve}
\boldsymbol{\mathrm{C}}_{2i} =
\left( \begin{array}{ccccccccc}
1 & 1 & 0 & 1 & 1 & 0 & 0 & 0 & 0 \\
0 & 1 & 1 & 0 & 1 & 1 & 0 & 0 & 0 \\
{\it 1} & 0 & 1 & {\it 1} & 0 & 1 & 0 & 0 & 0 \\
0 & 0 & 0 & 1 & 1 & 0 & 1 & 1 & 0 \\
0 & 0 & 0 & 0 & 1 & 1 & 0 & 1 & 1 \\
0 & 0 & 0 & {\it 1} & 0 & 1 & {\it 1} & 0 & 1 \\
{\it 1} & {\it 1} & 0 & 0 & 0 & 0 & 1 & 1 & 0 \\
0 & {\it 1} & {\it 1} & 0 & 0 & 0 & 0 & 1 & 1 \\
{\it 1} & 0 & {\it 1} & 0 & 0 & 0 & {\it 1} & 0 & 1 \end{array} \right).
\end{equation}
For the FCC lattice for odd and even layers
\begin{equation}
\label{eq-cifcc}
\boldsymbol{\mathrm{C}}_{i} =
\left( \begin{array}{ccccccccc}
1 & 1 & 0 & 0 & 0 & 0 & 0 & {\it 1} & 0 \\
0 & 1 & 1 & 0 & 0 & 0 & 0 & 0 & {\it 1} \\
{\it 1} & 0 & 1 & {\it 1} & 0 & 0 & 0 & 0 & 0 \\
0 & 1 & 0 & 1 & 1 & 0 & 0 & 0 & 0 \\
0 & 0 & 1 & 0 & 1 & 1 & 0 & 0 & 0 \\
0 & 0 & 0 & {\it 1} & 0 & 1 & {\it 1} & 0 & 0 \\
0 & 0 & 0 & 0 & 1 & 0 & 1 & 1 & 0 \\
0 & 0 & 0 & 0 & 0 & 1 & 0 & 1 & 1 \\
{\it 1} & 0 & 0 & 0 & 0 & 0 & {\it 1} & 0 & 1 \end{array} \right).
\end{equation}
In all cases, $i$ is a positive integer.

\chapter{Coulomb matrix elements for the 2DEG}
\label{app-mes}
As seen in Eqs.\ (\ref{eq-U2DEG}) and (\ref{eq-impmec}), graphene matrix elements of the Coulomb interaction can be expressed in terms of those for the 2DEG. The method for calculating such 2DEG matrix elements is explained in Ref.\ \cite{Dzyhab} and outlined here for convenience. Let us first examine the impurity matrix elements
\begin{equation}
V_{n \, m}=\langle nm|V_\mathrm{C}|nm\rangle=\int \int d \mathbf{r}
^2\psi_{nm}^\ast(\mathbf{r})V_\mathrm{C}(r)\psi_{nm}(\mathbf{r}).
\label{eq-2degimpme}
\end{equation}
The general technique is to express $V_\mathrm{C}(r)$ in terms of its Fourier transform
\begin{equation}
\tilde{V}_\mathrm{C}(q)=\frac{2\pi Z e^2}{\epsilon_\mathrm{imp} q}
\label{eq-ftvc} 
\end{equation}
and the operator $\mathrm{exp}\left(i \mathbf{q}\cdot\mathbf{r}\right)$ in terms of displacement operators
\begin{eqnarray}
\mathrm{exp}\left(i \mathbf{q}\cdot\mathbf{r}\right)&=&\mathrm{exp}\left(-\frac{\tilde{q}^\ast a^\dagger}{\sqrt{2}}+\frac{\tilde{q} a}{\sqrt{2}}\right)
\mathrm{exp}\left(\frac{i\tilde{q} b^\dagger}{\sqrt{2}}+\frac{i\tilde{q}^\ast b}{\sqrt{2}}\right)\\ \nonumber
&=&\hat{D}_n\left(-\frac{\tilde{q}^\ast}{\sqrt{2}}\right)\hat{D}_m\left(\frac{i\tilde{q}}{\sqrt{2}}\right),
\label{eq-disops} 
\end{eqnarray}
where $\tilde{q}=\left(q_x+iq_y\right)\ell_B$. Since the two different sets of Bose ladder operators commute,
\begin{equation}
\langle nm|\hat{D}_n\left(\alpha\right)\hat{D}_m\left(\beta\right)|nm\rangle=\langle n|\hat{D}_n\left(\alpha\right)|n\rangle \langle m|\hat{D}_m\left(\beta\right)|m\rangle.
\label{eq-dsplit} 
\end{equation}
One can show using the Baker-Campbell-Hausdorff formula and explicit expressions for Laguerre polynomials that performing the real space integrals gives
\begin{equation}
\langle n'|\hat{D}\left(\alpha\right)|n\rangle=
\Biggl\{\begin{array}{cc}
\sqrt{\frac{n!}{n'!}}\alpha^{n'-n}\mathrm{exp}\left(-\frac{|\alpha|^2}{2}\right)L_n^{n'-n}\left(|\alpha|^2\right) & n\le n'\\
\hspace{8mm}\sqrt{\frac{n'!}{n!}}\left(-\alpha^\ast\right)^{n-n'}\mathrm{exp}\left(-\frac{|\alpha|^2}{2}\right)L_{n'}^{n-n'}\left(|\alpha|^2\right) & n'\le n.
\end{array}
\label{eq-a9} 
\end{equation}
Using the above one can show
\begin{equation}
V_{n \, m}=\frac{Z E_0}{\sqrt{\pi}}\int_0^\infty x^{-1/2}e^{-x}L_n(x)L_m(x),
\label{eq-vnm} 
\end{equation}
which can be calculated explicitly for particular $n,m$ values. In fact the only elements used in our calculations are
\begin{equation}
V_{0 \, m}=\frac{Z E_0}{\sqrt{\pi}}\frac{\Gamma(m+1/2)}{m!},
\hspace{5 mm}
V_{1 \, m}=\frac{4m-1}{4m-2}V_{0 \, m}.
\label{eq-v01m} 
\end{equation}

Now consider the two-body Coulomb interaction matrix elements
\begin{equation}
\label{eq-u2degapp}
U_{n_1 m_1  n_2 m_2}^{n_1' m_1'  n_2' m_2'}=\int \int d \mathbf{r}_1 ^2 d
\mathbf{r}_2 ^2\psi^\ast_{ n_1'm_1'}(\mathbf{r}_1)\psi^\ast_{
n_2'm_2'}(\mathbf{r}_2)U\left( |\mathbf{r}_1-\mathbf{r}_2|\right)\psi_{ n_2 m_2}
(\mathbf{r}_2)\psi_{ n_1 m_1} (\mathbf{r}_1).
\end{equation}
The techniques used are the same as those for the impurity matrix elements.
Two important examples are the direct and exchange matrix elements
\begin{eqnarray}
\label{eq-u2degdir}
U_{r m_1  s m_2}^{r m_1'  s m_2'}&=&E_0 \sqrt{\frac{\mathrm{min}_1! \mathrm{min}_2!}{\pi(\mathrm{min}_1+t)!(\mathrm{min}_2+t)!}}\\ \nonumber
& & \times \int_0^\infty dx\hspace{1mm} x^{t-1/2}e^{-2x}L_r(x)L_s(x)L_{\mathrm{min}_1}^t(x) L_{\mathrm{min}_2}^t(x),
\end{eqnarray}
\begin{eqnarray}
\label{eq-u2degex}
U_{s m_1  r m_2}^{r m_1'  s m_2'} & = & E_0 \frac{r!}{s!}\sqrt{\frac{\mathrm{min}_1! \mathrm{min}_2!}{\pi(\mathrm{min}_1+t)!(\mathrm{min}_2+t)!}}\\ \nonumber
& & \times\int_0^\infty dx\hspace{1mm} x^{s-r+t-1/2}e^{-2x}(L_r^{s-r}(x))^2L_{\mathrm{min}_1}^t(x) L_{\mathrm{min}_2}^t(x),
\end{eqnarray}
where $t=|m_1-m_1'|=|m_2-m_2'|$, $\mathrm{min}_1=\mathrm{min}(m_1,m_1')$ and $\mathrm{min}_2=\mathrm{min}(m_2,m_2')$.

Eq.\ (\ref{eq-se}) gave an expression for the correction, $E_{\mathrm{SE}}\left( n,n'\right)$, to the energy of an
electron in the $n^{\rm th}$ LL of graphene due to exchange with electrons in the $n'^{\rm th}$ LL. By Eq.\ (\ref{eq-U2DEG}), $E_{\mathrm{SE}}\left( n,n'\right)$ can be expressed in terms of 2DEG matrix elements. Thus we calculate
\begin{eqnarray}
\label{eq-app3}
\sum_{m'}U_{n_1
m\hspace{2mm}   n_2 m'}^{n'_1  m' \hspace{1mm}   n'_2  m}&=&\hspace{-2mm} \sum_{m'}\int d\mathbf{q}^2 \biggl\{\tilde{U}(q)
\langle n'_1| \hat{D}_n\left(-\frac{\tilde{q}^\ast}{\sqrt{2}}\right)|n_1\rangle 
\langle n'_2| \hat{D}_n\left(\frac{\tilde{q}^\ast}{\sqrt{2}}\right)|n_2\rangle \\ \nonumber
& & \times \langle m| \hat{D}_m\left(-\frac{i\tilde{q}}{\sqrt{2}}\right)|m'\rangle
\langle m'| \hat{D}_m\left(\frac{i\tilde{q}}{\sqrt{2}}\right)|m\rangle\biggr\}/4\pi^2\\ \nonumber
& = & \int d\mathbf{q}^2 \tilde{U}(q)
\langle n'_1| \hat{D}_n\left(-\frac{\tilde{q}^\ast}{\sqrt{2}}\right)|n_1\rangle 
\langle n'_2| \hat{D}_n\left(\frac{\tilde{q}^\ast}{\sqrt{2}}\right)|n_2\rangle/4\pi^2,
\end{eqnarray}
where $\tilde{U}(q)=2\pi e^2/\epsilon q$ is the Fourier transform of the Coulomb potential, $U(r)$. Eq.\ (\ref{eq-a9}) may now be utilised to obtain results for particular $n_1,n'_1,n_2,n'_2$ values.

\chapter{Spin and pseudospin eigenstates}
\label{app-states}
The excitonic spin and pseudospin eigenstates are
\begin{equation}
\label{eq-appcs0t0}
|S=0, T=0; S_z=0, T_z=0\rangle =\frac{1}{2}\left(|\Uparrow \uparrow \hspace{1mm} \Uparrow \uparrow \rangle +|\Downarrow \uparrow \hspace{1mm} \Downarrow \uparrow \rangle
+ |\Uparrow \downarrow \hspace{1mm} \Uparrow \downarrow \rangle + |\Downarrow \downarrow \hspace{1mm} \Downarrow \downarrow \rangle
\right)
\end{equation}
\begin{equation}
\label{eq-appcs0t1tz-1}
|S=0, T=1; S_z=0, T_z=-1\rangle =\frac{1}{\sqrt{2}}\left(|\Downarrow \uparrow \hspace{1mm} \Uparrow \uparrow \rangle +|\Downarrow \downarrow \hspace{1mm} \Uparrow \downarrow \rangle\right),
\end{equation}
\begin{equation}
\label{eq-appcs0t1}
|S=0, T=1; S_z=0, T_z=0\rangle =\frac{1}{2}\left(|\Uparrow \uparrow \hspace{1mm} \Uparrow \uparrow \rangle -|\Downarrow \uparrow \hspace{1mm} \Downarrow \uparrow \rangle
+ |\Uparrow \downarrow \hspace{1mm} \Uparrow \downarrow \rangle - |\Downarrow \downarrow \hspace{1mm} \Downarrow \downarrow \rangle
\right),
\end{equation}
\begin{equation}
\label{eq-appcs0t1tz1}
|S=0, T=1; S_z=0, T_z=1\rangle =\frac{1}{\sqrt{2}}\left(|\Uparrow \uparrow \hspace{1mm} \Downarrow \uparrow \rangle +|\Uparrow \downarrow \hspace{1mm} \Downarrow \downarrow \rangle\right),
\end{equation}
\begin{equation}
\label{eq-appcs1t0sz-1}
|S=1, T=0; S_z=-1, T_z=0\rangle =\frac{1}{\sqrt{2}}\left(|\Uparrow \downarrow \hspace{1mm} \Uparrow \uparrow \rangle +|\Downarrow \downarrow \hspace{1mm} \Downarrow \uparrow \rangle\right),
\end{equation}
\begin{equation}
\label{eq-appcs1t0}
|S=1, T=0; S_z=0, T_z=0\rangle =\frac{1}{2}\left(|\Uparrow \uparrow \hspace{1mm} \Uparrow \uparrow \rangle +|\Downarrow \uparrow \hspace{1mm} \Downarrow \uparrow \rangle
- |\Uparrow \downarrow \hspace{1mm} \Uparrow \downarrow \rangle - |\Downarrow \downarrow \hspace{1mm} \Downarrow \downarrow \rangle
\right),
\end{equation}
\begin{equation}
\label{eq-appcs1t0sz1}
|S=1, T=0; S_z=1, T_z=0\rangle =\frac{1}{\sqrt{2}}\left(|\Uparrow \uparrow \hspace{1mm} \Uparrow \downarrow \rangle +|\Downarrow \uparrow \hspace{1mm} \Downarrow \downarrow \rangle\right),
\end{equation}
\begin{equation}
\label{eq-appcs1t1tz-1sz-1}
|S=1, T=1; S_z=-1, T_z=-1\rangle =|\Downarrow \downarrow \hspace{1mm} \Uparrow \uparrow \rangle,
\end{equation}
\begin{equation}
\label{eq-appcs1t1sz-1tz0}
|S=1, T=1; S_z=-1, T_z=0\rangle =\frac{1}{\sqrt{2}}\left(|\Uparrow \downarrow \hspace{1mm} \Uparrow \uparrow \rangle -|\Downarrow \downarrow \hspace{1mm} \Downarrow \uparrow \rangle\right),
\end{equation}
\begin{equation}
\label{eq-appcs1t1sz-1tz1}
|S=1, T=1; S_z=-1, T_z=1\rangle =|\Uparrow \downarrow \hspace{1mm} \Downarrow \uparrow \rangle,
\end{equation}
\begin{equation}
\label{eq-appcs1t1sz0tz-1}
|S=1, T=1; S_z=0, T_z=-1\rangle =\frac{1}{\sqrt{2}}\left(|\Downarrow \uparrow \hspace{1mm} \Uparrow \uparrow \rangle -|\Downarrow \downarrow \hspace{1mm} \Uparrow \downarrow \rangle\right),
\end{equation}
\begin{equation}
\label{eq-appcs1t1}
|S=1, T=1; S_z=0, T_z=0\rangle =\frac{1}{2}\left(|\Uparrow \uparrow \hspace{1mm} \Uparrow \uparrow \rangle -|\Downarrow \uparrow \hspace{1mm} \Downarrow \uparrow \rangle
- |\Uparrow \downarrow \hspace{1mm} \Uparrow \downarrow \rangle + |\Downarrow \downarrow \hspace{1mm} \Downarrow \downarrow \rangle
\right),
\end{equation}
\begin{equation}
\label{eq-appcs1t1sz0tz1}
|S=1, T=1; S_z=0, T_z=1\rangle =\frac{1}{\sqrt{2}}\left(|\Uparrow \uparrow \hspace{1mm} \Downarrow \uparrow \rangle -|\Uparrow \downarrow \hspace{1mm} \Downarrow \downarrow \rangle\right),
\end{equation}
\begin{equation}
\label{eq-appcs1t1sz1tz-1}
|S=1, T=1; S_z=1, T_z=-1\rangle =|\Downarrow \uparrow \hspace{1mm} \Uparrow \downarrow \rangle,
\end{equation}
\begin{equation}
\label{eq-appcs1t1sz1tz0}
|S=1, T=1; S_z=1, T_z=0\rangle =\frac{1}{\sqrt{2}}\left(|\Downarrow \uparrow \hspace{1mm} \Downarrow \downarrow \rangle -|\Uparrow \uparrow \hspace{1mm} \Uparrow \downarrow \rangle\right),
\end{equation}
\begin{equation}
\label{eq-appcs1t1sz1tz1}
|S=1, T=1; S_z=1, T_z=1\rangle =|\Uparrow \uparrow \hspace{1mm} \Downarrow \downarrow \rangle.
\end{equation}

{ \small
\begin{singlespace}
\bibliographystyle{unsrt}

\end{singlespace} 
}                                


\end{document}